%% file: paper.tex
\documentclass[twocolumn,11pt]{article}
\usepackage[colorlinks,allcolors=blue]{hyperref}

\usepackage{amsmath,lineno,authblk,natbib,relsize,abstract,dblfloatfix,type1cm,vmargin,latexsym,setspace,placeins,nicefrac}
\usepackage{subfig}
\date{}

\newcommand{\prog}[1]{{\rm\bf#1}}

\newenvironment{acknowledgements}{\section*{Acknowledgements}}{}
\setpapersize{A4}
\setmarginsrb{1.45cm}{1.45cm}{1.45cm}{1.45cm}{0mm}{0mm}{0mm}{11mm}
\usepackage{tikz-qtree}
\usepackage{color,pgf,float}
\usepackage{indentfirst}

\input{formulae_preamble}
\usepackage{tikz}
\usetikzlibrary{calc}

\usepackage{tikz-qtree}
\tikzset{sibling distance=8pt}
\tikzset{level distance=35pt}
\tikzset{edge from parent/.append style={very thick}}

\usepackage{fancyvrb}
\fvset{frame=single,framerule=.2mm,fontsize=\relscale{.86}}
\input{pygments}

\usepackage{float}
\newfloat{Listing}{hbtp}{lop}[section]
\newcommand*\FancyVerbStartString{}
\newcommand*\FancyVerbStopString{}

\usepackage{enumitem}
\setitemize{noitemsep,itemsep=1pt,topsep=1pt,parsep=0pt,partopsep=0pt}

\title{libmpdata++ 0.1: a library of parallel MPDATA solvers for~systems~of~generalised transport equations}

\author[1]{Anna Jaruga}
\author[1]{Sylwester Arabas}
\author[1,2]{Dorota Jarecka}
\author[1]{Hanna Pawlowska}
\author[3]{Piotr K. Smolarkiewicz\footnote{Affiliate Professor at the University of Warsaw}}
\author[1]{Maciej Waruszewski}

\affil[1]{Institute of Geophysics, Faculty of Physics, University of Warsaw, Warsaw, Poland}
\affil[2]{National Center for Atmospheric Research, Boulder, Colorado, USA}
\affil[3]{European Centre for Medium-Range Weather Forecasts, Reading, United Kingdom}

\begin{document}

  \maketitle 

  \begin{abstract}
    \noindent
    This paper accompanies first release of~\emph{libmpdata++}, a C++ library implementing  
      the Multidimensional Positive-Definite Advection Transport Algorithm (MPDATA).
    The library offers basic numerical solvers for systems of generalised transport equations.
    The solvers are forward-in-time, conservative and non-linearly stable. 
    The \emph{libmpdata++} library covers the 
      basic second-order-accurate formulation of MPDATA, its third-order variant,
      the infinite-gauge option for variable-sign fields and
      a flux-corrected transport extension to guarantee non-oscillatory solutions.
    The library is equipped with a non-symmetric variational elliptic solver
      for implicit evaluation of pressure gradient terms.
    All solvers offer parallelisation through domain decomposition 
      using shared-memory parallelisation. 

    The paper describes the library programming interface,
      and serves as a user guide.
    Supported options are illustrated with benchmarks discussed in the MPDATA literature.
    Benchmark descriptions include code snippets as well as quantitative representations 
      of simulation results.
    Examples of applications include:
      homogeneous transport in one, two and three dimensions in 
      Cartesian and spherical domains;
      shallow-water system compared with analytical solution (originally derived for a 2D case); and
      a buoyant convection problem in an incompressible Boussinesq fluid with interfacial instability.
    All the examples are implemented out of the library tree.
    Regardless of the differences in the problem dimensionality, right-hand-side terms,
      boundary conditions and parallelisation approach, all the examples use the same
      unmodified library, which is a key goal of \emph{libmpdata++} design.
    The design, based on the principle of separation of concerns, prioritises the user 
      and developer productivity.
    The \emph{libmpdata++} library is implemented in C++, 
      making use of the \emph{Blitz++} multi-dimensional array containers, and
      is released as free/libre and open-source software.
   
  \end{abstract}

  \begin{spacing}{0.9}
  \tableofcontents
  \end{spacing}

 \newpage
  \section{Introduction}

  The MPDATA advection scheme introduced in  \citet{Smolarkiewicz_1983} 
    has grown into a family of numerical algorithms 
    for geosciences and beyond
    \citep[see for example][]{Grabowski_and_Smolarkiewicz_2002, 
      Cotter_et_al_2002,                                        
      Smolarkiewicz_and_Szmelter_2009,                          
      Ortiz_and_Smolarkiewicz_2009,                             
      Hyman_et_al_2012,                                         
      Charbonneau_and_Smolarkiewicz_2013}.                      
  MPDATA stands for Multidimensional Positive-Definite Advection Transport 
    Algorithm\footnote{In fact, MPDATA is sign-preserving, rather than merely positive-definite, but for historical
    reasons the name remains unchanged}.
  It is a finite-difference/finite-volume algorithm for solving the generalised transport equation
  \begin{equation}
    \partial_t (G \psi) + \nabla \cdot (G \vec{u} \psi) = GR~.
    \label{gte}
  \end{equation}
  \noindent
  Equation~(\ref{gte}) describes the advection of a scalar field $\psi$ in a flow
    with velocity $\vec{u}$. 
  The field $R$ on the right-hand-side (rhs) is a total of source/sink terms.
  The scalar field $G$ can represent the fluid density, 
    the Jacobian of coordinate transformation or their product,
    and satisfies the equation
  \begin{equation}
    \partial_t (G) + \nabla \cdot (G \vec{u}) = 0~.
    \label{gte_G}
  \end{equation}
  \noindent
  In the homogeneous case ($R \equiv 0$), MPDATA is at least second-order-accurate in space and time, 
    conservative and non-linearly stable.

  The history of MPDATA spans three decades: \citet{Smolarkiewicz_1984} -- 
    \citet{Kuhnlein_et_al_2012}, \citet{Smolarkiewicz_et_al_2014} 
    and is widely documented in the literature - see \citet{Smolarkiewicz_and_Margolin_1998}, 
    \citet{Smolarkiewicz_2006} and \citet{Prusa_et_al_2008}
    for reviews.
  Notwithstanding, 
    from the authors' experience the software engineering aspects 
    still overshadow the benefits of MPDATA.
  To facilitate the use of MPDATA schemes, 
    hereby we present a new implementation
    of the MPDATA family of algorithms -- \emph{libmpdata++}.

  In the development of \emph{libmpdata++} we strive to comply with  
    the best practices sought-after among the scientific community \citep{Wilson_et_al_2014};
    in particular, with the paradigm of maximising code reuse.
  This paradigm is embodied in the ``{\it open source computational libraries -- 
    the main foundation upon which academic and also a
    significant part of industrial computational research rests}'' \citep{Bangerth_and_Heister_2013}.  

  The \emph{libmpdata++} has been developed in C++,\footnote{In the C++11 revision of the language} 
    making extensive use of object-oriented programming (OOP) and template programming.
  The primary goals when designing \emph{libmpdata++} were to maintain strict separation of concerns
    and to reproduce within the code 
    the mathematical ``blackboard abstractions'' used for documenting numerical algorithms.
  The adopted design contributes to the readability, maintainability and 
    conciseness of the code.
  The current development of \emph{libmpdata++} is an extension of the research on OOP implementation
    of the basic MPDATA scheme presented in \citet{Arabas_et_al_2014}.

  The goal of this article is twofold:
  first, to document the library interface by providing usage examples;
  and second, to validate the correctness of the implementation by verifying the results
    against published benchmarks.

  The structure of the paper is as follows.
  Section~\ref{sec:dizajn} outlines the library design.
  The four sections that follow correspond to four types of equation systems 
    solved by the implemented algorithms, namely:
    homogeneous advective transport; inhomogeneous transport;
    transport with prognosed velocity; systems featuring elliptic pressure equation.
  Each of these sections outlines the implemented algorithms, describes the library interface 
     and provides usage examples.
  Each example is accompanied with definition of the solved problem, 
    description of the program code and discussion of the results. 

  The paper structure reflects the solver inheritance hierarchy in \emph{libmpdata++}.
  All features discussed in preceding sections apply to the one that follow.
  The set of discussed problems was selected to match the tutorial structure of the paper.
  The presentation begins with simple examples focusing on the basic library interface.
  Subsequent examples use increasingly more complicated cases with the 
    most complex reflecting potential for advanced applications.

  The current version of \emph{libmpdata++} source code, including all examples presented herein, 
    can be found at \url{http://libmpdataxx.igf.fuw.edu.pl/}.

  \section{\label{sec:dizajn}Library design}

  \subsection{Dependencies}

  The \emph{libmpdata++} package is a header-only C++ library.
  It is built upon the \emph{Blitz++}\footnote{see \url{http://sf.net/projects/blitz/}} array containers.
  We refer the reader to the \emph{Blitz++} documentation \citep{Blitz_doc}
    for description of the \emph{Blitz++} interface, to which the user is exposed while
    working with \emph{libmpdata++}.
  The \emph{libmpdata++} library also depends on several components of the \emph{Boost}\footnote{see \url{http://boost.org/}}
    library collection, however these are used internally only.
  The library code requires a C++11-compliant compiler and has been
    tested to work with GNU \emph{g++}\footnote{see \url{http://gcc.gnu.org/}} and LLVM \emph{clang++}\footnote{see \url{http://llvm.org/}}.

  \subsection{\label{sec:comp}Components}

  \begin{figure*}
    \center
    \begin{tikzpicture}
      \tikzset{sibling distance=5pt}
      \tikzset{level distance=40pt}
      \tikzset{font=\scriptsize}
      \tikzset{edge from parent/.append style={very thick}}
      \tikzset{every tree node/.style={align=center,anchor=north}} 
      \Tree [.\fbox{\prog{mpdata}}
		[.\fbox{\prog{mpdata\_rhs}}
		    [.\fbox{\prog{mpdata\_rhs\_vip}}
			[.\fbox{\prog{mpdata\_rhs\_vip\_prs}} 
                            [.\prog{boussinesq} 
                                [.\fbox{\prog{output}}$<$\prog{boussinesq}$>$ ]
                            ]
                        ]
                        [.\prog{shallow\_water} 
                            [.\fbox{\prog{output}}$<$\prog{shallow\_water}$>$ ]
                        ]
		    ]
		    [.\prog{coupled\_harmosc} 
                        [.\fbox{\prog{output}}$<$\prog{coupled\_harmosc}$>$ ]
                    ]
		]
                [.\fbox{\prog{output}}$<$\fbox{\prog{mpdata}}$>$ ]
	    ]
    \end{tikzpicture}
    \caption{\label{fig:inherit}
      Inheritance diagram of classes mentioned in the paper.
      Classes defined within \emph{libmpdata++} have their 
        names surrounded with black frames.
      The user-defined classes \prog{coupled\_harmosc}, \prog{shallow\_water} and 
        \prog{boussinesq} are designed to solve a particular physical problem and are defined out of the library tree.
      The solid black lines show the inheritance relations.
      The \prog{output} label depicts any of the output handlers available
        in \emph{libmpdata++}.
    }
  \end{figure*}
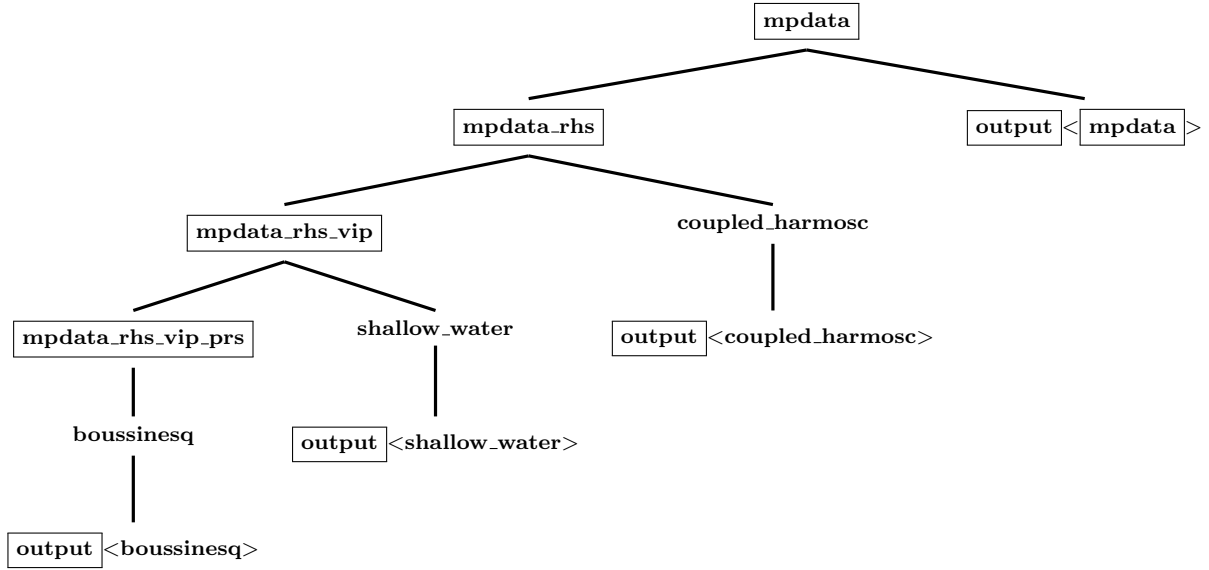

  \noindent
  Components of the library are grouped as follows:

  \begin{itemize}
    \item{solvers:
      \begin{itemize}
	\item{\prog{mpdata} intended for solving homogeneous transport problems, (section~\ref{sec:adv}),}
	\item{\prog{mpdata\_rhs} extending the above with rhs term handling, (section~\ref{sec:adv+rhs}),}
	\item{\prog{mpdata\_rhs\_vip} adding prognosed-velocity support, (section~\ref{sec:adv+rhs+vel}),}
	\item{\prog{mpdata\_rhs\_vip\_prs} further extending the above with elliptic pressure equation solvers, 
                    (section~\ref{sec:adv+rhs+vel+prs});}
      \end{itemize}
    }
    \item{output handlers:
      \begin{itemize}
	\item{\prog{gnuplot}  offering direct communication
	  with the \emph{gnuplot}\footnote{see \url{http://gnuplot.info/}} program with no intermediate output files,}
	\item{\prog{hdf5} offering basic \emph{HDF5}\footnote{see \url{http://hdfgroup.org/HDF5/}} 
          output compatible with netCDF\footnote{see \url{http://www.unidata.ucar.edu/software/netcdf/}} readers,}
	\item{\prog{hdf5\_xdmf} implementing the eXtensible Data Model and Format\footnote{see \url{http://xdmf.org/}} 
	  standard supported for instance by the \emph{Paraview}\footnote{see \url{http://paraview.org/}} visualisation tool;}
      \end{itemize}
    }
    \item{boundary conditions:
      \begin{itemize}
	\item{\prog{cyclic} implementing periodic boundaries,}
	\item{\prog{open} giving zero-divergence condition on domain edges,}
	\item{\prog{polar} applicable with spherical coordinates;}
      \end{itemize}
    }
    \item{concurrency handlers: 
      \begin{itemize} 
	\item{\prog{serial} for single-thread operation,}
	\item{\prog{openmp} for multi-thread operation using OpenMP,}
	\item{\prog{boost\_thread} for multi-thread operation using \emph{Boost.Thread},}
	\item{\prog{threads} that defaults to \prog{openmp} if supported by the compiler and falls back to \prog{boost\_thread} otherwise.}
      \end{itemize}
    }
  \end{itemize}
  Performing integration with \emph{libmpdata++} requires choosing one of the
    solvers, one output handler, one boundary condition per each domain
    edge and one concurrency handler.

  The inheritance diagram in Fig.~\ref{fig:inherit} shows relationships
    between \emph{libmpdata++} solvers and the classes defined in
    the examples discussed in the paper.
  The \textbf{mpdata} solver is displayed at the top, as it is the base class for all 
    other classes.
  The leftmost branch of the tree (solvers prefixed with \textbf{mpdata\_}) depicts the inheritance relationships among the
    solvers defined within \emph{libmpdata++}.
  The user-defined classes inherit from \emph{libmpdata++} solvers but are defined out
    of the library tree.

  \subsection{\label{sec:grid}Computational domain and grid}

  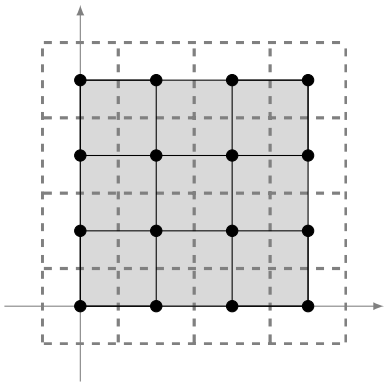
\begin{figure}[h!]
    \center
    \input{fig-domain}
    \caption{\label{fig:domain}
      Schematic of a 2D computational domain. Bullets mark the data points for the
        dependent variable $\psi$ in~(\ref{gte}), solid lines 
        depict edges of primary grid and dashed lines mark edges of dual grid in Fig.~\ref{fig:grid}.
    }
  \end{figure}

  The arrangement of the computational domain used in \emph{libmpdata++} is
    shown in Fig.~\ref{fig:domain}. 
  The initial condition for the dependent variable $\psi$ is assumed to be 
    known in $nx \times ny$ data points.
  The outermost data points are located at the boundaries of the domain.

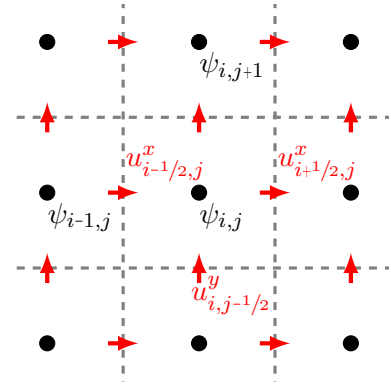
\begin{figure}[h!]
    \center
    \input{fig-grid}
    \caption{\label{fig:grid}
      A schematic of a 2D Arakawa-C grid. Bullets denote the cell centres and dashed 
      lines denote the cell walls corresponding to the dual grid in Fig.~\ref{fig:domain}.
    }
  \end{figure}

  The dual, staggered Arakawa-C grid \citep{Arakawa_and_Lamb_1977} used in \emph{libmpdata++} 
    is shown in Fig.~\ref{fig:grid}.
  In this spatial discretisation approach, the cell--mean values of the scalar fields $\psi$,
    and $G$ reside in the centres of computational cells, 
    --- corresponding to the data points of the primary grid in Fig.~\ref{fig:domain} --- 
    whereas the components of the velocity field $\vec{u}$ 
    are specified at the cell edges of the dual grid in Fig.~\ref{fig:domain}.

  \subsection{Error and progress reporting} 

  There are several error-handling mechanisms used within \emph{libmpdata++}.

  First, there are sanity checks within the code implemented using
    \prog{static\_assert()} calls.
  These are reported during compilation, for instance when invalid 
    values of compile-time parameters are supplied.
  
  Second, there are available numerous run-time sanity checks, implemented using 
    \prog{assert()} calls. 
  These are often time-consuming and are not intended to be executed 
    in production runs.
  To disable them, one needs to compile the program using \emph{libmpdata++}
    with the \mbox{\prog{-DNDEBUG}} compiler flag.
  Examples of such checks include detection of \prog{NaN} values within
    the model state variables, which may be useful to trace 
    origins of numerical instability problems.

  Third, the user may chose to activate the \emph{Blitz++} debug mode that
    enables run-time array range checks.
  Activating \emph{Blitz++} debug mode requires compiling the program using \emph{libmpdata++}
    with the \mbox{\prog{-DBZ\_DEBUG}} flag and linking with \prog{libblitz}.

  Finally, \emph{libmpdata++} reports run-time errors 
    by throwing \prog{std::runtime\_error} exceptions.

  Simulation progress is communicated to the user by
    continuously updating the process threads' name 
    with the percentage of work completed
    (can be observed e.g. by invoking \prog{top -H}).

  \section{Advective transport}\label{sec:adv}

  The focus of this section is on the advection algorithm used within \emph{libmpdata++}.
  Section \ref{sec:mpdata} provides a short introduction to the 
    implemented MPDATA scheme.
  Section \ref{sec:homo_api} describes the library interface needed for the 
    homogeneous transport cases.
  The following sections \ref{sec:1d} - \ref{sec:over_pole} show examples 
    of usage of \emph{libmpdata++} along with the 
    references to other MPDATA benchmarks.

  \subsection{Implemented algorithms}\label{sec:mpdata}

  This subsection is intended to provide the reader with an outline of selected MPDATA
    features that correspond to the options presently available in \emph{libmpdata++}.
  For the full derivation of the scheme and its 
    options see the reviews in \citet{Smolarkiewicz_and_Margolin_1998}
    and \citet{Smolarkiewicz_2006}; 
    whereas for an extended discussion of stability, positivity and convexity 
    see \citet{Smolarkiewicz_and_Szmelter_2005}.

  In the present implementation, it is assumed that $G$ is constant in time.
  Consequently, the governing homogeneous transport eq.~(\ref{gte}) can be written as 
  \begin{equation}
    \partial_t{\psi} + \frac{1}{G} \nabla \cdot (G\vec{u} \psi) = 0~.
    \label{gte_final}
  \end{equation}
  \noindent
  This particular form is solved by the \textbf{mpdata} solver of \emph{libmpdata++}.
  
  The following paragraphs will focus on the algorithms used for handling (\ref{gte_final}).
  The rules for applying source and sink terms are presented in section \ref{sec:adv+rhs}.

  \subsubsection{Basic MPDATA}

  MPDATA is an, at least, second-order-accurate iterative scheme 
    in which all iterations take the form of a first-order-accurate 
    donor-cell pass \citep[alias upwind, upstream; cf.][sec.~20.1.3]{Press_et_al_2007}.
  For the one-dimensional\footnote{one-dimensional case was chosen for simplicity, 
      multi-dimensional MPDATA formul\ae~can be found in 
      \citet[][sect.~2.2]{Smolarkiewicz_and_Margolin_1998}}
    case, after the discretisation in space (subscripts~$i$) 
    and time (superscripts~$n$), 
    the donor-cell pass applied to eq.~(\ref{gte_final}) yields
  \begin{equation}
    \begin{split}
      \psi_i^{n+1} = \psi_i^{n} - \frac{1}{G_i}[& F(\psi_i^n,     \psi_{i\pone}^n, G_{i\phlf}, u_{i\phlf}^{n\phlf}) - \\
				   & F(\psi_{i\mone}^n, \psi_i^n,     G_{i\mhlf},  u_{i\mhlf}^{n\phlf}) ]~.
    \label{upstream}
    \end{split}
  \end{equation}
  \noindent
  The flux function $F$ is defined as
  \begin{equation}
    \begin{split}
      F(\psi_{L}, \psi_{R}, G, u)  \equiv \left([u]^+ \psi_{L} + [u]^- \psi_{R}\right) G \frac{\Delta t}{\Delta x}~,
    \label{F}
    \end{split}
  \end{equation}
  \noindent  
  where $[u]^+ \equiv max(u,0)$ and $[u]^- \equiv min(u,0)$. 

  In the case of a time-varying velocity field, the velocity components are evaluated at 
    an intermediate time level denoted by the $n+\nicefrac{1}{2}$ superscript in eq.~(\ref{upstream}).
  Association of the velocity components with dual-cell edges is denoted by fractional 
    indices $i+\nicefrac{1}{2}$ and $i-\nicefrac{1}{2}$, see Fig.~\ref{fig:grid}.

  Hereafter, $G u\frac{\Delta t}{\Delta x}$ is written compactly
    as $GC$ where $C$ denotes the Courant number.
  $GC$ is referred to as the advector, while
    the scalar field $\psi$ as the advectee
    --- the nomenclature adopted after \citet{Dave_Randall}.  
 
  Evaluation of eq.~(\ref{upstream}) concludes the first pass of MPDATA.
  To compensate for the implicit diffusion of the donor-cell pass, the subsequent passes of MPDATA
    reuse eq.~(\ref{upstream}) and~(\ref{F}), but with
    $\psi$ replaced with the result of the preceding pass and 
    $\vec{u}$ replaced with the ``anti-diffusive'' pseudo-velocity.
  The pseudo-velocity is analytically derived by expanding eq.~(\ref{upstream}) in the second-order Taylor
    series about spatial point~$i$ and time level~$n$, 
    and representing the leading, dissipative truncation error as an advective flux; see
    \citet{Smolarkiewicz_1984} for a derivation.
  A single corrective pass ensures second-order accuracy in time and space.
  Subsequent corrective passes decrease the amplitude of the leading error, 
    within second-order accuracy.
  The one-dimensional formula for the basic antidiffusive advector is written as

  \begin{equation}
    GC_{i\phlf}^{k\pone} = 
                          \left[ |GC^{k}_{i\phlf}| - \frac{(GC^{k}_{i\phlf})^2}{0.5(G_{i\pone}+G_{i})} \right] 
                          \frac{\psi^{k}_{i\pone}-\psi^{k}_i}{\psi^{k}_{i\pone}+\psi^{k}_i}~,
    \label{adf_vel}
    \vspace{.666em}
  \end{equation} 
  where $k$ numbers MPDATA passes.
  For k=1, $C^k$ is the flow-velocity-based Courant number, whereas
    for k$>$1, $C^k$ is the pseudo-velocity-based Courant number.
  The number of corrective passes can be chosen within \emph{libmpdata++}. 


  The library features two implementations of the donor-cell algorithm defined by (\ref{upstream})~and~(\ref{F}).
  The default one employs the compensated summation algorithm of \citet{Kahan_1965} which
    reduces round-off error arising when summing numbers of different magnitudes.
  The alternative, slightly less resource-intensive one, 
    is a ``straightforward'' summation
    available as an option in {\emph{libmpdata++}}.
 
  \subsubsection{Third-order-accurate variant}\label{sec:tot}

  Accounting for third-order terms in the Taylor series expansion while deriving the
    pseudo-velocity improves the accuracy of MPDATA.
  When $G\equiv1$, $u=const$ and three or more corrective passes are applied,
    the procedure ensures third-order accuracy in time and space.
  The discretised formul\ae~for the third-order scheme, 
    derived analytically in \citet{Margolin_and_Smolarkiewicz_1998},
    can be found in \citet[][Eq.~36]{Smolarkiewicz_and_Margolin_1998}.

  \subsubsection{Divergent-flow variant}\label{sec:dfl}
 
  In case of a divergent flow, the pseudo-velocity formul\ae~ 
    are augmented with an additional term proportional to the flow divergence.
  This additional term is implemented in \emph{libmpdata++} following 
    \citet[][sec.~3.2(3)]{Smolarkiewicz_and_Margolin_1998}.

  \subsubsection{Non-oscillatory option}\label{sec:fct}

  Solutions obtained with the basic MPDATA are sign-preserving, and thus non-oscillatory near zero.
  Generally however, they feature dispersive ripples characteristic of higher-order numerical schemes.
  These can be suppressed by limiting the pseudo-velocities, in the spirit of flux-corrected transport.
  Application of the limiters reduces somewhat the accuracy of the scheme 
    \citep{Smolarkiewicz_and_Grabowski_1990},
    yet this loss is generally outweighed by ensuring non-oscillatory 
    (or ripple-free) solutions.
  Noteworthy, because MPDATA is built upon the donor-cell scheme characterised by 
    small phase error, the non-oscillatory corrections have to deal 
    with errors in signal amplitude only.
  The non-oscillatory option is a default option within the \emph{libmpdata++}.
  For the derivation and further discussion of the multi-dimensional non-oscillatory option  
    see \citet{Smolarkiewicz_and_Grabowski_1990}.

  \subsubsection{Variable-sign scalar fields}\label{sec:varsign}

  The basic MPDATA formulation assumes that the advected field $\psi$ 
    is exclusively either non-negative or non-positive.
  In particular, this assumption is evident in the $\psi$-fraction factor 
    $\frac{\psi^{k}_{i\pone}-\psi^{k}_i}{\psi^{k}_{i\pone}+\psi^{k}_i}$
    of eq.~(\ref{adf_vel}), which can become unbounded in case of variable-sign field. 
  The \emph{libmpdata++} library includes implementations of 
    two \mbox{MPDATA} options intended for 
    simulating advection of variable-sign field.

  The first method replaces $\psi$ with  $|\psi|$ in 
    all $\psi$-fraction factors that enter the pseudo-velocity expressions.
  This approach is robust but it
    reduces the solution quality where $\psi$ crosses through zero;
    see paragraph 3.2(4) in \citet{Smolarkiewicz_and_Margolin_1998}.
  
  The default method, is the ``infinite-gauge'' variant of the algorithm,
    a generalised one-step Lax-Wendroff (linear, oscillatory) limit of MPDATA at infinite constant background,
    discussed in \citet[][sec.~4.2]{Smolarkiewicz_2006}.
  In practice, the infinite-gauge option of MPDATA is used 
    with the non-oscillatory enhancement.
  

  \subsection{\label{sec:homo_api}Library interface}

  \subsubsection{Compile-time parameters}

  Compile-time parameters include
    number of dimensions, number of equations and algorithm options.
  Most of the compile-time parameters are declared by defining integer constants  
    within the compile-time parameter structure.
  Listing~\ref{lst:choice_1} depicts a minimal definition that
    inherits from the \prog{ct\_params\_default\_t} structure
    containing default values for numerous parameters.

  \begin{Listing}[h!]
  \fvset{gobble=2}%
  \renewcommand*\FancyVerbStartString{\PY{c+c1}{//\PYZlt{}listing\PYZhy{}1\PYZgt{}}}%
  \renewcommand*\FancyVerbStopString{\PY{c+c1}{//\PYZlt{}/listing\PYZhy{}1\PYZgt{}}}%
  \input{choice.cpp}
    \caption{\label{lst:choice_1} 
      Example definition of compile-time parameters structure.
    }
  \end{Listing}

  All solvers expect a structure with compile-time parameters as their 
    first template parameter, as exemplified in List.~\ref{lst:choice_2}.

  \begin{Listing}[h!]
  \fvset{gobble=4}%
  \renewcommand*\FancyVerbStartString{\PY{c+c1}{//\PYZlt{}listing\PYZhy{}2\PYZgt{}}}%
  \renewcommand*\FancyVerbStopString{\PY{c+c1}{//\PYZlt{}/listing\PYZhy{}2\PYZgt{}}}%

\input{choice.cpp}    \caption{\label{lst:choice_2} 
      Example alias declaration combining solver- and compile-time parameters choice.
    }
  \end{Listing}

  \subsubsection{Choosing library components}

  The library components listed in section \ref{sec:comp} are chosen through template parameters.
  First, the solver is equipped with an output mechanism by 
    passing the solver type as a template parameter to the output type, 
    as exemplified in List.~\ref{lst:choice_3}.
  The output classes inherit from solvers.

  \begin{Listing}[h!]
  \fvset{gobble=4}%
  \renewcommand*\FancyVerbStartString{\PY{c+c1}{//\PYZlt{}listing\PYZhy{}3\PYZgt{}}}%
  \renewcommand*\FancyVerbStopString{\PY{c+c1}{//\PYZlt{}/listing\PYZhy{}3\PYZgt{}}}%

\input{choice.cpp}    \caption{\label{lst:choice_3} 
      Example alias declaration of an output mechanism.
    }
  \end{Listing}

  Second, the concurrency handlers expect solver class (equipped with output) as the first template parameter.
  Subsequent template parameters control
    boundary condition types on each of the domain edges (see List.~\ref{lst:choice_4}).

  \begin{Listing}[h!]
  \fvset{gobble=4}%
  \renewcommand*\FancyVerbStartString{\PY{c+c1}{//\PYZlt{}listing\PYZhy{}4\PYZgt{}}}%
  \renewcommand*\FancyVerbStopString{\PY{c+c1}{//\PYZlt{}/listing\PYZhy{}4\PYZgt{}}}%

\input{choice.cpp}    \caption{\label{lst:choice_4} 
      Example alias declaration of a concurrency handler.  
    }
  \end{Listing}

  \subsubsection{\label{sec:run}Run-time parameters}

  Run-time parameters include the grid size, 
    number of MPDATA passes and output file name.
  The list of applicable run-time parameters is defined by fields
    of the \prog{rt\_params\_t} structure.
  This structure is defined 
    within each solver and extended when equipping the solver with
    an output mechanism.
  The concurrency handlers expect an instance of the run-time parameters
    structure as their constructor argument.
  Example code depicting how to set the run-time parameters and then instantiate
    a concurrency handler is presented in List.~\ref{lst:choice_5}.

  \begin{Listing}[h!]
  \fvset{gobble=4}%
  \renewcommand*\FancyVerbStartString{\PY{c+c1}{//\PYZlt{}listing\PYZhy{}5\PYZgt{}}}%
  \renewcommand*\FancyVerbStopString{\PY{c+c1}{//\PYZlt{}/listing\PYZhy{}5\PYZgt{}}}%

\input{choice.cpp}    \caption{\label{lst:choice_5} 
      Example run-time parameter structure declaration followed by 
      a concurrency handler instantiation. 
    }
  \end{Listing}

  \subsubsection{\label{sec:public}Public methods}

  The concurrency handlers act as controlling logic for the other components,
    and hence the user is exposed to the public interface of these handlers only.

  Listing~\ref{lst:api} contains signatures of methods implemented by each 
    of the concurrency handlers.
  \begin{Listing}[h!]
  \fvset{gobble=6}%
  \renewcommand*\FancyVerbStartString{\PY{c+c1}{//\PYZlt{}listing\PYZhy{}2\PYZgt{}}}%
  \renewcommand*\FancyVerbStopString{\PY{c+c1}{//\PYZlt{}/listing\PYZhy{}2\PYZgt{}}}%
  \input{any.cpp}%

  \fvset{gobble=6}%
  \renewcommand*\FancyVerbStartString{\PY{c+c1}{//\PYZlt{}listing\PYZhy{}3\PYZgt{}}}%
  \renewcommand*\FancyVerbStopString{\PY{c+c1}{//\PYZlt{}/listing\PYZhy{}3\PYZgt{}}}%
  \input{any.cpp}%

  \fvset{gobble=6}%
  \renewcommand*\FancyVerbStartString{\PY{c+c1}{//\PYZlt{}listing\PYZhy{}4\PYZgt{}}}%
  \renewcommand*\FancyVerbStopString{\PY{c+c1}{//\PYZlt{}/listing\PYZhy{}4\PYZgt{}}}%
  \input{any.cpp}%

  \fvset{gobble=6}%
  \renewcommand*\FancyVerbStartString{\PY{c+c1}{//\PYZlt{}listing\PYZhy{}1\PYZgt{}}}%
  \renewcommand*\FancyVerbStopString{\PY{c+c1}{//\PYZlt{}/listing\PYZhy{}1\PYZgt{}}}%
  \input{any.cpp}%

  \fvset{gobble=6}%
  \renewcommand*\FancyVerbStartString{\PY{c+c1}{//\PYZlt{}listing\PYZhy{}5\PYZgt{}}}%
  \renewcommand*\FancyVerbStopString{\PY{c+c1}{//\PYZlt{}/listing\PYZhy{}5\PYZgt{}}}%

\input{any.cpp}    \caption{\label{lst:api} 
      Signatures of all the methods within \emph{libmpdata++} application programming interface.}
  \end{Listing}
    
  The \prog{advectee()} is an accessor method for the advected scalar fields.
  It can be used for setting the initial condition 
    as well as for examining the solver state. 
  It expects an index of the requested advectee as the argument 
    (advected scalar fields are numbered from zero).
  This provides choice between different advected variables.
  The returned \prog{blitz::Array} is zero-base indexed and has the same size as
    the computational grid (set with the \prog{grid\_size} field of
    the run-time parameters structure, see List.~\ref{lst:choice_5}).

  The \prog{advector()} method allows to access the components of the 
    vector field of Courant numbers multiplied by the G factor (i.e., 
    a Jacobian of coordinate transformation, a fluid density field or their product).
  The argument selects the vector field components numbered from zero.
  The size of the returned array depends on the component.
  It equals the grid size in all but the selected dimension in which it is
    reduced by one (i.e. $nx\times(ny-1)$ for the ``y'' component and so forth, cf. Fig.~\ref{fig:grid}).
  
  The \prog{g\_factor()} is an accessor method for the $G$ field.
  The returned array has the same size as the one returned by \prog{advectee()}.
  The default value is set to $G\equiv1$, (for details, see Ex.~\ref{sec:over_pole}).

  The \prog{advance()} method launches the time-stepping logic of the solver advancing
    the solution by the number of time steps given as argument.
  This method can be called multiple times - the solvers maintain all information
    needed to resume the integration. 

  The \prog{panic\_ptr()} method returns a pointer to a Boolean variable that 
    if set to true will cause the solver to stop the computations 
    after the currently computed time step. 
  This method may be used, for instance, to implement signal handling
    within programs using \emph{libmpdata++}.

  All multi-dimensional arrays used in \emph{libmpdata++} use the default \emph{Blitz++}
    ``row-major'' memory layout with the last dimension varying fastest.
  Domain decomposition for parallel computations 
    is done over the first dimension only.

  \subsection{\label{sec:1d}Example: ``hello world''}

  The source code presented in this subsection is intended to serve as a minimal
    complete example on how to use \emph{libmpdata++}.
  In other examples presented throughout the paper, only the fragments of code that
    differ significantly from the minimal example will be presented.

  \begin{figure}[h]
    \pgfimage[width=.45\textwidth]{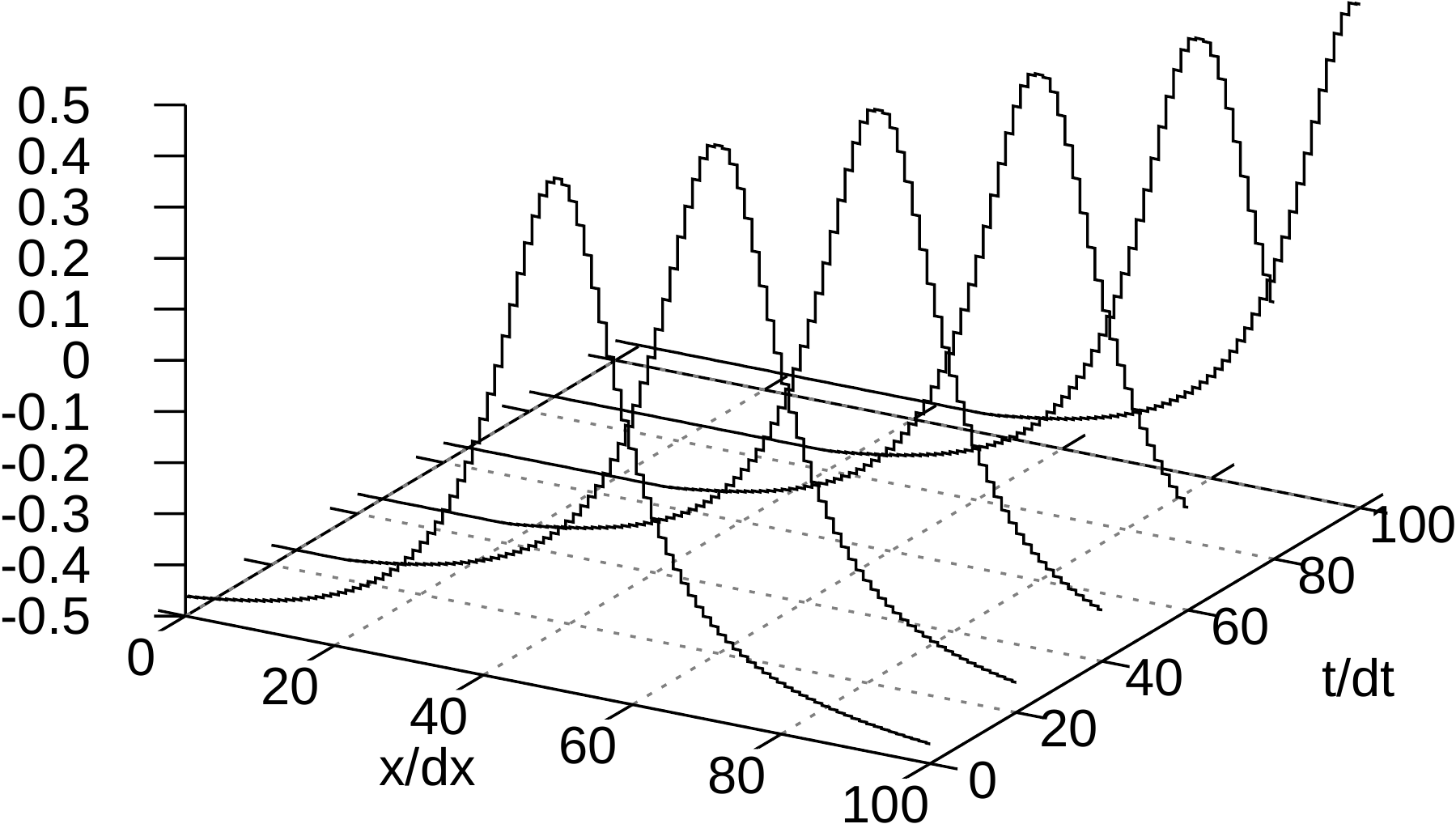}
    \caption{\label{fig:1} 
      Simulation results generated by the code in List.~\ref{lst:1}.
    }
  \end{figure}

  \begin{Listing}[th!]
  \fvset{gobble=0}%
  \renewcommand*\FancyVerbStartString{\PY{c+c1}{//\PYZlt{}listing\PYZhy{}1\PYZgt{}}}%
  \renewcommand*\FancyVerbStopString{\PY{c+c1}{//\PYZlt{}/listing\PYZhy{}1\PYZgt{}}}%
  \input{example_1.cpp.tex}
    \caption{\label{lst:1}
      A usage example of \emph{libmpdata++}.
      The listing contains the code needed to generate Fig.~\ref{fig:1}.
    }
  \end{Listing}

  The example consists of an elemental transport problem for a  
    one-dimensional, variable-sign field advected with a constant velocity. 
  The simulation results using code in List.~\ref{lst:1} are shown in Fig.~\ref{fig:1}.
  Spatial and temporal directions are depicted on the abscissa and ordinate, respectively.
  Cell-mean values of the transported field are shown on the applicate and are  
    presented in compliance with the assumption of data points
    representing grid-cell means of transported field.

  The code in List.~\ref{lst:1} begins with
    three include statements that
    reflect the choice of the library components: 
    solver, concurrency handler and output mechanism.
  All compile-time parameters are grouped into a structure passed as a template
    parameter to the solver.
  Here, this structure is named \textbf{ct\_params\_t}
    and inherits from \textbf{ct\_params\_default\_t} what results in assigning
    default values to parameters not defined within the inheriting class.
  The solvers expect the structure to contain a type \textbf{real\_t} 
    which controls the floating point format used.
  The two constants that do not have default values and need to be explicitly defined
    are \textbf{n\_dims} and \textbf{n\_eqns}. 
  They control the dimensionality of the problem and the number of equations to be solved, 
    respectively.

  Choice between different solver types, output mechanisms and concurrency handlers 
    is done via type alias declaration. 
  Here, the basic \prog{mpdata} solver is chosen which is then equipped with 
    the \prog{gnuplot} output mechanism.
  All output classes expect a solver class as their first template parameter,
    which is used to define the parent class (i.e., output classes inherit from
    solvers).

  Classes representing concurrency handlers 
    expect the output class and the boundary conditions as their template parameters.
  In the example, a basic serial handler is used and open boundary conditions
    on both ends of the domain are chosen.

  The choice of run-time parameters is done by assigning values to the member fields of
    the \textbf{rt\_params\_t} structure defined within the solver class and augmented
    with additional fields by the output class.
  In this example, the instance of \textbf{rt\_params\_t} structure is named \textbf{p},
    the grid size is set to 101 points
    and the output is set to be done every 20 time steps.
  An instance of the \textbf{rt\_params\_t} structure is expected as the constructor parameter
    for concurrency handlers. 

  The grid step \textbf{dx} is set to 0.1 and the number of time steps to 100. 
  Initial values of the Courant number and the transported scalar fields are set by assigning 
    to the arrays returned by the \prog{advector()} and \prog{advectee()} methods. 
  In this example, the Courant number equals $0.5$ and 
    the advected shape is described by the Witch of Agnesi formula $y(x) = 8a^3/(x^2+4a^2)$
    with the coefficient $a=0.5$.
  Initial shape is centred in the middle of computational domain and is shifted downwards by $0.5$.

  Finally, the actual integration is performed by calling the \prog{advance()} method with 
    the number of time steps as argument.

  \subsection{\label{sec:ex2}Example: advection scheme options}

  The following example is intended to present MPDATA advection scheme options 
    described in subsection \ref{sec:mpdata}.
  The way of choosing different options is discussed, and  
    the calling sequence of the library interface is shown
    for the case of advecting multiple scalar fields.

  The example consists of transporting two boxcar signals
    with different MPDATA options.
  In all tests, the first signal extends from 2 to 4 
    and the second signal extends from -1 to 1, 
    to observe the solution for 
    fixed-sign and variable-sign signals.

  Listing~\ref{lst:2_ct_params} shows the compile-time parameters structure fields
    common to all cases presented within this example.
  The number of dimensions is set to one 
    and the number of equations to solve is set to two.
  Consistent with List.~\ref{lst:1} from the ``hello world'' example, 
    \textbf{p} shown in List.~\ref{lst:2_rt_params} is 
    an instance of \textbf{rt\_params\_t} structure 
    with run-time parameters of the simulation.
  Setting the \textbf{outfreq} field to the number of time steps results in plotting 
    the initial condition and the final state.
  The \textbf{outvars} field contains a map 
    with a structure containing variable name, here left empty, 
    and unit defined for each of the advected scalar fields.
  Listing~\ref{lst:2_init} shows how to set initial values to multiple scalar fields
    using the \textbf{advectee()} method with an integer argument specifying
    the index of equation in the solved system.
 
 \begin{Listing}
  \fvset{gobble=4}%
  \renewcommand*\FancyVerbStartString{\PY{c+c1}{//\PYZlt{}listing\PYZhy{}1\PYZgt{}}}%
  \renewcommand*\FancyVerbStopString{\PY{c+c1}{//\PYZlt{}/listing\PYZhy{}1\PYZgt{}}}%
  \input{example_2.cpp}
    \caption{\label{lst:2_ct_params}
      Compile-time parameters for Ex.~\ref{sec:ex2}.
    }
  \end{Listing}
  \begin{Listing}
  \fvset{gobble=2}%
  \renewcommand*\FancyVerbStartString{\PY{c+c1}{//\PYZlt{}listing\PYZhy{}2\PYZgt{}}}%
  \renewcommand*\FancyVerbStopString{\PY{c+c1}{//\PYZlt{}/listing\PYZhy{}2\PYZgt{}}}%

\input{example_2.cpp}    \caption{\label{lst:2_rt_params}
      Run-time parameters for Ex.~\ref{sec:ex2}.
    }
  \end{Listing}
  \begin{Listing}
  \fvset{gobble=2}%
  \renewcommand*\FancyVerbStartString{\PY{c+c1}{//\PYZlt{}listing\PYZhy{}3\PYZgt{}}}%
  \renewcommand*\FancyVerbStopString{\PY{c+c1}{//\PYZlt{}/listing\PYZhy{}3\PYZgt{}}}%

\input{example_2.cpp}    \caption{\label{lst:2_init}
      Initial condition and velocity field for Ex.~\ref{sec:ex2}.
    }
  \end{Listing}

  \FloatBarrier
  \subsubsection{\label{sec:ex2_varsign}Variable-sign scalar fields}

  The \emph{libmpdata++} library is equipped with two options for handling 
    variable-sign fields; recall the discussion in paragraph \ref{sec:varsign}.
  The option using absolute values is named \textbf{abs}, 
    whereas the ``infinite-gauge'' option is dubbed \textbf{iga}.
  The option flags are defined in the \prog{opts} namespace.
  The option choice is made by defining the \prog{opts} field of the 
    compile-time parameters structure, in analogy to \prog{n\_dims} or \prog{n\_eqns}.

\begin{figure}[h!]
  \centering
    \pgfimage[width=.44\textwidth]{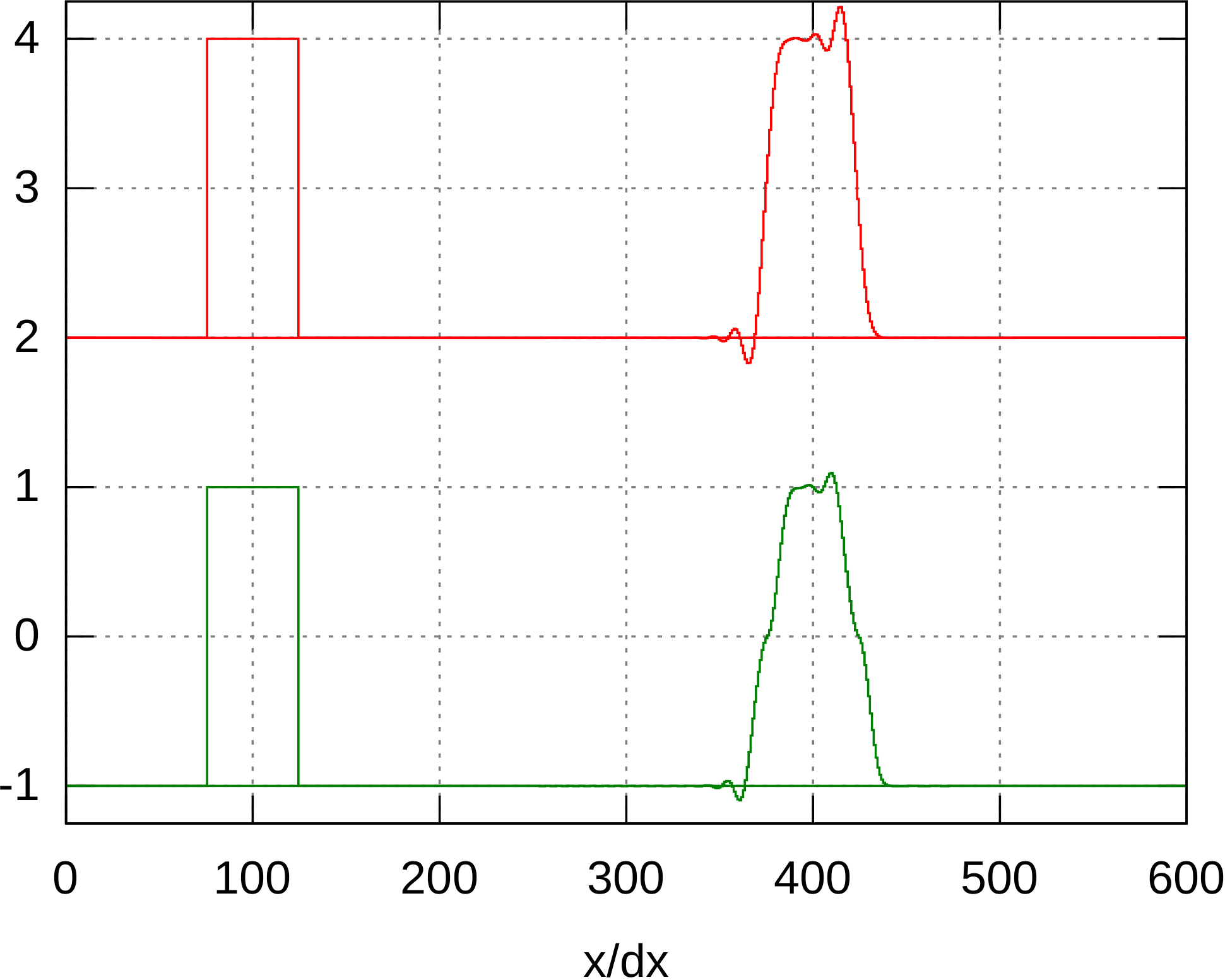}
    \caption{\label{fig:2_abs} 
      Result of the simulation with the advection scheme option for variable-sign signal set 
      to absolute value, cf.~List.~\ref{lst:2_abs}. 
    }
  \end{figure}

\begin{figure}[h!]
  \centering
    \pgfimage[width=.44\textwidth]{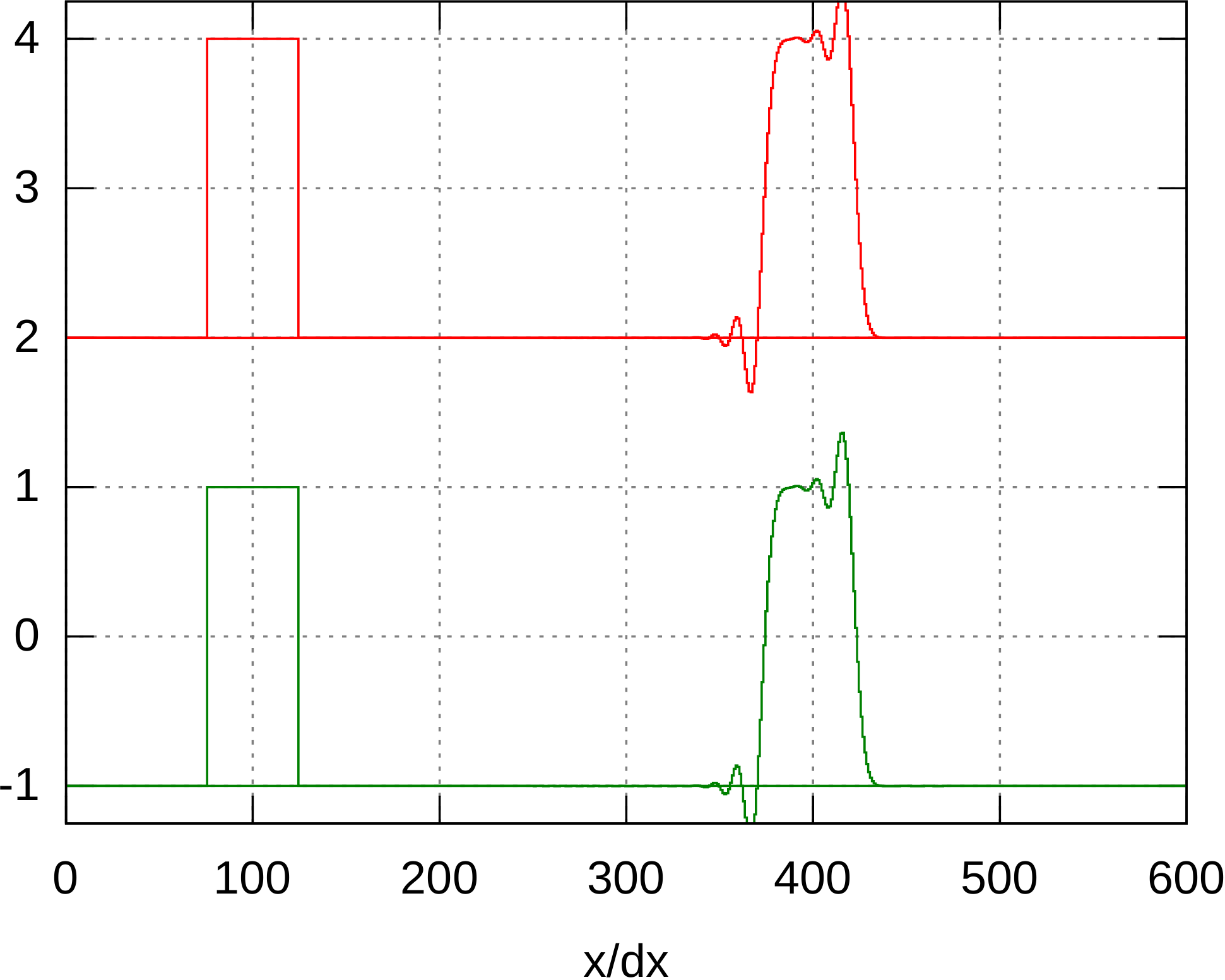}
    \caption{\label{fig:2_iga} 
      As in Fig.~\ref{fig:2_abs} but with variable-sign option set to ``infinite-gauge'', cf.~List.~\ref{lst:2_iga}.
    }
  \end{figure}

  In the first test, the choice of handling variable-sign signal 
    is set to \textbf{abs}, List.~\ref{lst:2_abs}.
  Figure~\ref{fig:2_abs} shows the result of simulation with parameters set 
    in List.~\ref{lst:2_ct_params}, \ref{lst:2_rt_params}, \ref{lst:2_init}  
     and \ref{lst:2_abs}.
  The final signal shows dispersive ripples characteristic of higher-order schemes. 
  It is also evident that the ripple magnitude depends on the constant background, 
    a manifestation of the scheme non-linearity.
  Furthermore, the final variable-sign signal features a bogus saddle point at the zero crossings
    (cf. paragraph \ref{sec:varsign}),
    and this can be eliminated by using the infinite-gauge (alias \prog{iga}) option.
  Listing~\ref{lst:2_iga} shows how to choose the \textbf{iga} option. 
  Figure~\ref{fig:2_iga} shows the result of simulation with parameters set 
    in List.~\ref{lst:2_ct_params}, \ref{lst:2_rt_params}, \ref{lst:2_init} 
     and \ref{lst:2_iga}.
  Although \textbf{iga} evinces more pronounced oscillations, their magnitude
    does not depend on the constant background.
  This, together with the robust behaviour of \textbf{iga} when crossing zero, substantiates
    the discussion of paragraph \ref{sec:varsign} on \textbf{iga} amounting to a linear limit of MPDATA.

  \begin{Listing}[h!]
  \fvset{gobble=4}%
  \renewcommand*\FancyVerbStartString{\PY{c+c1}{//\PYZlt{}listing\PYZhy{}4\PYZgt{}}}%
  \renewcommand*\FancyVerbStopString{\PY{c+c1}{//\PYZlt{}/listing\PYZhy{}4\PYZgt{}}}%

\input{example_2.cpp}    \caption{\label{lst:2_abs}
      Advection scheme options for Fig.~\ref{fig:2_abs}, variable-sign option is set to absolute value.
    }
  \end{Listing}

  \begin{Listing}[h!]
  \fvset{gobble=4}%
  \renewcommand*\FancyVerbStartString{\PY{c+c1}{//\PYZlt{}listing\PYZhy{}5\PYZgt{}}}%
  \renewcommand*\FancyVerbStopString{\PY{c+c1}{//\PYZlt{}/listing\PYZhy{}5\PYZgt{}}}%

\input{example_2.cpp}    \caption{\label{lst:2_iga}
      Advection scheme options for Fig.~\ref{fig:2_iga}, variable-sign option is set to ``infinite-gauge''.
    }
  \end{Listing}
 
  \FloatBarrier

  \subsubsection{\label{sec:ex2_third}Third-order-accurate variant}

\begin{figure}[h!]
  \centering
    \pgfimage[width=.44\textwidth]{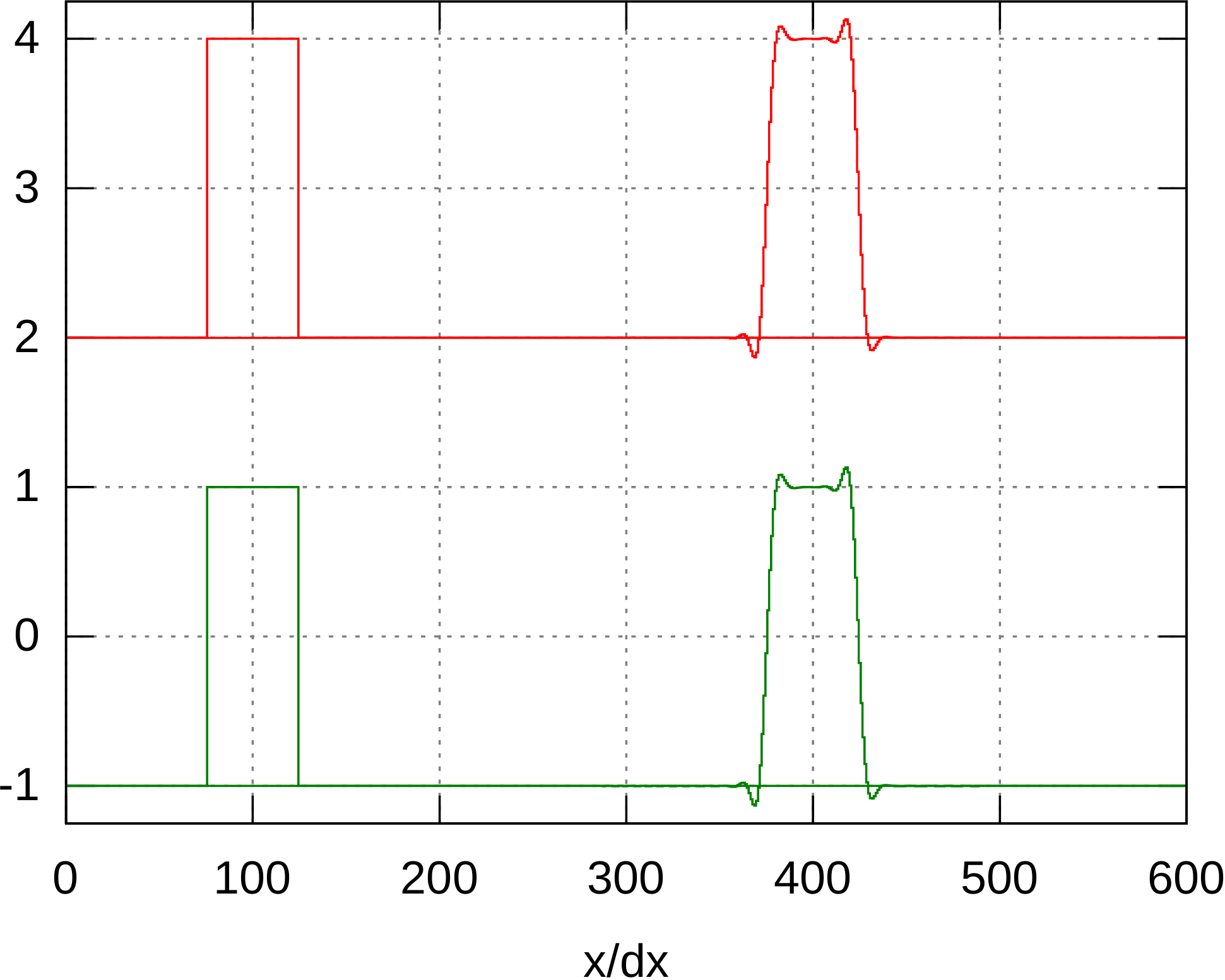}
    \caption{\label{fig:2_iga_tot}
      As in Fig.~\ref{fig:2_abs} but with variable-sign option set to ``infinite-gauge'' and 
      third-order-accurate variant, cf.~List.~\ref{lst:2_iga_tot}.
    }
  \end{figure}

  Choosing third-order variant enhances the accuracy of the scheme
    when used with more than two passes of MPDATA or with \textbf{iga}; recall paragraph \ref{sec:tot}. 
  Option \textbf{tot} enables the third-order variant of MPDATA scheme.
  Figure~\ref{fig:2_iga_tot} shows result of the same test as in Fig.~\ref{fig:2_abs} and~\ref{fig:2_iga}
    but with MPDATA options set as in List.~\ref{lst:2_iga_tot}.
  The resulting signal is evidently more accurate and symmetric, 
    but the oscillations are still present.

  \begin{Listing}[h!]
  \fvset{gobble=4}%
  \renewcommand*\FancyVerbStartString{\PY{c+c1}{//\PYZlt{}listing\PYZhy{}6\PYZgt{}}}%
  \renewcommand*\FancyVerbStopString{\PY{c+c1}{//\PYZlt{}/listing\PYZhy{}6\PYZgt{}}}%

\input{example_2.cpp}    \caption{\label{lst:2_iga_tot}
      Advection scheme options for Fig.~\ref{fig:2_iga_tot}, variable-sign option is set to ``infinite-gauge'' 
        and third-order accuracy variant is chosen.
    }
  \end{Listing}

  \FloatBarrier

  \subsubsection{\label{sec:ex2_nonosc}Non-oscillatory option}

  To eliminate oscillations apparent in the preceding tests,
    the non-oscillatory (\textbf{fct}) option (paragraph~\ref{sec:fct}) needs to be chosen.
  This option can be used together with all other MPDATA options, such as 
    basic scheme,
    variable-sign signals (\textbf{abs} or \textbf{iga})
    and the third-order-accurate variant (\textbf{tot}).

\begin{figure}[h!]
  \centering
    \pgfimage[width=.44\textwidth]{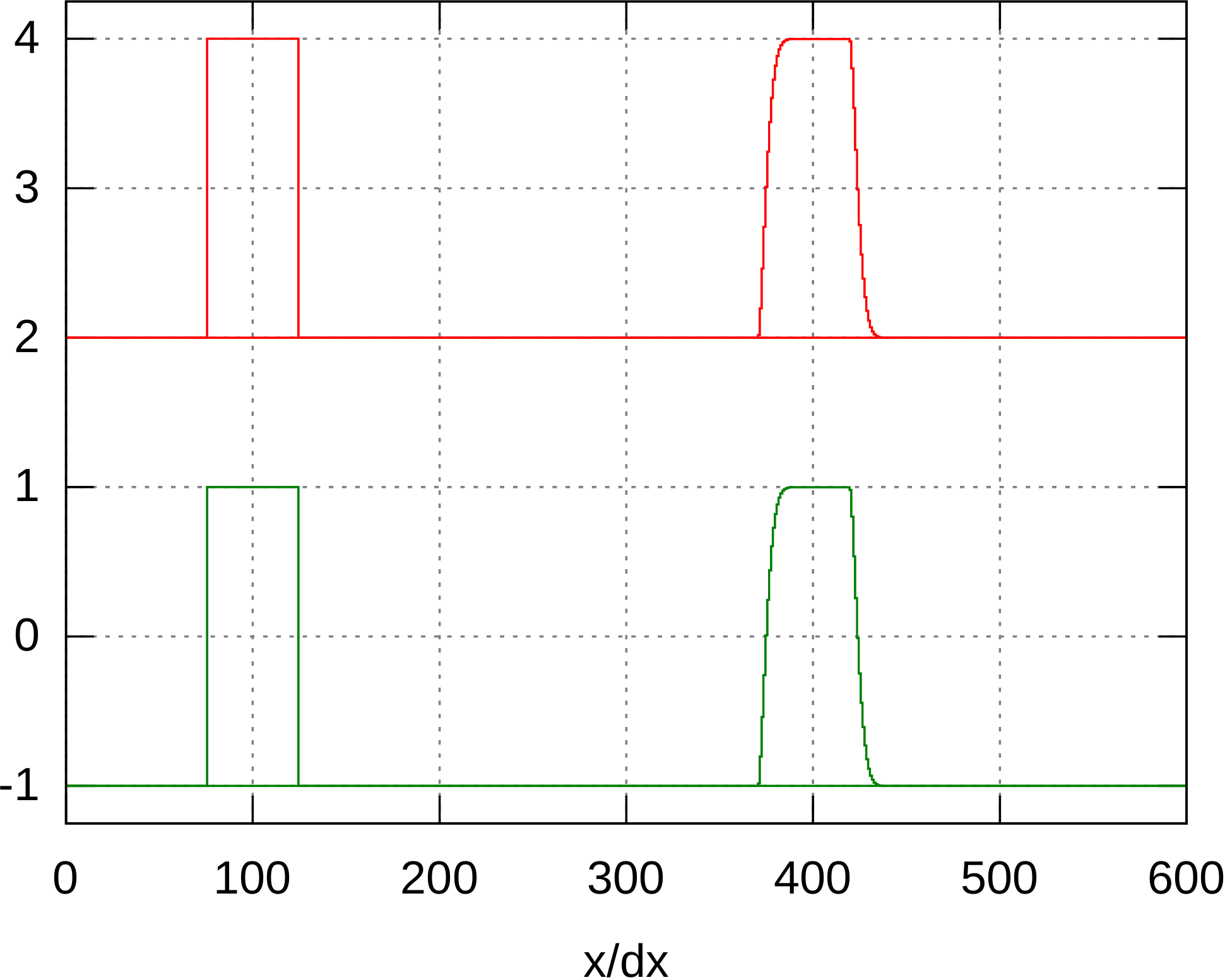}
    \caption{\label{fig:2_iga_fct}
      As in Fig.~\ref{fig:2_abs} but with options set to infinite-gauge treatment
      of variable-sign signal and flux corrections, cf. List.~\ref{lst:2_iga_fct}.
    }
  \end{figure}

  Here, \textbf{fct} is selected together with \prog{iga}, cf. List.~\ref{lst:2_iga_fct}.
  This is the default setting; i.e., when inheriting from the default parameters structure, 
    and not overriding the \prog{opts} setting, as illustrated 
    in List.~\ref{lst:1}.
  Figure~\ref{fig:2_iga_fct} shows the corresponding results.
  The solutions for both fixed-sign and variable-sign signals have indistinguishable profiles 
    and all of the dispersive ripples have been suppressed.
  \begin{Listing}[h!]
  \fvset{gobble=4}%
  \renewcommand*\FancyVerbStartString{\PY{c+c1}{//\PYZlt{}listing\PYZhy{}7\PYZgt{}}}%
  \renewcommand*\FancyVerbStopString{\PY{c+c1}{//\PYZlt{}/listing\PYZhy{}7\PYZgt{}}}%

\input{example_2.cpp}    \caption{\label{lst:2_iga_fct}
      Advection scheme options for Fig.~\ref{fig:2_iga_fct}, variable-sign option is set to ``infinite-gauge'' 
        and non-oscillatory option is enabled.
      This is the default setting in \emph{libmpdata++}.
    }
  \end{Listing}

  \begin{Listing}[h!]
  \fvset{gobble=4}%
  \renewcommand*\FancyVerbStartString{\PY{c+c1}{//\PYZlt{}listing\PYZhy{}8\PYZgt{}}}%
  \renewcommand*\FancyVerbStopString{\PY{c+c1}{//\PYZlt{}/listing\PYZhy{}8\PYZgt{}}}%

\input{example_2.cpp}    \caption{\label{lst:2_iga_tot_fct}
      Advection scheme options for Fig.~\ref{fig:2_iga_tot_fct}, variable-sign option is set to ``infinite-gauge'', 
        non-oscillatory option is enabled and third-order accuracy variant is chosen.
    }
  \end{Listing}

  \begin{figure}[h!]
  \centering
    \pgfimage[width=.44\textwidth]{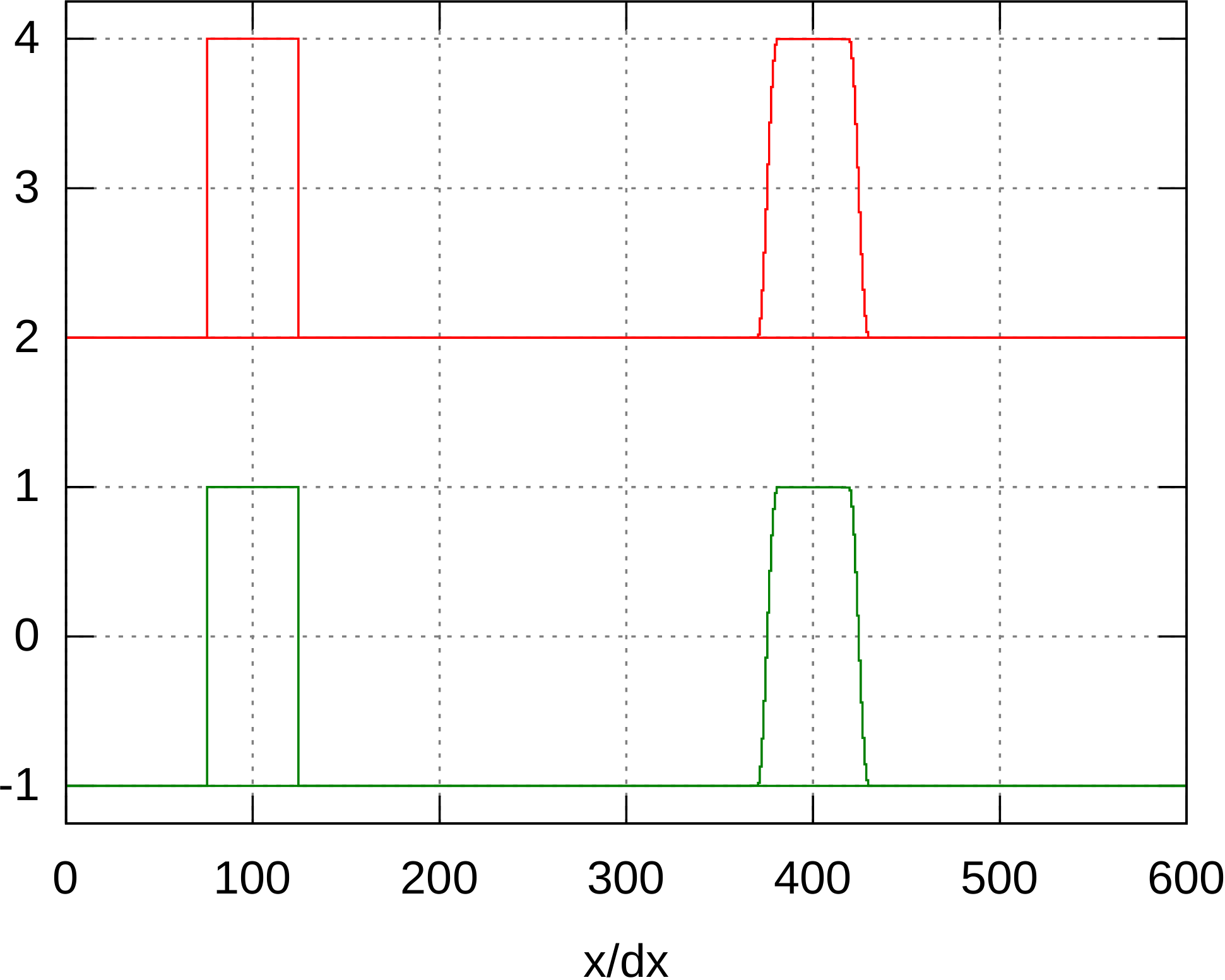}
    \caption{\label{fig:2_iga_tot_fct} 
      As in Fig.~\ref{fig:2_abs} but with options set to infinite-gauge treatment
      of variable-sign signal, non-oscillatory option and third-order accuracy variant, cf. List.~\ref{lst:2_iga_tot_fct}.
      }
  \end{figure}

  To further enhance the accuracy of the solution, \textbf{fct} and \textbf{iga} 
    can be combined with the \textbf{tot} variant; cf. List.~\ref{lst:2_iga_tot_fct}.
  The corresponding result is shown in Fig.~\ref{fig:2_iga_tot_fct}.
  Enabling the third-order-accurate variant improves the symmetry of the solution, as compared
    to the results presented in Fig.~\ref{fig:2_iga_fct}.

  \FloatBarrier

  \subsection{\label{sec:conv_test}Example: convergence tests in 1D}

  In this subsection the convergence test originated in \citet{Smolarkiewicz_and_Grabowski_1990} 
    is used to quantify the accuracy of various MPDATA options.

  The test consists of a series of one-dimensional simulations with Courant numbers
  \begin{eqnarray*}
    C \in (0.05, 0.1, 0.15, 0.2, \ldots, 0.85, 0.9, 0.95)~,
  \end{eqnarray*}
  and grid increments
  \begin{eqnarray*}
    \mathsmaller{\Delta} x \in 
    \left(\frac{\mathsmaller{\Delta x_m}}{2^0}, \frac{\mathsmaller{\Delta x_m}}{2^1}, 
     \frac{\mathsmaller{\Delta x_m}}{2^2}, \frac{\mathsmaller{\Delta x_m}}{2^3}, 
     \frac{\mathsmaller{\Delta x_m}}{2^4}, \frac{\mathsmaller{\Delta x_m}}{2^5}, 
     \frac{\mathsmaller{\Delta x_m}}{2^6}, \frac{\mathsmaller{\Delta x_m}}{2^7}\right),
  \end{eqnarray*}
  \noindent
  where $\Delta x_m = 1$ is the maximal increment.
  The series amounts to 152 simulations for each option. 
  In each simulation, the number of time steps $NT$ and the number of grid cells $NX$
    is adjusted so that the total time $T$ and total length of the domain $X$ remain 
    constant.
  The domain size $X = 44 \Delta x_m$ and simulation time $T = 1$ are selected. 
  The advective velocity is set to $u = \Delta x_m / T = 1$.

  In each simulation, a Gaussian profile 
  \begin{eqnarray}
    \psi_{ex}(x)_{t=0} = \frac{1}{\sigma \sqrt{2 \pi}} exp\left(-\frac{(x - x_0)^2}{2 \sigma^2}\right)
  \label{gauss}
  \end{eqnarray}
  \noindent
    is advected, and the result of the simulation is compared with the exact solution $\psi_{ex}$.
  The initial profiles and the exact solutions are calculated by analytically integrating 
    function~(\ref{gauss}) over the grid-cell extents, to comply with the inherent MPDATA assumption
    of a data point representing the grid-cell mean of transported field.
  The dispersion parameter of the initial profile (\ref{gauss}) is set to  $\sigma = 1.5 \Delta x_{m}$, 
    while the profile is centred in the middle of the domain $ x_{0} = 0.5 X$.

  As a measure of accuracy, a truncation-error function is introduced 
  \begin{equation}
    err(C, \Delta x) \equiv \frac{1}{T} \sqrt{\left. \sum_{i=1}^{NX} [\psi_{ex}(x_i) - \psi(x_i)]^2/NX \right|_{t=T}}~.
    \label{trerr}
    \vspace{.666em}
  \end{equation}
 
\begin{figure}[h!]
  \centering
  \subfloat[]{
    \pgfimage[width=.35\textwidth]{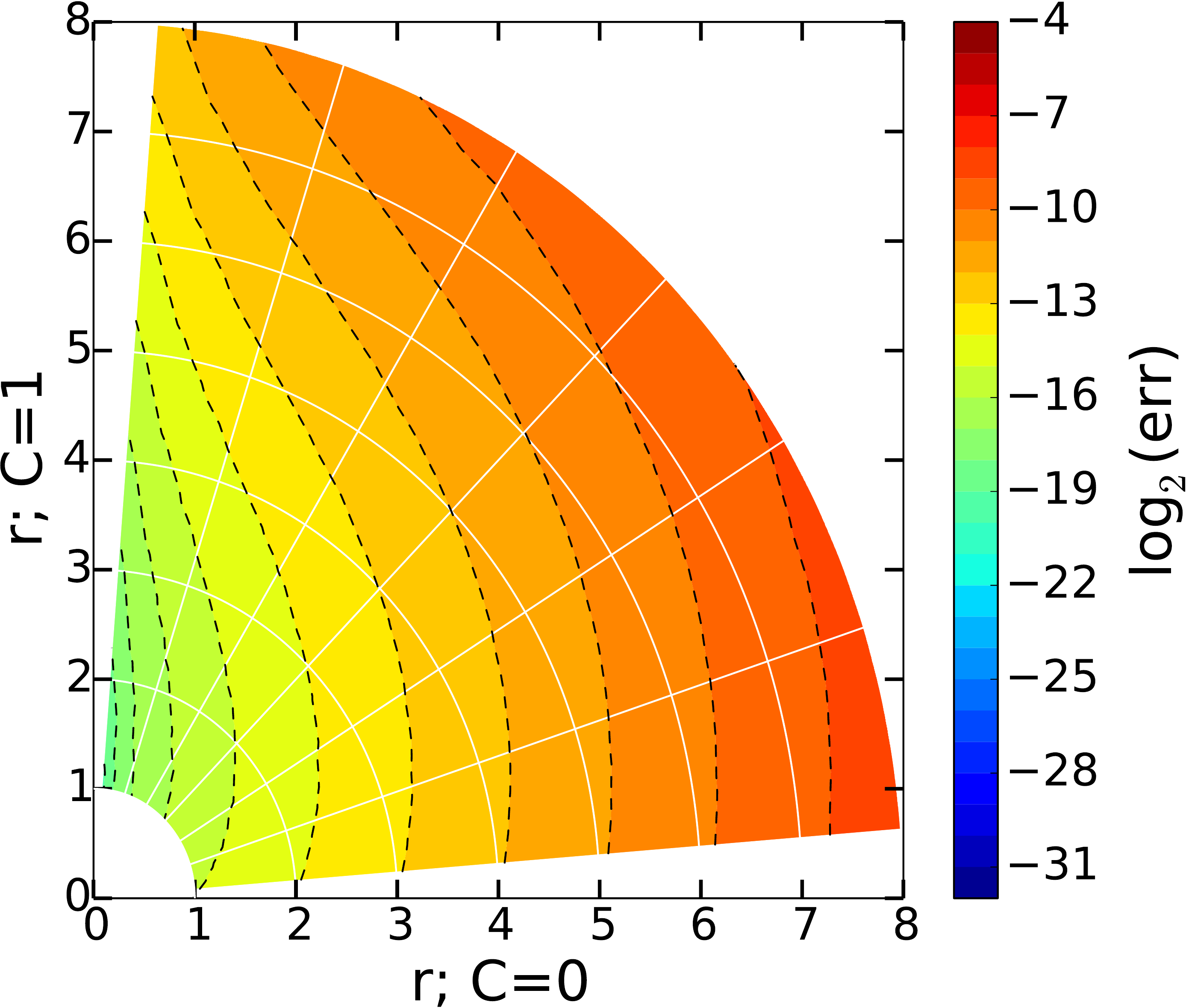}
    \label{fig:3_it1} 
  }
  \\
  \vspace{-0.9cm}
  \subfloat[]{
    \pgfimage[width=.35\textwidth]{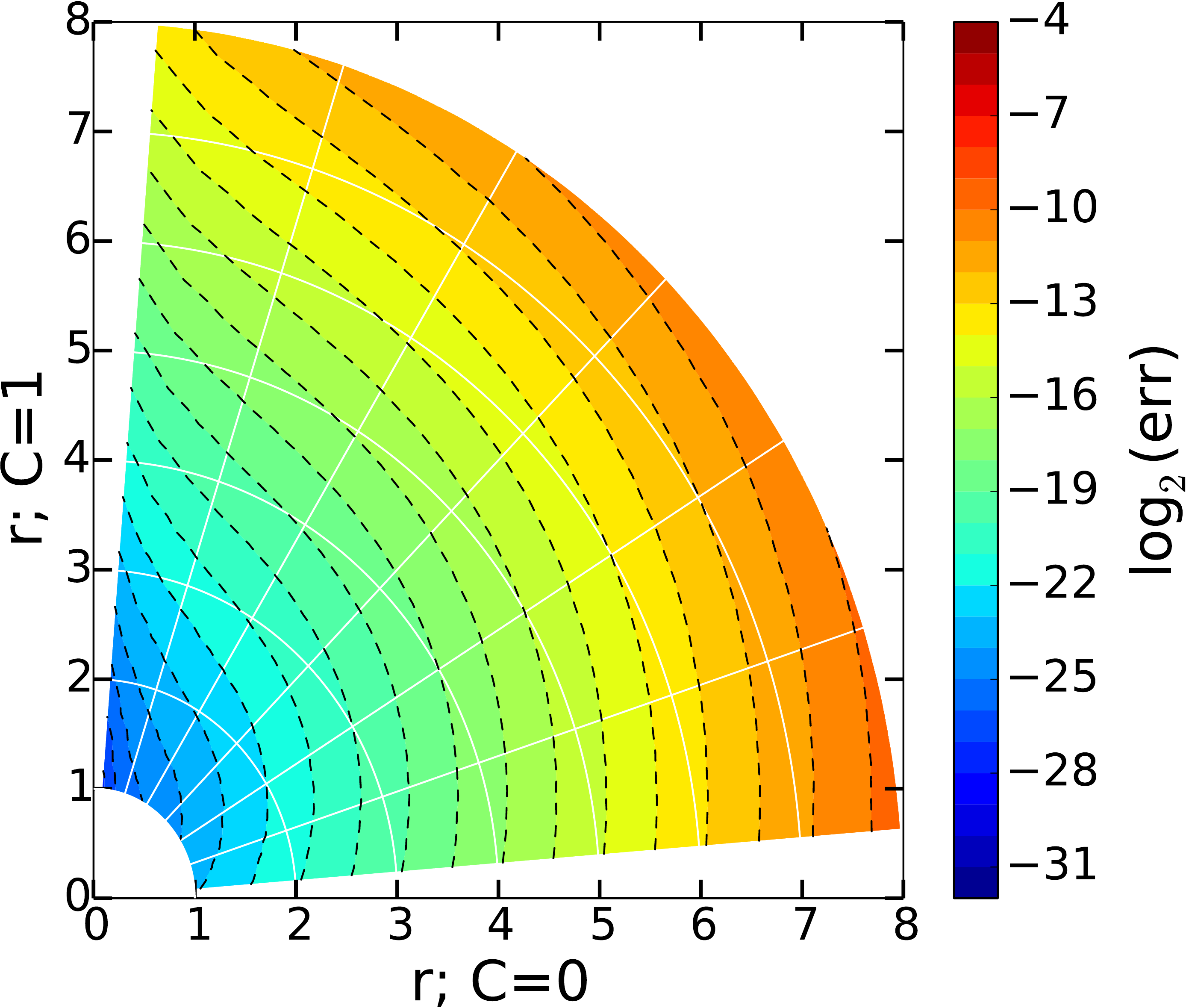}
    \label{fig:3_it2} 
  } 
  \\
  \vspace{-0.9cm}
  \subfloat[]{
    \pgfimage[width=.35\textwidth]{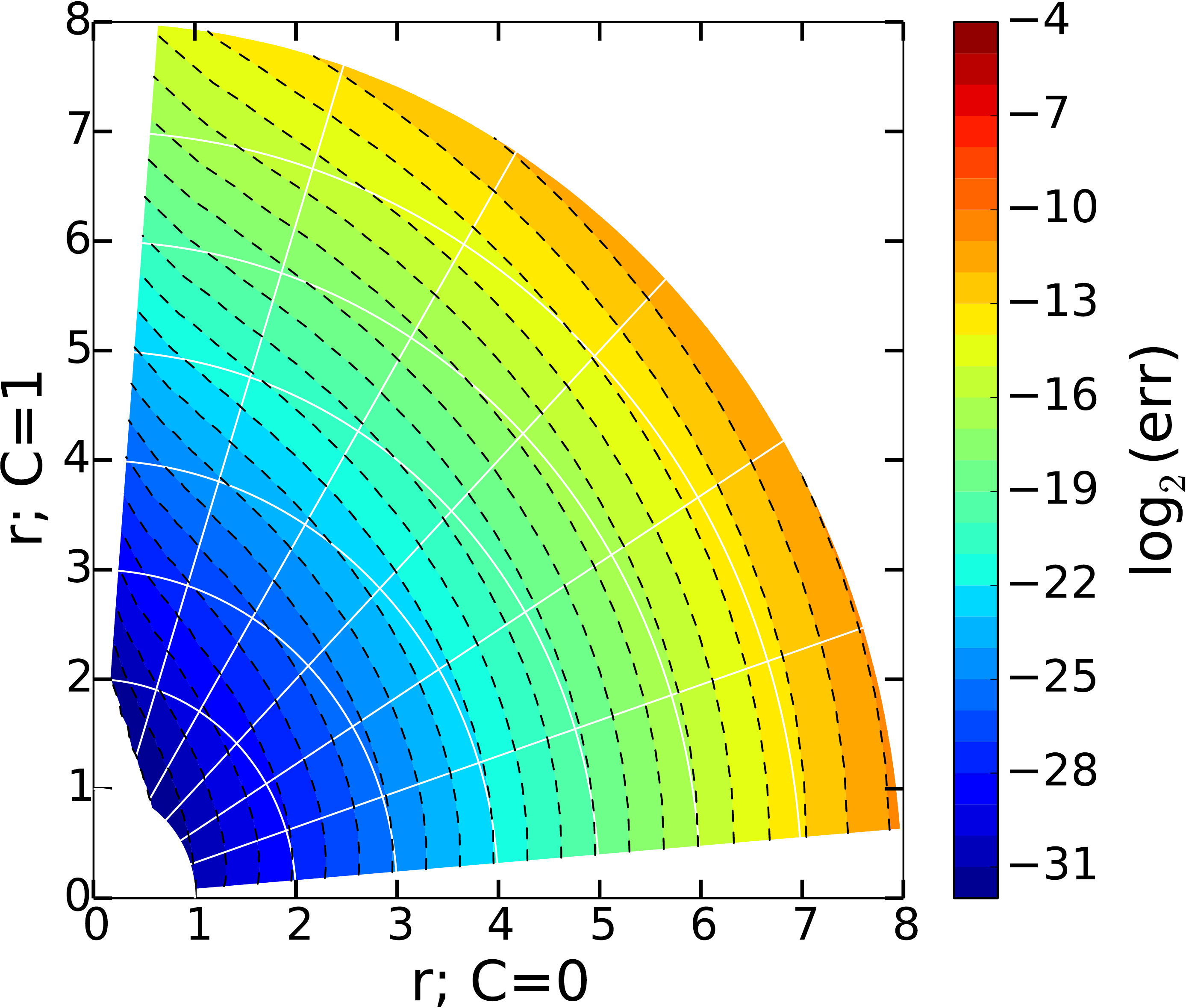}
    \label{fig:3_tot}
  }
  \centering
  \caption{\label{conv_panel_1} The result of the convergence test. \ref{fig:3_it1} for the donor-cell scheme, 
           \ref{fig:3_it2} for the basic MPDATA and \ref{fig:3_tot} for the third-order-accurate variant.}
  \end{figure}

  The results of the convergence test for the generic first-order-accurate donor-cell scheme, 
    the basic MPDATA and its third-order-accurate variant are shown in 
    Fig.~\ref{fig:3_it1}-\ref{fig:3_tot}. 
  Each figure displays, in polar coordinates, the base-two logarithm of the truncation-error function 
    (\ref{trerr}) for the entire series of 152 simulations.
  The radius and angle, respectively,    
    \begin{eqnarray}
    r=ln_2\left( \frac{\Delta x}{\Delta x_m}\right) + 8 ~, ~~~
    \phi = C \frac{\pi}{2}~,
  \end{eqnarray}
  indicate changes in grid increment and Courant number.
  Thus, closer to the origin are simulation results for finer grids, 
    closer to the abscissa are points for small Courant numbers, 
    and closer to the ordinate are points with Courant numbers approaching unity.
  The contour interval of dashed isolines and of the colour map is set to~1, 
    corresponding to error reduction by the factor of~2.
  Lines of constant grid-cell size and constant Courant number are overlaid with white contours.

  The figures contain information on the convergence rate of MPDATA options.
  When moving along the lines of constant Courant number towards the origin, 
    thus increasing the spatial and temporal resolution,
    the number of crossed dashed isolines determines the order of the scheme, cf. section 
    8.1 in \citet{Margolin_and_Smolarkiewicz_1998}. 
  Therefore, the results in Fig.~\ref{fig:3_it1}-\ref{fig:3_tot} attest to 
    the first-, second- and third-order asymptotic convergence rates, respectively.
  Furthermore, the shape of dashed isolines  
    conveys the dependency of the solution accuracy on the Courant number.
  In particular, they show that at fixed spatial resolution the solution accuracy
    increases with the Courant number.
  Moreover, as the order of the convergence increases the isolines become more circular 
    indicating more isotropic solution accuracy in the Courant number.

\begin{figure}[th!]
  \centering
  \subfloat[]{
    \pgfimage[width=.35\textwidth]{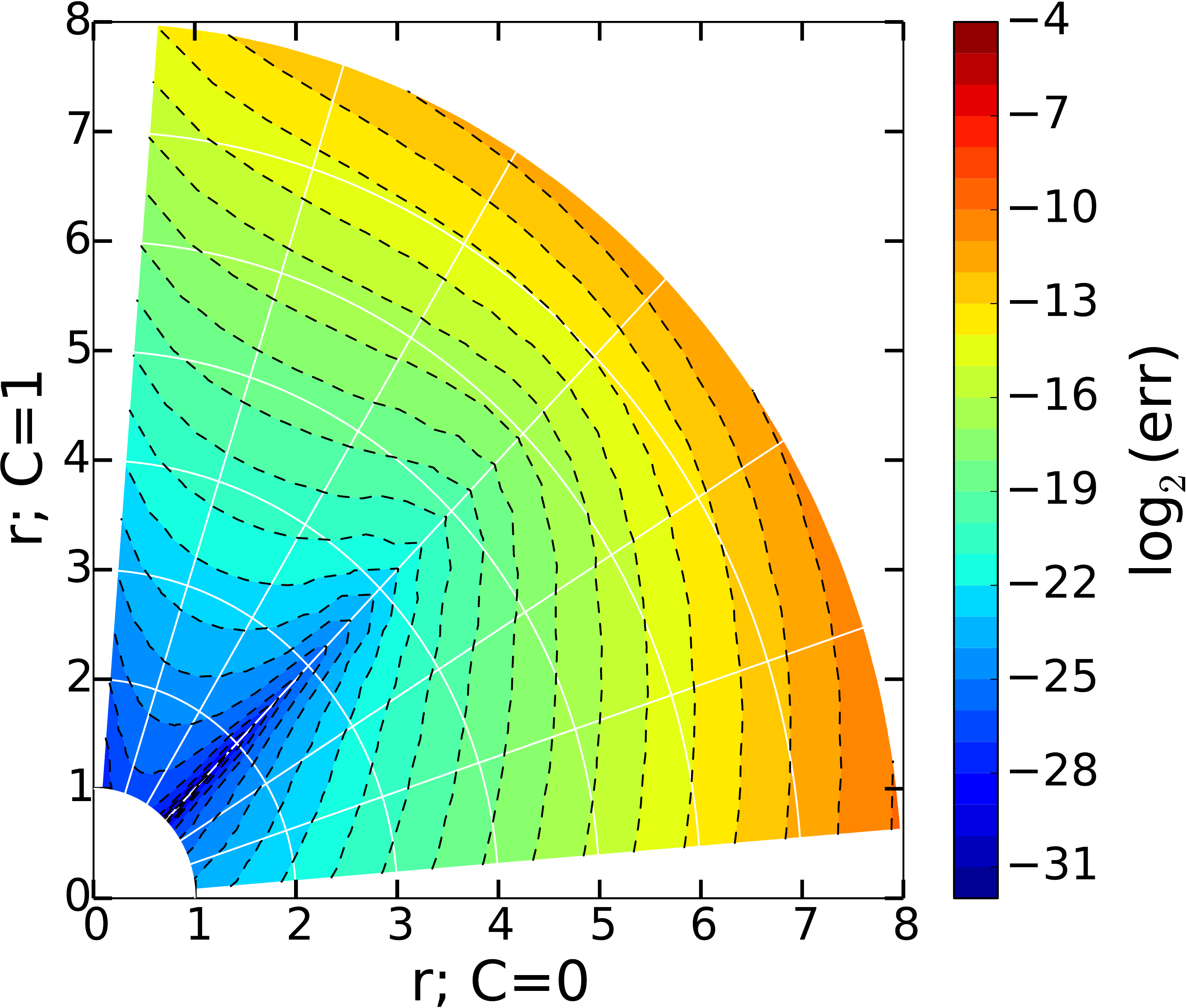}
    \label{fig:3_it3} 
  }
  \\
  \vspace{-0.95cm}
  \subfloat[]{
    \pgfimage[width=.35\textwidth]{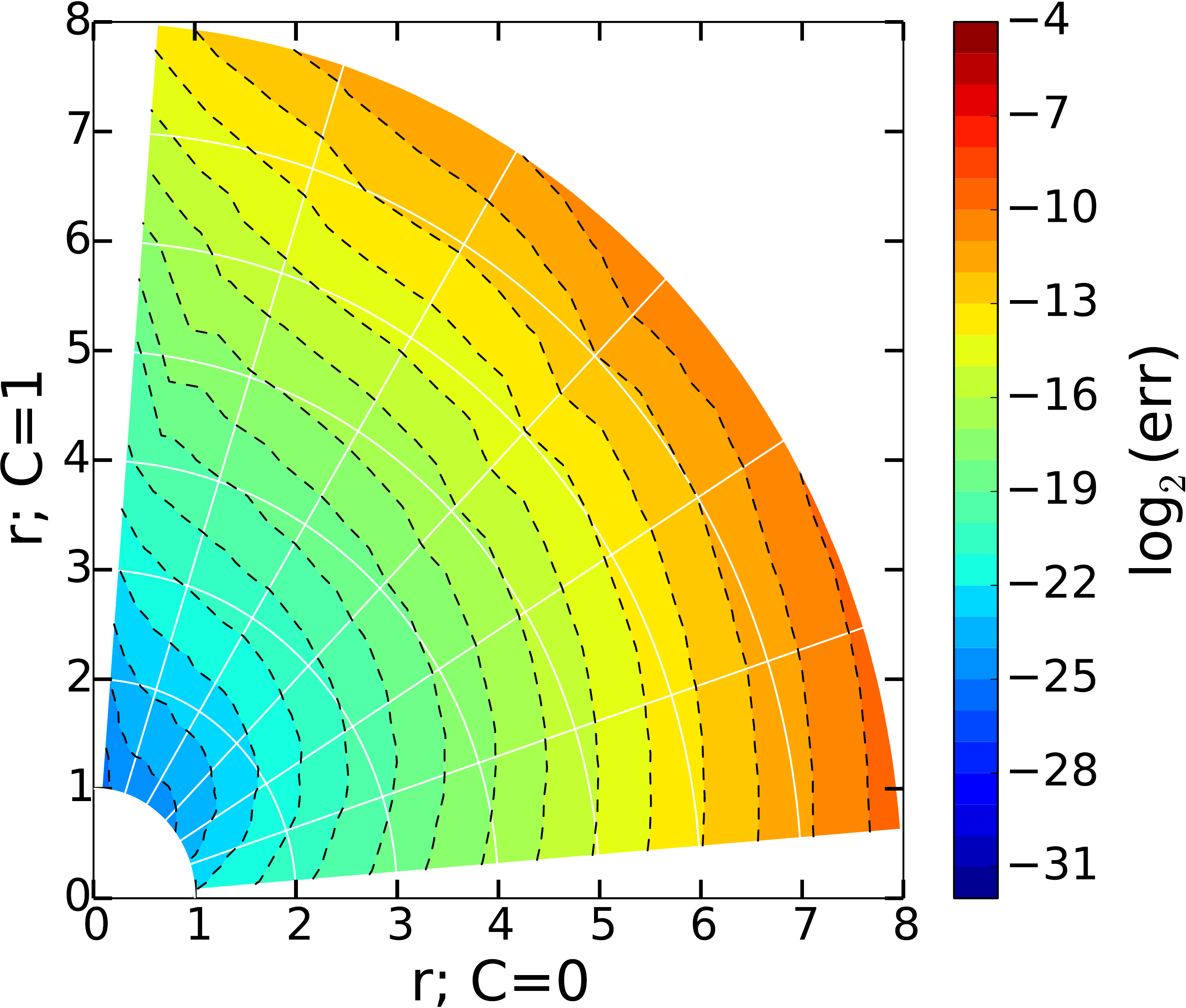}
    \label{fig:3_fct}
  }
  \\
  \vspace{-0.95cm}
  \subfloat[]{ 
    \pgfimage[width=.35\textwidth]{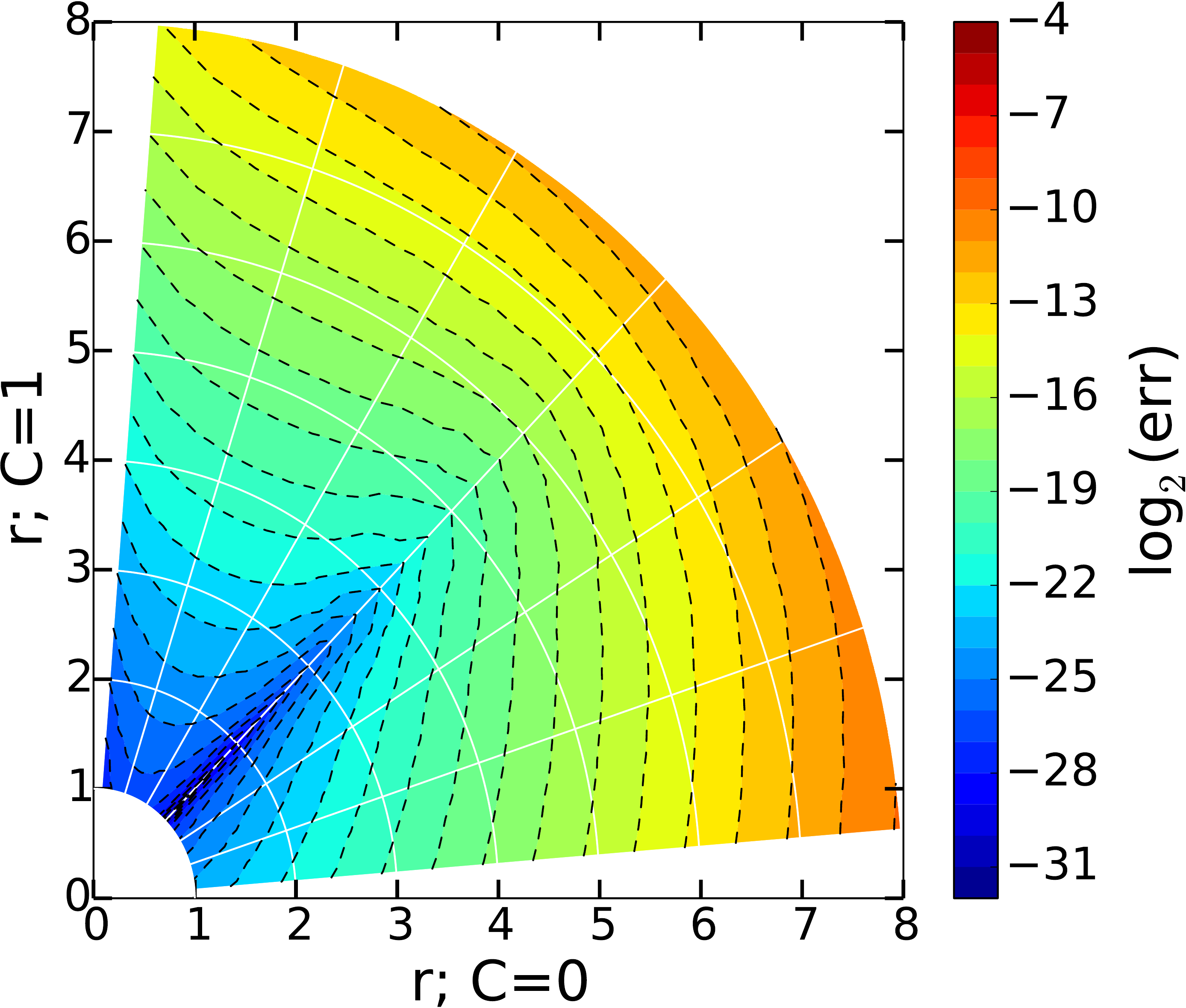}
    \label{fig:3_iga}
  }
  \\ 
  \vspace{-0.95cm}
  \subfloat[]{
    \pgfimage[width=.35\textwidth]{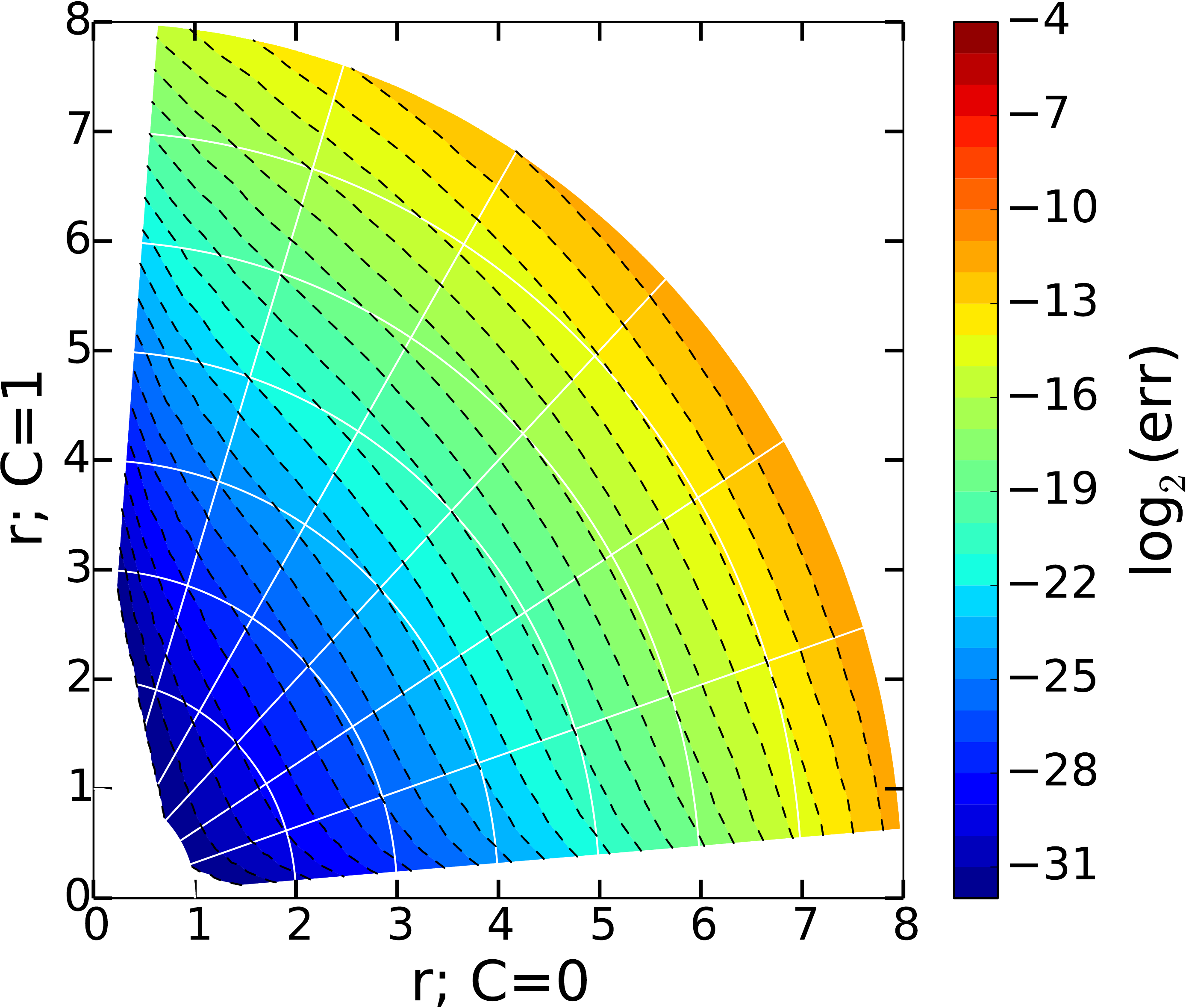}
    \label{fig:3_iga_tot}
  }
  \caption{\label{fig:conv_panel_2} As in Fig.~\ref{conv_panel_1}. 
           \ref{fig:3_it3} for three passes of MPDATA, 
           \ref{fig:3_fct} for two passes with non-oscillatory option, 
           \ref{fig:3_iga} for infinite-gauge option, 
           and \ref{fig:3_iga_tot} for infinite-gauge with third-order-accurate variant.}
  \end{figure}

  Figure~\ref{fig:3_it2} reproduces the solution in Fig.~1 
    of \citet{Smolarkiewicz_and_Grabowski_1990} and, thus, verifies the \emph{libmpdata++} implementation.
  For further verification 
    Fig.~\ref{fig:3_it3} and \ref{fig:3_fct} show results of the convergence test 
    for: i) three-pass MPDATA, (run-time solver parameter 	\textbf{n\_iters = 3}); 
    and ii) for two-pass MPDATA with \textbf{fct} option.
  These results reproduce Fig.~2 and 3 from \citet{Smolarkiewicz_and_Grabowski_1990}.
  Noteworthy, an interesting feature of Fig.~\ref{fig:3_it3} is the groove 
    of the third-order convergence rate formed
    around $\phi = 45^{\circ}$, characteristic of MPDATA with three or more passes
    \citep{Margolin_and_Smolarkiewicz_1998}.
  Next, comparing Fig.~\ref{fig:3_fct} with \ref{fig:3_it2} shows that 
    the price to be paid for an oscillation-free result is a reduction in the convergence rate
    \citep[from 2 to $\sim$1.8, section 4 in][]{Smolarkiewicz_and_Grabowski_1990}. 

  Figures~\ref{fig:3_iga} and \ref{fig:3_iga_tot} document original results for the convergence test 
    applied to the ``infinite-gauge'' limit of MPDATA.
  In particular, Fig.~\ref{fig:3_iga} shows that \textbf{iga} is as accurate as three-pass MPDATA, 
    \citep[cf. section 4 in][]{Smolarkiewicz_and_Clark_1986}; 
    whereas, Fig.~\ref{fig:3_iga_tot} reveals that the third-order-accurate \textbf{iga} 
    is more anisotropic in Courant number than the third-order-accurate standard 
    MPDATA in Fig.~\ref{fig:3_tot}.

  The convergence test results for the default setting of \emph{libmpdata++} 
    (\textbf{iga} plus \textbf{fct}) are not shown, because they resemble 
    results from Fig.~\ref{fig:3_fct} with somewhat enhanced accuracy
    for well-resolved fields (i.e., small grid-cells).

  \FloatBarrier
\subsection{\label{sec:cone}Example: rotating cone in 2D}


  This example introduces \emph{libmpdata++} programming interface  
    for two-dimensional simulations with the velocity field varying in space.
  Test results are compared with published MPDATA benchmarks.
  The example is based on the classical solid-body rotation test \citep{Molenkamp_1968}.
  The current setup follows \citet{Smolarkiewicz_and_Margolin_1998}.
  The initial condition features a cone 
    centred around the point $(x_0, y_0)=(50\Delta x, 75\Delta y)$.
  The grid interval is $\Delta x = \Delta y = 1$, and the domain size 
    is $100 \Delta x \times 100 \Delta y$ ---
    thus containing 101$\times$101 data points, cf. Fig.~\ref{fig:domain}.
  The height of the cone is set to 4, the radius to $15\Delta x$, and the background level to 1.
  The flow velocity is specified as $(u,v) = \omega \left(y-y_c, -(x-x_c) \right)$, 
    where angular velocity $\omega = 10^{-1}$ 
    and $(x_c, y_c)$ denotes coordinates of the domain centre. 
  With time interval $\Delta t = 0.1$, one full rotation requires 628 time steps. 
  The total integration time corresponds to six full rotations.

  Implementation of the set-up using the \emph{libmpdata++} interface begins with 
    definition of the compile-time parameters structure.
  The test features a single scalar field in a two-dimensional space, 
    what is reflected in the values of \prog{n\_dims} and \prog{n\_eqns} 
    set in List.~\ref{lst:cone1}.
  In one of the test runs, the number of MPDATA passes (\textbf{n\_iters}) 
    is set to 3, instead of the default value of 2.
  Corresponding field of run-time parameters structure is shown in List.~\ref{lst:cone4}.
  During instantiation of the concurrency handler, four boundary-condition settings 
    (two per each dimension) are passed as template arguments. 
  In this example, open boundary conditions (\prog{bcond::open}) are set in both dimensions 
    - see List.~\ref{lst:cone2}.

  \begin{Listing}
  \fvset{gobble=4}%
  \renewcommand*\FancyVerbStartString{\PY{c+c1}{//\PYZlt{}listing\PYZhy{}1\PYZgt{}}}%
  \renewcommand*\FancyVerbStopString{\PY{c+c1}{//\PYZlt{}/listing\PYZhy{}1\PYZgt{}}}%
  \input{rotating_cone.cpp.tex}
    \caption{\label{lst:cone1}
      Compile-time parameter settings for the rotating-cone test.
    }
  \end{Listing}

  \begin{Listing}
  \fvset{gobble=6}%
  \renewcommand*\FancyVerbStartString{\PY{c+c1}{//\PYZlt{}listing\PYZhy{}4\PYZgt{}}}%
  \renewcommand*\FancyVerbStopString{\PY{c+c1}{//\PYZlt{}/listing\PYZhy{}4\PYZgt{}}}%

\input{rotating_cone.cpp.tex}    \caption{\label{lst:cone4}
      Run-time parameter responsible for setting the number of MPDATA passes in Fig.~\ref{fig:cone_acc2}.
    }
  \end{Listing}

  The choice of the \prog{threads} concurrency handler in List.~\ref{lst:cone2} results in multi-threaded calculations
    -- using OpenMP if the compiler supports it, or using \emph{Boost.Thread} otherwise.
  The number of computational subdomains (and hence threads) is controlled by the 
    \prog{OMP\_NUM\_THREADS} environment
    variable, regardless if OpenMP or \emph{Boost.Thread} implementation is used.
  The default is to use all CPUs/cores available in the system.
  Notably, replacing \prog{concurr::serial} from the previous examples with \prog{concurr::threads} 
    is the only modification needed to enable domain decomposition via shared-memory parallelism.

  \begin{Listing}
  \fvset{gobble=2}%
  \renewcommand*\FancyVerbStartString{\PY{c+c1}{//\PYZlt{}listing\PYZhy{}2\PYZgt{}}}%
  \renewcommand*\FancyVerbStopString{\PY{c+c1}{//\PYZlt{}/listing\PYZhy{}2\PYZgt{}}}%

\input{rotating_cone.cpp.tex}    \caption{\label{lst:cone2}
      Concurrency handler instantiation for the rotating-cone test.}
  \end{Listing}

  The way the initial condition and
    the velocity field are set is shown in List.~\ref{lst:cone3}.
  The Courant number components are specified using calls
    to the \prog{advector()} method with the argument defining
    the component index.

  \begin{Listing}
  \fvset{gobble=4}%
  \renewcommand*\FancyVerbStartString{\PY{c+c1}{//\PYZlt{}listing\PYZhy{}3\PYZgt{}}}%
  \renewcommand*\FancyVerbStopString{\PY{c+c1}{//\PYZlt{}/listing\PYZhy{}3\PYZgt{}}}%

\input{rotating_cone.cpp.tex}    \caption{\label{lst:cone3}
      Initial condition for the rotating-cone test.}
  \end{Listing}

 \begin{figure}[t!]
   \centering
   \subfloat[]{
     \pgfimage[width=.36\textwidth]{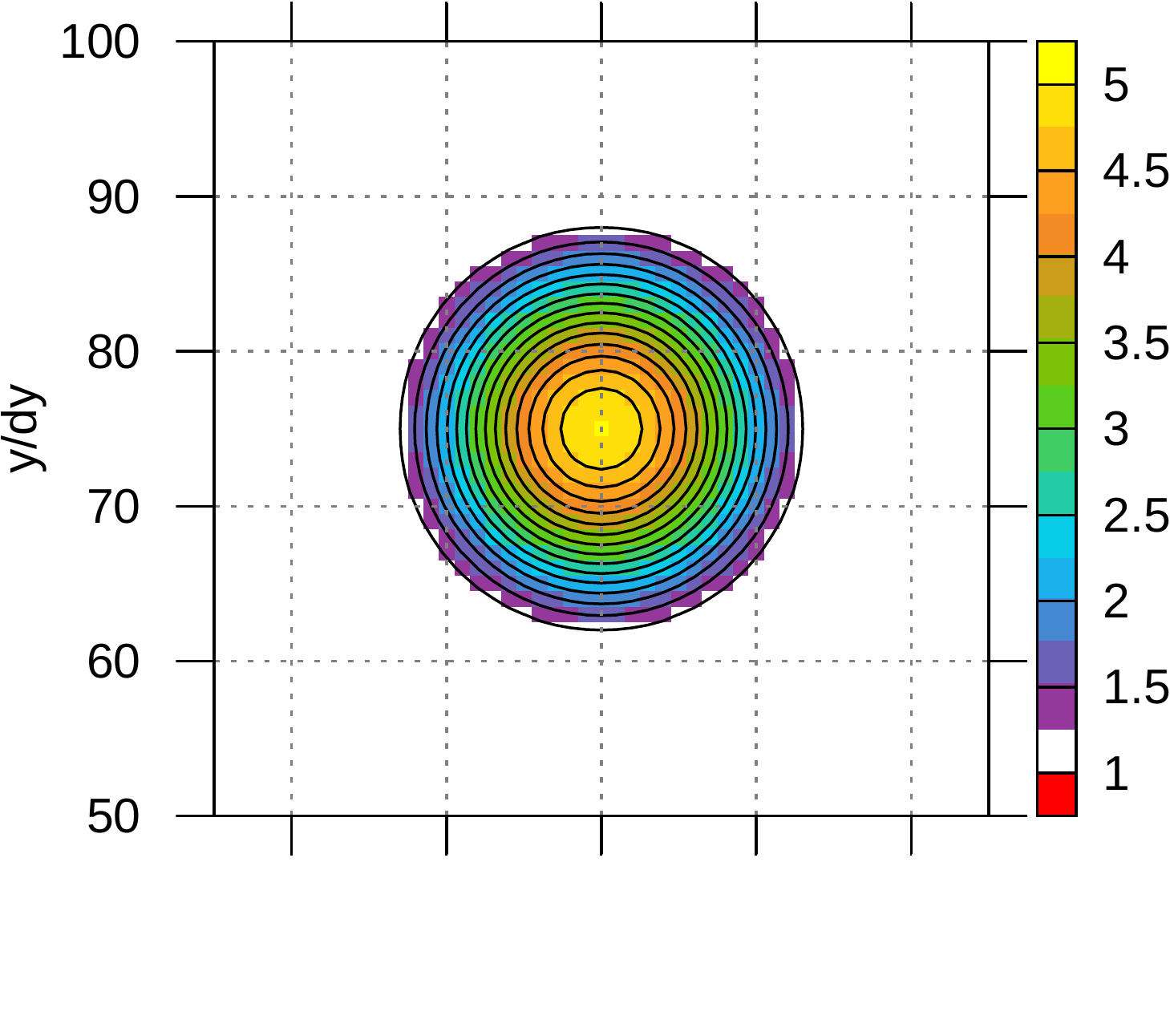}
     \label{fig:cone_init}
   }
   \\
   \vspace{-1.725cm}
   \subfloat[]{
     \pgfimage[width=.36\textwidth]{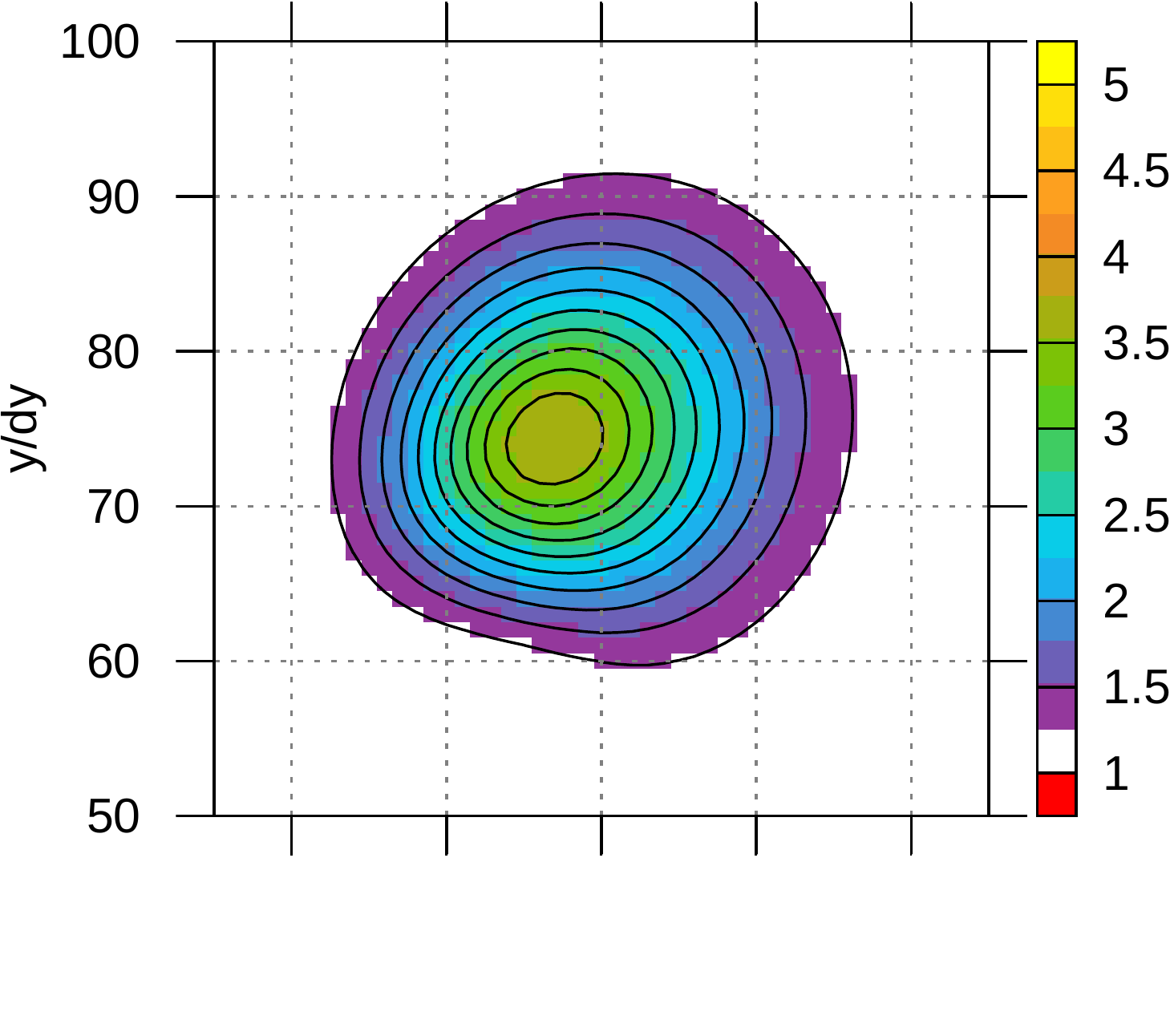}
     \label{fig:cone_fct} 
   }
   \\
   \vspace{-1.725cm}
   \subfloat[]{
     \pgfimage[width=.36\textwidth]{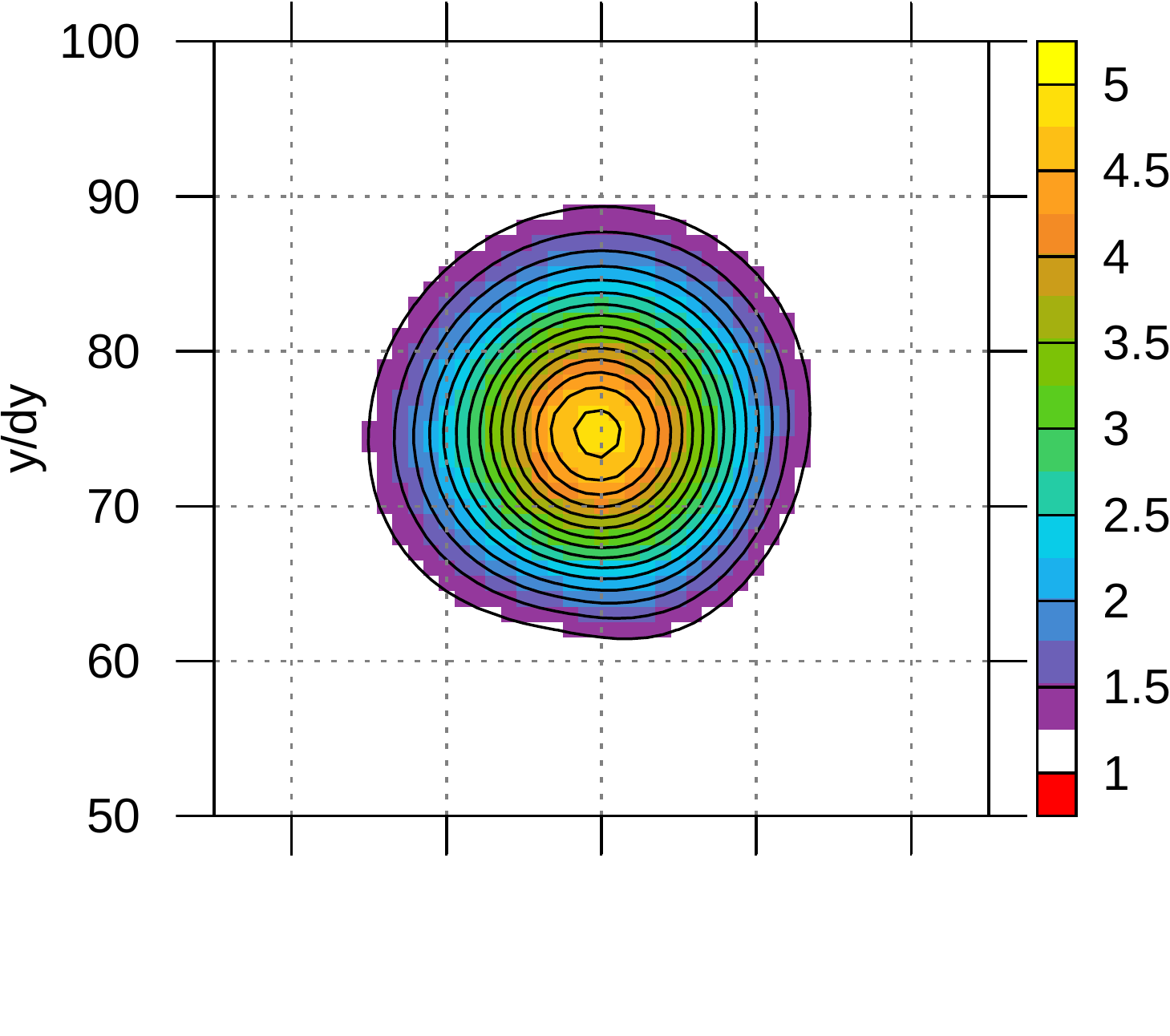}
     \label{fig:cone_acc2}
   }
   \\
   \vspace{-1.725cm}
   \subfloat[]{
     \pgfimage[width=.36\textwidth]{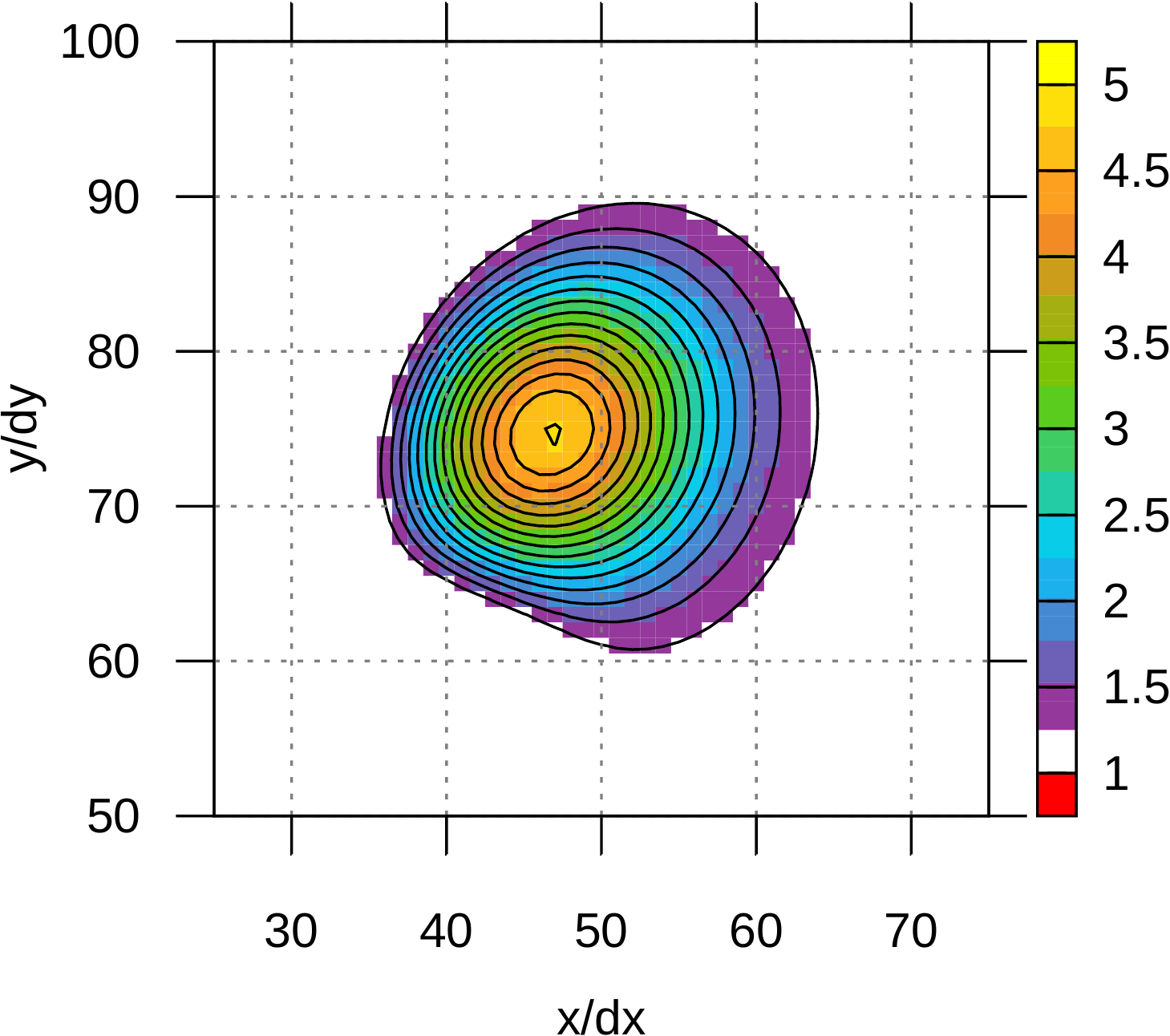}
     \label{fig:cone_default}
   }
   \centering
   \caption{\label{cone_panel} The results of Ex.~\ref{sec:cone}; only a quarter of the domain is shown: 
        \ref{fig:cone_init} shows initial condition of Ex.~\ref{sec:cone},
        \ref{fig:cone_fct} results for basic MPDATA with \textbf{fct}, 
        \ref{fig:cone_acc2} for MPDATA with three passes with \textbf{fct} and \textbf{tot}
        and \ref{fig:cone_default} for the default setting of \emph{libmpdata++} (\textbf{iga} and \textbf{fct}).
   }
   \end{figure}

  The initial condition is displayed in Fig.~\ref{fig:cone_init}, and 
    the results after total integration time are shown in Fig.~\ref{fig:cone_fct}--\ref{fig:cone_default}.
  All plots are centred around cone's initial location 
    and show only a quarter of the computational domain.
  The isolines of the advected cone are plotted with 0.25 interval. 
  The results in Figs.~\ref{fig:cone_fct} and \ref{fig:cone_acc2} were
    obtained with the \textbf{fct} and the three-pass \textbf{tot}+\textbf{fct} MPDATA, respectively.
  They match those presented by
    \citet[][Fig.~1 therein]{Smolarkiewicz_and_Margolin_1998}
    and \citet[][Fig.~4 therein]{Smolarkiewicz_and_Szmelter_2005}.
  Figure~\ref{fig:cone_default} shows test result for the default setting of \emph{libmpdata++}.

\FloatBarrier

  \subsection{\label{sec:sphere}Example: revolving sphere in 3D}
  
  This example extends Ex.~\ref{sec:cone} to three spatial dimensions.
  It exemplifies how to specify a three-dimensional set-up using \emph{libmpdata++}.
  Furthermore, the option is described for saving the simulation results to \emph{HDF5} files with
    \emph{XDMF} annotations.
   
  The setup follows \citet{Smolarkiewicz_1984}:
    the domain size is $40\Delta x \times 40\Delta y \times 40\Delta z$,
    with uniform grid spacing $\Delta x = \Delta y = \Delta z = 2.5$.
  The initial condition is a sphere of radius $7 \Delta x$ centred around the
    point $(x_0,~y_0,~z_0) = (20\Delta x -7 \cdot 6^{-1/2}\Delta x,~20\Delta y - 7 \cdot 6^{-1/2} \Delta
    y,~20\Delta z + 14 \cdot 6^{-1/2} \Delta z)$
    with density linearly varying from 4 at the centre to 0 at the edge.
  The sphere is rotating with constant angular velocity $\vec{\Omega} = \omega/\sqrt{3}(1,~1,~1)$ 
    of magnitude $\omega = 0.1$.
  The components of the advecting velocity field are 
    $(u, v, w) = (-\Omega_z (y - y_c) + \Omega_y (z - z_c),\;
    \Omega_z (x - x_c) - \Omega_x (z - z_c),\; -\Omega_y (x - x_c) + \Omega_x(y -
    y_c))$, where the coordinates of the rotation centre are 
    $(x_c, y_c, z_c) = (20\Delta x,\; 20\Delta y,\; 20 \Delta z)$.
  One full revolution takes 314 time-steps, and the test lasts for five revolutions. 

  \begin{Listing}
  \fvset{gobble=2}%
  \renewcommand*\FancyVerbStartString{\PY{c+c1}{//\PYZlt{}listing\PYZhy{}1\PYZgt{}}}%
  \renewcommand*\FancyVerbStopString{\PY{c+c1}{//\PYZlt{}/listing\PYZhy{}1\PYZgt{}}}%
  \input{rotating_sphere.cpp.tex}
    \caption{\label{lst:sphere1}
      Compile time parameter setting for the revolving-sphere test.
    }
  \end{Listing}

  Specifying the 3D setup with the \emph{libmpdata++} programming interface calls 
     starts by setting the \prog{n\_dims} field to 3, List.~\ref{lst:sphere1}.
  Listing~\ref{lst:sphere2} shows the choice of recommended three dimensional output handler
    \prog{hdf5\_xdmf}.
  This results in output consisting of \emph{HDF5}
    files with \emph{XDMF}
    annotation that can be viewed, for example, with the \emph{Paraview}
    visualisation software.
  This output is saved in a directory specified by the \prog{outdir} field of
    the run-time parameters, see List.~\ref{lst:sphere3}.
  
  \begin{Listing}
  \fvset{gobble=2}%
  \renewcommand*\FancyVerbStartString{\PY{c+c1}{//\PYZlt{}listing\PYZhy{}2\PYZgt{}}}%
  \renewcommand*\FancyVerbStopString{\PY{c+c1}{//\PYZlt{}/listing\PYZhy{}2\PYZgt{}}}%

\input{rotating_sphere.cpp.tex}    \caption{\label{lst:sphere2}
      Alias declaration of an output mechanism for the revolving-sphere test.
    }
  \end{Listing}

  \begin{Listing}
  \fvset{gobble=2}%
  \renewcommand*\FancyVerbStartString{\PY{c+c1}{//\PYZlt{}listing\PYZhy{}3\PYZgt{}}}%
  \renewcommand*\FancyVerbStopString{\PY{c+c1}{//\PYZlt{}/listing\PYZhy{}3\PYZgt{}}}%

\input{rotating_sphere.cpp.tex}    \caption{\label{lst:sphere3}
      Run-time parameters field specifying output directory for the revolving-sphere test.
    }
  \end{Listing}

\begin{figure}[h!]
   \centering
   \subfloat[]{
     \pgfimage[width=.295\textwidth]{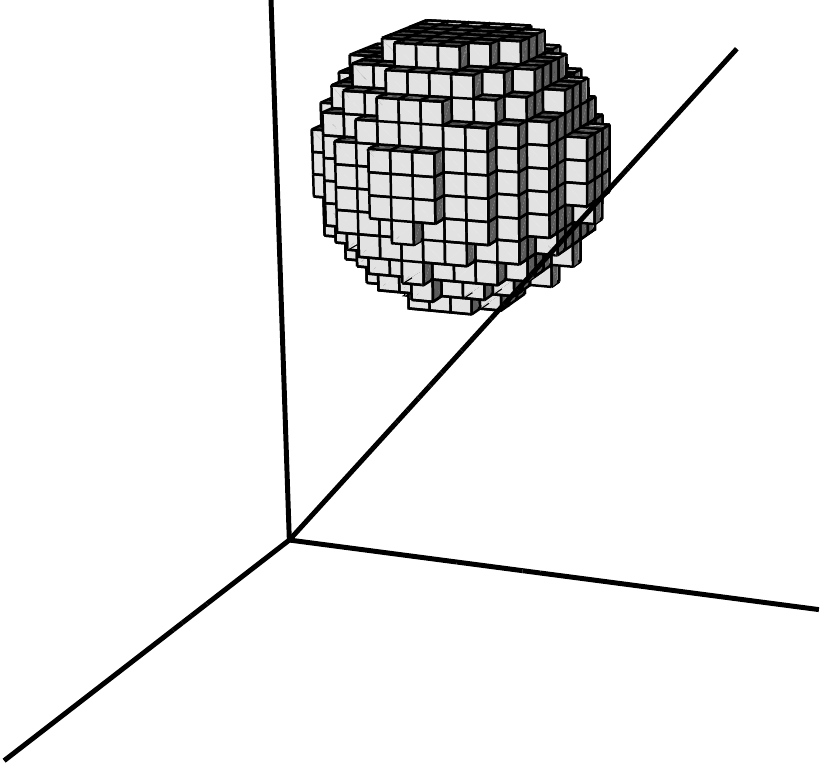}
     \label{fig:rot_sph_init}
   }
   \\
   \vspace{-1.2cm}
   \subfloat[]{
     \pgfimage[width=.295\textwidth]{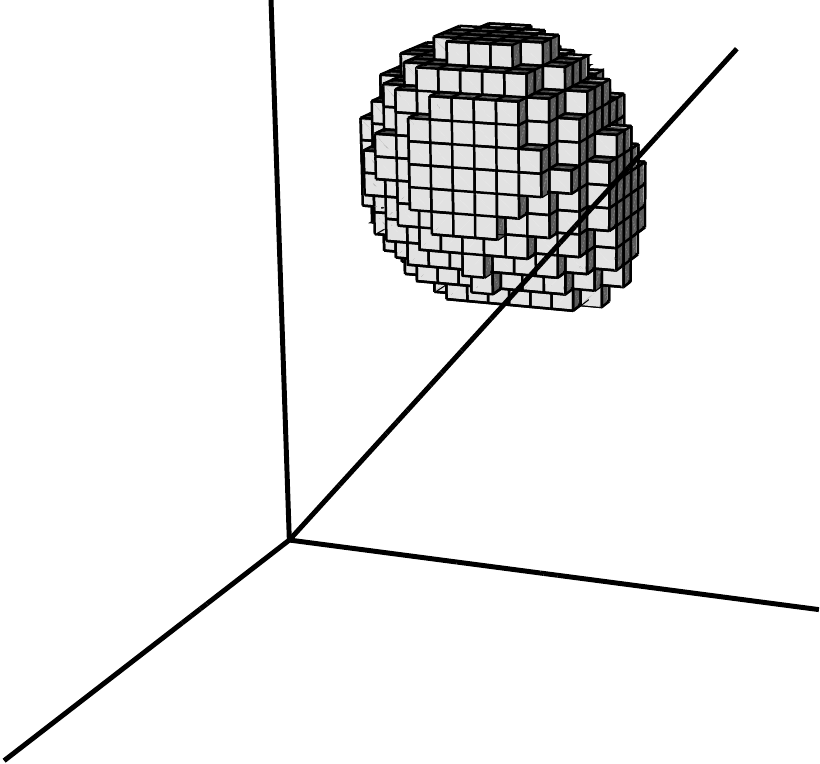}
     \label{fig:rot_sph_mpd4}
   }
   \\
   \vspace{-1.2cm}
   \subfloat[]{
     \pgfimage[width=.295\textwidth]{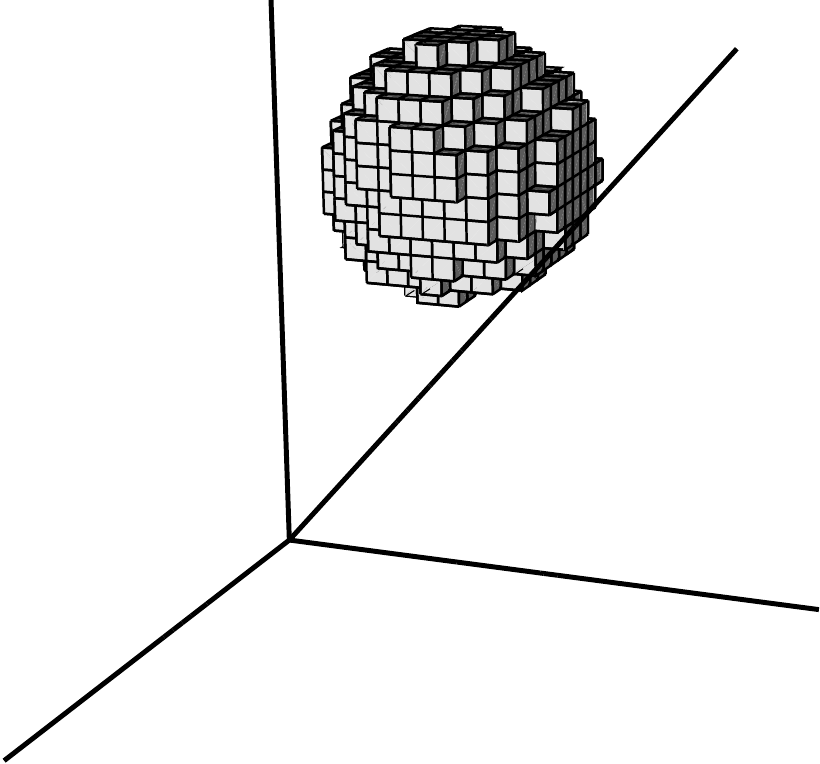}
     \label{fig:rot_sph_mpd4_tot}
   }
   \centering
   \caption{The results of Ex.~\ref{sec:sphere}; only a part of the domain is shown. 
           \ref{fig:rot_sph_init} shows initial condition, 
           \ref{fig:rot_sph_mpd4} results for the four-pass MPDATA, 
           \ref{fig:rot_sph_mpd4_tot} results for third-order-accurate variant with four passes.
   }
   \end{figure}

  Figure~\ref{fig:rot_sph_init} shows the initial condition,  
    Fig.~\ref{fig:rot_sph_mpd4} and \ref{fig:rot_sph_mpd4_tot} show the results
    after five revolutions
    for the four-pass MPDATA without and with \textbf{tot}.
  Only a portion of the computational domain centred at the sphere is shown.
  The black line crossing the XY plane is the axis of rotation. 
  The grey volume is composed of dual-grid cells (section \ref{sec:grid}) encompassing 
    data points with cell-mean values of density greater or equal $0.5$.

  The solution in Fig.~\ref{fig:rot_sph_mpd4} is deformed, 
    but this deformation is significantly reduced when the third-order-accurate variant is set, 
    Fig.~\ref{fig:rot_sph_mpd4_tot}.
  Obtained results can be compared with those presented by 
    \citet[][Fig.~13-16]{Smolarkiewicz_1984}.

  \FloatBarrier

  \subsection{\label{sec:over_pole}Example: 2D advection on a sphere} 

  This subsection concludes homogeneous transport examples with a 2D solid-body
    rotation test on a spherical surface \citep{Williamson_And_Rasch_1989}.
  The purpose of this example is to present methods for setting up the simulations 
    in spherical coordinates.\footnote{
      The same method, used here to specify a Jacobian of coordinate
      transformation, can be applied to prescribe a variable-in-space fluid density.}

  Following \citet{Smolarkiewicz_and_Rasch_1991} only the case when the initial field
    rotates over the poles is presented.
  The initial condition is a cone centred around the point $(3\pi / 2, 0)$ with 
    height and radius equal to 1 and $7\pi / 64$, respectively.
  The wind field is given by
    \begin{equation}
      \begin{aligned}
        u &= - U  \sin{\phi} \cos{\lambda}~, \\
        v &= ~~  U  \sin{\lambda}~,
      \end{aligned}
    \end{equation}
    where $\lambda$ and $\phi$ denote respectively longitude and latitude, 
    and $U = \pi / 128$.
  The computational domain $[0, 2\pi] \times [-\pi/2, \pi/2]$  
    is resolved with $128 \times 64$ grid increments $\Delta \lambda = \Delta \phi$
    and is shifted by $0.5 \Delta \phi$ so that there are no data points on the poles. 
  The test is run for 5120 time-steps corresponding to one revolution around the globe.

  The advection equation in spherical
    coordinates has the form of the generalised transport eq.~(\ref{gte}) with 
    the Jacobian of coordinate transformation
    \begin{equation}
      G = \cos{\phi}.
    \end{equation}
  In order to solve the generalised transport equation with $G \not\equiv  1$
    the \prog{nug} option has to be set, see List.~\ref{lst:pole1}.

  \begin{Listing}
  \fvset{gobble=4}%
  \renewcommand*\FancyVerbStartString{\PY{c+c1}{//\PYZlt{}listing\PYZhy{}1\PYZgt{}}}%
  \renewcommand*\FancyVerbStopString{\PY{c+c1}{//\PYZlt{}/listing\PYZhy{}1\PYZgt{}}}%
  \input{over_the_pole.cpp.tex}
    \caption{\label{lst:pole1}
      Compile-time parameter field for Ex.~\ref{sec:over_pole}.
    }
  \end{Listing}

  \begin{Listing}
  \fvset{gobble=2}%
  \renewcommand*\FancyVerbStartString{\PY{c+c1}{//\PYZlt{}listing\PYZhy{}2\PYZgt{}}}%
  \renewcommand*\FancyVerbStopString{\PY{c+c1}{//\PYZlt{}/listing\PYZhy{}2\PYZgt{}}}%

\input{over_the_pole.cpp.tex}    \caption{\label{lst:pole2}
      Concurrency handler for Ex.~\ref{sec:over_pole}.
    }
  \end{Listing}

  \begin{Listing}
  \fvset{gobble=2}%
  \renewcommand*\FancyVerbStartString{\PY{c+c1}{//\PYZlt{}listing\PYZhy{}3\PYZgt{}}}%
  \renewcommand*\FancyVerbStopString{\PY{c+c1}{//\PYZlt{}/listing\PYZhy{}3\PYZgt{}}}%

\input{over_the_pole.cpp.tex}    \caption{\label{lst:pole3}The Jacobian setting for Ex.~\ref{sec:over_pole}.}
  \end{Listing}

  Boundary conditions in this example incorporate principles of
    differential geometry \citep[cf. chapter XIV in][]{Maurin_1980}
    in the classical spherical latitude-longitude framework
    \citep{Szmelter_and_Smolarkiewicz_2010}.
  They are cyclic (\prog{bcond::cyclic}) in the zonal direction, whereas in the
     meridional direction they represent two degenerated charts (of the atlas composed of three)
     defining differentiation of dependent variables in vicinity of the poles
     (\prog{bcond::polar}), List.~\ref{lst:pole2}.
  The setting of $G$ is done using the \prog{g\_factor()} accessor method 
    as shown in List.~\ref{lst:pole3}; 
    note the shift in latitude by $\Delta \phi / 2$. 

   \begin{figure}[h!]
   \centering
   \subfloat[]{
     \pgfimage[height=.15\textheight]{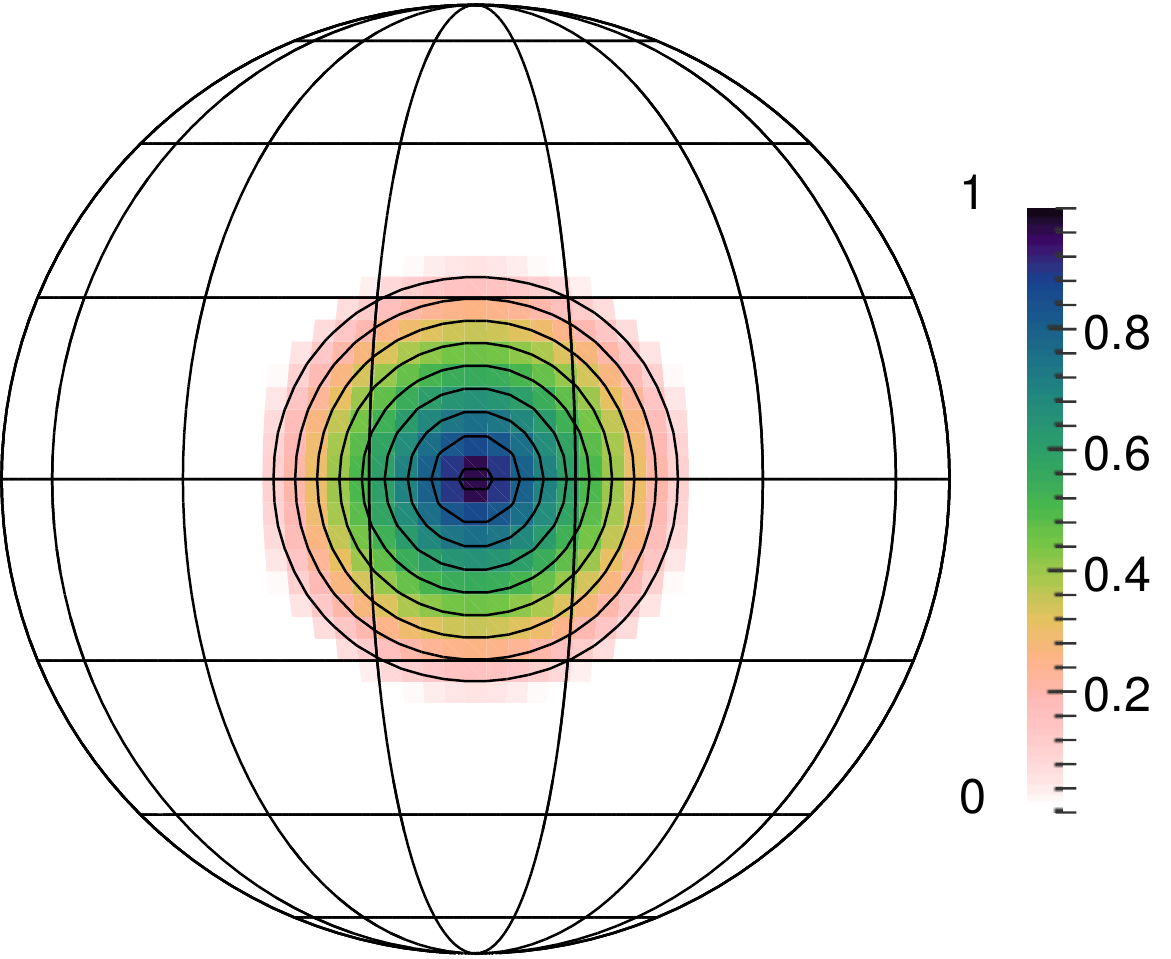}
     \label{fig:pole_ini} 
   }
   \\
   \vspace{-0.93cm}
    \subfloat[]{
      \pgfimage[height=.15\textheight]{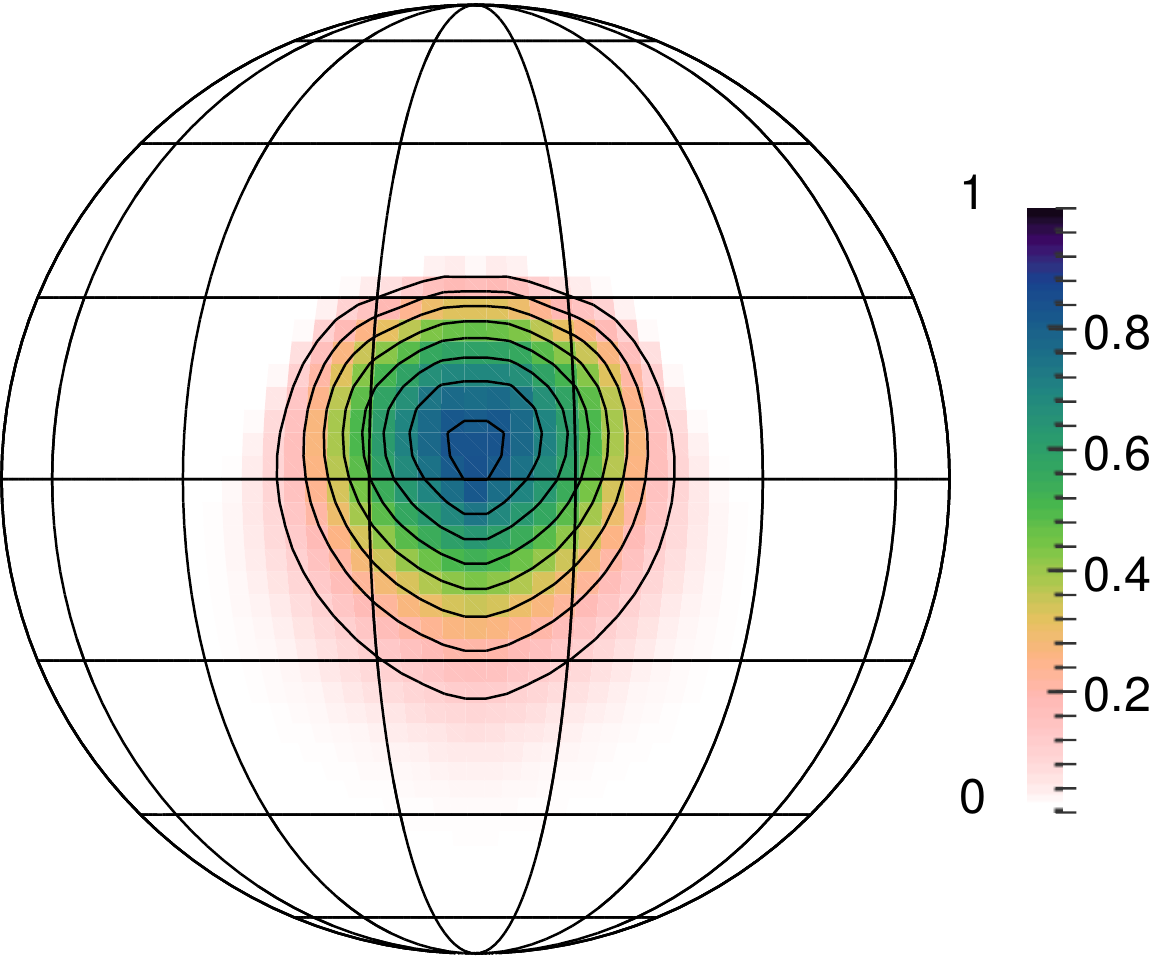}
      \label{fig:pole_def} 
   }
   \\
   \vspace{-0.93cm}
    \subfloat[]{
      \pgfimage[height=.15\textheight]{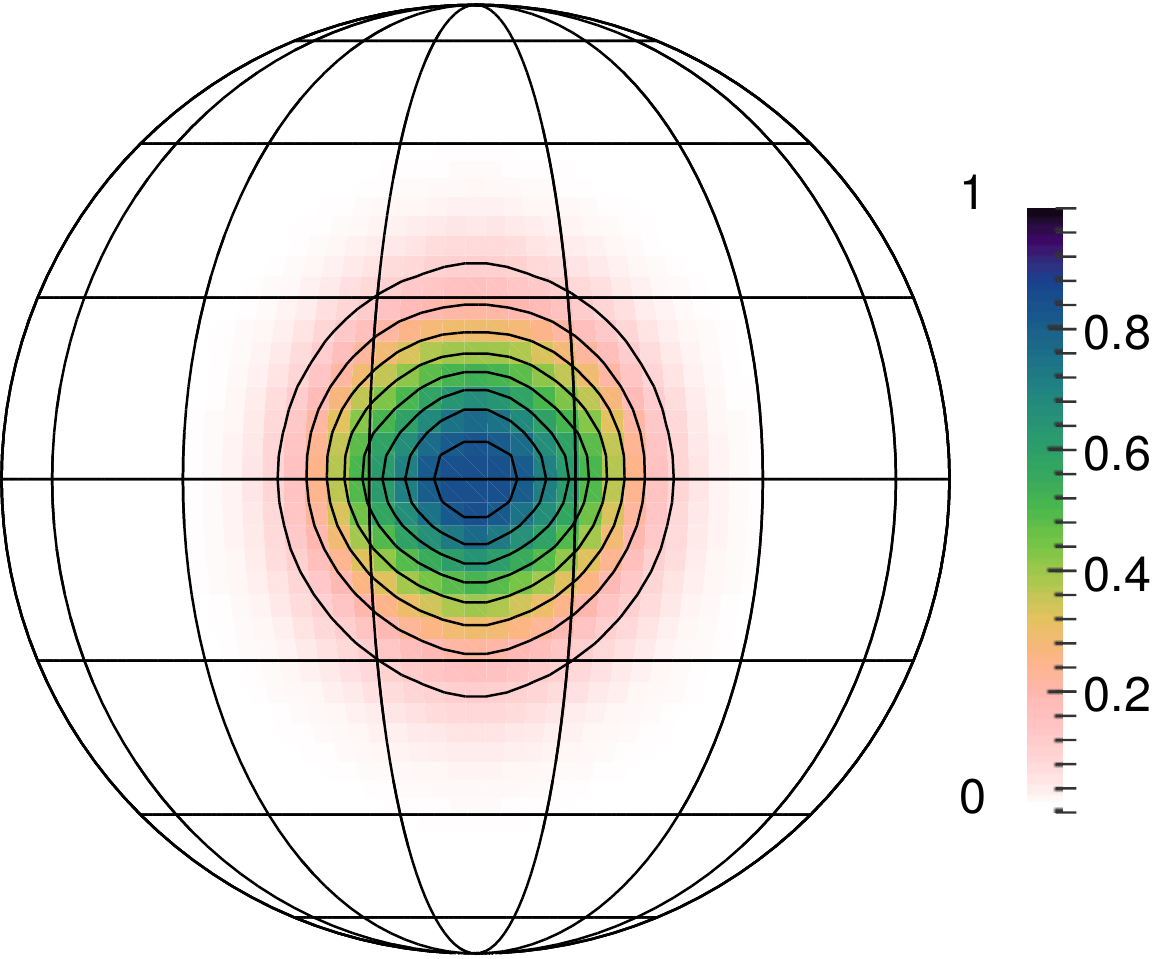}
      \label{fig:pole_bst} 
   }
   \centering
   \caption{\label{pole_panel} The results of Ex.~\ref{sec:over_pole}:
           \ref{fig:pole_ini} shows the initial condition, 
           \ref{fig:pole_def} results for the default \emph{libmpdata++} options and
           \ref{fig:pole_bst} results for the three-pass MPDATA with \textbf{fct} and \textbf{tot}.
   }
   \end{figure}
  \FloatBarrier
   The initial condition for the test is plotted in Fig.~\ref{fig:pole_ini}, 
    whereas the results are displayed in Fig.~\ref{fig:pole_def} and \ref{fig:pole_bst}. 
  All figures use orthographic projection, with the perspective centred at
    the initial condition (the true solution), with the contour interval 0.1.
  Figure~\ref{fig:pole_def} shows the result for the default \emph{libmpdata++} options.
  There is a visible deformation in the direction of motion, 
    consistent with earlier Cartesian rotational tests.
  The result in Fig.~\ref{fig:pole_bst},
    obtained using three passes of MPDATA with \textbf{fct} and \textbf{tot}, 
    shows reduced deformation
    and reproduces Fig.~6 in \citet{Smolarkiewicz_and_Rasch_1991}.

  \section{Inhomogeneous advective transport}\label{sec:adv+rhs}

  \subsection{Implemented algorithms}

  As of the current release, \emph{libmpdata++} provides three ways of handling
    source terms in the inhomogeneous extension of eq.~(\ref{gte_final}) 
  \begin{equation}
    \partial_t{\psi} + \frac{1}{G} \nabla \cdot (G\vec{u} \psi) = R~.
    \label{gte_final_rhs}
  \end{equation}
  The available time integration schemes include: 
    the two variants of the first-order-accurate Euler-forward scheme 
    (hereafter referred to as \textbf{euler\_a}  and \textbf{euler\_b}); 
    and the second-order-accurate Crank-Nicolson scheme (\textbf{trapez}).
  The Euler schemes are implemented to account for parameterised forcings (e.g., 
    due to cloud microphysics), whereas
    the Crank-Nicolson scheme is standard for basic dynamics 
    (e.g., pressure gradient, Coriolis and  buoyancy forces).
  In both Euler schemes, while calculating the solver state at the time level \emph{n+1},
    the right-hand-side at the time level \emph{n} is only needed.
  In the \textbf{euler\_a} option (eq.~\ref{euler_a}), the source 
    terms are computed and applied standardly after the advection
  \begin{equation}
    \psi^{n+1} = ADV(\psi^{n}) + \Delta t R^{n}~.
    \label{euler_a}
  \end{equation}
  In the \textbf{euler\_b} option (eq.~\ref{euler_b}), the source terms are computed and applied  
    \citep[arguably in the Lagrangian spirit; section 3.2 in][]{Smolarkiewicz_and_Szmelter_2009} 
    before the advection
  \begin{equation}
    \psi^{n+1} = ADV(\psi^{n} + \Delta t R^{n})~.
    \label{euler_b}
  \end{equation}
  In the \textbf{trapez} option (eq.~\ref{trapez}), half of the sources terms are computed and applied 
    as in the \textbf{euler\_a} and half as in the \textbf{euler\_b} 
    \citep[arguably in the spirit of the Lagrangian trapezoidal rule; section 2.2 in][]{Smolarkiewicz_and_Szmelter_2009}
  \begin{equation}
    \psi^{n+1} = ADV(\psi^{n} + 0.5 \Delta t R^{n}) + 0.5 \Delta t R^{n+1}~.
    \label{trapez}
  \end{equation} 

  \subsection{Library interface}

  The logic for handling source terms is implemented in the \textbf{mpdata\_rhs}
    solver that inherits from the \textbf{mpdata} class, Fig.~\ref{fig:inherit}.
  Consequently, all options discussed in the preceding section apply.
  The choice of the source-term integration scheme is controlled by the
    \textbf{rhs\_scheme} compile-time parameter with the valid values of 
    \textbf{euler\_a}, \textbf{euler\_b} or \textbf{trapez}.

  The user is expected to provide information on the source terms by 
    defining a derived class of \textbf{mpdata\_rhs} with the
    \textbf{update\_rhs()} method overloaded.
  The \textbf{update\_rhs()} signature is given in List.~\ref{lst:update_rhs},
    whereas the usage example is given in subsection~\ref{sec:harmosc}.
  The method is called by the solver with the following arguments:
    \begin{itemize}
      \item{a vector of arrays \textbf{rhs} storing the source terms for each 
            equation of the integrated system,}
      \item{a floating-point value \textbf{dt} with the time-step value,}
      \item{an integer number \textbf{at} indicating if the source terms
	    are to be computed at time level \emph{n} (if at=0) or \emph{n+1} (if at=1).}
    \end{itemize}

  \begin{Listing}
  \fvset{gobble=6}%
  \renewcommand*\FancyVerbStartString{\PY{c+c1}{//\PYZlt{}listing\PYZhy{}1\PYZgt{}}}%
  \renewcommand*\FancyVerbStopString{\PY{c+c1}{//\PYZlt{}/listing\PYZhy{}1\PYZgt{}}}%
  \input{mpdata_rhs.cpp.tex}
    \caption{\label{lst:update_rhs}Signature of the method used for defining source-terms.}
  \end{Listing}

  Calculation of forcings at the \emph{n+1} time level
    is needed if \textbf{rhs\_scheme=trapez} option is chosen.
  The case of \textbf{at} equal zero is used in the Euler schemes and in the very first time step 
    when using the \textbf{trapez} option (i.e., once per simulation).
  When the \textbf{trapez} option is used, the \textbf{dt} passed to the \textbf{update\_rhs} method
    equals half of the original time-step.

  The \textbf{update\_rhs} method is expected to first call \textbf{parent\_t::update\_rhs()}
    to zero out the source and sink terms stored in \textbf{rhs}.
  Later, it is expected to calculate the \textbf{rhs} terms in a given time-step by 
    summing all sources and sinks and ''augment assign'' them to the \textbf{rhs} field
    (e.g., using the += operator).

  All elements of the \textbf{rhs} vector corresponding to subsequent equations in the system 
    are expected to be modified in a single \textbf{update\_rhs()} call.

  \subsection{\label{sec:harmosc}Example: translating oscillator}

  The purpose of this example is to show how to include rhs terms in \emph{libmpdata++},
    by creating a user-defined class out of the library tree.

  A system of two one-dimensional advection equations 
  \begin{equation}
    \begin{aligned}
      \partial_t \psi + \partial_x (u_o \psi) &= ~~ \omega \phi \\
      \partial_t \phi + \partial_x (u_o \phi) &= -\omega \psi
    \label{harm1}
    \end{aligned}
  \end{equation}
    represents a harmonic oscillator translating with $u_o = {\rm const.}$;
    see section 4.1 in \citet{Smolarkiewicz_2006} for a discussion.\footnote{The 
      implicit manner of prescribing forcings, 
      similar to the one presented herein,
      is an archetype for integrating Coriolis force in \citet{Prusa_et_al_2008}.}
  Applying the trapezoidal rule to integrate the PDE system (\ref{harm1}) leads to following
    system of coupled implicit algebraic equations  
  
  \begin{equation}
    \begin{aligned}
      \psi_i^{n+1} &= \psi_i^{*} + 0.5 \: \Delta t \: \omega \: \phi_i^{n+1} \\       
      \phi_i^{n+1} &= \phi_i^{*} - 0.5 \: \Delta t \: \omega \: \psi_i^{n+1}~, 
    \label{osc_disc}      
    \end{aligned}
  \end{equation} 
  where $\psi_i^{*}$ and $\phi_i^{*}$ stand for
  
  \begin{eqnarray}
      \psi_i^{*} &=& MPDATA(\psi_i^n + 0.5 \: \Delta t \: \omega \: \phi_i^n, C) \\
      \phi_i^{*} &=& MPDATA(\phi_i^n - 0.5 \: \Delta t \: \omega \: \phi_i^n, C)~.
    \label{trapez_osc}
  \end{eqnarray}

  Substituting in (\ref{osc_disc}) $\psi_i^{n+1}$ with $\phi_i^{n+1}$ and vice versa
    and then regrouping leads to:
  \begin{equation}
    \begin{aligned}
      \psi_i^{n+1} &= \frac{\psi_i^{*} + 0.5 \: \Delta t \: \omega \: \phi_i^{*}}
			  {1+(0.5 \: \Delta t \: \omega)^2} \\
      \phi_i^{n+1} &= \frac{\phi_i^{*} - 0.5 \: \Delta t \: \omega \: \psi_i^{*}}
			  {1+(0.5 \: \Delta t \: \omega)^2}~.
    \label{final_osc}
     \end{aligned}
  \end{equation} 

  Implementation of forcing terms prescribed in eq.~(\ref{final_osc}) 
    is presented in List.~\ref{lst:rhs}.
  A new solver \textbf{coupled\_harmosc} is defined
    by inheriting from the \textbf{mpdata\_rhs} class.
  A member field \textbf{omega} is defined to store the frequency of oscillations.

  The rhs terms  are defined for both variables,
    \textbf{ix::psi} and \textbf{ix::phi}
    within the \textbf{update\_rhs()} method.
  The method implements both implicit and explicit formul\ae,
    the two cases are switched by the \textbf{at} argument.
  Defining forcings for both \emph{n} and \emph{n+1} cases allows to use
    this class with both \textbf{euler} and \textbf{trapez} options.
  The current state of the model is obtained via a call to the \textbf{state()} method.
  Note how the formul\ae~defined in \textbf{update\_rhs()} in case(1) loosely resemble
    the mathematical notation presented in eq.~(\ref{final_osc}).
  The $0.5$ is absent because the $\Delta t$ passed as argument in \textbf{trapez}
    option is already divided by $2$.

  \begin{Listing}[th!]
  \fvset{gobble=0}%
  \renewcommand*\FancyVerbStartString{\PY{c+c1}{//\PYZlt{}listing\PYZhy{}1\PYZgt{}}}%
  \renewcommand*\FancyVerbStopString{\PY{c+c1}{//\PYZlt{}/listing\PYZhy{}1\PYZgt{}}}%
  \input{example_7.cpp}
    \caption{\label{lst:rhs}
      Definition of the solver used in Ex.~\ref{sec:harmosc}.
    }
    \vspace{-1em}
  \end{Listing}

  Next, the \textbf{rt\_params\_t} structure is augmented (by inheriting from
    parent's \textbf{rt\_params\_t}) with the \textbf{omega}. 
  Last, the \textbf{coupled\_harmosc} constructor is defined.
  Within it, the choice of \textbf{omega} 
    is handled by copying its value from 
    the \textbf{p.omega} to \textbf{omega} member field and then checking if the 
    user has altered the default value of~0.

  \begin{Listing}
  \fvset{gobble=2}%
  \renewcommand*\FancyVerbStartString{\PY{c+c1}{//\PYZlt{}listing\PYZhy{}1\PYZgt{}}}%
  \renewcommand*\FancyVerbStopString{\PY{c+c1}{//\PYZlt{}/listing\PYZhy{}1\PYZgt{}}}%
  \input{example_71.cpp}
    \caption{\label{lst:ct_params_osc}
      Compile-time parameter structure for Ex.~~\ref{sec:harmosc}.
    }
  \end{Listing}

  For inhomogeneous transport the \textbf{rhs\_scheme} within the \textbf{ct\_params\_t}
    structure needs to be defined.
  In this example it is set to \textbf{trapez}, List.~\ref{lst:ct_params_osc}.
  MPDATA advection scheme options are set to default by inheriting from the 
    \textbf{ct\_params\_t\_default} structure.
  The structure \textbf{ix} allows to call advected variables by their labels, 
    \textbf{phi} and \textbf{psi}, rather than integer numbers.
  Last, when defining the \textbf{rt\_params\_t} structure a value is assigned to the 
    member field \textbf{p.omega}, see List.~\ref{lst:rt_params_osc}.

  \begin{Listing}
  \fvset{gobble=2}%
  \renewcommand*\FancyVerbStartString{\PY{c+c1}{//\PYZlt{}listing\PYZhy{}2\PYZgt{}}}%
  \renewcommand*\FancyVerbStopString{\PY{c+c1}{//\PYZlt{}/listing\PYZhy{}2\PYZgt{}}}%

\input{example_71.cpp}    \caption{\label{lst:rt_params_osc}
      Run-time parameter structure for Ex.~\ref{sec:harmosc}.
    }
  \end{Listing}

  In the present example, the initial condition for $\psi$ is defined as 
    $\psi(x) = 0.5[1+cos(2\pi x/ 100)]$ for $x \in (50, 150)$ and zero elsewhere.
  The initial condition for $\phi$ is set to zero.

\begin{figure}[H]
    \pgfimage[width=.45\textwidth]{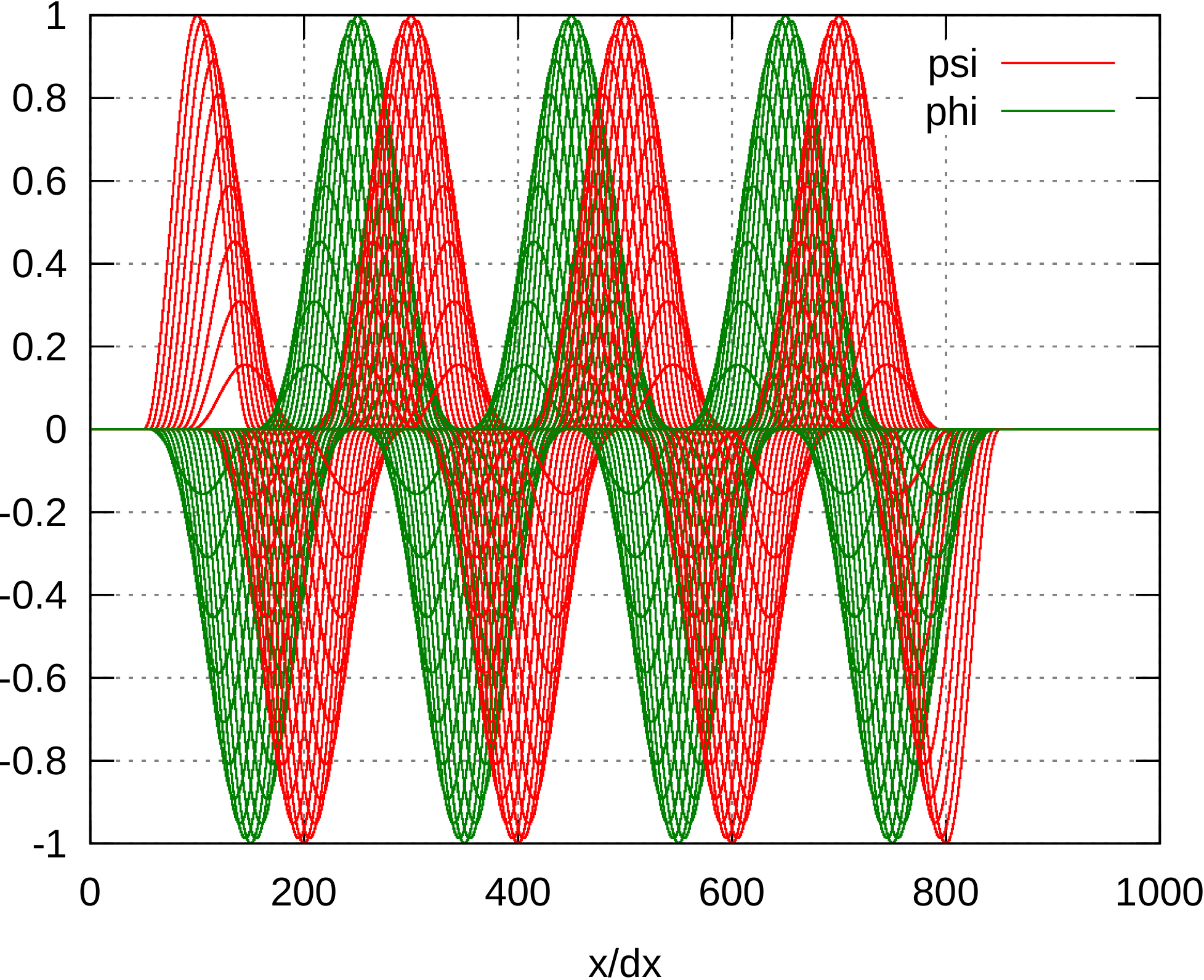}
    \caption{\label{fig:harm_osc} Simulation results for Ex.~\ref{sec:harmosc}.}
  \end{figure}

  The result of 1400~s of simulated time are shown in Fig.~\ref{fig:harm_osc}.
  Note that the solutions for both $\psi$ and $\phi$ remain in phase
    and feature no substantial amplitude error.
  This contrasts with calculations using Euler-forward schemes (not shown).


  \section{Transport with prognosed velocity}\label{sec:adv+rhs+vel}

  \subsection{Implemented algorithms}

  Whenever the velocity field changes in time, the second-order accuracy of the solution 
    at \emph{n+1} requires estimate of the advector at \emph{n+1/2}.
  This is provided by linear extrapolations from \emph{n} and \emph{n-1} values 
    \citep{Smolarkiewicz_and_Margolin_1998}.
  Furthermore, when the velocity is a dependent variable of the model, eq.~(\ref{gte_final_rhs})
    embodies equations of motion.
  Then the velocity (or momentum) components are treated as 
    advected scalars (i.e. advectees) and are predicted at the centres 
    of the dual-grid cells, Fig.~\ref{fig:grid}. 
  The advector field is then interpolated linearly to the centres of the cell walls.

  \subsection{\label{sec:vip_lib}Library interface}

  The algorithms for interpolating in space and extrapolating in time the advector 
    field from the model variables are defined in the \prog{mpdata\_rhs\_vip} class 
    and all user-created solvers with time-varying velocity must 
    inherit from this class.

  The transported fields may represent either velocity or momenta.
  In the latter case  
    the prognosed velocity components are calculated as ratios of two advectee fields 
    (e.g. momentum components and density).
  The index of the advectee that forms the common
    denominator for all velocity components should be assigned to \prog{vip\_den}. 
  The \prog{vip\_i}, \prog{vip\_j} and \prog{vip\_k} store
    the index of the advected fields appearing in the numerators for each velocity component.
  These velocity components are then used to calculate the advector field. 
  In case when the velocity components are model variables 
    (as in the example of section~\ref{sec:bombel}),
    the common denominator is redundant and value -1 should be assigned to \textbf{vip\_den}.

  For systems where numerators and denominators can uniformly approach zeros,
    the \prog{vip\_eps} value is defined to prevent divisions by zero.
  Then, if the denominator at a given grid-point is less than the \prog{vip\_eps},
    the resulting advector is set to zero therein.
  The default value of \prog{vip\_eps}
    depends on the precision chosen for the simulation. 
  Namely, it is set to be the smallest number that added to 1 produces
    a result that is not equal~1.

  The \prog{vip\_i}, \prog{vip\_j}, \prog{vip\_k} and \prog{vip\_den}  
    are expected to be members of the compile-time parameters structure 
    \prog{ct\_params\_t} of the \prog{mpdata\_rhs\_vip} class.
  The \prog{vip\_eps} value is a run-time parameter.

  As of the current release, the prognosed-velocity features of \emph{libmpdata++}
    are implemented for constant $G\equiv1$ only.

  \subsection{\label{sec:shw1}Example: 1D shallow-water system}

  The aim of this example is to show how to define simulations with prognosed
    velocity field.
  The necessary compile-time and run-time parameters as well as the user-defined class
    with source and sink terms are discussed.
  The obtained results are compared with the analytical solution and a published MPDATA benchmark.

  The idealised system of 1D inviscid shallow-water equations is considered,
    with both the surface friction and background rotation neglected.
  The simulated physical scenario is a slab-symmetric parabolic drop spreading under gravity;
    see \citet{Frei_1993} for a general context
    and \citet{Schar_and_Smolarkiewicz_1996} for bespoke analytical solutions. 
  The corresponding governing equations take the dimensionless form
  \noindent
  \begin{equation}
    \begin{aligned}
      \partial_t h + \partial_x (uh) &=& 0~,  \\
      \partial_t (uh) + \partial_x ( uuh) &=& - h \partial_x h~, 
    \label{shw1d}
    \end{aligned}
  \end{equation}
  \noindent
  where $h$ is a normalised depth of the fluid layer and $u$ is a normalised velocity.
  Following \citet{Schar_and_Smith_1993} the selected velocity scale is 
     $u_o = (gh_o)^{1/2}$ where $h_o$ is the initial height 
    of the drop and $g$ denotes the gravitational acceleration.
  The characteristic time-scale is $t_o = a/u_o$, where $a$ denotes the initial half-width of the drop.
  At the initial time a drop is confined to $|x| \le 1$ 
    and centred about $x=0$,
  \begin{equation}
    h(x,t=0) = 
       \begin{cases}
         1-x^2  ,& for \;\;\; |x| \le 1  \\
         0      ,& for \;\;\; |x| > 1~. 
       \end{cases}
    \label{shw1_init}
  \end{equation}
  \noindent
  The time-step is set to 0.01 and the grid spacing is set to 0.05.
  The crux of the test is a synchronous solution for 
    the depth and momentum near the drop edge that accurately diagnoses 
    the velocity.
  
  The definition of the rhs terms for Ex.~\ref{sec:shw1}
    is presented in List.~\ref{lst:shw1_1}.
  Only the method for calculating the forcing terms is shown; for the full
    out-of-the-library-tree definition of source-terms see List.~\ref{lst:rhs}.
  As in the List.~\ref{lst:rhs}, the definition in List.~\ref{lst:shw1_1} 
    attempts to follow the mathematical notation.
  Because of the use of the \prog{grad} function, the \prog{nabla} namespace 
    is included.

  \begin{Listing}[h!]
  \fvset{gobble=2}%
  \renewcommand*\FancyVerbStartString{\PY{c+c1}{//\PYZlt{}listing\PYZhy{}1\PYZgt{}}}%
  \renewcommand*\FancyVerbStopString{\PY{c+c1}{//\PYZlt{}/listing\PYZhy{}1\PYZgt{}}}%
  \input{shallow_water.cpp}
    \caption{\label{lst:shw1_1}Method for calculating source and sink terms for Ex.~\ref{sec:shw1}.}
  \end{Listing}

  \begin{Listing}[h!]
  \fvset{gobble=0}%
  \renewcommand*\FancyVerbStartString{\PY{c+c1}{//\PYZlt{}listing\PYZhy{}1\PYZgt{}}}%
  \renewcommand*\FancyVerbStopString{\PY{c+c1}{//\PYZlt{}/listing\PYZhy{}1\PYZgt{}}}%
  \input{shallow_water_params.cpp}
    \caption{\label{lst:shw1_2}Compile-time parameters for Ex.~\ref{sec:shw1}.}
  \end{Listing}

  Listing~\ref{lst:shw1_2} specifies the compile-time parameters structure.
  Because fluid flow in this example is divergent the \textbf{opts::dfl} 
    correction is enabled, cf. sec.~\ref{sec:dfl}.
  The  \prog{rhs\_scheme} is set to \prog{trapez}.\footnote{Because the equation for $h$ 
      is homogeneous, the momentum forcing at \emph{n+1} time level can be readily evaluated
      after advecting~$h$.}
  Within the \prog{ix} structure, the equation indices are assigned.
  Furthermore, the recipe for calculating the velocity is defined by 
    assigning the indices to \prog{vip\_i} and \prog{vip\_den}.
  Lack of the rhs terms 
    is specified by toggling n-th bit of the \textbf{hints\_norhs} field, 
    where n is the index of the homogeneous equation.
  This prevents the unnecessary summation of zeros.

  Listing~\ref{lst:shw1_3} shows the run-time parameters structure. 
  The value of gravitational acceleration \prog{p.g} is set to 1 to follow the dimensionless 
    notation of~(\ref{shw1d}), and 
    the \prog{vip\_eps} is set arbitrarily to $10^{-8}$.
 
  \begin{Listing}[h!]
  \fvset{gobble=2}%
  \renewcommand*\FancyVerbStartString{\PY{c+c1}{//\PYZlt{}listing\PYZhy{}2\PYZgt{}}}%
  \renewcommand*\FancyVerbStopString{\PY{c+c1}{//\PYZlt{}/listing\PYZhy{}2\PYZgt{}}}%

\input{shallow_water_params.cpp}    \caption{\label{lst:shw1_3}Run-time parameters for Ex.~\ref{sec:shw1}.}
  \end{Listing}
 
\begin{figure}[h!]
    \centering
    \subfloat[]{
      \pgfimage[width=.44\textwidth]{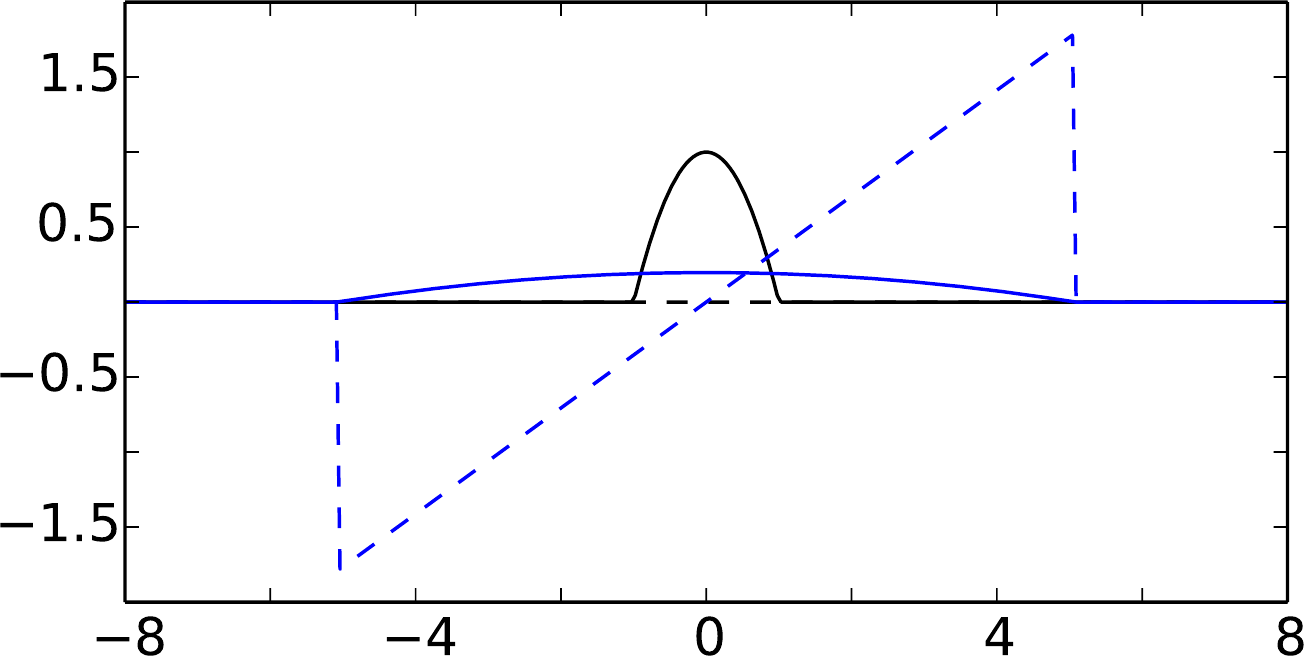}
      \label{fig:shw1_1}
    }
    \\
    \vspace{-0.7cm}
    \subfloat[]{
      \pgfimage[width=.44\textwidth]{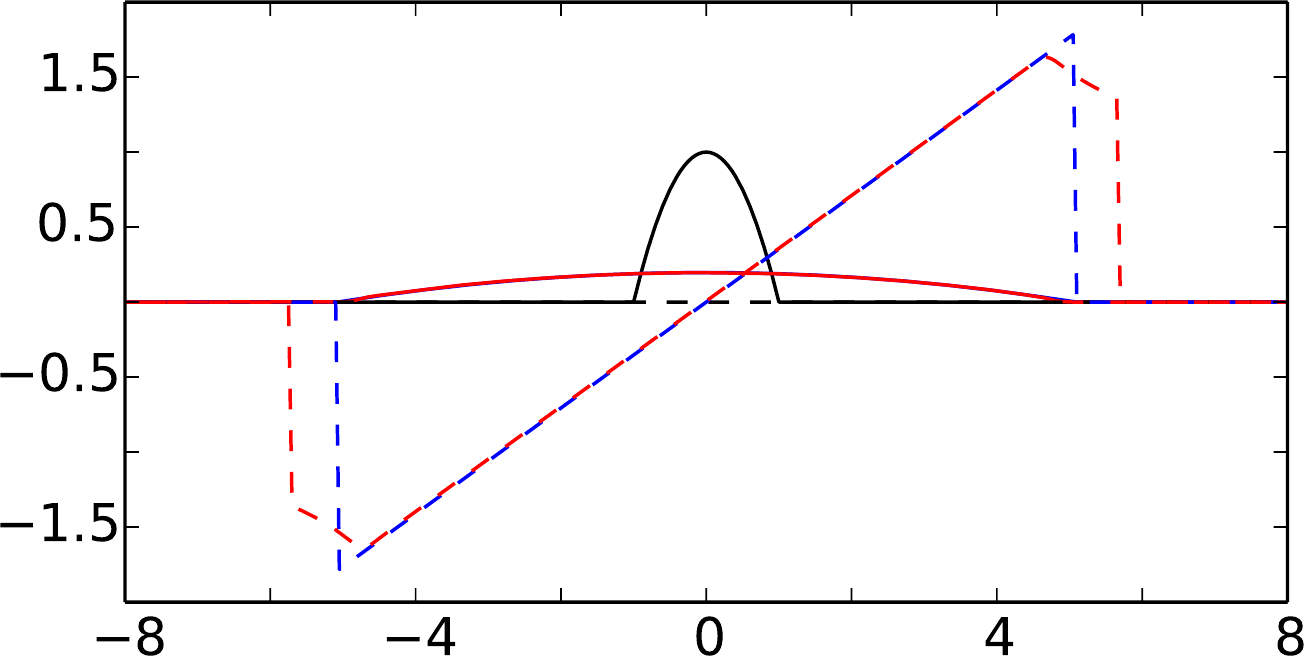}
      \label{fig:shw1_2}
    }
    \\
    \vspace{-0.7cm}
    \subfloat[]{
      \pgfimage[width=.44\textwidth]{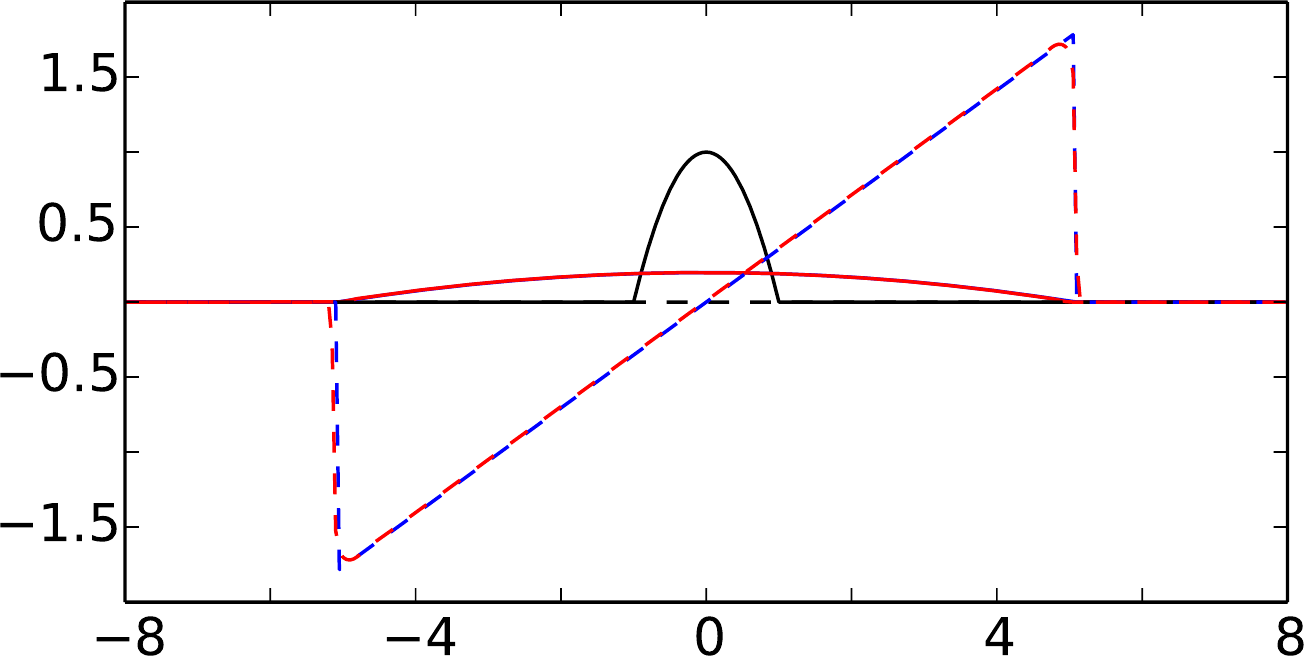}
      \label{fig:shw1_3}
    }
    \centering
    \caption{\label{fig:shw1} Simulation results for Ex.~\ref{sec:shw1}.
        Solid lines represent fluid height and dashed lines represent fluid velocity.
        Initial condition is plotted in black, analytical solution in blue and test results in red.
        \ref{fig:shw1_1} shows the initial condition and analytical solution at t=3.
        \ref{fig:shw1_2} and \ref{fig:shw1_3} show numerical results plotted over \ref{fig:shw1_1}
        obtained with options \textbf{abs} + \textbf{fct} and \textbf{iga} + \textbf{fct}, respectively.}
    \end{figure}

  The results of the test are plotted in Fig.~\ref{fig:shw1}.
  Figure~\ref{fig:shw1_1} shows the initial condition (black) and the analytical solution for t=3 (blue).
  Solid lines mark the fluid depth and the dashed line the velocity.
  The remaining two panels show numerical results\footnote{Similar to advector field evaluation 
    discussed in Sec.~\ref{sec:vip_lib} the \prog{vip\_eps} value was used 
    as cutoff value to prevent divisions by zero when calculating velocity field} 
    at t=3 for different MPDATA options (red) plotted over the top panel.
  Figure~\ref{fig:shw1_2} shows the solution with options \prog{abs} and \prog{fct}, 
    whereas Fig.~\ref{fig:shw1_3} shows the solution obtained with options \prog{iga} and \prog{fct}.

  All presented results are free of apparent artefacts near the drop edge.
  The \textbf{abs}+\textbf{fct} in the central panel compares well with Fig.~7b
    in \citet{Schar_and_Smolarkiewicz_1996};
  whereas, the \textbf{iga}+\textbf{fct} solution in the bottom panel 
    closely reproduces the analytical result.
 
    \FloatBarrier

  \subsection{\label{sec:shw2}Example: 2D shallow-water system}
  
  The 2D shallow-water test discussed here is
    an original axis-symmetric extension of the 1D slab-symmetric test in Ex.~\ref{sec:shw1}.
  The corresponding dimensionless equations take the form
  \noindent
  \begin{equation}
    \begin{aligned}
      \partial_t h + \partial_x (uh) + \partial_y (vh) &=& 0~,  \\
      \partial_t (uh) + \partial_x ( uuh) + \partial_y (vuh) &=& - h \partial_x h~, \\
      \partial_t (vh) + \partial_x ( uvh) + \partial_y (vvh) &=& - h \partial_y h~. 
    \label{shw2d}
    \end{aligned}
  \end{equation}

  \noindent
  As in 1D case, $h$ stands for the fluid height and $(u,v)$ are the velocity components
    in $x$ and $y$ directions, respectively.
  Again, the initial condition consists of a parabolic drop centred at the origin and
    confined to $x^2 + y^2 \le 1$,
    
  \begin{equation}
    h(x, y, t=0) = 
       \begin{cases}
         1-x^2 -y^2  ,& for \; \sqrt{x^2 + y^2} \le 1  \\
         0           ,& for \; \sqrt{x^2 + y^2} > 1~. 
       \end{cases}
    \label{shw2_init}
  \end{equation}

  \noindent
  Following the method presented by \citet{Frei_1993} and \citet{Schar_and_Smolarkiewicz_1996}
    the analytical solution of the system (\ref{shw2d}) can be obtained as
  \begin{equation}
    \begin{aligned} 
      h(x,y,t) &=& \frac{1}{\lambda^2} - \frac{x^2 + y^2}{\lambda^4}~,\\
      u(x,t)   &=& x \frac{\lambda_t}{\lambda}~, \\
      v(y,t)   &=& y \frac{\lambda_t}{\lambda}~.
    \end{aligned}
  \end{equation}
  \noindent
  Here $\lambda(t)$ is half-width of the drop, evolving according~to
  \begin{equation}
      \lambda(t) = \sqrt{2t^2 +1}
  \end{equation}
  \noindent
  and $\lambda_t = \partial \lambda / \partial t$ is the velocity
    of the leading edge.
  \newline
  \newline

\begin{figure}[h!]
    \centering
    \pgfimage[width=.43\textwidth]{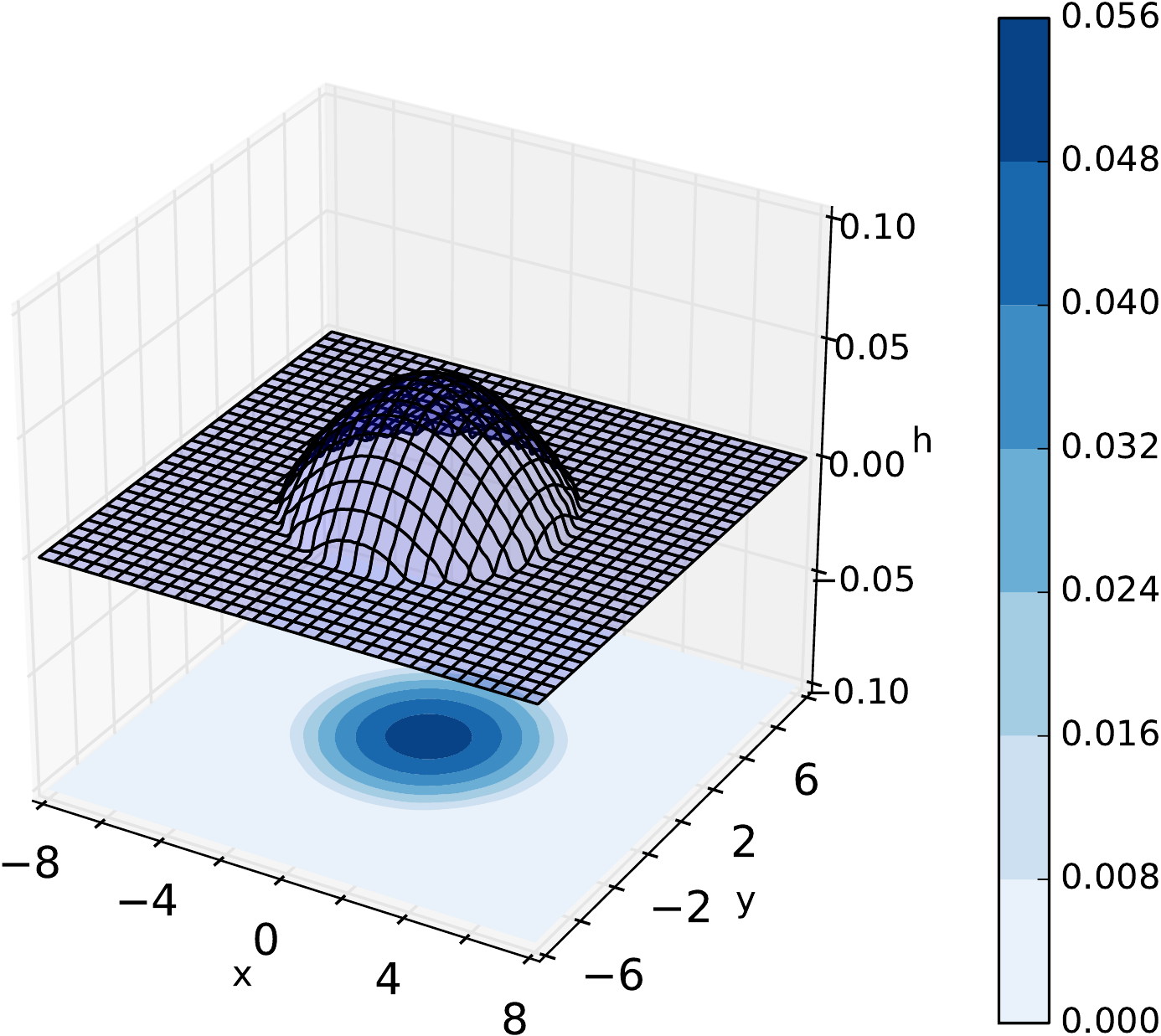}
    \caption{\label{fig:shw2d} Drop height at t=3 of the Ex.~\ref{sec:shw2}.}
  \end{figure}

\begin{figure}[h!]
  \centering
  \subfloat[]{
    \pgfimage[width=.44\textwidth]{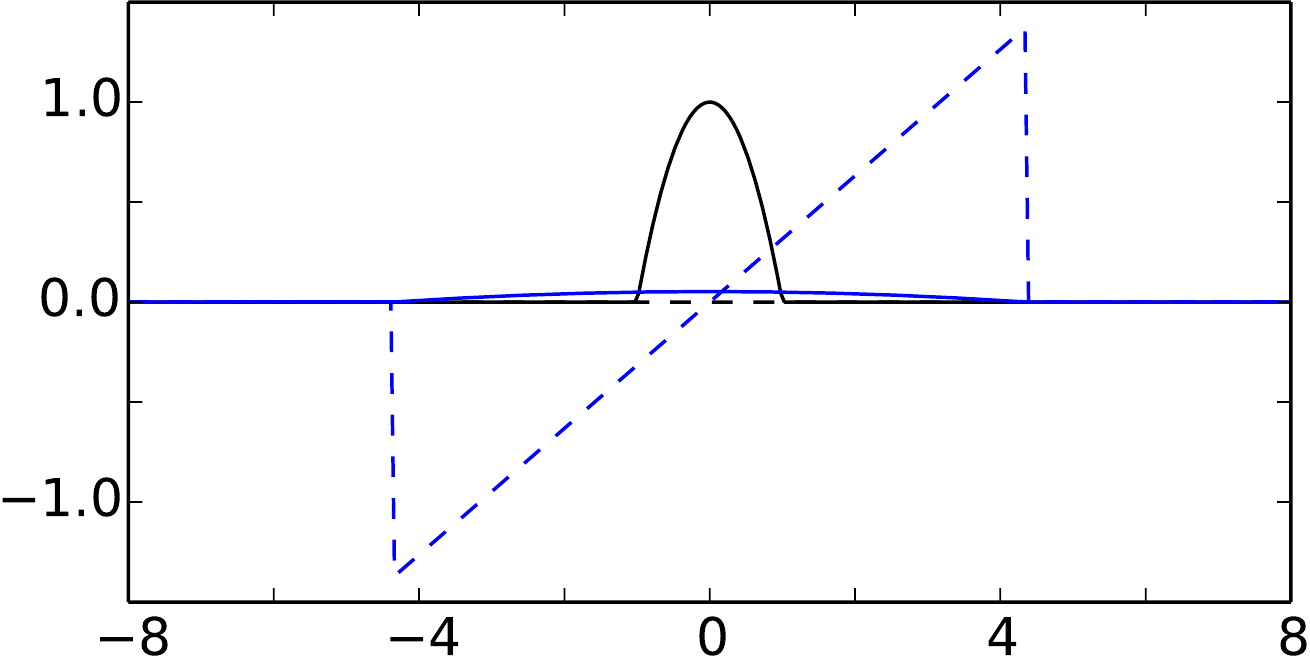}
    \label{fig:shw2_1}
  }
  \\
  \vspace{-0.7cm}
  \subfloat[]{
    \pgfimage[width=.44\textwidth]{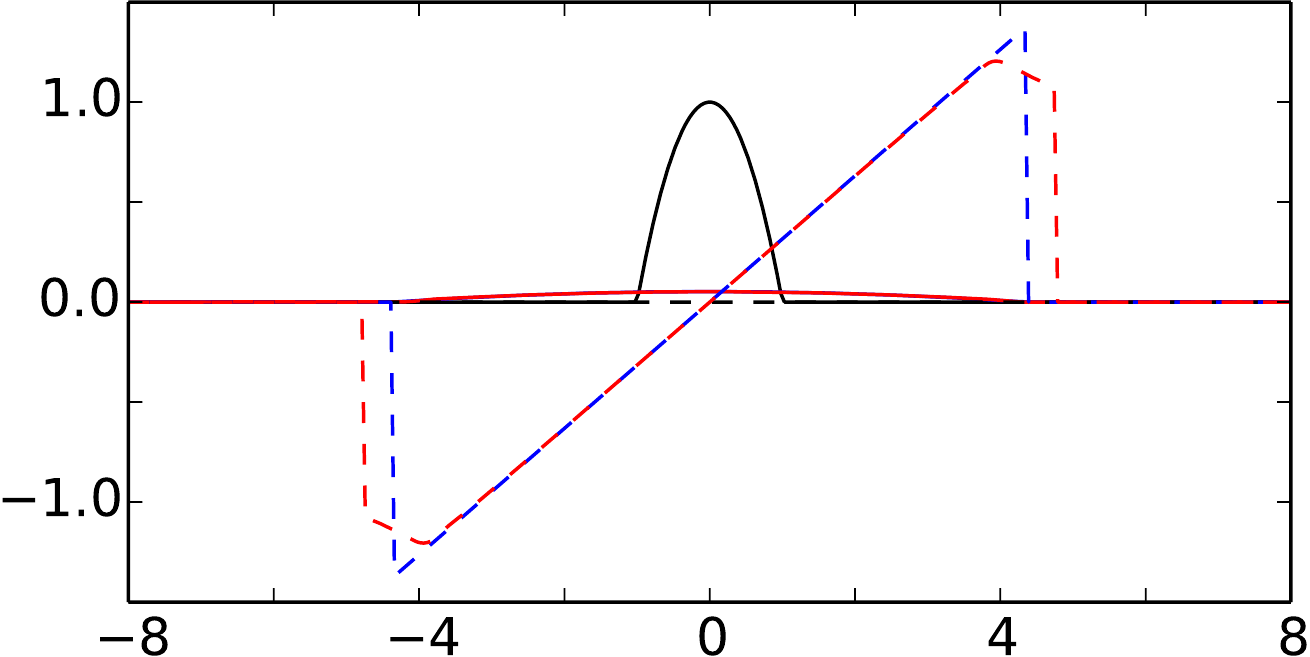}
    \label{fig:shw2_2}
  }
  \\
  \vspace{-0.7cm}
  \subfloat[]{
    \pgfimage[width=.44\textwidth]{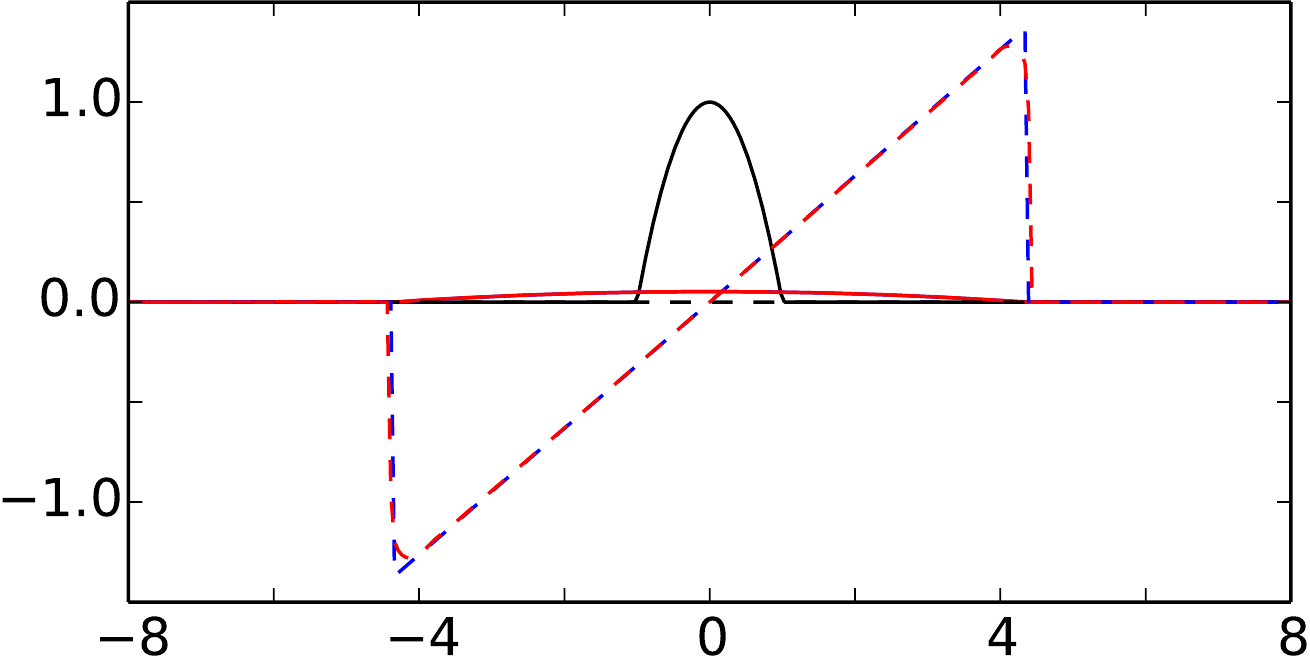}
    \label{fig:shw2_3}
  }
  \centering
  \caption{\label{fig:shw2} The same as in Fig.~\ref{fig:shw1} but for the two-dimensional case}
  \end{figure}

  Figure~\ref{fig:shw2d} shows a perspective display of drop height at $t=3$,
    whereas Fig.~\ref{fig:shw2} shows the profiles of velocity and height of the drop.
  Similarly to Fig.~\ref{fig:shw1}, the top panel shows the initial condition (black) 
    and analytical solution for $t=3$ (blue).
  Central and bottom panels show corresponding numerical results at $t=3$ (red).
  Solid lines represent the fluid height and the dashed lines the velocity.
  The central panel shows the solution with options \prog{abs} and \prog{fct}, whereas the bottom panel
    shows the solution with options \prog{iga} and \prog{fct}.
  As in the 1D case, the velocity field near the drop edge is regular 
    and the \textbf{iga}+\textbf{fct} result closely follows the analytical solution.
  
  \FloatBarrier

  \FloatBarrier

  \section{Systems with elliptic pressure equation}\label{sec:adv+rhs+vel+prs}

  \subsection{Implemented algorithms}

  The \emph{libmpdata++} library includes an implicit representation of pressure gradient terms
    for incompressible fluid equations.
  This necessitates the solution of an elliptic Poisson problem for pressure.
  The elliptic problem is solved after applying all explicitly known forcings 
    to ensure a non-divergent velocity field at the end of each time step.
  As of the current release, the library is equipped with 
    the minimal- and conjugate-residual variational iterative solvers.
  For the derivation of used schemes and further discussion of the elliptic problem see
     \citet{Smolarkiewicz_and_Margolin_1994}, \citet{Smolarkiewicz_and_Szmelter_2011}
     and references therein.

  \subsection{Library interface}
 
  The methods for solving the elliptic problem are implemented in the 
    \prog{mpdata\_rhs\_vip\_prs} class, Fig.~\ref{fig:inherit}.
  This class inherits from the \prog{mpdata\_rhs\_vip} class. 
  Therefore the way to specify other source terms as well as time-varying velocity field
    remains unchanged.

  The choice of elliptic solver is controlled by setting the compile-time parameter \prog{prs\_scheme}
    to \prog{mr} and \prog{cr} for the minimal-residual and  conjugate-residual solver, respectively.
  The iterations within the elliptic solver stop when the divergence of the velocity
    field is lower than a threshold tolerance set by a run-time parameter \prog{prs\_tol}, 
    cf \citep{Smolarkiewicz_et_al_1997}.
  
  \subsection{\label{sec:bombel}Example: Boussinesq convection}

  The goal of this example is to show the user interface for simulations
    featuring elliptic pressure equation.
  The governing PDE system consists of momentum, potential temperature, and mass-continuity
    equations for an ideal, 2D, incompressible Boussinesq fluid

  \begin{eqnarray}
    \partial_t \vec{v} + \nabla \cdot ( \vec{v} \otimes \vec{v}) &=& - \nabla \pi - \vec{g} \frac{\theta^{'}}{\theta_o}~, \\
    \partial_t \theta  + \nabla \cdot (\vec{v} \theta) &=& 0~, \label{eq:bombel_th} \\
    \nabla \cdot \vec{v} &=& 0~. \label{eq:bombel_div}
  \end{eqnarray}
  \noindent
    Here,
  $\vec{v} = (u,w)$ denotes the velocity field,
  $\pi$ is the pressure perturbation about the hydrostatic reference state normalised
    by the reference density $\rho_o$, constant in the Boussinesq model.
    Furthermore,
  $\theta^{'}$ represents the potential temperature perturbation about the reference state
  $\theta_o = {\rm const}$, and $\otimes$ denotes the tensor product.

  \noindent
  Combining the velocity prediction from the momentum equation, according to eq.~(\ref{trapez}),
    with the mass continuity eq.~(\ref{eq:bombel_div}) leads to the elliptic Poisson problem
  \begin{equation}
    \label{eq:bombel_ellip}
    -\frac{1}{\rho_o} \nabla \cdot \left(\rho_o \left(\widehat{\vec{v}} - 0.5 \Delta t \nabla \pi\right)\right) = 0~,
  \end{equation}
  \noindent
  where
  $\widehat{\vec{v}}$ is the velocity field after the advection
    summed with all the explicitly known source terms at time level \emph{n+1},
    namely buoyancy in this example.\footnote{Because
      the potential temperature equation (\ref{eq:bombel_th}) is homogeneous,
      the buoyancy at \emph{n+1} time level can be readily evaluated after advecting $\theta$.
    }
  In eq.~(\ref{eq:bombel_ellip}) the pressure perturbation field $\pi$ is unknown,
    and it needs to be adjusted such
    that the final velocity field $\widehat{\vec{v}} - 0.5 \Delta t \nabla \pi$
    satisfies the mass continuity equation (\ref{eq:bombel_div}).
  Denoting $0.5 \Delta t \pi$ as $\phi$ allows to symbolise eq.~(\ref{eq:bombel_ellip})
    using standard notation for linear sparse problems, \citep{Smolarkiewicz_and_Margolin_1994}

  \begin{equation}
    \mathcal{L} ({\phi}) - \mathcal{R} = 0~. 
  \end{equation}

  The setup of the test follows \citet{Smolarkiewicz_and_Pudykiewicz_1992}.
  It consists of a circular potential temperature anomaly
    of radius 250~m, embedded in a neutrally stratified quiescent environment,
    with $\theta_o = 300K$,
    in the domain resolved with 200 $\times$ 200 grid-cells of the size dx=dy=10m.
  The initial anomaly $\theta ' = 0.5K$
    is centred in the horizontal, 260 metres above the lower boundary.
  The time-step is set to $\Delta t=0.75$~s and the simulation takes 800 time-steps.

  Listing~\ref{lst:bombel_params} shows the compile-time parameters structure.
  The time integration scheme for the buoyancy forcing is set to \prog{trapez},
    as the user has a choice of the algorithm.
  However, as of the current release, the elliptic problem formulation requires
    forcings to be independent of velocity if handled using the \prog{trapez} scheme.
  The implicit pressure gradient terms
     are always integrated with the trapezoidal rule (\ref{trapez}),
     regardless of the \textbf{rhs\_scheme} setting. 
  In List.~\ref{lst:bombel_params} the elliptic solver option is set to the
    conjugate-residual scheme \prog{cr}.
  The \prog{vip\_den} is set to -1,
    because here the velocity components are the model kinematic variables,
    cf. the discussion in second paragraph of section \ref{sec:vip_lib}.

  \begin{Listing}
  \fvset{gobble=2}%
  \renewcommand*\FancyVerbStartString{\PY{c+c1}{//\PYZlt{}listing\PYZhy{}1\PYZgt{}}}%
  \renewcommand*\FancyVerbStopString{\PY{c+c1}{//\PYZlt{}/listing\PYZhy{}1\PYZgt{}}}%
  \input{bombel_params.cpp}
    \caption{\label{lst:bombel_params} Compile-time parameters for Ex.~\ref{sec:bombel}.}
  \end{Listing}

  \begin{Listing}
  \fvset{gobble=2}%
  \renewcommand*\FancyVerbStartString{\PY{c+c1}{//\PYZlt{}listing\PYZhy{}2\PYZgt{}}}%
  \renewcommand*\FancyVerbStopString{\PY{c+c1}{//\PYZlt{}/listing\PYZhy{}2\PYZgt{}}}%

\input{bombel_params.cpp}    \caption{\label{lst:bombel_tol}Run-time parameter field setting the accuracy of the pressure solver.}
  \end{Listing}

  The convergence threshold of the elliptic solver, $\nabla\cdot(\vec{v})\le\varepsilon$,
    is set to 10\textsuperscript{-7} via the run-time parameter \prog{prs\_tol},
    List.~\ref{lst:bombel_tol}.

  Listing~\ref{lst:bombel_rhs} shows the buoyancy forcing definition.\\

  \begin{Listing}
  \fvset{gobble=2}%
  \renewcommand*\FancyVerbStartString{\PY{c+c1}{//\PYZlt{}listing\PYZhy{}1\PYZgt{}}}%
  \renewcommand*\FancyVerbStopString{\PY{c+c1}{//\PYZlt{}/listing\PYZhy{}1\PYZgt{}}}%
  \input{bombel_rhs.cpp}
    \caption{\label{lst:bombel_rhs}Method for calculating source and sink terms for Ex.~\ref{sec:bombel}.}
  \end{Listing}
 
  \vspace{-1cm}

  \begin{figure}[h!]
  \centering
  \subfloat[]{
    \pgfimage[width=.38\textwidth]{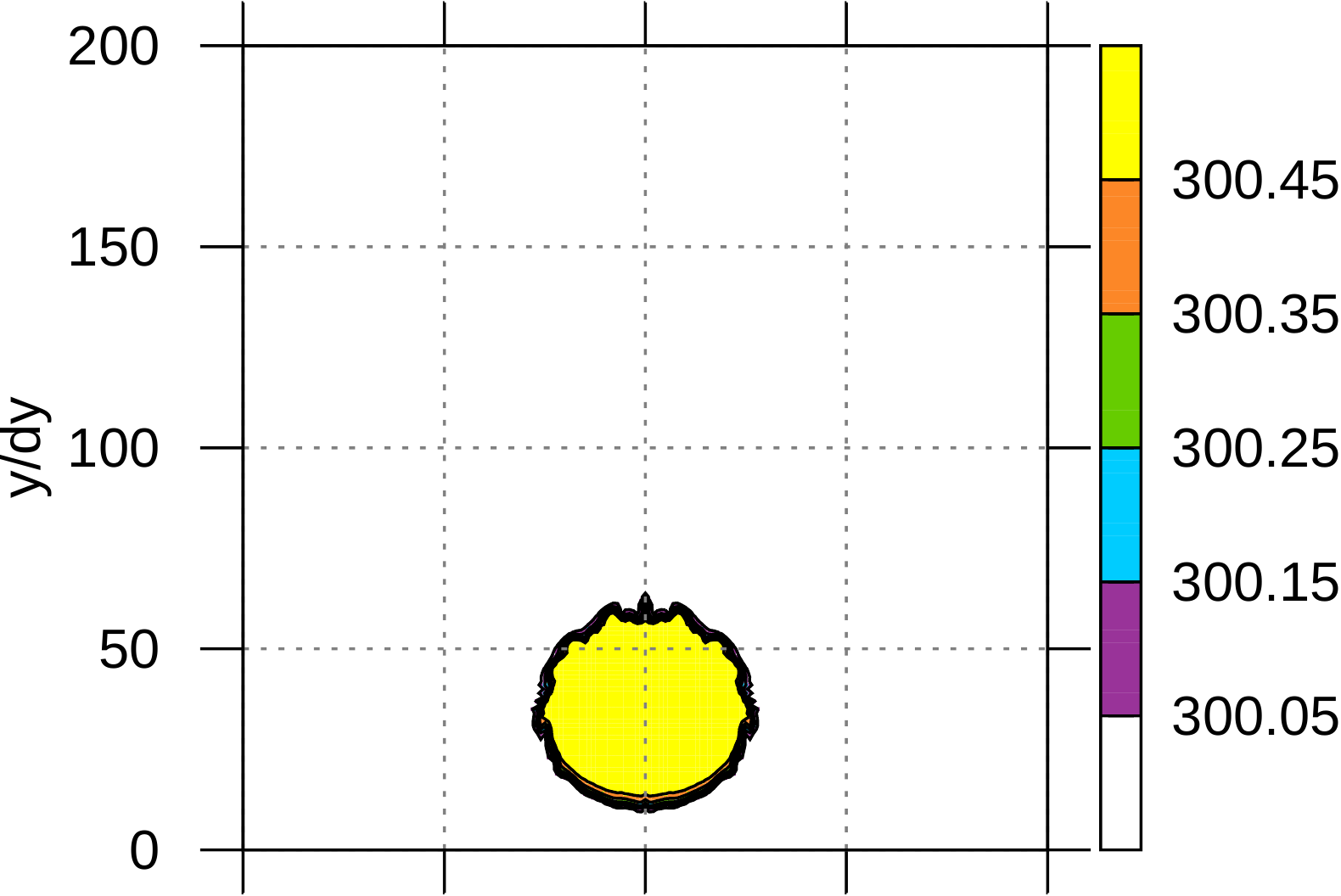}
    \label{fig:bom1}
  }
  \\
  \vspace{-1.1cm}
  \subfloat[]{
     \pgfimage[width=.38\textwidth]{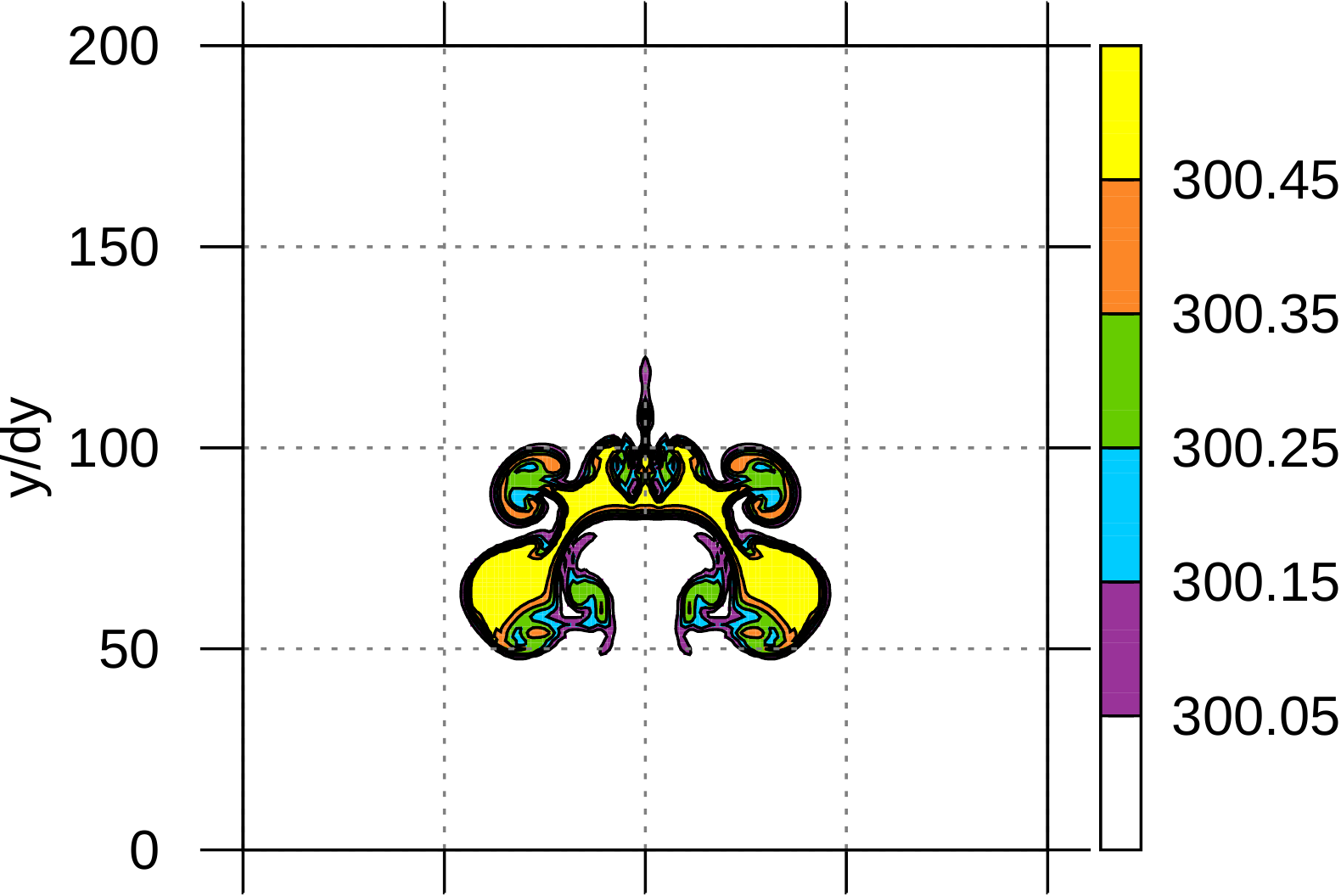}
     \label{fig:bom2}
  }
  \\
  \vspace{-1.1cm}
  \subfloat[]{
     \pgfimage[width=.38\textwidth]{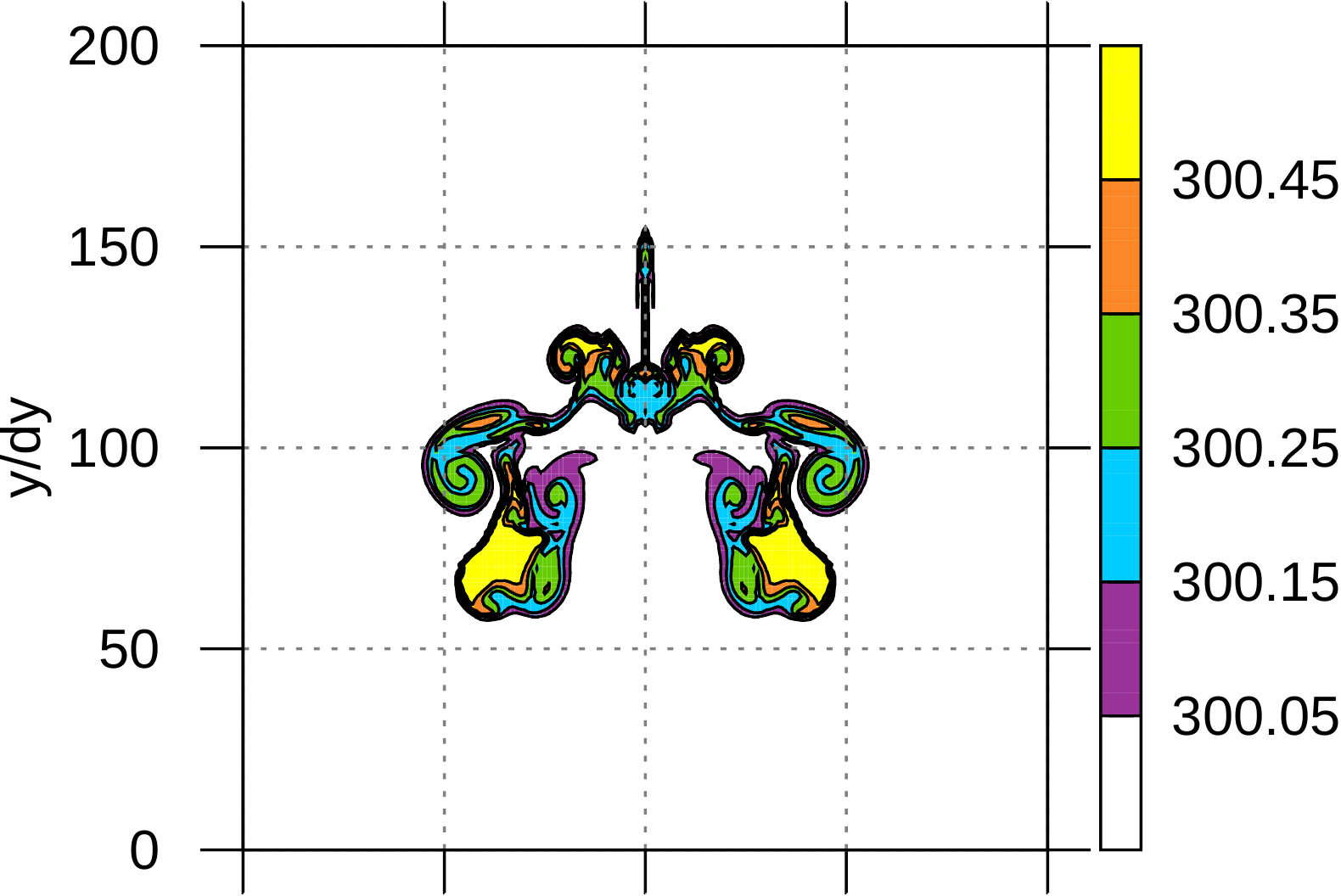}
     \label{fig:bom3}
  }
  \centering
  \caption{\label{fig:bom_panel} The results of Ex.~\ref{sec:bombel}, 
     \ref{fig:bom1} from the 200$^{th}$, 
     \ref{fig:bom2} from the 600$^{th}$ and
     \ref{fig:bom3} from the 800$^{th}$ time step.
  }
  \end{figure}
  The evolved $\theta$ fields after $200$, $600$ and $800$ time steps
    are shown in Figs.~\ref{fig:bom1}, \ref{fig:bom2} and \ref{fig:bom3}.
  These results correspond to plots from Fig.~3 in
    \citet{Smolarkiewicz_and_Pudykiewicz_1992} and illustrate that
    \emph{libmpdata++} captures the interfacial instabilities and sharp gradients,
    including small turbulent structures in Fig.~\ref{fig:bom3}.
  Yet, the solutions contain small (imperceptible in the plots) under- and over-shoots,
   developing at the rate of $\Delta \theta / \Delta t \sim\Delta t\theta_o\nabla\cdot(\vec{v})$.
  These oscillations depend on the magnitude of the residual errors, $\nabla \cdot \vec{v} \ne 0$,
   controlled by the convergence threshold \prog{prs\_tol}.
  For substantiation, Tab.~\ref{tab:bombel} displays the magnitude of such spurious extrema $\delta \theta_{max}$ --- 
    defined as the larger from the maximal magnitudes of normalised under- and over-shoots 
    with respects to their initial values --- 
    against \prog{prs\_tol} at the time of Fig.~\ref{fig:bom3}. 
  Note that $\delta\theta_{max}$ is bounded by \prog{prs\_tol}($\times 800\Delta t$).

  \begin{table}[h!]
  \centering
  \begin{tabular}{ c | c | c | c }
  \prog{prs\_tol} & $10^{-5}$ & $10^{-7}$ & $10^{-9}$ \\ \hline
  $\delta\theta_{max}$& $3 \cdot 10^{-4}$ & $8 \cdot 10^{-6}$ & $1 \cdot 10^{-7}$
  \end{tabular}
  \caption{Maximal spurious extrema of $\theta$ field after 800 time steps for
   various values of the convergence threshold \prog{prs\_tol}.}
  \label{tab:bombel}
  \end{table}

  \FloatBarrier

\section{Remarks}  

In this paper the first release of \emph{libmpdata++} was introduced.
Versatility of the user interface as well as 
  the correctness of the implementation were illustrated with a series of examples
  with increasing degree of physical, mathematical and programming complexity.
Starting from elementary advection in Cartesian domain,
  through passive advection on the sphere, 
  through slab- and axis-symmetric water drop spreading under gravity,
  to buoyant convection in an incompressible Boussinesq fluid, 
  the accompanying discussions included code snippets, 
  description of the user interface and 
  comparison with previously published benchmarks.

Our priority in the development of \emph{libmpdata++} is the researcher productivity.
In case of scientific software such as \emph{libmpdata++},
  the researchers are both users and developers of the library.
The adherence to the principle of separation of concerns and employment of programming
  techniques that promote code conciseness --- e.g. the current release consists of less than 10k lines of code ---
  contribute to the developers productivity.
The user productivity is amplified by ensuring that the release of the library is accompanied 
  with example-rich documentation. 
Both the users and developers benefit from the free/libre open-source software release 
  of the library.

Work is under way on several new features for the subsequent release 
  of \emph{libmpdata++}, including distributed-memory parallelisation.

  \FloatBarrier
  \relscale{.95}
  \begin{acknowledgements}
  Personal reviews from Christian K{\"u}hnlein 
    and Willem Deconinck helped to improve the presentation.
  Development of \emph{libmpdata++} was funded by the Polish National
    Science Centre -- projects no. 2011/01/N/ST10/01483 (PRELUDIUM)
    and no. 2012/06/M/ST10/00434 (HARMONIA).
  PKS acknowledges support by funding received from the European 
    Research Council under the European Union's Seventh Framework Programme 
    (FP7/2012/ERC Grant agreement no. 320375).
  Part of the work was carried out during visits of AJ to the National
    Center for Atmospheric Research (NCAR) in Boulder, Colorado, USA
    and to the European Centre for Medium-Range Weather Forecasts (ECMWF), Reading, UK.
  Part of the work was carried out during visits of DJ to NCAR 
    funded by Polish Ministry of Science and Higher Education - project no. 1119/MOB/13/2014/0.
  NCAR is operated by the University Corporation for Atmospheric Research.
  Figures were generated using \emph{gnuplot} (\url{http://gnuplot.info/}), \emph{Paraview}
    (\url{http://paraview.org/}) and \emph{matplotlib} (\url{http://matplotlib.org/}).
  \end{acknowledgements}

  \bibliography{paper}

\end{document}

%% file: formulae_preamble.tex
\usepackage{amsmath}

\usepackage{xfrac}

\newcommand{\minus}{\scalebox{.4}[.8]{\( - \)}}
\newcommand{\plus}{\scalebox{.4}[.6]{\( + \)}}
\newcommand{\pone}{\plus1}
\newcommand{\mone}{\minus1}

\newcommand{\phlf}{\plus\sfrac{1}{2}}
\newcommand{\mhlf}{\minus\sfrac{1}{2}}

%% file: fig-domain.tex
\begin{tikzpicture}
  \coordinate (Origin)   at (0,0);
  \coordinate (XAxisMin) at (-2.5,-1.5);
  \coordinate (XAxisMax) at (2.5,-1.5);
  \coordinate (YAxisMin) at (-1.5,-2.5);
  \coordinate (YAxisMax) at (-1.5,2.5);

  \draw [thin, gray,-latex] (XAxisMin) -- (XAxisMax);
  \draw [thin, gray,-latex] (YAxisMin) -- (YAxisMax);

  \clip (-2.01,-2.01) rectangle (2.01cm,2.01cm); 
	
  \draw[style=help lines,dashed, very thick] (-14,-14) grid[step=1cm] (14,14);
	
  \filldraw[fill=gray, fill opacity=0.3, draw=black] (-1.5,-1.5) rectangle (1.5,1.5);
  \filldraw[fill=gray, fill opacity=0, draw=black] (-.5,-.5) rectangle (.5, .5);
  \filldraw[fill=gray, fill opacity=0, draw=black] (-1.5,-1.5) rectangle (-.5, -.5);
  \filldraw[fill=gray, fill opacity=0, draw=black] (.5, -1.5) rectangle (1.5,-.5);
  \filldraw[fill=gray, fill opacity=0, draw=black] (-1.5, .5) rectangle (-.5, 1.5);
  \filldraw[fill=gray, fill opacity=0, draw=black] (.5, .5) rectangle (1.5,1.5);

  \foreach \x in {0,1,...,3}{
    \foreach \y in {0,1,...,3}{
      \node[draw,circle,inner sep=1.5pt,fill] at (-1.5+\x,-1.5+\y) {};
    }
  }

\end{tikzpicture}

%% file: fig-grid.tex
\begin{tikzpicture}

  \clip (-1.5,-1.5) rectangle (3.5cm,3.5cm); 

  \draw[style=help lines,dashed, very thick] (-13,-13) grid[step=2cm] (13,13);

  \foreach \x in {-7,-6,...,7}{
    \foreach \y in {-7,-6,...,7}{
      \node[draw,circle,inner sep=2pt,fill] at (1+2*\x,1+2*\y) {};
    }
  }
  \draw [black] (1,1) -- (1,1) node [below right] {$\!\!\psi_{i,j}$};
  \draw [black] (-1,1) -- (-1,1) node [below right] {$\!\!\psi_{i\mone,j}$};
  \draw [black] (1,3) -- (1,3) node [below right] {$\!\!\psi_{i,j\pone}$};

  \draw [ultra thick,-latex,red] (1.8,1) -- (2.2,1) node [above right] {$\!\!\!\!\!u^{x}_{i\phlf,j}$};
  \draw [ultra thick,-latex,red] (-.2,1) -- (.2,1) node [above right] {$\!\!\!\!\!u^{x}_{i\mhlf,j}$};
  \draw [ultra thick,-latex,red] (1.8,3) -- (2.2,3) node [above] {};
  \draw [ultra thick,-latex,red] (-.2,3) -- (.2,3) node [above] {};
  \draw [ultra thick,-latex,red] (1.8,-1) -- (2.2,-1) node [above] {};
  \draw [ultra thick,-latex,red] (-.2,-1) -- (.2,-1) node [above] {};

  \draw [ultra thick,-latex,red] (1,1.8) -- (1,2.2) node [above] {};
  \draw [ultra thick,-latex,red] (1,-.2) -- (1,.2) node at (1.42,.-.37) {$u^{y}_{i,j\mhlf}$};
  \draw [ultra thick,-latex,red] (-1,1.8) -- (-1,2.2) node [above] {};
  \draw [ultra thick,-latex,red] (-1,-.2) -- (-1,.2) node [above] {};
  \draw [ultra thick,-latex,red] (3,1.8) -- (3,2.2) node [above] {};
  \draw [ultra thick,-latex,red] (3,-.2) -- (3,.2) node [above] {};
\end{tikzpicture}

%% file: choice.cpp.tex
\begin{Verbatim}[commandchars=\\\{\}]
\PY{c+cm}{/** }
\PY{c+cm}{ * @file}
\PY{c+cm}{ * @copyright University of Warsaw}
\PY{c+cm}{ * @section LICENSE}
\PY{c+cm}{ * GPLv3+ (see the COPYING file or http://www.gnu.org/licenses/)}
\PY{c+cm}{ *}
\PY{c+cm}{ * \PYZbs{}include \PYZdq{}concurrent\PYZus{}1d/test\PYZus{}concurrent\PYZus{}1d.cpp\PYZdq{}}
\PY{c+cm}{ */}

\PY{c+cp}{\PYZsh{}}\PY{c+cp}{include \PYZlt{}libmpdata++}\PY{c+cp}{/}\PY{c+cp}{concurr}\PY{c+cp}{/}\PY{c+cp}{openmp.hpp\PYZgt{}}
\PY{c+cp}{\PYZsh{}}\PY{c+cp}{include \PYZlt{}libmpdata++}\PY{c+cp}{/}\PY{c+cp}{concurr}\PY{c+cp}{/}\PY{c+cp}{boost\PYZus{}thread.hpp\PYZgt{}}
\PY{c+cp}{\PYZsh{}}\PY{c+cp}{include \PYZlt{}libmpdata++}\PY{c+cp}{/}\PY{c+cp}{concurr}\PY{c+cp}{/}\PY{c+cp}{serial.hpp\PYZgt{}}
\PY{c+cp}{\PYZsh{}}\PY{c+cp}{include \PYZlt{}libmpdata++}\PY{c+cp}{/}\PY{c+cp}{concurr}\PY{c+cp}{/}\PY{c+cp}{threads.hpp\PYZgt{}}

\PY{c+cp}{\PYZsh{}}\PY{c+cp}{include \PYZlt{}libmpdata++}\PY{c+cp}{/}\PY{c+cp}{solvers}\PY{c+cp}{/}\PY{c+cp}{mpdata.hpp\PYZgt{}}

\PY{c+cp}{\PYZsh{}}\PY{c+cp}{include \PYZlt{}libmpdata++}\PY{c+cp}{/}\PY{c+cp}{output}\PY{c+cp}{/}\PY{c+cp}{gnuplot.hpp\PYZgt{}}

\PY{k+kt}{int} \PY{n+nf}{main}\PY{p}{(}\PY{p}{)}
\PY{p}{\PYZob{}}
  \PY{k}{using} \PY{k}{namespace} \PY{n}{libmpdataxx}\PY{p}{;}

  \PY{n}{std}\PY{o}{:}\PY{o}{:}\PY{n}{cerr} \PY{o}{\PYZlt{}}\PY{o}{\PYZlt{}} \PY{l+s}{\PYZdq{}}\PY{l+s}{OpenMP: }\PY{l+s}{\PYZdq{}}\PY{p}{;}
\PY{c+cp}{\PYZsh{}}\PY{c+cp}{if defined(\PYZus{}OPENMP)}
  \PY{n}{std}\PY{o}{:}\PY{o}{:}\PY{n}{cerr} \PY{o}{\PYZlt{}}\PY{o}{\PYZlt{}} \PY{l+s}{\PYZdq{}}\PY{l+s}{on}\PY{l+s}{\PYZdq{}} \PY{o}{\PYZlt{}}\PY{o}{\PYZlt{}} \PY{n}{std}\PY{o}{:}\PY{o}{:}\PY{n}{endl}\PY{p}{;}
\PY{c+cp}{\PYZsh{}}\PY{c+cp}{else }
  \PY{n}{std}\PY{o}{:}\PY{o}{:}\PY{n}{cerr} \PY{o}{\PYZlt{}}\PY{o}{\PYZlt{}} \PY{l+s}{\PYZdq{}}\PY{l+s}{off}\PY{l+s}{\PYZdq{}} \PY{o}{\PYZlt{}}\PY{o}{\PYZlt{}} \PY{n}{std}\PY{o}{:}\PY{o}{:}\PY{n}{endl}\PY{p}{;}
\PY{c+cp}{\PYZsh{}}\PY{c+cp}{endif}

\PY{c+c1}{//\PYZlt{}listing\PYZhy{}1\PYZgt{}}
  \PY{k}{struct} \PY{k+kt}{ct\PYZus{}params\PYZus{}t} \PY{o}{:} \PY{k+kt}{ct\PYZus{}params\PYZus{}default\PYZus{}t}
  \PY{p}{\PYZob{}} 
    \PY{k}{using} \PY{k+kt}{real\PYZus{}t} \PY{o}{=} \PY{k+kt}{double}\PY{p}{;} 
    \PY{k}{enum} \PY{p}{\PYZob{}} \PY{n}{n\PYZus{}dims} \PY{o}{=} \PY{l+m+mi}{1} \PY{p}{\PYZcb{}}\PY{p}{;} 
    \PY{k}{enum} \PY{p}{\PYZob{}} \PY{n}{n\PYZus{}eqns} \PY{o}{=} \PY{l+m+mi}{1} \PY{p}{\PYZcb{}}\PY{p}{;}
  \PY{p}{\PYZcb{}}\PY{p}{;}
\PY{c+c1}{//\PYZlt{}/listing\PYZhy{}1\PYZgt{}}

  \PY{k}{const} \PY{k+kt}{int} \PY{n}{nx} \PY{o}{=} \PY{l+m+mi}{10}\PY{p}{,} \PY{n}{nt} \PY{o}{=} \PY{l+m+mi}{1000}\PY{p}{;}
   
  \PY{c+c1}{// OpenMP}
  \PY{n}{std}\PY{o}{:}\PY{o}{:}\PY{n}{cerr} \PY{o}{\PYZlt{}}\PY{o}{\PYZlt{}} \PY{l+s}{\PYZdq{}}\PY{l+s}{OpenMP run}\PY{l+s}{\PYZdq{}} \PY{o}{\PYZlt{}}\PY{o}{\PYZlt{}} \PY{n}{std}\PY{o}{:}\PY{o}{:}\PY{n}{endl}\PY{p}{;}
  \PY{p}{\PYZob{}}
\PY{c+c1}{//\PYZlt{}listing\PYZhy{}2\PYZgt{}}
    \PY{k}{using} \PY{k+kt}{slv\PYZus{}t} \PY{o}{=} \PY{n}{solvers}\PY{o}{:}\PY{o}{:}\PY{n}{mpdata}\PY{o}{\PYZlt{}}\PY{k+kt}{ct\PYZus{}params\PYZus{}t}\PY{o}{\PYZgt{}}\PY{p}{;}
\PY{c+c1}{//\PYZlt{}/listing\PYZhy{}2\PYZgt{}}
\PY{c+c1}{//\PYZlt{}listing\PYZhy{}3\PYZgt{}}
    \PY{k}{using} \PY{k+kt}{slv\PYZus{}out\PYZus{}t} \PY{o}{=} \PY{n}{output}\PY{o}{:}\PY{o}{:}\PY{n}{gnuplot}\PY{o}{\PYZlt{}}\PY{k+kt}{slv\PYZus{}t}\PY{o}{\PYZgt{}}\PY{p}{;}
\PY{c+c1}{//\PYZlt{}/listing\PYZhy{}3\PYZgt{}}
\PY{c+c1}{//\PYZlt{}listing\PYZhy{}4\PYZgt{}}
    \PY{k}{using} \PY{k+kt}{run\PYZus{}t} \PY{o}{=} \PY{n}{concurr}\PY{o}{:}\PY{o}{:}\PY{n}{openmp}\PY{o}{\PYZlt{}}
      \PY{k+kt}{slv\PYZus{}out\PYZus{}t}\PY{p}{,} 
      \PY{n}{bcond}\PY{o}{:}\PY{o}{:}\PY{n}{cyclic}\PY{p}{,} \PY{n}{bcond}\PY{o}{:}\PY{o}{:}\PY{n}{cyclic}\PY{p}{,}
      \PY{n}{bcond}\PY{o}{:}\PY{o}{:}\PY{n}{open}\PY{p}{,}   \PY{n}{bcond}\PY{o}{:}\PY{o}{:}\PY{n}{open}
    \PY{o}{\PYZgt{}}\PY{p}{;}
\PY{c+c1}{//\PYZlt{}/listing\PYZhy{}4\PYZgt{}}
\PY{c+c1}{//\PYZlt{}listing\PYZhy{}5\PYZgt{}}
    \PY{k}{typename} \PY{k+kt}{slv\PYZus{}out\PYZus{}t}\PY{o}{:}\PY{o}{:}\PY{k+kt}{rt\PYZus{}params\PYZus{}t} \PY{n}{p}\PY{p}{;}
    \PY{n}{p}\PY{p}{.}\PY{n}{grid\PYZus{}size} \PY{o}{=} \PY{p}{\PYZob{}} \PY{n}{nx} \PY{p}{\PYZcb{}}\PY{p}{;}
    \PY{k+kt}{run\PYZus{}t} \PY{n}{run}\PY{p}{(}\PY{n}{p}\PY{p}{)}\PY{p}{;}
\PY{c+c1}{//\PYZlt{}/listing\PYZhy{}5\PYZgt{}}
    \PY{n}{run}\PY{p}{.}\PY{n}{advectee}\PY{p}{(}\PY{p}{)} \PY{o}{=} \PY{l+m+mi}{0}\PY{p}{;}
    \PY{n}{run}\PY{p}{.}\PY{n}{advector}\PY{p}{(}\PY{p}{)} \PY{o}{=} \PY{l+m+mi}{0}\PY{p}{;}
    \PY{n}{run}\PY{p}{.}\PY{n}{advance}\PY{p}{(}\PY{n}{nt}\PY{p}{)}\PY{p}{;}
  \PY{p}{\PYZcb{}}

  \PY{c+c1}{// Boost.Thread}
  \PY{n}{std}\PY{o}{:}\PY{o}{:}\PY{n}{cerr} \PY{o}{\PYZlt{}}\PY{o}{\PYZlt{}} \PY{l+s}{\PYZdq{}}\PY{l+s}{Boost.Thread run}\PY{l+s}{\PYZdq{}} \PY{o}{\PYZlt{}}\PY{o}{\PYZlt{}} \PY{n}{std}\PY{o}{:}\PY{o}{:}\PY{n}{endl}\PY{p}{;}
  \PY{p}{\PYZob{}}
    \PY{k}{using} \PY{k+kt}{solver\PYZus{}t} \PY{o}{=} \PY{n}{solvers}\PY{o}{:}\PY{o}{:}\PY{n}{mpdata}\PY{o}{\PYZlt{}}\PY{k+kt}{ct\PYZus{}params\PYZus{}t}\PY{o}{\PYZgt{}}\PY{p}{;}
    \PY{k}{typename} \PY{k+kt}{solver\PYZus{}t}\PY{o}{:}\PY{o}{:}\PY{k+kt}{rt\PYZus{}params\PYZus{}t} \PY{n}{p}\PY{p}{;}
    \PY{n}{p}\PY{p}{.}\PY{n}{grid\PYZus{}size} \PY{o}{=} \PY{p}{\PYZob{}}\PY{n}{nx}\PY{p}{\PYZcb{}}\PY{p}{;}
    \PY{n}{concurr}\PY{o}{:}\PY{o}{:}\PY{n}{boost\PYZus{}thread}\PY{o}{\PYZlt{}}\PY{k+kt}{solver\PYZus{}t}\PY{p}{,} \PY{n}{bcond}\PY{o}{:}\PY{o}{:}\PY{n}{cyclic}\PY{p}{,} \PY{n}{bcond}\PY{o}{:}\PY{o}{:}\PY{n}{cyclic}\PY{o}{\PYZgt{}} \PY{n}{run}\PY{p}{(}\PY{n}{p}\PY{p}{)}\PY{p}{;}
    \PY{n}{run}\PY{p}{.}\PY{n}{advectee}\PY{p}{(}\PY{p}{)} \PY{o}{=} \PY{l+m+mi}{0}\PY{p}{;}
    \PY{n}{run}\PY{p}{.}\PY{n}{advector}\PY{p}{(}\PY{p}{)} \PY{o}{=} \PY{l+m+mi}{0}\PY{p}{;}
    \PY{n}{run}\PY{p}{.}\PY{n}{advance}\PY{p}{(}\PY{n}{nt}\PY{p}{)}\PY{p}{;}
  \PY{p}{\PYZcb{}}

  \PY{c+c1}{// trheads (i.e. auto)}
  \PY{n}{std}\PY{o}{:}\PY{o}{:}\PY{n}{cerr} \PY{o}{\PYZlt{}}\PY{o}{\PYZlt{}} \PY{l+s}{\PYZdq{}}\PY{l+s}{threads run}\PY{l+s}{\PYZdq{}} \PY{o}{\PYZlt{}}\PY{o}{\PYZlt{}} \PY{n}{std}\PY{o}{:}\PY{o}{:}\PY{n}{endl}\PY{p}{;}
  \PY{p}{\PYZob{}}
    \PY{k}{using} \PY{k+kt}{solver\PYZus{}t} \PY{o}{=} \PY{n}{solvers}\PY{o}{:}\PY{o}{:}\PY{n}{mpdata}\PY{o}{\PYZlt{}}\PY{k+kt}{ct\PYZus{}params\PYZus{}t}\PY{o}{\PYZgt{}}\PY{p}{;}
    \PY{k}{typename} \PY{k+kt}{solver\PYZus{}t}\PY{o}{:}\PY{o}{:}\PY{k+kt}{rt\PYZus{}params\PYZus{}t} \PY{n}{p}\PY{p}{;}
    \PY{n}{p}\PY{p}{.}\PY{n}{grid\PYZus{}size} \PY{o}{=} \PY{p}{\PYZob{}}\PY{n}{nx}\PY{p}{\PYZcb{}}\PY{p}{;}
    \PY{n}{concurr}\PY{o}{:}\PY{o}{:}\PY{n}{threads}\PY{o}{\PYZlt{}}\PY{k+kt}{solver\PYZus{}t}\PY{p}{,} \PY{n}{bcond}\PY{o}{:}\PY{o}{:}\PY{n}{cyclic}\PY{p}{,} \PY{n}{bcond}\PY{o}{:}\PY{o}{:}\PY{n}{cyclic}\PY{o}{\PYZgt{}} \PY{n}{run}\PY{p}{(}\PY{n}{p}\PY{p}{)}\PY{p}{;}
    \PY{n}{run}\PY{p}{.}\PY{n}{advectee}\PY{p}{(}\PY{p}{)} \PY{o}{=} \PY{l+m+mi}{0}\PY{p}{;}
    \PY{n}{run}\PY{p}{.}\PY{n}{advector}\PY{p}{(}\PY{p}{)} \PY{o}{=} \PY{l+m+mi}{0}\PY{p}{;}
    \PY{n}{run}\PY{p}{.}\PY{n}{advance}\PY{p}{(}\PY{n}{nt}\PY{p}{)}\PY{p}{;}
  \PY{p}{\PYZcb{}}

  \PY{c+c1}{// serial}
  \PY{n}{std}\PY{o}{:}\PY{o}{:}\PY{n}{cerr} \PY{o}{\PYZlt{}}\PY{o}{\PYZlt{}} \PY{l+s}{\PYZdq{}}\PY{l+s}{serial run}\PY{l+s}{\PYZdq{}} \PY{o}{\PYZlt{}}\PY{o}{\PYZlt{}} \PY{n}{std}\PY{o}{:}\PY{o}{:}\PY{n}{endl}\PY{p}{;}
  \PY{p}{\PYZob{}}
    \PY{k}{using} \PY{k+kt}{solver\PYZus{}t} \PY{o}{=} \PY{n}{solvers}\PY{o}{:}\PY{o}{:}\PY{n}{mpdata}\PY{o}{\PYZlt{}}\PY{k+kt}{ct\PYZus{}params\PYZus{}t}\PY{o}{\PYZgt{}}\PY{p}{;}
    \PY{k}{typename} \PY{k+kt}{solver\PYZus{}t}\PY{o}{:}\PY{o}{:}\PY{k+kt}{rt\PYZus{}params\PYZus{}t} \PY{n}{p}\PY{p}{;}
    \PY{n}{p}\PY{p}{.}\PY{n}{grid\PYZus{}size} \PY{o}{=} \PY{p}{\PYZob{}}\PY{n}{nx}\PY{p}{\PYZcb{}}\PY{p}{;}
    \PY{n}{concurr}\PY{o}{:}\PY{o}{:}\PY{n}{serial}\PY{o}{\PYZlt{}}\PY{k+kt}{solver\PYZus{}t}\PY{p}{,} \PY{n}{bcond}\PY{o}{:}\PY{o}{:}\PY{n}{cyclic}\PY{p}{,} \PY{n}{bcond}\PY{o}{:}\PY{o}{:}\PY{n}{cyclic}\PY{o}{\PYZgt{}} \PY{n}{run}\PY{p}{(}\PY{n}{p}\PY{p}{)}\PY{p}{;}
    \PY{n}{run}\PY{p}{.}\PY{n}{advectee}\PY{p}{(}\PY{p}{)} \PY{o}{=} \PY{l+m+mi}{0}\PY{p}{;}
    \PY{n}{run}\PY{p}{.}\PY{n}{advector}\PY{p}{(}\PY{p}{)} \PY{o}{=} \PY{l+m+mi}{0}\PY{p}{;}
    \PY{n}{run}\PY{p}{.}\PY{n}{advance}\PY{p}{(}\PY{n}{nt}\PY{p}{)}\PY{p}{;}
  \PY{p}{\PYZcb{}}
\PY{p}{\PYZcb{}}
\end{Verbatim}

%% file: any.cpp.tex
\begin{Verbatim}[commandchars=\\\{\}]
\PY{c+cm}{/** @file}
\PY{c+cm}{ * @copyright University of Warsaw}
\PY{c+cm}{ * @section LICENSE}
\PY{c+cm}{ * GPLv3+ (see the COPYING file or http://www.gnu.org/licenses/)}
\PY{c+cm}{ */}

\PY{c+cp}{\PYZsh{}}\PY{c+cp}{pragma once}

\PY{c+cp}{\PYZsh{}}\PY{c+cp}{include \PYZlt{}boost}\PY{c+cp}{/}\PY{c+cp}{ptr\PYZus{}container}\PY{c+cp}{/}\PY{c+cp}{ptr\PYZus{}vector.hpp\PYZgt{}}
\PY{c+cp}{\PYZsh{}}\PY{c+cp}{include \PYZlt{}libmpdata++}\PY{c+cp}{/}\PY{c+cp}{blitz.hpp\PYZgt{}}

\PY{c+cp}{\PYZsh{}}\PY{c+cp}{include \PYZlt{}libmpdata++}\PY{c+cp}{/}\PY{c+cp}{bcond}\PY{c+cp}{/}\PY{c+cp}{cyclic\PYZus{}1d.hpp\PYZgt{}}
\PY{c+cp}{\PYZsh{}}\PY{c+cp}{include \PYZlt{}libmpdata++}\PY{c+cp}{/}\PY{c+cp}{bcond}\PY{c+cp}{/}\PY{c+cp}{cyclic\PYZus{}2d.hpp\PYZgt{}}
\PY{c+cp}{\PYZsh{}}\PY{c+cp}{include \PYZlt{}libmpdata++}\PY{c+cp}{/}\PY{c+cp}{bcond}\PY{c+cp}{/}\PY{c+cp}{cyclic\PYZus{}3d.hpp\PYZgt{}}
\PY{c+cp}{\PYZsh{}}\PY{c+cp}{include \PYZlt{}libmpdata++}\PY{c+cp}{/}\PY{c+cp}{bcond}\PY{c+cp}{/}\PY{c+cp}{open\PYZus{}1d.hpp\PYZgt{}}
\PY{c+cp}{\PYZsh{}}\PY{c+cp}{include \PYZlt{}libmpdata++}\PY{c+cp}{/}\PY{c+cp}{bcond}\PY{c+cp}{/}\PY{c+cp}{open\PYZus{}2d.hpp\PYZgt{}}
\PY{c+cp}{\PYZsh{}}\PY{c+cp}{include \PYZlt{}libmpdata++}\PY{c+cp}{/}\PY{c+cp}{bcond}\PY{c+cp}{/}\PY{c+cp}{open\PYZus{}3d.hpp\PYZgt{}}
\PY{c+cp}{\PYZsh{}}\PY{c+cp}{include \PYZlt{}libmpdata++}\PY{c+cp}{/}\PY{c+cp}{bcond}\PY{c+cp}{/}\PY{c+cp}{polar\PYZus{}2d.hpp\PYZgt{}}

\PY{c+cp}{\PYZsh{}}\PY{c+cp}{include \PYZlt{}libmpdata++}\PY{c+cp}{/}\PY{c+cp}{concurr}\PY{c+cp}{/}\PY{c+cp}{detail}\PY{c+cp}{/}\PY{c+cp}{sharedmem.hpp\PYZgt{}}
\PY{c+cp}{\PYZsh{}}\PY{c+cp}{include \PYZlt{}libmpdata++}\PY{c+cp}{/}\PY{c+cp}{concurr}\PY{c+cp}{/}\PY{c+cp}{detail}\PY{c+cp}{/}\PY{c+cp}{timer.hpp\PYZgt{}}

\PY{c+c1}{// TODO: simplify 1d/2d/3d logic below or split into separate files?}

\PY{k}{namespace} \PY{n}{libmpdataxx}
\PY{p}{\PYZob{}}
  \PY{c+c1}{/// @brief concurr namespace}
  \PY{k}{namespace} \PY{n}{concurr}
  \PY{p}{\PYZob{}}
    \PY{k}{template} \PY{o}{\PYZlt{}}\PY{k}{typename} \PY{k+kt}{real\PYZus{}t}\PY{p}{,} \PY{k+kt}{int} \PY{n}{n\PYZus{}dims}\PY{o}{\PYZgt{}}
    \PY{k}{struct} \PY{n}{any}
    \PY{p}{\PYZob{}}
      \PY{k}{virtual} 
\PY{c+c1}{//\PYZlt{}listing\PYZhy{}1\PYZgt{}}
      \PY{k+kt}{void} \PY{n}{advance}\PY{p}{(}\PY{k+kt}{int}\PY{p}{)} 
\PY{c+c1}{//\PYZlt{}/listing\PYZhy{}1\PYZgt{}}
      \PY{p}{\PYZob{}} \PY{n}{assert}\PY{p}{(}\PY{n+nb}{false}\PY{p}{)}\PY{p}{;} \PY{k}{throw}\PY{p}{;} \PY{p}{\PYZcb{}}  

      \PY{k}{virtual} 
\PY{c+c1}{//\PYZlt{}listing\PYZhy{}2\PYZgt{}}
      \PY{n}{blitz}\PY{o}{:}\PY{o}{:}\PY{n}{Array}\PY{o}{\PYZlt{}}\PY{k+kt}{real\PYZus{}t}\PY{p}{,} \PY{n}{n\PYZus{}dims}\PY{o}{\PYZgt{}} \PY{n}{advectee}\PY{p}{(}\PY{k+kt}{int} \PY{n}{eqn} \PY{o}{=} \PY{l+m+mi}{0}\PY{p}{)}
\PY{c+c1}{//\PYZlt{}/listing\PYZhy{}2\PYZgt{}}
      \PY{p}{\PYZob{}} \PY{n}{assert}\PY{p}{(}\PY{n+nb}{false}\PY{p}{)}\PY{p}{;} \PY{k}{throw}\PY{p}{;} \PY{p}{\PYZcb{}}

      \PY{k}{virtual} 
\PY{c+c1}{//\PYZlt{}listing\PYZhy{}3\PYZgt{}}
      \PY{n}{blitz}\PY{o}{:}\PY{o}{:}\PY{n}{Array}\PY{o}{\PYZlt{}}\PY{k+kt}{real\PYZus{}t}\PY{p}{,} \PY{n}{n\PYZus{}dims}\PY{o}{\PYZgt{}} \PY{n}{advector}\PY{p}{(}\PY{k+kt}{int} \PY{n}{dim} \PY{o}{=} \PY{l+m+mi}{0}\PY{p}{)} 
\PY{c+c1}{//\PYZlt{}/listing\PYZhy{}3\PYZgt{}}
      \PY{p}{\PYZob{}} \PY{n}{assert}\PY{p}{(}\PY{n+nb}{false}\PY{p}{)}\PY{p}{;} \PY{k}{throw}\PY{p}{;} \PY{p}{\PYZcb{}}

      \PY{k}{virtual} 
\PY{c+c1}{//\PYZlt{}listing\PYZhy{}4\PYZgt{}}
      \PY{n}{blitz}\PY{o}{:}\PY{o}{:}\PY{n}{Array}\PY{o}{\PYZlt{}}\PY{k+kt}{real\PYZus{}t}\PY{p}{,} \PY{n}{n\PYZus{}dims}\PY{o}{\PYZgt{}} \PY{n}{g\PYZus{}factor}\PY{p}{(}\PY{p}{)} 
\PY{c+c1}{//\PYZlt{}/listing\PYZhy{}4\PYZgt{}}
      \PY{p}{\PYZob{}} \PY{n}{assert}\PY{p}{(}\PY{n+nb}{false}\PY{p}{)}\PY{p}{;} \PY{k}{throw}\PY{p}{;} \PY{p}{\PYZcb{}}

      \PY{k}{virtual} 
\PY{c+c1}{//\PYZlt{}listing\PYZhy{}5\PYZgt{}}
      \PY{k+kt}{bool} \PY{o}{*}\PY{n}{panic\PYZus{}ptr}\PY{p}{(}\PY{p}{)} 
\PY{c+c1}{//\PYZlt{}/listing\PYZhy{}5\PYZgt{}}
      \PY{p}{\PYZob{}} \PY{n}{assert}\PY{p}{(}\PY{n+nb}{false} \PY{o}{\PYZam{}}\PY{o}{\PYZam{}} \PY{l+s}{\PYZdq{}}\PY{l+s}{unimplemented!}\PY{l+s}{\PYZdq{}}\PY{p}{)}\PY{p}{;} \PY{k}{throw}\PY{p}{;} \PY{p}{\PYZcb{}}

      \PY{c+c1}{// dtor}
      \PY{k}{virtual} \PY{o}{\PYZti{}}\PY{n}{any}\PY{p}{(}\PY{p}{)} \PY{p}{\PYZob{}}\PY{p}{\PYZcb{}}
    \PY{p}{\PYZcb{}}\PY{p}{;}

    \PY{k}{namespace} \PY{n}{detail}
    \PY{p}{\PYZob{}}
      \PY{k}{template}\PY{o}{\PYZlt{}}
        \PY{k}{class} \PY{n+nc}{solver\PYZus{}t\PYZus{}}\PY{p}{,} 
        \PY{n}{bcond}\PY{o}{:}\PY{o}{:}\PY{n}{bcond\PYZus{}e} \PY{n}{bcxl}\PY{p}{,} \PY{n}{bcond}\PY{o}{:}\PY{o}{:}\PY{n}{bcond\PYZus{}e} \PY{n}{bcxr}\PY{p}{,}
        \PY{n}{bcond}\PY{o}{:}\PY{o}{:}\PY{n}{bcond\PYZus{}e} \PY{n}{bcyl}\PY{p}{,} \PY{n}{bcond}\PY{o}{:}\PY{o}{:}\PY{n}{bcond\PYZus{}e} \PY{n}{bcyr}\PY{p}{,}
        \PY{n}{bcond}\PY{o}{:}\PY{o}{:}\PY{n}{bcond\PYZus{}e} \PY{n}{bczl}\PY{p}{,} \PY{n}{bcond}\PY{o}{:}\PY{o}{:}\PY{n}{bcond\PYZus{}e} \PY{n}{bczr}
      \PY{o}{\PYZgt{}}
      \PY{k}{class} \PY{n+nc}{concurr\PYZus{}common} \PY{o}{:} \PY{k}{public} \PY{n}{any}\PY{o}{\PYZlt{}}\PY{k}{typename} \PY{n}{solver\PYZus{}t\PYZus{}}\PY{o}{:}\PY{o}{:}\PY{k+kt}{real\PYZus{}t}\PY{p}{,} \PY{n}{solver\PYZus{}t\PYZus{}}\PY{o}{:}\PY{o}{:}\PY{n}{n\PYZus{}dims}\PY{o}{\PYZgt{}}
      \PY{p}{\PYZob{}}
        \PY{n+nl}{public:}

        \PY{k}{typedef} \PY{n}{solver\PYZus{}t\PYZus{}} \PY{k+kt}{solver\PYZus{}t}\PY{p}{;}

        \PY{n+nl}{private:}

        \PY{c+c1}{// helper methods to define subdomain ranges}
	\PY{k+kt}{int} \PY{n+nf}{min}\PY{p}{(}\PY{k}{const} \PY{k+kt}{int} \PY{o}{\PYZam{}}\PY{n}{grid\PYZus{}size}\PY{p}{,} \PY{k}{const} \PY{k+kt}{int} \PY{o}{\PYZam{}}\PY{n}{rank}\PY{p}{,} \PY{k}{const} \PY{k+kt}{int} \PY{o}{\PYZam{}}\PY{n}{size}\PY{p}{)} 
	\PY{p}{\PYZob{}} 
	  \PY{k}{return} \PY{n}{rank} \PY{o}{*} \PY{n}{grid\PYZus{}size} \PY{o}{/} \PY{n}{size}\PY{p}{;} 
	\PY{p}{\PYZcb{}}

	\PY{k+kt}{int} \PY{n+nf}{max}\PY{p}{(}\PY{k}{const} \PY{k+kt}{int} \PY{o}{\PYZam{}}\PY{n}{grid\PYZus{}size}\PY{p}{,} \PY{k}{const} \PY{k+kt}{int} \PY{o}{\PYZam{}}\PY{n}{rank}\PY{p}{,} \PY{k}{const} \PY{k+kt}{int} \PY{o}{\PYZam{}}\PY{n}{size}\PY{p}{)} 
	\PY{p}{\PYZob{}} 
          \PY{k}{return} \PY{n}{min}\PY{p}{(}\PY{n}{grid\PYZus{}size}\PY{p}{,} \PY{n}{rank} \PY{o}{+} \PY{l+m+mi}{1}\PY{p}{,} \PY{n}{size}\PY{p}{)} \PY{o}{\PYZhy{}} \PY{l+m+mi}{1}\PY{p}{;} 
	\PY{p}{\PYZcb{}}

	\PY{n+nl}{protected:}

        \PY{c+c1}{// (cannot be nested due to tempaltes)}
        \PY{k}{typedef} \PY{n}{sharedmem}\PY{o}{\PYZlt{}}
          \PY{k}{typename} \PY{k+kt}{solver\PYZus{}t}\PY{o}{:}\PY{o}{:}\PY{k+kt}{real\PYZus{}t}\PY{p}{,}
          \PY{k+kt}{solver\PYZus{}t}\PY{o}{:}\PY{o}{:}\PY{n}{n\PYZus{}dims}\PY{p}{,}
          \PY{k+kt}{solver\PYZus{}t}\PY{o}{:}\PY{o}{:}\PY{n}{n\PYZus{}tlev}
        \PY{o}{\PYZgt{}} \PY{k+kt}{mem\PYZus{}t}\PY{p}{;}

	\PY{c+c1}{// member fields}
	\PY{n}{boost}\PY{o}{:}\PY{o}{:}\PY{n}{ptr\PYZus{}vector}\PY{o}{\PYZlt{}}\PY{k+kt}{solver\PYZus{}t}\PY{o}{\PYZgt{}} \PY{n}{algos}\PY{p}{;} 
        \PY{n}{std}\PY{o}{:}\PY{o}{:}\PY{n}{unique\PYZus{}ptr}\PY{o}{\PYZlt{}}\PY{k+kt}{mem\PYZus{}t}\PY{o}{\PYZgt{}} \PY{n}{mem}\PY{p}{;}
        \PY{n}{timer} \PY{n}{tmr}\PY{p}{;}

	\PY{n+nl}{public:}

        \PY{k}{typedef} \PY{k}{typename} \PY{k+kt}{solver\PYZus{}t}\PY{o}{:}\PY{o}{:}\PY{k+kt}{real\PYZus{}t} \PY{k+kt}{real\PYZus{}t}\PY{p}{;}

        \PY{c+c1}{// dtor}
	\PY{k}{virtual} \PY{o}{\PYZti{}}\PY{n}{concurr\PYZus{}common}\PY{p}{(}\PY{p}{)}
        \PY{p}{\PYZob{}}
          \PY{n}{tmr}\PY{p}{.}\PY{n}{print}\PY{p}{(}\PY{p}{)}\PY{p}{;}
        \PY{p}{\PYZcb{}}

	\PY{c+c1}{// ctor}
	\PY{n}{concurr\PYZus{}common}\PY{p}{(}
	  \PY{k}{const} \PY{k}{typename} \PY{k+kt}{solver\PYZus{}t}\PY{o}{:}\PY{o}{:}\PY{k+kt}{rt\PYZus{}params\PYZus{}t} \PY{o}{\PYZam{}}\PY{n}{p}\PY{p}{,}
          \PY{k+kt}{mem\PYZus{}t} \PY{o}{*}\PY{n}{mem\PYZus{}p}\PY{p}{,}
	  \PY{k}{const} \PY{k+kt}{int} \PY{o}{\PYZam{}}\PY{n}{size}
	\PY{p}{)} \PY{p}{\PYZob{}}
          \PY{c+c1}{// allocate the memory to be shared by multiple threads}
          \PY{n}{mem}\PY{p}{.}\PY{n}{reset}\PY{p}{(}\PY{n}{mem\PYZus{}p}\PY{p}{)}\PY{p}{;}
	  \PY{k+kt}{solver\PYZus{}t}\PY{o}{:}\PY{o}{:}\PY{n}{alloc}\PY{p}{(}\PY{n}{mem}\PY{p}{.}\PY{n}{get}\PY{p}{(}\PY{p}{)}\PY{p}{,} \PY{n}{p}\PY{p}{)}\PY{p}{;}

          \PY{c+c1}{// allocate per\PYZhy{}thread structures}
          \PY{n}{init}\PY{p}{(}\PY{n}{p}\PY{p}{,} \PY{n}{p}\PY{p}{.}\PY{n}{grid\PYZus{}size}\PY{p}{,} \PY{n}{size}\PY{p}{)}\PY{p}{;} 
        \PY{p}{\PYZcb{}}

        \PY{n+nl}{private:}
 
        \PY{c+c1}{// 1D version}
        \PY{k+kt}{void} \PY{n}{init}\PY{p}{(}
          \PY{k}{const} \PY{k}{typename} \PY{k+kt}{solver\PYZus{}t}\PY{o}{:}\PY{o}{:}\PY{k+kt}{rt\PYZus{}params\PYZus{}t} \PY{o}{\PYZam{}}\PY{n}{p}\PY{p}{,}
          \PY{k}{const} \PY{n}{std}\PY{o}{:}\PY{o}{:}\PY{n}{array}\PY{o}{\PYZlt{}}\PY{k+kt}{int}\PY{p}{,} \PY{l+m+mi}{1}\PY{o}{\PYZgt{}} \PY{o}{\PYZam{}}\PY{n}{grid\PYZus{}size}\PY{p}{,} \PY{k}{const} \PY{k+kt}{int} \PY{o}{\PYZam{}}\PY{n}{n0}
        \PY{p}{)}
        \PY{p}{\PYZob{}}
          \PY{k}{typename} \PY{k+kt}{solver\PYZus{}t}\PY{o}{:}\PY{o}{:}\PY{k+kt}{bcp\PYZus{}t} \PY{n}{bxl}\PY{p}{,} \PY{n}{bxr}\PY{p}{,} \PY{n}{shrdl}\PY{p}{,} \PY{n}{shrdr}\PY{p}{;}

          \PY{k}{switch} \PY{p}{(}\PY{n}{bcxl}\PY{p}{)} \PY{c+c1}{// TODO: make a function that does it}
          \PY{p}{\PYZob{}}
            \PY{k}{case} \PY{n}{bcond}:\PY{o}{:}\PY{n}{cyclic}\PY{o}{:} 
              \PY{n}{bxl}\PY{p}{.}\PY{n}{reset}\PY{p}{(}\PY{k}{new} \PY{n}{bcond}\PY{o}{:}\PY{o}{:}\PY{n}{cyclic\PYZus{}left\PYZus{}1d}\PY{o}{\PYZlt{}}\PY{k+kt}{real\PYZus{}t}\PY{o}{\PYZgt{}}\PY{p}{(}\PY{k+kt}{rng\PYZus{}t}\PY{p}{(}\PY{l+m+mi}{0}\PY{p}{,} \PY{n}{grid\PYZus{}size}\PY{p}{[}\PY{l+m+mi}{0}\PY{p}{]}\PY{o}{\PYZhy{}}\PY{l+m+mi}{1}\PY{p}{)}\PY{p}{,} \PY{k+kt}{solver\PYZus{}t}\PY{o}{:}\PY{o}{:}\PY{n}{halo}\PY{p}{)}\PY{p}{)}\PY{p}{;}
              \PY{k}{break}\PY{p}{;}
            \PY{k}{case} \PY{n}{bcond}:\PY{o}{:}\PY{n}{open}\PY{o}{:} 
              \PY{n}{bxl}\PY{p}{.}\PY{n}{reset}\PY{p}{(}\PY{k}{new} \PY{n}{bcond}\PY{o}{:}\PY{o}{:}\PY{n}{open\PYZus{}left\PYZus{}1d}\PY{o}{\PYZlt{}}\PY{k+kt}{real\PYZus{}t}\PY{o}{\PYZgt{}}\PY{p}{(}\PY{k+kt}{rng\PYZus{}t}\PY{p}{(}\PY{l+m+mi}{0}\PY{p}{,} \PY{n}{grid\PYZus{}size}\PY{p}{[}\PY{l+m+mi}{0}\PY{p}{]}\PY{o}{\PYZhy{}}\PY{l+m+mi}{1}\PY{p}{)}\PY{p}{,} \PY{k+kt}{solver\PYZus{}t}\PY{o}{:}\PY{o}{:}\PY{n}{halo}\PY{p}{)}\PY{p}{)}\PY{p}{;}
              \PY{k}{break}\PY{p}{;}
            \PY{n+nl}{default:} \PY{n}{assert}\PY{p}{(}\PY{n+nb}{false}\PY{p}{)}\PY{p}{;}
          \PY{p}{\PYZcb{}}

          \PY{k}{switch} \PY{p}{(}\PY{n}{bcxr}\PY{p}{)} \PY{c+c1}{// TODO: make a function that does it}
          \PY{p}{\PYZob{}}
            \PY{k}{case} \PY{n}{bcond}:\PY{o}{:}\PY{n}{cyclic}\PY{o}{:}
	      \PY{n}{bxr}\PY{p}{.}\PY{n}{reset}\PY{p}{(}\PY{k}{new} \PY{n}{bcond}\PY{o}{:}\PY{o}{:}\PY{n}{cyclic\PYZus{}rght\PYZus{}1d}\PY{o}{\PYZlt{}}\PY{k+kt}{real\PYZus{}t}\PY{o}{\PYZgt{}}\PY{p}{(}\PY{k+kt}{rng\PYZus{}t}\PY{p}{(}\PY{l+m+mi}{0}\PY{p}{,} \PY{n}{grid\PYZus{}size}\PY{p}{[}\PY{l+m+mi}{0}\PY{p}{]}\PY{o}{\PYZhy{}}\PY{l+m+mi}{1}\PY{p}{)}\PY{p}{,} \PY{k+kt}{solver\PYZus{}t}\PY{o}{:}\PY{o}{:}\PY{n}{halo}\PY{p}{)}\PY{p}{)}\PY{p}{;}
              \PY{k}{break}\PY{p}{;}
            \PY{k}{case} \PY{n}{bcond}:\PY{o}{:}\PY{n}{open}\PY{o}{:} 
              \PY{n}{bxr}\PY{p}{.}\PY{n}{reset}\PY{p}{(}\PY{k}{new} \PY{n}{bcond}\PY{o}{:}\PY{o}{:}\PY{n}{open\PYZus{}rght\PYZus{}1d}\PY{o}{\PYZlt{}}\PY{k+kt}{real\PYZus{}t}\PY{o}{\PYZgt{}}\PY{p}{(}\PY{k+kt}{rng\PYZus{}t}\PY{p}{(}\PY{l+m+mi}{0}\PY{p}{,} \PY{n}{grid\PYZus{}size}\PY{p}{[}\PY{l+m+mi}{0}\PY{p}{]}\PY{o}{\PYZhy{}}\PY{l+m+mi}{1}\PY{p}{)}\PY{p}{,} \PY{k+kt}{solver\PYZus{}t}\PY{o}{:}\PY{o}{:}\PY{n}{halo}\PY{p}{)}\PY{p}{)}\PY{p}{;}
              \PY{k}{break}\PY{p}{;}
            \PY{n+nl}{default:} \PY{n}{assert}\PY{p}{(}\PY{n+nb}{false}\PY{p}{)}\PY{p}{;}
          \PY{p}{\PYZcb{}}

	  \PY{k}{for} \PY{p}{(}\PY{k+kt}{int} \PY{n}{i0} \PY{o}{=} \PY{l+m+mi}{0}\PY{p}{;} \PY{n}{i0} \PY{o}{\PYZlt{}} \PY{n}{n0}\PY{p}{;} \PY{o}{+}\PY{o}{+}\PY{n}{i0}\PY{p}{)} 
          \PY{p}{\PYZob{}}
            \PY{n}{shrdl}\PY{p}{.}\PY{n}{reset}\PY{p}{(}\PY{k}{new} \PY{n}{bcond}\PY{o}{:}\PY{o}{:}\PY{n}{shared}\PY{o}{\PYZlt{}}\PY{k+kt}{real\PYZus{}t}\PY{o}{\PYZgt{}}\PY{p}{(}\PY{p}{)}\PY{p}{)}\PY{p}{;}
            \PY{n}{shrdr}\PY{p}{.}\PY{n}{reset}\PY{p}{(}\PY{k}{new} \PY{n}{bcond}\PY{o}{:}\PY{o}{:}\PY{n}{shared}\PY{o}{\PYZlt{}}\PY{k+kt}{real\PYZus{}t}\PY{o}{\PYZgt{}}\PY{p}{(}\PY{p}{)}\PY{p}{)}\PY{p}{;}
            \PY{k}{const} \PY{k+kt}{rng\PYZus{}t} \PY{n+nf}{i}\PY{p}{(}\PY{n}{min}\PY{p}{(}\PY{n}{grid\PYZus{}size}\PY{p}{[}\PY{l+m+mi}{0}\PY{p}{]}\PY{p}{,} \PY{n}{i0}\PY{p}{,} \PY{n}{n0}\PY{p}{)}\PY{p}{,} \PY{n}{max}\PY{p}{(}\PY{n}{grid\PYZus{}size}\PY{p}{[}\PY{l+m+mi}{0}\PY{p}{]}\PY{p}{,} \PY{n}{i0}\PY{p}{,} \PY{n}{n0}\PY{p}{)}\PY{p}{)}\PY{p}{;} 
	    \PY{n}{algos}\PY{p}{.}\PY{n}{push\PYZus{}back}\PY{p}{(}
              \PY{k}{new} \PY{k+kt}{solver\PYZus{}t}\PY{p}{(}
                \PY{k}{typename} \PY{k+kt}{solver\PYZus{}t}\PY{o}{:}\PY{o}{:}\PY{k+kt}{ctor\PYZus{}args\PYZus{}t}\PY{p}{(}\PY{p}{\PYZob{}}
		  \PY{n}{mem}\PY{p}{.}\PY{n}{get}\PY{p}{(}\PY{p}{)}\PY{p}{,} 
		  \PY{n}{i0} \PY{o}{=}\PY{o}{=} \PY{l+m+mi}{0}      \PY{o}{?} \PY{n}{bxl} \PY{o}{:} \PY{n}{shrdl}\PY{p}{,}
		  \PY{n}{i0} \PY{o}{=}\PY{o}{=} \PY{n}{n0} \PY{o}{\PYZhy{}} \PY{l+m+mi}{1} \PY{o}{?} \PY{n}{bxr} \PY{o}{:} \PY{n}{shrdr}\PY{p}{,}
		  \PY{n}{i}
                \PY{p}{\PYZcb{}}\PY{p}{)}\PY{p}{,} 
                \PY{n}{p}
              \PY{p}{)}
            \PY{p}{)}\PY{p}{;}
          \PY{p}{\PYZcb{}}
	\PY{p}{\PYZcb{}}

        \PY{c+c1}{// 2D version}
        \PY{k+kt}{void} \PY{n}{init}\PY{p}{(}
          \PY{k}{const} \PY{k}{typename} \PY{k+kt}{solver\PYZus{}t}\PY{o}{:}\PY{o}{:}\PY{k+kt}{rt\PYZus{}params\PYZus{}t} \PY{o}{\PYZam{}}\PY{n}{p}\PY{p}{,}
	  \PY{k}{const} \PY{n}{std}\PY{o}{:}\PY{o}{:}\PY{n}{array}\PY{o}{\PYZlt{}}\PY{k+kt}{int}\PY{p}{,} \PY{l+m+mi}{2}\PY{o}{\PYZgt{}} \PY{o}{\PYZam{}}\PY{n}{grid\PYZus{}size}\PY{p}{,} 
          \PY{k}{const} \PY{k+kt}{int} \PY{o}{\PYZam{}}\PY{n}{n0}\PY{p}{,} \PY{k}{const} \PY{k+kt}{int} \PY{o}{\PYZam{}}\PY{n}{n1} \PY{o}{=} \PY{l+m+mi}{1}
        \PY{p}{)} \PY{p}{\PYZob{}}
\PY{c+c1}{// TODO: assert parallelisation in the right dimensions! (blitz::assertContiguous)}
          \PY{k}{for} \PY{p}{(}\PY{k+kt}{int} \PY{n}{i0} \PY{o}{=} \PY{l+m+mi}{0}\PY{p}{;} \PY{n}{i0} \PY{o}{\PYZlt{}} \PY{n}{n0}\PY{p}{;} \PY{o}{+}\PY{o}{+}\PY{n}{i0}\PY{p}{)} 
          \PY{p}{\PYZob{}}
            \PY{k}{for} \PY{p}{(}\PY{k+kt}{int} \PY{n}{i1} \PY{o}{=} \PY{l+m+mi}{0}\PY{p}{;} \PY{n}{i1} \PY{o}{\PYZlt{}} \PY{n}{n1}\PY{p}{;} \PY{o}{+}\PY{o}{+}\PY{n}{i1}\PY{p}{)} 
            \PY{p}{\PYZob{}}
	      \PY{k}{typename} \PY{k+kt}{solver\PYZus{}t}\PY{o}{:}\PY{o}{:}\PY{k+kt}{bcp\PYZus{}t} \PY{n}{bxl}\PY{p}{,} \PY{n}{bxr}\PY{p}{,} \PY{n}{byl}\PY{p}{,} \PY{n}{byr}\PY{p}{,} \PY{n}{shrdl}\PY{p}{,} \PY{n}{shrdr}\PY{p}{;}

              \PY{c+c1}{// dim 1, left}
              \PY{k}{switch} \PY{p}{(}\PY{n}{bcxl}\PY{p}{)} \PY{c+c1}{// TODO: make a function taht does it}
              \PY{p}{\PYZob{}}
	        \PY{k}{case} \PY{n}{bcond}:\PY{o}{:}\PY{n}{cyclic}\PY{o}{:}
		  \PY{n}{bxl}\PY{p}{.}\PY{n}{reset}\PY{p}{(}\PY{k}{new} \PY{n}{bcond}\PY{o}{:}\PY{o}{:}\PY{n}{cyclic\PYZus{}left\PYZus{}2d}\PY{o}{\PYZlt{}}\PY{l+m+mi}{0}\PY{p}{,} \PY{k+kt}{real\PYZus{}t}\PY{o}{\PYZgt{}}\PY{p}{(}\PY{k+kt}{rng\PYZus{}t}\PY{p}{(}\PY{l+m+mi}{0}\PY{p}{,} \PY{n}{grid\PYZus{}size}\PY{p}{[}\PY{l+m+mi}{0}\PY{p}{]}\PY{o}{\PYZhy{}}\PY{l+m+mi}{1}\PY{p}{)}\PY{p}{,} \PY{k+kt}{solver\PYZus{}t}\PY{o}{:}\PY{o}{:}\PY{n}{halo}\PY{p}{)}\PY{p}{)}\PY{p}{;}
                  \PY{k}{break}\PY{p}{;}
                \PY{k}{case} \PY{n}{bcond}:\PY{o}{:}\PY{n}{open}\PY{o}{:}
	          \PY{n}{bxl}\PY{p}{.}\PY{n}{reset}\PY{p}{(}\PY{k}{new} \PY{n}{bcond}\PY{o}{:}\PY{o}{:}\PY{n}{open\PYZus{}left\PYZus{}2d}\PY{o}{\PYZlt{}}\PY{l+m+mi}{0}\PY{p}{,} \PY{k+kt}{real\PYZus{}t}\PY{o}{\PYZgt{}}\PY{p}{(}\PY{k+kt}{rng\PYZus{}t}\PY{p}{(}\PY{l+m+mi}{0}\PY{p}{,} \PY{n}{grid\PYZus{}size}\PY{p}{[}\PY{l+m+mi}{0}\PY{p}{]}\PY{o}{\PYZhy{}}\PY{l+m+mi}{1}\PY{p}{)}\PY{p}{,} \PY{k+kt}{solver\PYZus{}t}\PY{o}{:}\PY{o}{:}\PY{n}{halo}\PY{p}{)}\PY{p}{)}\PY{p}{;}
                  \PY{k}{break}\PY{p}{;}
	        \PY{n+nl}{default:} \PY{n}{assert}\PY{p}{(}\PY{n+nb}{false}\PY{p}{)}\PY{p}{;}
              \PY{p}{\PYZcb{}}

              \PY{c+c1}{// dim 1, rght}
	      \PY{k}{switch} \PY{p}{(}\PY{n}{bcxr}\PY{p}{)} \PY{c+c1}{// TODO: make a function taht does it}
              \PY{p}{\PYZob{}}
                \PY{k}{case} \PY{n}{bcond}:\PY{o}{:}\PY{n}{cyclic}\PY{o}{:}
                  \PY{n}{bxr}\PY{p}{.}\PY{n}{reset}\PY{p}{(}\PY{k}{new} \PY{n}{bcond}\PY{o}{:}\PY{o}{:}\PY{n}{cyclic\PYZus{}rght\PYZus{}2d}\PY{o}{\PYZlt{}}\PY{l+m+mi}{0}\PY{p}{,} \PY{k+kt}{real\PYZus{}t}\PY{o}{\PYZgt{}}\PY{p}{(}\PY{k+kt}{rng\PYZus{}t}\PY{p}{(}\PY{l+m+mi}{0}\PY{p}{,} \PY{n}{grid\PYZus{}size}\PY{p}{[}\PY{l+m+mi}{0}\PY{p}{]}\PY{o}{\PYZhy{}}\PY{l+m+mi}{1}\PY{p}{)}\PY{p}{,} \PY{k+kt}{solver\PYZus{}t}\PY{o}{:}\PY{o}{:}\PY{n}{halo}\PY{p}{)}\PY{p}{)}\PY{p}{;}
                  \PY{k}{break}\PY{p}{;}
                \PY{k}{case} \PY{n}{bcond}:\PY{o}{:}\PY{n}{open}\PY{o}{:}
                  \PY{n}{bxr}\PY{p}{.}\PY{n}{reset}\PY{p}{(}\PY{k}{new} \PY{n}{bcond}\PY{o}{:}\PY{o}{:}\PY{n}{open\PYZus{}rght\PYZus{}2d}\PY{o}{\PYZlt{}}\PY{l+m+mi}{0}\PY{p}{,} \PY{k+kt}{real\PYZus{}t}\PY{o}{\PYZgt{}}\PY{p}{(}\PY{k+kt}{rng\PYZus{}t}\PY{p}{(}\PY{l+m+mi}{0}\PY{p}{,} \PY{n}{grid\PYZus{}size}\PY{p}{[}\PY{l+m+mi}{0}\PY{p}{]}\PY{o}{\PYZhy{}}\PY{l+m+mi}{1}\PY{p}{)}\PY{p}{,} \PY{k+kt}{solver\PYZus{}t}\PY{o}{:}\PY{o}{:}\PY{n}{halo}\PY{p}{)}\PY{p}{)}\PY{p}{;}
                  \PY{k}{break}\PY{p}{;}
	        \PY{n+nl}{defalt:} \PY{n}{assert}\PY{p}{(}\PY{n+nb}{false}\PY{p}{)}\PY{p}{;}
              \PY{p}{\PYZcb{}}

              \PY{c+c1}{// dim 2, left}
	      \PY{k}{switch} \PY{p}{(}\PY{n}{bcyl}\PY{p}{)} \PY{c+c1}{// TODO: make a function taht does it}
              \PY{p}{\PYZob{}}
                \PY{k}{case} \PY{n}{bcond}:\PY{o}{:}\PY{n}{cyclic}\PY{o}{:}
                  \PY{n}{byl}\PY{p}{.}\PY{n}{reset}\PY{p}{(}\PY{k}{new} \PY{n}{bcond}\PY{o}{:}\PY{o}{:}\PY{n}{cyclic\PYZus{}left\PYZus{}2d}\PY{o}{\PYZlt{}}\PY{l+m+mi}{1}\PY{p}{,} \PY{k+kt}{real\PYZus{}t}\PY{o}{\PYZgt{}}\PY{p}{(}\PY{k+kt}{rng\PYZus{}t}\PY{p}{(}\PY{l+m+mi}{0}\PY{p}{,} \PY{n}{grid\PYZus{}size}\PY{p}{[}\PY{l+m+mi}{1}\PY{p}{]}\PY{o}{\PYZhy{}}\PY{l+m+mi}{1}\PY{p}{)}\PY{p}{,} \PY{k+kt}{solver\PYZus{}t}\PY{o}{:}\PY{o}{:}\PY{n}{halo}\PY{p}{)}\PY{p}{)}\PY{p}{;}
                  \PY{k}{break}\PY{p}{;}
                \PY{k}{case} \PY{n}{bcond}:\PY{o}{:}\PY{n}{polar}\PY{o}{:}
                  \PY{n}{byl}\PY{p}{.}\PY{n}{reset}\PY{p}{(}\PY{k}{new} \PY{n}{bcond}\PY{o}{:}\PY{o}{:}\PY{n}{polar\PYZus{}left\PYZus{}2d}\PY{o}{\PYZlt{}}\PY{l+m+mi}{1}\PY{p}{,} \PY{k+kt}{real\PYZus{}t}\PY{o}{\PYZgt{}}\PY{p}{(}\PY{k+kt}{rng\PYZus{}t}\PY{p}{(}\PY{l+m+mi}{0}\PY{p}{,} \PY{n}{grid\PYZus{}size}\PY{p}{[}\PY{l+m+mi}{1}\PY{p}{]}\PY{o}{\PYZhy{}}\PY{l+m+mi}{1}\PY{p}{)}\PY{p}{,}
                            \PY{k+kt}{solver\PYZus{}t}\PY{o}{:}\PY{o}{:}\PY{n}{halo}\PY{p}{,}
                            \PY{p}{(}\PY{n}{grid\PYZus{}size}\PY{p}{[}\PY{l+m+mi}{0}\PY{p}{]} \PY{o}{\PYZhy{}} \PY{l+m+mi}{1}\PY{p}{)} \PY{o}{/} \PY{l+m+mi}{2}\PY{p}{)}\PY{p}{)}\PY{p}{;}
                  \PY{k}{break}\PY{p}{;}
                \PY{k}{case} \PY{n}{bcond}:\PY{o}{:}\PY{n}{open}\PY{o}{:}
                  \PY{n}{byl}\PY{p}{.}\PY{n}{reset}\PY{p}{(}\PY{k}{new} \PY{n}{bcond}\PY{o}{:}\PY{o}{:}\PY{n}{open\PYZus{}left\PYZus{}2d}\PY{o}{\PYZlt{}}\PY{l+m+mi}{1}\PY{p}{,} \PY{k+kt}{real\PYZus{}t}\PY{o}{\PYZgt{}}\PY{p}{(}\PY{k+kt}{rng\PYZus{}t}\PY{p}{(}\PY{l+m+mi}{0}\PY{p}{,} \PY{n}{grid\PYZus{}size}\PY{p}{[}\PY{l+m+mi}{1}\PY{p}{]}\PY{o}{\PYZhy{}}\PY{l+m+mi}{1}\PY{p}{)}\PY{p}{,} \PY{k+kt}{solver\PYZus{}t}\PY{o}{:}\PY{o}{:}\PY{n}{halo}\PY{p}{)}\PY{p}{)}\PY{p}{;}
                  \PY{k}{break}\PY{p}{;}
                \PY{n+nl}{default:} \PY{n}{assert}\PY{p}{(}\PY{n+nb}{false}\PY{p}{)}\PY{p}{;}
              \PY{p}{\PYZcb{}}

              \PY{c+c1}{// dim 2, rght}
	      \PY{k}{switch} \PY{p}{(}\PY{n}{bcyr}\PY{p}{)} \PY{c+c1}{// TODO: make a function taht does it}
              \PY{p}{\PYZob{}}
                \PY{k}{case} \PY{n}{bcond}:\PY{o}{:}\PY{n}{cyclic}\PY{o}{:}
                  \PY{n}{byr}\PY{p}{.}\PY{n}{reset}\PY{p}{(}\PY{k}{new} \PY{n}{bcond}\PY{o}{:}\PY{o}{:}\PY{n}{cyclic\PYZus{}rght\PYZus{}2d}\PY{o}{\PYZlt{}}\PY{l+m+mi}{1}\PY{p}{,} \PY{k+kt}{real\PYZus{}t}\PY{o}{\PYZgt{}}\PY{p}{(}\PY{k+kt}{rng\PYZus{}t}\PY{p}{(}\PY{l+m+mi}{0}\PY{p}{,} \PY{n}{grid\PYZus{}size}\PY{p}{[}\PY{l+m+mi}{1}\PY{p}{]}\PY{o}{\PYZhy{}}\PY{l+m+mi}{1}\PY{p}{)}\PY{p}{,} \PY{k+kt}{solver\PYZus{}t}\PY{o}{:}\PY{o}{:}\PY{n}{halo}\PY{p}{)}\PY{p}{)}\PY{p}{;}
                  \PY{k}{break}\PY{p}{;}
                \PY{k}{case} \PY{n}{bcond}:\PY{o}{:}\PY{n}{polar}\PY{o}{:}
                  \PY{n}{byr}\PY{p}{.}\PY{n}{reset}\PY{p}{(}\PY{k}{new} \PY{n}{bcond}\PY{o}{:}\PY{o}{:}\PY{n}{polar\PYZus{}rght\PYZus{}2d}\PY{o}{\PYZlt{}}\PY{l+m+mi}{1}\PY{p}{,} \PY{k+kt}{real\PYZus{}t}\PY{o}{\PYZgt{}}\PY{p}{(}\PY{k+kt}{rng\PYZus{}t}\PY{p}{(}\PY{l+m+mi}{0}\PY{p}{,} \PY{n}{grid\PYZus{}size}\PY{p}{[}\PY{l+m+mi}{1}\PY{p}{]}\PY{o}{\PYZhy{}}\PY{l+m+mi}{1}\PY{p}{)}\PY{p}{,}
                            \PY{k+kt}{solver\PYZus{}t}\PY{o}{:}\PY{o}{:}\PY{n}{halo}\PY{p}{,}
                            \PY{p}{(}\PY{n}{grid\PYZus{}size}\PY{p}{[}\PY{l+m+mi}{0}\PY{p}{]} \PY{o}{\PYZhy{}} \PY{l+m+mi}{1}\PY{p}{)} \PY{o}{/} \PY{l+m+mi}{2}\PY{p}{)}\PY{p}{)}\PY{p}{;}
                  \PY{k}{break}\PY{p}{;}
                \PY{k}{case} \PY{n}{bcond}:\PY{o}{:}\PY{n}{open}\PY{o}{:}
                  \PY{n}{byr}\PY{p}{.}\PY{n}{reset}\PY{p}{(}\PY{k}{new} \PY{n}{bcond}\PY{o}{:}\PY{o}{:}\PY{n}{open\PYZus{}rght\PYZus{}2d}\PY{o}{\PYZlt{}}\PY{l+m+mi}{1}\PY{p}{,} \PY{k+kt}{real\PYZus{}t}\PY{o}{\PYZgt{}}\PY{p}{(}\PY{k+kt}{rng\PYZus{}t}\PY{p}{(}\PY{l+m+mi}{0}\PY{p}{,} \PY{n}{grid\PYZus{}size}\PY{p}{[}\PY{l+m+mi}{1}\PY{p}{]}\PY{o}{\PYZhy{}}\PY{l+m+mi}{1}\PY{p}{)}\PY{p}{,} \PY{k+kt}{solver\PYZus{}t}\PY{o}{:}\PY{o}{:}\PY{n}{halo}\PY{p}{)}\PY{p}{)}\PY{p}{;}
                  \PY{k}{break}\PY{p}{;}
                \PY{n+nl}{default:} \PY{n}{assert}\PY{p}{(}\PY{n+nb}{false}\PY{p}{)}\PY{p}{;}
              \PY{p}{\PYZcb{}}

              \PY{n}{shrdl}\PY{p}{.}\PY{n}{reset}\PY{p}{(}\PY{k}{new} \PY{n}{bcond}\PY{o}{:}\PY{o}{:}\PY{n}{shared}\PY{o}{\PYZlt{}}\PY{k+kt}{real\PYZus{}t}\PY{o}{\PYZgt{}}\PY{p}{(}\PY{p}{)}\PY{p}{)}\PY{p}{;} \PY{c+c1}{// TODO: shrdy if n1 != 1}
              \PY{n}{shrdr}\PY{p}{.}\PY{n}{reset}\PY{p}{(}\PY{k}{new} \PY{n}{bcond}\PY{o}{:}\PY{o}{:}\PY{n}{shared}\PY{o}{\PYZlt{}}\PY{k+kt}{real\PYZus{}t}\PY{o}{\PYZgt{}}\PY{p}{(}\PY{p}{)}\PY{p}{)}\PY{p}{;} \PY{c+c1}{// TODO: shrdy if n1 != 1}

              \PY{k}{const} \PY{k+kt}{rng\PYZus{}t} 
                \PY{n+nf}{i}\PY{p}{(} \PY{n}{min}\PY{p}{(}\PY{n}{grid\PYZus{}size}\PY{p}{[}\PY{l+m+mi}{0}\PY{p}{]}\PY{p}{,} \PY{n}{i0}\PY{p}{,} \PY{n}{n0}\PY{p}{)}\PY{p}{,} \PY{n}{max}\PY{p}{(}\PY{n}{grid\PYZus{}size}\PY{p}{[}\PY{l+m+mi}{0}\PY{p}{]}\PY{p}{,} \PY{n}{i0}\PY{p}{,} \PY{n}{n0}\PY{p}{)} \PY{p}{)}\PY{p}{,}
                \PY{n}{j}\PY{p}{(} \PY{n}{min}\PY{p}{(}\PY{n}{grid\PYZus{}size}\PY{p}{[}\PY{l+m+mi}{1}\PY{p}{]}\PY{p}{,} \PY{n}{i1}\PY{p}{,} \PY{n}{n1}\PY{p}{)}\PY{p}{,} \PY{n}{max}\PY{p}{(}\PY{n}{grid\PYZus{}size}\PY{p}{[}\PY{l+m+mi}{1}\PY{p}{]}\PY{p}{,} \PY{n}{i1}\PY{p}{,} \PY{n}{n1}\PY{p}{)} \PY{p}{)}\PY{p}{;}
              \PY{n}{algos}\PY{p}{.}\PY{n}{push\PYZus{}back}\PY{p}{(}
                \PY{k}{new} \PY{k+kt}{solver\PYZus{}t}\PY{p}{(}
                  \PY{k}{typename} \PY{k+kt}{solver\PYZus{}t}\PY{o}{:}\PY{o}{:}\PY{k+kt}{ctor\PYZus{}args\PYZus{}t}\PY{p}{(}\PY{p}{\PYZob{}}
		    \PY{n}{mem}\PY{p}{.}\PY{n}{get}\PY{p}{(}\PY{p}{)}\PY{p}{,} 
		    \PY{n}{i0} \PY{o}{=}\PY{o}{=} \PY{l+m+mi}{0}      \PY{o}{?} \PY{n}{bxl} \PY{o}{:} \PY{n}{shrdl}\PY{p}{,}
		    \PY{n}{i0} \PY{o}{=}\PY{o}{=} \PY{n}{n0} \PY{o}{\PYZhy{}} \PY{l+m+mi}{1} \PY{o}{?} \PY{n}{bxr} \PY{o}{:} \PY{n}{shrdr}\PY{p}{,}
		    \PY{n}{byl}\PY{p}{,} \PY{n}{byr}\PY{p}{,} 
		    \PY{n}{i}\PY{p}{,} \PY{n}{j}
                  \PY{p}{\PYZcb{}}\PY{p}{)}\PY{p}{,} 
                  \PY{n}{p}
                \PY{p}{)}
              \PY{p}{)}\PY{p}{;}
            \PY{p}{\PYZcb{}}
          \PY{p}{\PYZcb{}}
	\PY{p}{\PYZcb{}}

        \PY{c+c1}{// 3D version}
        \PY{k+kt}{void} \PY{n}{init}\PY{p}{(}
          \PY{k}{const} \PY{k}{typename} \PY{k+kt}{solver\PYZus{}t}\PY{o}{:}\PY{o}{:}\PY{k+kt}{rt\PYZus{}params\PYZus{}t} \PY{o}{\PYZam{}}\PY{n}{p}\PY{p}{,}
	  \PY{k}{const} \PY{n}{std}\PY{o}{:}\PY{o}{:}\PY{n}{array}\PY{o}{\PYZlt{}}\PY{k+kt}{int}\PY{p}{,} \PY{l+m+mi}{3}\PY{o}{\PYZgt{}} \PY{o}{\PYZam{}}\PY{n}{grid\PYZus{}size}\PY{p}{,} 
          \PY{k}{const} \PY{k+kt}{int} \PY{o}{\PYZam{}}\PY{n}{n0}\PY{p}{,} \PY{k}{const} \PY{k+kt}{int} \PY{o}{\PYZam{}}\PY{n}{n1} \PY{o}{=} \PY{l+m+mi}{1}\PY{p}{,} \PY{k}{const} \PY{k+kt}{int} \PY{o}{\PYZam{}}\PY{n}{n2} \PY{o}{=} \PY{l+m+mi}{1}
        \PY{p}{)} \PY{p}{\PYZob{}}
          \PY{k}{typename} \PY{k+kt}{solver\PYZus{}t}\PY{o}{:}\PY{o}{:}\PY{k+kt}{bcp\PYZus{}t} \PY{n}{bxl}\PY{p}{,} \PY{n}{bxr}\PY{p}{,} \PY{n}{byl}\PY{p}{,} \PY{n}{byr}\PY{p}{,} \PY{n}{bzl}\PY{p}{,} \PY{n}{bzr}\PY{p}{,} \PY{n}{shrdl}\PY{p}{,} \PY{n}{shrdr}\PY{p}{;}

\PY{c+c1}{// TODO: renew pointers only if invalid ?}
	  \PY{k}{for} \PY{p}{(}\PY{k+kt}{int} \PY{n}{i0} \PY{o}{=} \PY{l+m+mi}{0}\PY{p}{;} \PY{n}{i0} \PY{o}{\PYZlt{}} \PY{n}{n0}\PY{p}{;} \PY{o}{+}\PY{o}{+}\PY{n}{i0}\PY{p}{)} 
          \PY{p}{\PYZob{}}
	    \PY{k}{for} \PY{p}{(}\PY{k+kt}{int} \PY{n}{i1} \PY{o}{=} \PY{l+m+mi}{0}\PY{p}{;} \PY{n}{i1} \PY{o}{\PYZlt{}} \PY{n}{n1}\PY{p}{;} \PY{o}{+}\PY{o}{+}\PY{n}{i1}\PY{p}{)} 
            \PY{p}{\PYZob{}}
	      \PY{k}{for} \PY{p}{(}\PY{k+kt}{int} \PY{n}{i2} \PY{o}{=} \PY{l+m+mi}{0}\PY{p}{;} \PY{n}{i2} \PY{o}{\PYZlt{}} \PY{n}{n2}\PY{p}{;} \PY{o}{+}\PY{o}{+}\PY{n}{i2}\PY{p}{)} 
              \PY{p}{\PYZob{}}
                \PY{c+c1}{// dim 1, left}
                \PY{k}{switch} \PY{p}{(}\PY{n}{bcxl}\PY{p}{)} \PY{c+c1}{// TODO: make a function that does it}
                \PY{p}{\PYZob{}}
                  \PY{k}{case} \PY{n}{bcond}:\PY{o}{:}\PY{n}{cyclic}\PY{o}{:}
                    \PY{n}{bxl}\PY{p}{.}\PY{n}{reset}\PY{p}{(}\PY{k}{new} \PY{n}{bcond}\PY{o}{:}\PY{o}{:}\PY{n}{cyclic\PYZus{}left\PYZus{}3d}\PY{o}{\PYZlt{}}\PY{l+m+mi}{0}\PY{p}{,} \PY{k+kt}{real\PYZus{}t}\PY{o}{\PYZgt{}}\PY{p}{(}\PY{k+kt}{rng\PYZus{}t}\PY{p}{(}\PY{l+m+mi}{0}\PY{p}{,} \PY{n}{grid\PYZus{}size}\PY{p}{[}\PY{l+m+mi}{0}\PY{p}{]}\PY{o}{\PYZhy{}}\PY{l+m+mi}{1}\PY{p}{)}\PY{p}{,} \PY{k+kt}{solver\PYZus{}t}\PY{o}{:}\PY{o}{:}\PY{n}{halo}\PY{p}{)}\PY{p}{)}\PY{p}{;}
                    \PY{k}{break}\PY{p}{;}
                  \PY{k}{case} \PY{n}{bcond}:\PY{o}{:}\PY{n}{open}\PY{o}{:}
                    \PY{n}{bxl}\PY{p}{.}\PY{n}{reset}\PY{p}{(}\PY{k}{new} \PY{n}{bcond}\PY{o}{:}\PY{o}{:}\PY{n}{open\PYZus{}left\PYZus{}3d}\PY{o}{\PYZlt{}}\PY{l+m+mi}{0}\PY{p}{,} \PY{k+kt}{real\PYZus{}t}\PY{o}{\PYZgt{}}\PY{p}{(}\PY{k+kt}{rng\PYZus{}t}\PY{p}{(}\PY{l+m+mi}{0}\PY{p}{,} \PY{n}{grid\PYZus{}size}\PY{p}{[}\PY{l+m+mi}{0}\PY{p}{]}\PY{o}{\PYZhy{}}\PY{l+m+mi}{1}\PY{p}{)}\PY{p}{,} \PY{k+kt}{solver\PYZus{}t}\PY{o}{:}\PY{o}{:}\PY{n}{halo}\PY{p}{)}\PY{p}{)}\PY{p}{;}
                    \PY{k}{break}\PY{p}{;}
                  \PY{n+nl}{default:} \PY{n}{assert}\PY{p}{(}\PY{n+nb}{false}\PY{p}{)}\PY{p}{;}
                \PY{p}{\PYZcb{}}

                \PY{c+c1}{// dim 1, rght}
                \PY{k}{switch} \PY{p}{(}\PY{n}{bcxr}\PY{p}{)} \PY{c+c1}{// TODO: make a function that does it}
                \PY{p}{\PYZob{}}
                  \PY{k}{case} \PY{n}{bcond}:\PY{o}{:}\PY{n}{cyclic}\PY{o}{:}
                    \PY{n}{bxr}\PY{p}{.}\PY{n}{reset}\PY{p}{(}\PY{k}{new} \PY{n}{bcond}\PY{o}{:}\PY{o}{:}\PY{n}{cyclic\PYZus{}rght\PYZus{}3d}\PY{o}{\PYZlt{}}\PY{l+m+mi}{0}\PY{p}{,} \PY{k+kt}{real\PYZus{}t}\PY{o}{\PYZgt{}}\PY{p}{(}\PY{k+kt}{rng\PYZus{}t}\PY{p}{(}\PY{l+m+mi}{0}\PY{p}{,} \PY{n}{grid\PYZus{}size}\PY{p}{[}\PY{l+m+mi}{0}\PY{p}{]}\PY{o}{\PYZhy{}}\PY{l+m+mi}{1}\PY{p}{)}\PY{p}{,} \PY{k+kt}{solver\PYZus{}t}\PY{o}{:}\PY{o}{:}\PY{n}{halo}\PY{p}{)}\PY{p}{)}\PY{p}{;}
                    \PY{k}{break}\PY{p}{;}
                  \PY{k}{case} \PY{n}{bcond}:\PY{o}{:}\PY{n}{open}\PY{o}{:}
                    \PY{n}{bxr}\PY{p}{.}\PY{n}{reset}\PY{p}{(}\PY{k}{new} \PY{n}{bcond}\PY{o}{:}\PY{o}{:}\PY{n}{open\PYZus{}rght\PYZus{}3d}\PY{o}{\PYZlt{}}\PY{l+m+mi}{0}\PY{p}{,} \PY{k+kt}{real\PYZus{}t}\PY{o}{\PYZgt{}}\PY{p}{(}\PY{k+kt}{rng\PYZus{}t}\PY{p}{(}\PY{l+m+mi}{0}\PY{p}{,} \PY{n}{grid\PYZus{}size}\PY{p}{[}\PY{l+m+mi}{0}\PY{p}{]}\PY{o}{\PYZhy{}}\PY{l+m+mi}{1}\PY{p}{)}\PY{p}{,} \PY{k+kt}{solver\PYZus{}t}\PY{o}{:}\PY{o}{:}\PY{n}{halo}\PY{p}{)}\PY{p}{)}\PY{p}{;}
                    \PY{k}{break}\PY{p}{;}
                  \PY{n+nl}{default:} \PY{n}{assert}\PY{p}{(}\PY{n+nb}{false}\PY{p}{)}\PY{p}{;}
                \PY{p}{\PYZcb{}}

                \PY{c+c1}{// dim 2, left}
                \PY{k}{switch} \PY{p}{(}\PY{n}{bcyl}\PY{p}{)} \PY{c+c1}{// TODO: make a function taht does it}
                \PY{p}{\PYZob{}}
                  \PY{k}{case} \PY{n}{bcond}:\PY{o}{:}\PY{n}{cyclic}\PY{o}{:}
                    \PY{n}{byl}\PY{p}{.}\PY{n}{reset}\PY{p}{(}\PY{k}{new} \PY{n}{bcond}\PY{o}{:}\PY{o}{:}\PY{n}{cyclic\PYZus{}left\PYZus{}3d}\PY{o}{\PYZlt{}}\PY{l+m+mi}{1}\PY{p}{,} \PY{k+kt}{real\PYZus{}t}\PY{o}{\PYZgt{}}\PY{p}{(}\PY{k+kt}{rng\PYZus{}t}\PY{p}{(}\PY{l+m+mi}{0}\PY{p}{,} \PY{n}{grid\PYZus{}size}\PY{p}{[}\PY{l+m+mi}{1}\PY{p}{]}\PY{o}{\PYZhy{}}\PY{l+m+mi}{1}\PY{p}{)}\PY{p}{,} \PY{k+kt}{solver\PYZus{}t}\PY{o}{:}\PY{o}{:}\PY{n}{halo}\PY{p}{)}\PY{p}{)}\PY{p}{;}
                    \PY{k}{break}\PY{p}{;}
                  \PY{k}{case} \PY{n}{bcond}:\PY{o}{:}\PY{n}{open}\PY{o}{:}
                    \PY{n}{byl}\PY{p}{.}\PY{n}{reset}\PY{p}{(}\PY{k}{new} \PY{n}{bcond}\PY{o}{:}\PY{o}{:}\PY{n}{open\PYZus{}left\PYZus{}3d}\PY{o}{\PYZlt{}}\PY{l+m+mi}{1}\PY{p}{,} \PY{k+kt}{real\PYZus{}t}\PY{o}{\PYZgt{}}\PY{p}{(}\PY{k+kt}{rng\PYZus{}t}\PY{p}{(}\PY{l+m+mi}{0}\PY{p}{,} \PY{n}{grid\PYZus{}size}\PY{p}{[}\PY{l+m+mi}{1}\PY{p}{]}\PY{o}{\PYZhy{}}\PY{l+m+mi}{1}\PY{p}{)}\PY{p}{,} \PY{k+kt}{solver\PYZus{}t}\PY{o}{:}\PY{o}{:}\PY{n}{halo}\PY{p}{)}\PY{p}{)}\PY{p}{;}
                    \PY{k}{break}\PY{p}{;}
                  \PY{n+nl}{default:} \PY{n}{assert}\PY{p}{(}\PY{n+nb}{false}\PY{p}{)}\PY{p}{;}
                \PY{p}{\PYZcb{}}

                \PY{c+c1}{// dim 2, rght}
                \PY{k}{switch} \PY{p}{(}\PY{n}{bcyr}\PY{p}{)} \PY{c+c1}{// TODO: make a function taht does it}
                \PY{p}{\PYZob{}}
                  \PY{k}{case} \PY{n}{bcond}:\PY{o}{:}\PY{n}{cyclic}\PY{o}{:}
		    \PY{n}{byr}\PY{p}{.}\PY{n}{reset}\PY{p}{(}\PY{k}{new} \PY{n}{bcond}\PY{o}{:}\PY{o}{:}\PY{n}{cyclic\PYZus{}rght\PYZus{}3d}\PY{o}{\PYZlt{}}\PY{l+m+mi}{1}\PY{p}{,} \PY{k+kt}{real\PYZus{}t}\PY{o}{\PYZgt{}}\PY{p}{(}\PY{k+kt}{rng\PYZus{}t}\PY{p}{(}\PY{l+m+mi}{0}\PY{p}{,} \PY{n}{grid\PYZus{}size}\PY{p}{[}\PY{l+m+mi}{1}\PY{p}{]}\PY{o}{\PYZhy{}}\PY{l+m+mi}{1}\PY{p}{)}\PY{p}{,} \PY{k+kt}{solver\PYZus{}t}\PY{o}{:}\PY{o}{:}\PY{n}{halo}\PY{p}{)}\PY{p}{)}\PY{p}{;}
                    \PY{k}{break}\PY{p}{;}
                  \PY{k}{case} \PY{n}{bcond}:\PY{o}{:}\PY{n}{open}\PY{o}{:}
		    \PY{n}{byr}\PY{p}{.}\PY{n}{reset}\PY{p}{(}\PY{k}{new} \PY{n}{bcond}\PY{o}{:}\PY{o}{:}\PY{n}{open\PYZus{}rght\PYZus{}3d}\PY{o}{\PYZlt{}}\PY{l+m+mi}{1}\PY{p}{,} \PY{k+kt}{real\PYZus{}t}\PY{o}{\PYZgt{}}\PY{p}{(}\PY{k+kt}{rng\PYZus{}t}\PY{p}{(}\PY{l+m+mi}{0}\PY{p}{,} \PY{n}{grid\PYZus{}size}\PY{p}{[}\PY{l+m+mi}{1}\PY{p}{]}\PY{o}{\PYZhy{}}\PY{l+m+mi}{1}\PY{p}{)}\PY{p}{,} \PY{k+kt}{solver\PYZus{}t}\PY{o}{:}\PY{o}{:}\PY{n}{halo}\PY{p}{)}\PY{p}{)}\PY{p}{;}
                    \PY{k}{break}\PY{p}{;}
                  \PY{n+nl}{default:} \PY{n}{assert}\PY{p}{(}\PY{n+nb}{false}\PY{p}{)}\PY{p}{;}
                \PY{p}{\PYZcb{}}

                \PY{c+c1}{// dim 3, left}
                \PY{k}{switch} \PY{p}{(}\PY{n}{bczl}\PY{p}{)} \PY{c+c1}{// TODO: make a function taht does it}
                \PY{p}{\PYZob{}}
                  \PY{k}{case} \PY{n}{bcond}:\PY{o}{:}\PY{n}{cyclic}\PY{o}{:}
		    \PY{n}{bzl}\PY{p}{.}\PY{n}{reset}\PY{p}{(}\PY{k}{new} \PY{n}{bcond}\PY{o}{:}\PY{o}{:}\PY{n}{cyclic\PYZus{}left\PYZus{}3d}\PY{o}{\PYZlt{}}\PY{l+m+mi}{2}\PY{p}{,} \PY{k+kt}{real\PYZus{}t}\PY{o}{\PYZgt{}}\PY{p}{(}\PY{k+kt}{rng\PYZus{}t}\PY{p}{(}\PY{l+m+mi}{0}\PY{p}{,} \PY{n}{grid\PYZus{}size}\PY{p}{[}\PY{l+m+mi}{2}\PY{p}{]}\PY{o}{\PYZhy{}}\PY{l+m+mi}{1}\PY{p}{)}\PY{p}{,} \PY{k+kt}{solver\PYZus{}t}\PY{o}{:}\PY{o}{:}\PY{n}{halo}\PY{p}{)}\PY{p}{)}\PY{p}{;}
                    \PY{k}{break}\PY{p}{;}
                  \PY{k}{case} \PY{n}{bcond}:\PY{o}{:}\PY{n}{open}\PY{o}{:}
		    \PY{n}{bzl}\PY{p}{.}\PY{n}{reset}\PY{p}{(}\PY{k}{new} \PY{n}{bcond}\PY{o}{:}\PY{o}{:}\PY{n}{open\PYZus{}left\PYZus{}3d}\PY{o}{\PYZlt{}}\PY{l+m+mi}{2}\PY{p}{,} \PY{k+kt}{real\PYZus{}t}\PY{o}{\PYZgt{}}\PY{p}{(}\PY{k+kt}{rng\PYZus{}t}\PY{p}{(}\PY{l+m+mi}{0}\PY{p}{,} \PY{n}{grid\PYZus{}size}\PY{p}{[}\PY{l+m+mi}{2}\PY{p}{]}\PY{o}{\PYZhy{}}\PY{l+m+mi}{1}\PY{p}{)}\PY{p}{,} \PY{k+kt}{solver\PYZus{}t}\PY{o}{:}\PY{o}{:}\PY{n}{halo}\PY{p}{)}\PY{p}{)}\PY{p}{;}
                    \PY{k}{break}\PY{p}{;}
                  \PY{n+nl}{default:} \PY{n}{assert}\PY{p}{(}\PY{n+nb}{false}\PY{p}{)}\PY{p}{;}
                \PY{p}{\PYZcb{}}

                \PY{c+c1}{// dim 3, rght}
                \PY{k}{switch} \PY{p}{(}\PY{n}{bczr}\PY{p}{)} \PY{c+c1}{// TODO: make a function taht does it}
                \PY{p}{\PYZob{}}
                  \PY{k}{case} \PY{n}{bcond}:\PY{o}{:}\PY{n}{cyclic}\PY{o}{:}
		    \PY{n}{bzr}\PY{p}{.}\PY{n}{reset}\PY{p}{(}\PY{k}{new} \PY{n}{bcond}\PY{o}{:}\PY{o}{:}\PY{n}{cyclic\PYZus{}rght\PYZus{}3d}\PY{o}{\PYZlt{}}\PY{l+m+mi}{2}\PY{p}{,} \PY{k+kt}{real\PYZus{}t}\PY{o}{\PYZgt{}}\PY{p}{(}\PY{k+kt}{rng\PYZus{}t}\PY{p}{(}\PY{l+m+mi}{0}\PY{p}{,} \PY{n}{grid\PYZus{}size}\PY{p}{[}\PY{l+m+mi}{2}\PY{p}{]}\PY{o}{\PYZhy{}}\PY{l+m+mi}{1}\PY{p}{)}\PY{p}{,} \PY{k+kt}{solver\PYZus{}t}\PY{o}{:}\PY{o}{:}\PY{n}{halo}\PY{p}{)}\PY{p}{)}\PY{p}{;}
                    \PY{k}{break}\PY{p}{;}
                  \PY{k}{case} \PY{n}{bcond}:\PY{o}{:}\PY{n}{open}\PY{o}{:}
		    \PY{n}{bzr}\PY{p}{.}\PY{n}{reset}\PY{p}{(}\PY{k}{new} \PY{n}{bcond}\PY{o}{:}\PY{o}{:}\PY{n}{open\PYZus{}rght\PYZus{}3d}\PY{o}{\PYZlt{}}\PY{l+m+mi}{2}\PY{p}{,} \PY{k+kt}{real\PYZus{}t}\PY{o}{\PYZgt{}}\PY{p}{(}\PY{k+kt}{rng\PYZus{}t}\PY{p}{(}\PY{l+m+mi}{0}\PY{p}{,} \PY{n}{grid\PYZus{}size}\PY{p}{[}\PY{l+m+mi}{2}\PY{p}{]}\PY{o}{\PYZhy{}}\PY{l+m+mi}{1}\PY{p}{)}\PY{p}{,} \PY{k+kt}{solver\PYZus{}t}\PY{o}{:}\PY{o}{:}\PY{n}{halo}\PY{p}{)}\PY{p}{)}\PY{p}{;}
                    \PY{k}{break}\PY{p}{;}
                  \PY{n+nl}{default:} \PY{n}{assert}\PY{p}{(}\PY{n+nb}{false}\PY{p}{)}\PY{p}{;}
                \PY{p}{\PYZcb{}}

                \PY{n}{shrdl}\PY{p}{.}\PY{n}{reset}\PY{p}{(}\PY{k}{new} \PY{n}{bcond}\PY{o}{:}\PY{o}{:}\PY{n}{shared}\PY{o}{\PYZlt{}}\PY{k+kt}{real\PYZus{}t}\PY{o}{\PYZgt{}}\PY{p}{(}\PY{p}{)}\PY{p}{)}\PY{p}{;} \PY{c+c1}{// TODO: shrdy if n1 != 1}
                \PY{n}{shrdr}\PY{p}{.}\PY{n}{reset}\PY{p}{(}\PY{k}{new} \PY{n}{bcond}\PY{o}{:}\PY{o}{:}\PY{n}{shared}\PY{o}{\PYZlt{}}\PY{k+kt}{real\PYZus{}t}\PY{o}{\PYZgt{}}\PY{p}{(}\PY{p}{)}\PY{p}{)}\PY{p}{;} \PY{c+c1}{// TODO: shrdy if n1 != 1}

                \PY{k+kt}{rng\PYZus{}t}
                  \PY{n+nf}{i}\PY{p}{(} \PY{n}{min}\PY{p}{(}\PY{n}{grid\PYZus{}size}\PY{p}{[}\PY{l+m+mi}{0}\PY{p}{]}\PY{p}{,} \PY{n}{i0}\PY{p}{,} \PY{n}{n0}\PY{p}{)}\PY{p}{,} \PY{n}{max}\PY{p}{(}\PY{n}{grid\PYZus{}size}\PY{p}{[}\PY{l+m+mi}{0}\PY{p}{]}\PY{p}{,} \PY{n}{i0}\PY{p}{,} \PY{n}{n0}\PY{p}{)} \PY{p}{)}\PY{p}{,}
                  \PY{n}{j}\PY{p}{(} \PY{n}{min}\PY{p}{(}\PY{n}{grid\PYZus{}size}\PY{p}{[}\PY{l+m+mi}{1}\PY{p}{]}\PY{p}{,} \PY{n}{i1}\PY{p}{,} \PY{n}{n1}\PY{p}{)}\PY{p}{,} \PY{n}{max}\PY{p}{(}\PY{n}{grid\PYZus{}size}\PY{p}{[}\PY{l+m+mi}{1}\PY{p}{]}\PY{p}{,} \PY{n}{i1}\PY{p}{,} \PY{n}{n1}\PY{p}{)} \PY{p}{)}\PY{p}{,}
                  \PY{n}{k}\PY{p}{(} \PY{n}{min}\PY{p}{(}\PY{n}{grid\PYZus{}size}\PY{p}{[}\PY{l+m+mi}{2}\PY{p}{]}\PY{p}{,} \PY{n}{i2}\PY{p}{,} \PY{n}{n2}\PY{p}{)}\PY{p}{,} \PY{n}{max}\PY{p}{(}\PY{n}{grid\PYZus{}size}\PY{p}{[}\PY{l+m+mi}{2}\PY{p}{]}\PY{p}{,} \PY{n}{i2}\PY{p}{,} \PY{n}{n2}\PY{p}{)} \PY{p}{)}\PY{p}{;}

		\PY{n}{algos}\PY{p}{.}\PY{n}{push\PYZus{}back}\PY{p}{(}
                  \PY{k}{new} \PY{k+kt}{solver\PYZus{}t}\PY{p}{(}
                    \PY{k}{typename} \PY{k+kt}{solver\PYZus{}t}\PY{o}{:}\PY{o}{:}\PY{k+kt}{ctor\PYZus{}args\PYZus{}t}\PY{p}{(}\PY{p}{\PYZob{}}
                      \PY{n}{mem}\PY{p}{.}\PY{n}{get}\PY{p}{(}\PY{p}{)}\PY{p}{,} 
		      \PY{n}{i0} \PY{o}{=}\PY{o}{=} \PY{l+m+mi}{0}      \PY{o}{?} \PY{n}{bxl} \PY{o}{:} \PY{n}{shrdl}\PY{p}{,}
		      \PY{n}{i0} \PY{o}{=}\PY{o}{=} \PY{n}{n0} \PY{o}{\PYZhy{}} \PY{l+m+mi}{1} \PY{o}{?} \PY{n}{bxr} \PY{o}{:} \PY{n}{shrdr}\PY{p}{,}
                      \PY{n}{byl}\PY{p}{,} \PY{n}{byr}\PY{p}{,} 
                      \PY{n}{bzl}\PY{p}{,} \PY{n}{bzr}\PY{p}{,} 
                      \PY{n}{i}\PY{p}{,} \PY{n}{j}\PY{p}{,} \PY{n}{k}
                    \PY{p}{\PYZcb{}}\PY{p}{)}\PY{p}{,} 
                    \PY{n}{p}
                  \PY{p}{)}
                \PY{p}{)}\PY{p}{;}
              \PY{p}{\PYZcb{}}
            \PY{p}{\PYZcb{}}
          \PY{p}{\PYZcb{}}
        \PY{p}{\PYZcb{}}

        \PY{k}{virtual} \PY{k+kt}{void} \PY{n}{solve}\PY{p}{(}\PY{k+kt}{int} \PY{n}{nt}\PY{p}{)} \PY{o}{=} \PY{l+m+mi}{0}\PY{p}{;}

        \PY{n+nl}{public:}
    
        \PY{k+kt}{void} \PY{n+nf}{advance}\PY{p}{(}\PY{k+kt}{int} \PY{n}{nt}\PY{p}{)} \PY{n}{final}
        \PY{p}{\PYZob{}}   
          \PY{n}{tmr}\PY{p}{.}\PY{n}{resume}\PY{p}{(}\PY{p}{)}\PY{p}{;}
          \PY{n}{solve}\PY{p}{(}\PY{n}{nt}\PY{p}{)}\PY{p}{;}
          \PY{n}{tmr}\PY{p}{.}\PY{n}{stop}\PY{p}{(}\PY{p}{)}\PY{p}{;}
        \PY{p}{\PYZcb{}}  

	\PY{k}{typename} \PY{k+kt}{solver\PYZus{}t}\PY{o}{:}\PY{o}{:}\PY{k+kt}{arr\PYZus{}t} \PY{n}{advectee}\PY{p}{(}\PY{k+kt}{int} \PY{n}{e} \PY{o}{=} \PY{l+m+mi}{0}\PY{p}{)} \PY{n}{final}
	\PY{p}{\PYZob{}}
	  \PY{k}{return} \PY{n}{mem}\PY{o}{\PYZhy{}}\PY{o}{\PYZgt{}}\PY{n}{advectee}\PY{p}{(}\PY{n}{e}\PY{p}{)}\PY{p}{;}
	\PY{p}{\PYZcb{}}

	\PY{k}{typename} \PY{k+kt}{solver\PYZus{}t}\PY{o}{:}\PY{o}{:}\PY{k+kt}{arr\PYZus{}t} \PY{n}{advector}\PY{p}{(}\PY{k+kt}{int} \PY{n}{d} \PY{o}{=} \PY{l+m+mi}{0}\PY{p}{)} \PY{n}{final}
	\PY{p}{\PYZob{}}
	  \PY{k}{return} \PY{n}{mem}\PY{o}{\PYZhy{}}\PY{o}{\PYZgt{}}\PY{n}{advector}\PY{p}{(}\PY{n}{d}\PY{p}{)}\PY{p}{;}
	\PY{p}{\PYZcb{}}

	\PY{k}{typename} \PY{k+kt}{solver\PYZus{}t}\PY{o}{:}\PY{o}{:}\PY{k+kt}{arr\PYZus{}t} \PY{n}{g\PYZus{}factor}\PY{p}{(}\PY{p}{)} \PY{n}{final}
	\PY{p}{\PYZob{}}
	  \PY{k}{return} \PY{n}{mem}\PY{o}{\PYZhy{}}\PY{o}{\PYZgt{}}\PY{n}{g\PYZus{}factor}\PY{p}{(}\PY{p}{)}\PY{p}{;}
	\PY{p}{\PYZcb{}}

        \PY{k+kt}{bool} \PY{o}{*}\PY{n}{panic\PYZus{}ptr}\PY{p}{(}\PY{p}{)} \PY{n}{final}
        \PY{p}{\PYZob{}}
          \PY{k}{return} \PY{o}{\PYZam{}}\PY{k}{this}\PY{o}{\PYZhy{}}\PY{o}{\PYZgt{}}\PY{n}{mem}\PY{o}{\PYZhy{}}\PY{o}{\PYZgt{}}\PY{n}{panic}\PY{p}{;}
        \PY{p}{\PYZcb{}}
      \PY{p}{\PYZcb{}}\PY{p}{;}
    \PY{p}{\PYZcb{}}\PY{p}{;} \PY{c+c1}{// namespace detail}
  \PY{p}{\PYZcb{}}\PY{p}{;} \PY{c+c1}{// namespace concurr}
\PY{p}{\PYZcb{}}\PY{p}{;} \PY{c+c1}{// namespace libmpdataxx}
\end{Verbatim}

%% file: example_1.cpp.tex
\begin{Verbatim}[commandchars=\\\{\}]
\PY{c+c1}{//\PYZlt{}listing\PYZhy{}1\PYZgt{}}
\PY{c+cp}{\PYZsh{}}\PY{c+cp}{include \PYZlt{}libmpdata++}\PY{c+cp}{/}\PY{c+cp}{solvers}\PY{c+cp}{/}\PY{c+cp}{mpdata.hpp\PYZgt{}}
\PY{c+cp}{\PYZsh{}}\PY{c+cp}{include \PYZlt{}libmpdata++}\PY{c+cp}{/}\PY{c+cp}{concurr}\PY{c+cp}{/}\PY{c+cp}{serial.hpp\PYZgt{}}
\PY{c+cp}{\PYZsh{}}\PY{c+cp}{include \PYZlt{}libmpdata++}\PY{c+cp}{/}\PY{c+cp}{output}\PY{c+cp}{/}\PY{c+cp}{gnuplot.hpp\PYZgt{}}

\PY{k}{using} \PY{k}{namespace} \PY{n}{libmpdataxx}\PY{p}{;}

\PY{k+kt}{int} \PY{n+nf}{main}\PY{p}{(}\PY{p}{)}
\PY{p}{\PYZob{}}
  \PY{c+c1}{// compile\PYZhy{}time parameters}
  \PY{k}{struct} \PY{k+kt}{ct\PYZus{}params\PYZus{}t} \PY{o}{:} \PY{k+kt}{ct\PYZus{}params\PYZus{}default\PYZus{}t}
  \PY{p}{\PYZob{}}
    \PY{k}{using} \PY{k+kt}{real\PYZus{}t} \PY{o}{=} \PY{k+kt}{double}\PY{p}{;}
    \PY{k}{enum} \PY{p}{\PYZob{}} \PY{n}{n\PYZus{}dims} \PY{o}{=} \PY{l+m+mi}{1} \PY{p}{\PYZcb{}}\PY{p}{;}
    \PY{k}{enum} \PY{p}{\PYZob{}} \PY{n}{n\PYZus{}eqns} \PY{o}{=} \PY{l+m+mi}{1} \PY{p}{\PYZcb{}}\PY{p}{;}
  \PY{p}{\PYZcb{}}\PY{p}{;}

  \PY{c+c1}{// solver choice}
  \PY{k}{using} \PY{k+kt}{slv\PYZus{}t} \PY{o}{=} \PY{n}{solvers}\PY{o}{:}\PY{o}{:}\PY{n}{mpdata}\PY{o}{\PYZlt{}}\PY{k+kt}{ct\PYZus{}params\PYZus{}t}\PY{o}{\PYZgt{}}\PY{p}{;}

  \PY{c+c1}{// output choice}
  \PY{k}{using} \PY{k+kt}{slv\PYZus{}out\PYZus{}t} \PY{o}{=} \PY{n}{output}\PY{o}{:}\PY{o}{:}\PY{n}{gnuplot}\PY{o}{\PYZlt{}}\PY{k+kt}{slv\PYZus{}t}\PY{o}{\PYZgt{}}\PY{p}{;}
  
  \PY{c+c1}{// concurency choice}
  \PY{k}{using} \PY{k+kt}{run\PYZus{}t} \PY{o}{=} \PY{n}{concurr}\PY{o}{:}\PY{o}{:}\PY{n}{serial}\PY{o}{\PYZlt{}}
    \PY{k+kt}{slv\PYZus{}out\PYZus{}t}\PY{p}{,} \PY{n}{bcond}\PY{o}{:}\PY{o}{:}\PY{n}{open}\PY{p}{,} \PY{n}{bcond}\PY{o}{:}\PY{o}{:}\PY{n}{open}
  \PY{o}{\PYZgt{}}\PY{p}{;}         \PY{c+c1}{//left bcond   //right bcond}

  \PY{c+c1}{// run\PYZhy{}time parameters}
  \PY{k}{typename} \PY{k+kt}{slv\PYZus{}out\PYZus{}t}\PY{o}{:}\PY{o}{:}\PY{k+kt}{rt\PYZus{}params\PYZus{}t} \PY{n}{p}\PY{p}{;}

  \PY{k+kt}{int} \PY{n}{nx} \PY{o}{=} \PY{l+m+mi}{101}\PY{p}{,} \PY{n}{nt} \PY{o}{=} \PY{l+m+mi}{100}\PY{p}{;}
  \PY{k+kt}{ct\PYZus{}params\PYZus{}t}\PY{o}{:}\PY{o}{:}\PY{k+kt}{real\PYZus{}t} \PY{n}{dx} \PY{o}{=} \PY{l+m+mf}{0.1}\PY{p}{;}

  \PY{n}{p}\PY{p}{.}\PY{n}{grid\PYZus{}size} \PY{o}{=} \PY{p}{\PYZob{}} \PY{n}{nx} \PY{p}{\PYZcb{}}\PY{p}{;}
  \PY{n}{p}\PY{p}{.}\PY{n}{outfreq} \PY{o}{=} \PY{l+m+mi}{20}\PY{p}{;} 
 
  \PY{c+c1}{// instantiation}
  \PY{k+kt}{run\PYZus{}t} \PY{n}{run}\PY{p}{(}\PY{n}{p}\PY{p}{)}\PY{p}{;}

  \PY{c+c1}{// initial condition}
  \PY{n}{blitz}\PY{o}{:}\PY{o}{:}\PY{n}{firstIndex} \PY{n}{i}\PY{p}{;}
  \PY{c+c1}{// Witch of Agnesi with a=.5 }
  \PY{n}{run}\PY{p}{.}\PY{n}{advectee}\PY{p}{(}\PY{p}{)} \PY{o}{=} \PY{o}{\PYZhy{}}\PY{l+m+mf}{.5} \PY{o}{+} \PY{l+m+mi}{1} \PY{o}{/} \PY{p}{(}
    \PY{n}{pow}\PY{p}{(}\PY{n}{dx}\PY{o}{*}\PY{p}{(}\PY{n}{i} \PY{o}{\PYZhy{}} \PY{p}{(}\PY{n}{nx}\PY{o}{\PYZhy{}}\PY{l+m+mi}{1}\PY{p}{)}\PY{o}{/}\PY{l+m+mf}{2.}\PY{p}{)}\PY{p}{,} \PY{l+m+mi}{2}\PY{p}{)} \PY{o}{+} \PY{l+m+mi}{1}
  \PY{p}{)}\PY{p}{;}
  \PY{c+c1}{// Courant number}
  \PY{n}{run}\PY{p}{.}\PY{n}{advector}\PY{p}{(}\PY{p}{)} \PY{o}{=} \PY{l+m+mf}{.5}\PY{p}{;}

  \PY{c+c1}{// integration}
  \PY{n}{run}\PY{p}{.}\PY{n}{advance}\PY{p}{(}\PY{n}{nt}\PY{p}{)}\PY{p}{;}
\PY{p}{\PYZcb{}}
\PY{c+c1}{//\PYZlt{}/listing\PYZhy{}1\PYZgt{}}
\end{Verbatim}

%% file: example_2.cpp.tex
\begin{Verbatim}[commandchars=\\\{\}]
\PY{c+cp}{\PYZsh{}}\PY{c+cp}{include \PYZlt{}libmpdata++}\PY{c+cp}{/}\PY{c+cp}{solvers}\PY{c+cp}{/}\PY{c+cp}{mpdata.hpp\PYZgt{}}
\PY{c+cp}{\PYZsh{}}\PY{c+cp}{include \PYZlt{}libmpdata++}\PY{c+cp}{/}\PY{c+cp}{concurr}\PY{c+cp}{/}\PY{c+cp}{serial.hpp\PYZgt{}}
\PY{c+cp}{\PYZsh{}}\PY{c+cp}{include \PYZlt{}libmpdata++}\PY{c+cp}{/}\PY{c+cp}{output}\PY{c+cp}{/}\PY{c+cp}{gnuplot.hpp\PYZgt{}}

\PY{k}{using} \PY{k}{namespace} \PY{n}{libmpdataxx}\PY{p}{;}

\PY{k}{template} \PY{o}{\PYZlt{}}\PY{k+kt}{int} \PY{n}{opts\PYZus{}arg}\PY{o}{\PYZgt{}}
\PY{k+kt}{void} \PY{n}{test}\PY{p}{(}\PY{k}{const} \PY{n}{std}\PY{o}{:}\PY{o}{:}\PY{n}{string} \PY{n}{filename}\PY{p}{)}
\PY{p}{\PYZob{}}
  \PY{c+c1}{// compile\PYZhy{}time parameters}
  \PY{k}{struct} \PY{k+kt}{ct\PYZus{}params\PYZus{}t} \PY{o}{:} \PY{k+kt}{ct\PYZus{}params\PYZus{}default\PYZus{}t}
  \PY{p}{\PYZob{}}
    \PY{k}{using} \PY{k+kt}{real\PYZus{}t} \PY{o}{=} \PY{k+kt}{double}\PY{p}{;}
\PY{c+c1}{//\PYZlt{}listing\PYZhy{}1\PYZgt{}}
    \PY{k}{enum} \PY{p}{\PYZob{}} \PY{n}{n\PYZus{}dims} \PY{o}{=} \PY{l+m+mi}{1} \PY{p}{\PYZcb{}}\PY{p}{;}
    \PY{k}{enum} \PY{p}{\PYZob{}} \PY{n}{n\PYZus{}eqns} \PY{o}{=} \PY{l+m+mi}{2} \PY{p}{\PYZcb{}}\PY{p}{;}
\PY{c+c1}{//\PYZlt{}/listing\PYZhy{}1\PYZgt{}}
    \PY{k}{enum} \PY{p}{\PYZob{}} \PY{n}{opts} \PY{o}{=} \PY{n}{opts\PYZus{}arg} \PY{p}{\PYZcb{}}\PY{p}{;}
  \PY{p}{\PYZcb{}}\PY{p}{;}

  \PY{k}{using} \PY{k+kt}{sim\PYZus{}t} \PY{o}{=} \PY{n}{output}\PY{o}{:}\PY{o}{:}\PY{n}{gnuplot}\PY{o}{\PYZlt{}}
    \PY{n}{solvers}\PY{o}{:}\PY{o}{:}\PY{n}{mpdata}\PY{o}{\PYZlt{}}\PY{k+kt}{ct\PYZus{}params\PYZus{}t}\PY{o}{\PYZgt{}}
  \PY{o}{\PYZgt{}}\PY{p}{;}
  \PY{k}{typename} \PY{k+kt}{sim\PYZus{}t}\PY{o}{:}\PY{o}{:}\PY{k+kt}{rt\PYZus{}params\PYZus{}t} \PY{n}{p}\PY{p}{;}

\PY{c+c1}{//\PYZlt{}listing\PYZhy{}2\PYZgt{}}
  \PY{k+kt}{int} \PY{n}{nx} \PY{o}{=} \PY{l+m+mi}{601}\PY{p}{,} \PY{n}{nt} \PY{o}{=} \PY{l+m+mi}{1200}\PY{p}{;}
  \PY{c+c1}{// run\PYZhy{}time parameters}
  \PY{n}{p}\PY{p}{.}\PY{n}{grid\PYZus{}size} \PY{o}{=} \PY{p}{\PYZob{}} \PY{n}{nx} \PY{p}{\PYZcb{}}\PY{p}{;}
  \PY{n}{p}\PY{p}{.}\PY{n}{outfreq} \PY{o}{=} \PY{n}{nt}\PY{p}{;} 
  \PY{n}{p}\PY{p}{.}\PY{n}{outvars} \PY{o}{=} \PY{p}{\PYZob{}}
    \PY{p}{\PYZob{}}\PY{l+m+mi}{0}\PY{p}{,} \PY{p}{\PYZob{}}\PY{p}{.}\PY{n}{name} \PY{o}{=} \PY{l+s}{\PYZdq{}}\PY{l+s}{\PYZdq{}}\PY{p}{,} \PY{p}{.}\PY{n}{unit} \PY{o}{=} \PY{l+s}{\PYZdq{}}\PY{l+s}{1}\PY{l+s}{\PYZdq{}}\PY{p}{\PYZcb{}}\PY{p}{\PYZcb{}}\PY{p}{,}
    \PY{p}{\PYZob{}}\PY{l+m+mi}{1}\PY{p}{,} \PY{p}{\PYZob{}}\PY{p}{.}\PY{n}{name} \PY{o}{=} \PY{l+s}{\PYZdq{}}\PY{l+s}{\PYZdq{}}\PY{p}{,} \PY{p}{.}\PY{n}{unit} \PY{o}{=} \PY{l+s}{\PYZdq{}}\PY{l+s}{1}\PY{l+s}{\PYZdq{}}\PY{p}{\PYZcb{}}\PY{p}{\PYZcb{}}
  \PY{p}{\PYZcb{}}\PY{p}{;}
\PY{c+c1}{//\PYZlt{}/listing\PYZhy{}2\PYZgt{}}
  \PY{n}{p}\PY{p}{.}\PY{n}{gnuplot\PYZus{}output} \PY{o}{=} \PY{n}{filename}\PY{p}{;} 
  \PY{n}{p}\PY{p}{.}\PY{n}{gnuplot\PYZus{}command} \PY{o}{=} \PY{l+s}{\PYZdq{}}\PY{l+s}{plot}\PY{l+s}{\PYZdq{}}\PY{p}{;}
  \PY{n}{p}\PY{p}{.}\PY{n}{gnuplot\PYZus{}with} \PY{o}{=} \PY{l+s}{\PYZdq{}}\PY{l+s}{histeps}\PY{l+s}{\PYZdq{}}\PY{p}{;}
  \PY{n}{p}\PY{p}{.}\PY{n}{gnuplot\PYZus{}yrange} \PY{o}{=} \PY{l+s}{\PYZdq{}}\PY{l+s}{[\PYZhy{}1.25:4.25]}\PY{l+s}{\PYZdq{}}\PY{p}{;}
  \PY{n}{p}\PY{p}{.}\PY{n}{gnuplot\PYZus{}fontsize} \PY{o}{=} \PY{l+s}{\PYZdq{}}\PY{l+s}{17}\PY{l+s}{\PYZdq{}}\PY{p}{;}

  \PY{c+c1}{// instantiation}
  \PY{n}{concurr}\PY{o}{:}\PY{o}{:}\PY{n}{serial}\PY{o}{\PYZlt{}}\PY{k+kt}{sim\PYZus{}t}\PY{p}{,} \PY{n}{bcond}\PY{o}{:}\PY{o}{:}\PY{n}{cyclic}\PY{p}{,} \PY{n}{bcond}\PY{o}{:}\PY{o}{:}\PY{n}{cyclic}\PY{o}{\PYZgt{}} \PY{n}{run}\PY{p}{(}\PY{n}{p}\PY{p}{)}\PY{p}{;}

\PY{c+c1}{//\PYZlt{}listing\PYZhy{}3\PYZgt{}}
  \PY{c+c1}{// initial condition}
  \PY{n}{blitz}\PY{o}{:}\PY{o}{:}\PY{n}{firstIndex} \PY{n}{i}\PY{p}{;}
  \PY{n}{run}\PY{p}{.}\PY{n}{advectee}\PY{p}{(}\PY{l+m+mi}{0}\PY{p}{)} \PY{o}{=} \PY{n}{where}\PY{p}{(}
    \PY{n}{i} \PY{o}{\PYZlt{}}\PY{o}{=} \PY{l+m+mi}{75} \PY{o}{|}\PY{o}{|} \PY{n}{i} \PY{o}{\PYZgt{}}\PY{o}{=} \PY{l+m+mi}{125}\PY{p}{,}   \PY{c+c1}{// if}
    \PY{l+m+mi}{2}\PY{p}{,}                     \PY{c+c1}{// then}
    \PY{l+m+mi}{4}                      \PY{c+c1}{// else}
  \PY{p}{)}\PY{p}{;} 
  \PY{n}{run}\PY{p}{.}\PY{n}{advectee}\PY{p}{(}\PY{l+m+mi}{1}\PY{p}{)} \PY{o}{=} \PY{n}{where}\PY{p}{(}
    \PY{n}{i} \PY{o}{\PYZlt{}}\PY{o}{=} \PY{l+m+mi}{75} \PY{o}{|}\PY{o}{|} \PY{n}{i} \PY{o}{\PYZgt{}}\PY{o}{=} \PY{l+m+mi}{125}\PY{p}{,}   \PY{c+c1}{// if }
    \PY{o}{\PYZhy{}}\PY{l+m+mi}{1}\PY{p}{,}                    \PY{c+c1}{// then}
    \PY{l+m+mi}{1}                      \PY{c+c1}{// else}
  \PY{p}{)}\PY{p}{;} 
  \PY{n}{run}\PY{p}{.}\PY{n}{advector}\PY{p}{(}\PY{p}{)} \PY{o}{=} \PY{o}{\PYZhy{}}\PY{l+m+mf}{.75}\PY{p}{;}  \PY{c+c1}{// Courant}
\PY{c+c1}{//\PYZlt{}/listing\PYZhy{}3\PYZgt{}}

  \PY{c+c1}{// integration}
  \PY{n}{run}\PY{p}{.}\PY{n}{advance}\PY{p}{(}\PY{n}{nt}\PY{p}{)}\PY{p}{;}
\PY{p}{\PYZcb{}}

\PY{k+kt}{int} \PY{n}{main}\PY{p}{(}\PY{p}{)}
\PY{p}{\PYZob{}}
  \PY{p}{\PYZob{}}
\PY{c+c1}{//\PYZlt{}listing\PYZhy{}4\PYZgt{}}
    \PY{k}{enum} \PY{p}{\PYZob{}} \PY{n}{opts} \PY{o}{=} \PY{n}{opts}\PY{o}{:}\PY{o}{:}\PY{n}{abs} \PY{p}{\PYZcb{}}\PY{p}{;}
\PY{c+c1}{//\PYZlt{}/listing\PYZhy{}4\PYZgt{}}
    \PY{n}{test}\PY{o}{\PYZlt{}}\PY{n}{opts}\PY{o}{\PYZgt{}}\PY{p}{(}\PY{l+s}{\PYZdq{}}\PY{l+s}{out\PYZus{}abs.svg}\PY{l+s}{\PYZdq{}}\PY{p}{)}\PY{p}{;}
  \PY{p}{\PYZcb{}}
  \PY{p}{\PYZob{}}
\PY{c+c1}{//\PYZlt{}listing\PYZhy{}5\PYZgt{}}
    \PY{k}{enum} \PY{p}{\PYZob{}} \PY{n}{opts} \PY{o}{=} \PY{n}{opts}\PY{o}{:}\PY{o}{:}\PY{n}{iga} \PY{p}{\PYZcb{}}\PY{p}{;}
\PY{c+c1}{//\PYZlt{}/listing\PYZhy{}5\PYZgt{}}
    \PY{n}{test}\PY{o}{\PYZlt{}}\PY{n}{opts}\PY{o}{\PYZgt{}}\PY{p}{(}\PY{l+s}{\PYZdq{}}\PY{l+s}{out\PYZus{}iga.svg}\PY{l+s}{\PYZdq{}}\PY{p}{)}\PY{p}{;}
  \PY{p}{\PYZcb{}}
  \PY{p}{\PYZob{}}
\PY{c+c1}{//\PYZlt{}listing\PYZhy{}6\PYZgt{}}
    \PY{k}{enum} \PY{p}{\PYZob{}} \PY{n}{opts} \PY{o}{=} \PY{n}{opts}\PY{o}{:}\PY{o}{:}\PY{n}{iga} \PY{o}{|} \PY{n}{opts}\PY{o}{:}\PY{o}{:}\PY{n}{tot} \PY{p}{\PYZcb{}}\PY{p}{;}
\PY{c+c1}{//\PYZlt{}/listing\PYZhy{}6\PYZgt{}}
    \PY{n}{test}\PY{o}{\PYZlt{}}\PY{n}{opts}\PY{o}{\PYZgt{}}\PY{p}{(}\PY{l+s}{\PYZdq{}}\PY{l+s}{out\PYZus{}iga\PYZus{}tot.svg}\PY{l+s}{\PYZdq{}}\PY{p}{)}\PY{p}{;}
  \PY{p}{\PYZcb{}}
  \PY{p}{\PYZob{}}
\PY{c+c1}{//\PYZlt{}listing\PYZhy{}7\PYZgt{}}
    \PY{k}{enum} \PY{p}{\PYZob{}} \PY{n}{opts} \PY{o}{=} \PY{n}{opts}\PY{o}{:}\PY{o}{:}\PY{n}{iga} \PY{o}{|} \PY{n}{opts}\PY{o}{:}\PY{o}{:}\PY{n}{fct} \PY{p}{\PYZcb{}}\PY{p}{;}
\PY{c+c1}{//\PYZlt{}/listing\PYZhy{}7\PYZgt{}}
    \PY{n}{test}\PY{o}{\PYZlt{}}\PY{n}{opts}\PY{o}{\PYZgt{}}\PY{p}{(}\PY{l+s}{\PYZdq{}}\PY{l+s}{out\PYZus{}iga\PYZus{}fct.svg}\PY{l+s}{\PYZdq{}}\PY{p}{)}\PY{p}{;}
  \PY{p}{\PYZcb{}}
  \PY{p}{\PYZob{}}
\PY{c+c1}{//\PYZlt{}listing\PYZhy{}8\PYZgt{}}
    \PY{k}{enum} \PY{p}{\PYZob{}} \PY{n}{opts} \PY{o}{=} \PY{n}{opts}\PY{o}{:}\PY{o}{:}\PY{n}{iga} \PY{o}{|} \PY{n}{opts}\PY{o}{:}\PY{o}{:}\PY{n}{tot} \PY{o}{|} \PY{n}{opts}\PY{o}{:}\PY{o}{:}\PY{n}{fct} \PY{p}{\PYZcb{}}\PY{p}{;}
\PY{c+c1}{//\PYZlt{}/listing\PYZhy{}8\PYZgt{}}
    \PY{n}{test}\PY{o}{\PYZlt{}}\PY{n}{opts}\PY{o}{\PYZgt{}}\PY{p}{(}\PY{l+s}{\PYZdq{}}\PY{l+s}{out\PYZus{}iga\PYZus{}tot\PYZus{}fct.svg}\PY{l+s}{\PYZdq{}}\PY{p}{)}\PY{p}{;}
  \PY{p}{\PYZcb{}}
\PY{p}{\PYZcb{}}
\end{Verbatim}

%% file: rotating_cone.cpp.tex
\begin{Verbatim}[commandchars=\\\{\}]
\PY{c+cm}{/* }
\PY{c+cm}{ * @file}
\PY{c+cm}{ * @copyright University of Warsaw}
\PY{c+cm}{ * @section LICENSE}
\PY{c+cm}{ * GPLv3+ (see the COPYING file or http://www.gnu.org/licenses/)}
\PY{c+cm}{ *}
\PY{c+cm}{ * \PYZbs{}include \PYZdq{}rotating\PYZus{}cone/test\PYZus{}rotating\PYZus{}cone.cpp\PYZdq{}}
\PY{c+cm}{ * \PYZbs{}image html \PYZdq{}../../tests/rotating\PYZus{}cone/figure.svg\PYZdq{}}
\PY{c+cm}{ */}

\PY{c+cp}{\PYZsh{}}\PY{c+cp}{include \PYZlt{}cmath\PYZgt{}}

\PY{c+cp}{\PYZsh{}}\PY{c+cp}{include \PYZlt{}boost}\PY{c+cp}{/}\PY{c+cp}{math}\PY{c+cp}{/}\PY{c+cp}{constants}\PY{c+cp}{/}\PY{c+cp}{constants.hpp\PYZgt{}}
\PY{k}{using} \PY{n}{boost}\PY{o}{:}\PY{o}{:}\PY{n}{math}\PY{o}{:}\PY{o}{:}\PY{n}{constants}\PY{o}{:}\PY{o}{:}\PY{n}{pi}\PY{p}{;}

\PY{c+cp}{\PYZsh{}}\PY{c+cp}{include \PYZlt{}libmpdata++}\PY{c+cp}{/}\PY{c+cp}{solvers}\PY{c+cp}{/}\PY{c+cp}{mpdata.hpp\PYZgt{}}
\PY{c+cp}{\PYZsh{}}\PY{c+cp}{include \PYZlt{}libmpdata++}\PY{c+cp}{/}\PY{c+cp}{concurr}\PY{c+cp}{/}\PY{c+cp}{threads.hpp\PYZgt{}}
\PY{c+cp}{\PYZsh{}}\PY{c+cp}{include \PYZlt{}libmpdata++}\PY{c+cp}{/}\PY{c+cp}{output}\PY{c+cp}{/}\PY{c+cp}{gnuplot.hpp\PYZgt{}}
\PY{k}{using} \PY{k}{namespace} \PY{n}{libmpdataxx}\PY{p}{;}

\PY{k}{template} \PY{o}{\PYZlt{}}\PY{k+kt}{int} \PY{n}{opts\PYZus{}arg}\PY{p}{,} \PY{k+kt}{int} \PY{n}{opts\PYZus{}iters}\PY{o}{\PYZgt{}}
\PY{k+kt}{void} \PY{n}{test}\PY{p}{(}\PY{k}{const} \PY{n}{std}\PY{o}{:}\PY{o}{:}\PY{n}{string} \PY{n}{filename}\PY{p}{)}
\PY{p}{\PYZob{}}
  \PY{k}{struct} \PY{k+kt}{ct\PYZus{}params\PYZus{}t} \PY{o}{:} \PY{k+kt}{ct\PYZus{}params\PYZus{}default\PYZus{}t}
  \PY{p}{\PYZob{}}
    \PY{k}{using} \PY{k+kt}{real\PYZus{}t} \PY{o}{=} \PY{k+kt}{double}\PY{p}{;}
\PY{c+c1}{//\PYZlt{}listing\PYZhy{}1\PYZgt{}}
    \PY{k}{enum} \PY{p}{\PYZob{}} \PY{n}{n\PYZus{}dims} \PY{o}{=} \PY{l+m+mi}{2} \PY{p}{\PYZcb{}}\PY{p}{;}
    \PY{k}{enum} \PY{p}{\PYZob{}} \PY{n}{n\PYZus{}eqns} \PY{o}{=} \PY{l+m+mi}{1} \PY{p}{\PYZcb{}}\PY{p}{;}
\PY{c+c1}{//\PYZlt{}/listing\PYZhy{}1\PYZgt{}}
    \PY{k}{enum} \PY{p}{\PYZob{}} \PY{n}{opts} \PY{o}{=} \PY{n}{opts\PYZus{}arg} \PY{p}{\PYZcb{}}\PY{p}{;}
  \PY{p}{\PYZcb{}}\PY{p}{;}

  \PY{k}{typename} \PY{k+kt}{ct\PYZus{}params\PYZus{}t}\PY{o}{:}\PY{o}{:}\PY{k+kt}{real\PYZus{}t} 
    \PY{n}{dt} \PY{o}{=} \PY{l+m+mf}{.1}\PY{p}{,}
    \PY{n}{dx} \PY{o}{=} \PY{l+m+mi}{1}\PY{p}{,}
    \PY{n}{dy} \PY{o}{=} \PY{l+m+mi}{1}\PY{p}{,}
    \PY{n}{omg} \PY{o}{=} \PY{l+m+mf}{.1}\PY{p}{,}
    \PY{n}{h} \PY{o}{=} \PY{l+m+mf}{4.}\PY{p}{,} 
    \PY{n}{h0} \PY{o}{=} \PY{l+m+mi}{1}\PY{p}{;}

\PY{c+c1}{/// @brief settings from @copybrief Anderson\PYZus{}and\PYZus{}Fattahi\PYZus{}1974}
\PY{c+c1}{//    dt = 10 * pi\PYZlt{}real\PYZus{}t\PYZgt{}(),}
\PY{c+c1}{//    omg = \PYZhy{}.001,// / (2 * pi\PYZlt{}real\PYZus{}t\PYZgt{}()),}
\PY{c+c1}{//    r = 4. * dx,}
\PY{c+c1}{//    h0 = \PYZhy{}.5,}
\PY{c+c1}{//    x0 = 21. * dx,}
\PY{c+c1}{//    y0 = 15. * dy;}

  \PY{k+kt}{int} \PY{n}{nt} \PY{o}{=} \PY{l+m+mi}{628} \PY{o}{*} \PY{l+m+mi}{6}\PY{p}{;}

  \PY{k}{using} \PY{k+kt}{slv\PYZus{}out\PYZus{}t} \PY{o}{=} \PY{n}{output}\PY{o}{:}\PY{o}{:}\PY{n}{gnuplot}\PY{o}{\PYZlt{}}\PY{n}{solvers}\PY{o}{:}\PY{o}{:}\PY{n}{mpdata}\PY{o}{\PYZlt{}}\PY{k+kt}{ct\PYZus{}params\PYZus{}t}\PY{o}{\PYZgt{}}\PY{o}{\PYZgt{}}\PY{p}{;}
  \PY{k}{typename} \PY{k+kt}{slv\PYZus{}out\PYZus{}t}\PY{o}{:}\PY{o}{:}\PY{k+kt}{rt\PYZus{}params\PYZus{}t} \PY{n}{p}\PY{p}{;}

  \PY{c+c1}{// pre instantiation}
  \PY{k}{switch} \PY{p}{(}\PY{n}{opts\PYZus{}iters}\PY{p}{)} \PY{c+c1}{// the crazy logic below is just for prettying the listing!}
  \PY{p}{\PYZob{}}
    \PY{k}{case} \PY{l+m+mi}{3}: 
\PY{c+c1}{//\PYZlt{}listing\PYZhy{}4\PYZgt{}}
      \PY{n}{p}\PY{p}{.}\PY{n}{n\PYZus{}iters} \PY{o}{=} \PY{l+m+mi}{3}\PY{p}{;}
\PY{c+c1}{//\PYZlt{}/listing\PYZhy{}4\PYZgt{}}
      \PY{k}{break}\PY{p}{;}
    \PY{n+nl}{default:}
      \PY{n}{p}\PY{p}{.}\PY{n}{n\PYZus{}iters} \PY{o}{=} \PY{n}{opts\PYZus{}iters}\PY{p}{;} 
  \PY{p}{\PYZcb{}}
  \PY{n}{p}\PY{p}{.}\PY{n}{grid\PYZus{}size} \PY{o}{=} \PY{p}{\PYZob{}}\PY{l+m+mi}{101}\PY{p}{,} \PY{l+m+mi}{101}\PY{p}{\PYZcb{}}\PY{p}{;}

  \PY{n}{p}\PY{p}{.}\PY{n}{outfreq} \PY{o}{=} \PY{n}{nt}\PY{p}{;} 
  \PY{n}{p}\PY{p}{.}\PY{n}{outvars}\PY{p}{[}\PY{l+m+mi}{0}\PY{p}{]}\PY{p}{.}\PY{n}{name} \PY{o}{=} \PY{l+s}{\PYZdq{}}\PY{l+s}{psi}\PY{l+s}{\PYZdq{}}\PY{p}{;}
  \PY{p}{\PYZob{}}
    \PY{n}{std}\PY{o}{:}\PY{o}{:}\PY{n}{ostringstream} \PY{n}{tmp}\PY{p}{;}
    \PY{n}{tmp} \PY{o}{\PYZlt{}}\PY{o}{\PYZlt{}} \PY{n}{filename} \PY{o}{\PYZlt{}}\PY{o}{\PYZlt{}} \PY{l+s}{\PYZdq{}}\PY{l+s}{\PYZus{}\PYZpc{}s\PYZus{}\PYZpc{}d.svg}\PY{l+s}{\PYZdq{}}\PY{p}{;}
    \PY{n}{p}\PY{p}{.}\PY{n}{gnuplot\PYZus{}output} \PY{o}{=} \PY{n}{tmp}\PY{p}{.}\PY{n}{str}\PY{p}{(}\PY{p}{)}\PY{p}{;}    
  \PY{p}{\PYZcb{}}
  \PY{n}{p}\PY{p}{.}\PY{n}{gnuplot\PYZus{}view} \PY{o}{=} \PY{l+s}{\PYZdq{}}\PY{l+s}{map}\PY{l+s}{\PYZdq{}}\PY{p}{;}
  \PY{n}{p}\PY{p}{.}\PY{n}{gnuplot\PYZus{}with} \PY{o}{=} \PY{l+s}{\PYZdq{}}\PY{l+s}{lines}\PY{l+s}{\PYZdq{}}\PY{p}{;}
  \PY{n}{p}\PY{p}{.}\PY{n}{gnuplot\PYZus{}surface} \PY{o}{=} \PY{n+nb}{false}\PY{p}{;}
  \PY{n}{p}\PY{p}{.}\PY{n}{gnuplot\PYZus{}contour} \PY{o}{=} \PY{n+nb}{true}\PY{p}{;}
  \PY{p}{\PYZob{}}
    \PY{n}{std}\PY{o}{:}\PY{o}{:}\PY{n}{ostringstream} \PY{n}{tmp}\PY{p}{;}
    \PY{n}{tmp} \PY{o}{\PYZlt{}}\PY{o}{\PYZlt{}} \PY{l+s}{\PYZdq{}}\PY{l+s}{[}\PY{l+s}{\PYZdq{}} \PY{o}{\PYZlt{}}\PY{o}{\PYZlt{}} \PY{n}{h0} \PY{o}{\PYZhy{}}\PY{l+m+mf}{.5} \PY{o}{\PYZlt{}}\PY{o}{\PYZlt{}} \PY{l+s}{\PYZdq{}}\PY{l+s}{ : }\PY{l+s}{\PYZdq{}} \PY{o}{\PYZlt{}}\PY{o}{\PYZlt{}} \PY{n}{h0} \PY{o}{+} \PY{n}{h} \PY{o}{+} \PY{l+m+mf}{.5} \PY{o}{\PYZlt{}}\PY{o}{\PYZlt{}} \PY{l+s}{\PYZdq{}}\PY{l+s}{]}\PY{l+s}{\PYZdq{}}\PY{p}{;}
    \PY{n}{p}\PY{p}{.}\PY{n}{gnuplot\PYZus{}cbrange} \PY{o}{=} \PY{n}{tmp}\PY{p}{.}\PY{n}{str}\PY{p}{(}\PY{p}{)}\PY{p}{;}
  \PY{p}{\PYZcb{}}
  \PY{n}{p}\PY{p}{.}\PY{n}{gnuplot\PYZus{}xrange} \PY{o}{=} \PY{l+s}{\PYZdq{}}\PY{l+s}{[25 : 75]}\PY{l+s}{\PYZdq{}}\PY{p}{;}
  \PY{n}{p}\PY{p}{.}\PY{n}{gnuplot\PYZus{}yrange} \PY{o}{=} \PY{l+s}{\PYZdq{}}\PY{l+s}{[50 : 100]}\PY{l+s}{\PYZdq{}}\PY{p}{;}
\PY{c+c1}{//  p.gnuplot\PYZus{}xrange = \PYZdq{}[0 : 100]\PYZdq{};}
\PY{c+c1}{//  p.gnuplot\PYZus{}yrange = \PYZdq{}[0 : 100]\PYZdq{};}
  \PY{p}{\PYZob{}}
    \PY{n}{std}\PY{o}{:}\PY{o}{:}\PY{n}{ostringstream} \PY{n}{tmp}\PY{p}{;}
    \PY{n}{tmp} \PY{o}{\PYZlt{}}\PY{o}{\PYZlt{}} \PY{l+s}{\PYZdq{}}\PY{l+s}{levels incremental }\PY{l+s}{\PYZdq{}} \PY{o}{\PYZlt{}}\PY{o}{\PYZlt{}} \PY{n}{h0} \PY{o}{\PYZhy{}}\PY{l+m+mf}{.25} \PY{o}{\PYZlt{}}\PY{o}{\PYZlt{}} \PY{l+s}{\PYZdq{}}\PY{l+s}{, .25,}\PY{l+s}{\PYZdq{}} \PY{o}{\PYZlt{}}\PY{o}{\PYZlt{}} \PY{n}{h0} \PY{o}{+} \PY{n}{h} \PY{o}{+} \PY{l+m+mf}{.25}\PY{p}{;}
    \PY{n}{p}\PY{p}{.}\PY{n}{gnuplot\PYZus{}cntrparam} \PY{o}{=} \PY{n}{tmp}\PY{p}{.}\PY{n}{str}\PY{p}{(}\PY{p}{)}\PY{p}{;}
  \PY{p}{\PYZcb{}}
  \PY{n}{p}\PY{p}{.}\PY{n}{gnuplot\PYZus{}fontsize} \PY{o}{=} \PY{l+s}{\PYZdq{}}\PY{l+s}{14}\PY{l+s}{\PYZdq{}}\PY{p}{;}
  \PY{n}{p}\PY{p}{.}\PY{n}{gnuplot\PYZus{}cbrange} \PY{o}{=} \PY{l+s}{\PYZdq{}}\PY{l+s}{[.75 : 5.25]}\PY{l+s}{\PYZdq{}}\PY{p}{;}
  \PY{n}{p}\PY{p}{.}\PY{n}{gnuplot\PYZus{}palette} \PY{o}{=} \PY{l+s}{\PYZdq{}}\PY{l+s}{defined (}\PY{l+s}{\PYZdq{}} 
    \PY{l+s}{\PYZdq{}}\PY{l+s}{0.75 \PYZsq{}\PYZsh{}ff0000\PYZsq{},}\PY{l+s}{\PYZdq{}}
    \PY{l+s}{\PYZdq{}}\PY{l+s}{1.00 \PYZsq{}\PYZsh{}ff0000\PYZsq{},}\PY{l+s}{\PYZdq{}}
    \PY{l+s}{\PYZdq{}}\PY{l+s}{1.00 \PYZsq{}\PYZsh{}ffffff\PYZsq{},}\PY{l+s}{\PYZdq{}}
    \PY{l+s}{\PYZdq{}}\PY{l+s}{1.25 \PYZsq{}\PYZsh{}ffffff\PYZsq{},}\PY{l+s}{\PYZdq{}}
    \PY{l+s}{\PYZdq{}}\PY{l+s}{1.25 \PYZsq{}\PYZsh{}993399\PYZsq{},}\PY{l+s}{\PYZdq{}}
    \PY{l+s}{\PYZdq{}}\PY{l+s}{2.25 \PYZsq{}\PYZsh{}00CCFF\PYZsq{},}\PY{l+s}{\PYZdq{}}
    \PY{l+s}{\PYZdq{}}\PY{l+s}{3.25 \PYZsq{}\PYZsh{}66CC00\PYZsq{},}\PY{l+s}{\PYZdq{}}
    \PY{l+s}{\PYZdq{}}\PY{l+s}{4.25 \PYZsq{}\PYZsh{}FC8727\PYZsq{},}\PY{l+s}{\PYZdq{}}
    \PY{l+s}{\PYZdq{}}\PY{l+s}{5.25 \PYZsq{}\PYZsh{}FFFF00\PYZsq{}) maxcolors 18}\PY{l+s}{\PYZdq{}}\PY{p}{;}
  \PY{n}{p}\PY{p}{.}\PY{n}{gnuplot\PYZus{}term} \PY{o}{=} \PY{l+s}{\PYZdq{}}\PY{l+s}{svg}\PY{l+s}{\PYZdq{}}\PY{p}{;}
  \PY{n}{p}\PY{p}{.}\PY{n}{gnuplot\PYZus{}title} \PY{o}{=} \PY{l+s}{\PYZdq{}}\PY{l+s}{notitle}\PY{l+s}{\PYZdq{}}\PY{p}{;}

\PY{c+c1}{//\PYZlt{}listing\PYZhy{}2\PYZgt{}}
  \PY{c+c1}{// instantiation}
  \PY{n}{concurr}\PY{o}{:}\PY{o}{:}\PY{n}{threads}\PY{o}{\PYZlt{}}
    \PY{k+kt}{slv\PYZus{}out\PYZus{}t}\PY{p}{,} 
    \PY{n}{bcond}\PY{o}{:}\PY{o}{:}\PY{n}{open}\PY{p}{,} \PY{n}{bcond}\PY{o}{:}\PY{o}{:}\PY{n}{open}\PY{p}{,}
    \PY{n}{bcond}\PY{o}{:}\PY{o}{:}\PY{n}{open}\PY{p}{,} \PY{n}{bcond}\PY{o}{:}\PY{o}{:}\PY{n}{open}
  \PY{o}{\PYZgt{}} \PY{n}{run}\PY{p}{(}\PY{n}{p}\PY{p}{)}\PY{p}{;} 
\PY{c+c1}{//\PYZlt{}/listing\PYZhy{}2\PYZgt{}}
  \PY{p}{\PYZob{}}

\PY{c+c1}{//TODO \PYZhy{} dawniej listing 3 zaczynal się tutaj \PYZhy{} może tak zostać?}
    \PY{c+c1}{// constants used in the set\PYZhy{}up definition}
    \PY{k}{enum} \PY{p}{\PYZob{}}\PY{n}{x}\PY{p}{,} \PY{n}{y}\PY{p}{\PYZcb{}}\PY{p}{;}
    \PY{k}{const} \PY{k}{typename} \PY{k+kt}{ct\PYZus{}params\PYZus{}t}\PY{o}{:}\PY{o}{:}\PY{k+kt}{real\PYZus{}t}
      \PY{n}{r} \PY{o}{=} \PY{l+m+mf}{15.} \PY{o}{*} \PY{n}{dx}\PY{p}{,}
      \PY{n}{x0} \PY{o}{=} \PY{l+m+mi}{50} \PY{o}{*} \PY{n}{dx}\PY{p}{,}
      \PY{n}{y0} \PY{o}{=} \PY{l+m+mi}{75} \PY{o}{*} \PY{n}{dy}\PY{p}{,}
      \PY{n}{xc} \PY{o}{=} \PY{l+m+mf}{.5} \PY{o}{*} \PY{p}{(}\PY{n}{p}\PY{p}{.}\PY{n}{grid\PYZus{}size}\PY{p}{[}\PY{n}{x}\PY{p}{]}\PY{o}{\PYZhy{}}\PY{l+m+mi}{1}\PY{p}{)} \PY{o}{*} \PY{n}{dx}\PY{p}{,}
      \PY{n}{yc} \PY{o}{=} \PY{l+m+mf}{.5} \PY{o}{*} \PY{p}{(}\PY{n}{p}\PY{p}{.}\PY{n}{grid\PYZus{}size}\PY{p}{[}\PY{n}{y}\PY{p}{]}\PY{o}{\PYZhy{}}\PY{l+m+mi}{1}\PY{p}{)} \PY{o}{*} \PY{n}{dy}\PY{p}{;}

\PY{c+c1}{//\PYZlt{}listing\PYZhy{}3\PYZgt{}}
    \PY{c+c1}{// temporary array of the same ...}
    \PY{n}{decltype}\PY{p}{(}\PY{n}{run}\PY{p}{.}\PY{n}{advectee}\PY{p}{(}\PY{p}{)}\PY{p}{)}        \PY{c+c1}{// type }
      \PY{n}{tmp}\PY{p}{(}\PY{n}{run}\PY{p}{.}\PY{n}{advectee}\PY{p}{(}\PY{p}{)}\PY{p}{.}\PY{n}{extent}\PY{p}{(}\PY{p}{)}\PY{p}{)}\PY{p}{;} \PY{c+c1}{// and size }
    \PY{c+c1}{// ... as the one returned by advectee()}

    \PY{c+c1}{// helper vars for Blitz++ tensor notation}
    \PY{n}{blitz}\PY{o}{:}\PY{o}{:}\PY{n}{firstIndex} \PY{n}{i}\PY{p}{;}
    \PY{n}{blitz}\PY{o}{:}\PY{o}{:}\PY{n}{secondIndex} \PY{n}{j}\PY{p}{;}

    \PY{c+c1}{// cone shape ...}
    \PY{n}{tmp} \PY{o}{=} \PY{n}{blitz}\PY{o}{:}\PY{o}{:}\PY{n}{pow}\PY{p}{(}\PY{n}{i} \PY{o}{*} \PY{n}{dx} \PY{o}{\PYZhy{}} \PY{n}{x0}\PY{p}{,} \PY{l+m+mi}{2}\PY{p}{)} \PY{o}{+} 
          \PY{n}{blitz}\PY{o}{:}\PY{o}{:}\PY{n}{pow}\PY{p}{(}\PY{n}{j} \PY{o}{*} \PY{n}{dy} \PY{o}{\PYZhy{}} \PY{n}{y0}\PY{p}{,} \PY{l+m+mi}{2}\PY{p}{)}\PY{p}{;}

    \PY{c+c1}{// ... cut off at zero}
    \PY{n}{run}\PY{p}{.}\PY{n}{advectee}\PY{p}{(}\PY{p}{)} \PY{o}{=} \PY{n}{h0} \PY{o}{+} \PY{n}{where}\PY{p}{(}
      \PY{n}{tmp} \PY{o}{\PYZhy{}} \PY{n}{pow}\PY{p}{(}\PY{n}{r}\PY{p}{,} \PY{l+m+mi}{2}\PY{p}{)} \PY{o}{\PYZlt{}}\PY{o}{=} \PY{l+m+mi}{0}\PY{p}{,}                  \PY{c+c1}{//if}
      \PY{n}{h} \PY{o}{*} \PY{n}{blitz}\PY{o}{:}\PY{o}{:}\PY{n}{sqr}\PY{p}{(}\PY{l+m+mi}{1} \PY{o}{\PYZhy{}} \PY{n}{tmp} \PY{o}{/} \PY{n}{pow}\PY{p}{(}\PY{n}{r}\PY{p}{,} \PY{l+m+mi}{2}\PY{p}{)}\PY{p}{)}\PY{p}{,}   \PY{c+c1}{//then}
      \PY{l+m+mf}{0.}                                     \PY{c+c1}{//else}
    \PY{p}{)}\PY{p}{;}

    \PY{c+c1}{// constant\PYZhy{}angular\PYZhy{}velocity rotational field}
    \PY{n}{run}\PY{p}{.}\PY{n}{advector}\PY{p}{(}\PY{n}{x}\PY{p}{)} \PY{o}{=}  \PY{n}{omg} \PY{o}{*} \PY{p}{(}\PY{n}{j} \PY{o}{*} \PY{n}{dy} \PY{o}{\PYZhy{}} \PY{n}{yc}\PY{p}{)} \PY{o}{*} \PY{n}{dt}\PY{o}{/}\PY{n}{dx}\PY{p}{;}
    \PY{n}{run}\PY{p}{.}\PY{n}{advector}\PY{p}{(}\PY{n}{y}\PY{p}{)} \PY{o}{=} \PY{o}{\PYZhy{}}\PY{n}{omg} \PY{o}{*} \PY{p}{(}\PY{n}{i} \PY{o}{*} \PY{n}{dx} \PY{o}{\PYZhy{}} \PY{n}{xc}\PY{p}{)} \PY{o}{*} \PY{n}{dt}\PY{o}{/}\PY{n}{dy}\PY{p}{;}
\PY{c+c1}{//\PYZlt{}/listing\PYZhy{}3\PYZgt{}}
  \PY{p}{\PYZcb{}}
    \PY{c+c1}{// TODO: an assert confirming that the above did what it should have done}
    \PY{c+c1}{//       (in context of the advector()\PYZsq{}s use of blitz::Array::reindex())}

  \PY{c+c1}{// time stepping}
  \PY{n}{run}\PY{p}{.}\PY{n}{advance}\PY{p}{(}\PY{n}{nt}\PY{p}{)}\PY{p}{;}
  
  \PY{n}{std}\PY{o}{:}\PY{o}{:}\PY{n}{cout}\PY{o}{\PYZlt{}}\PY{o}{\PYZlt{}}\PY{l+s}{\PYZdq{}}\PY{l+s}{min(psi) = }\PY{l+s}{\PYZdq{}} \PY{o}{\PYZlt{}}\PY{o}{\PYZlt{}} \PY{n}{min}\PY{p}{(}\PY{n}{run}\PY{p}{.}\PY{n}{advectee}\PY{p}{(}\PY{p}{)}\PY{p}{)} \PY{o}{\PYZlt{}}\PY{o}{\PYZlt{}} \PY{n}{std}\PY{o}{:}\PY{o}{:}\PY{n}{endl}\PY{p}{;}
\PY{p}{\PYZcb{}}

\PY{k+kt}{int} \PY{n}{main}\PY{p}{(}\PY{p}{)}
\PY{p}{\PYZob{}}
  \PY{p}{\PYZob{}}
    \PY{k}{enum} \PY{p}{\PYZob{}} \PY{n}{opts} \PY{o}{=} \PY{l+m+mi}{0} \PY{p}{\PYZcb{}}\PY{p}{;}
    \PY{k}{enum} \PY{p}{\PYZob{}} \PY{n}{opts\PYZus{}iters} \PY{o}{=} \PY{l+m+mi}{2}\PY{p}{\PYZcb{}}\PY{p}{;}
    \PY{n}{test}\PY{o}{\PYZlt{}}\PY{n}{opts}\PY{p}{,} \PY{n}{opts\PYZus{}iters}\PY{o}{\PYZgt{}}\PY{p}{(}\PY{l+s}{\PYZdq{}}\PY{l+s}{basic}\PY{l+s}{\PYZdq{}}\PY{p}{)}\PY{p}{;}
  \PY{p}{\PYZcb{}}
  \PY{p}{\PYZob{}}
    \PY{k}{enum} \PY{p}{\PYZob{}} \PY{n}{opts} \PY{o}{=} \PY{n}{opts}\PY{o}{:}\PY{o}{:}\PY{n}{fct} \PY{p}{\PYZcb{}}\PY{p}{;}
    \PY{k}{enum} \PY{p}{\PYZob{}} \PY{n}{opts\PYZus{}iters} \PY{o}{=} \PY{l+m+mi}{2}\PY{p}{\PYZcb{}}\PY{p}{;}
    \PY{n}{test}\PY{o}{\PYZlt{}}\PY{n}{opts}\PY{p}{,} \PY{n}{opts\PYZus{}iters}\PY{o}{\PYZgt{}}\PY{p}{(}\PY{l+s}{\PYZdq{}}\PY{l+s}{fct}\PY{l+s}{\PYZdq{}}\PY{p}{)}\PY{p}{;}
  \PY{p}{\PYZcb{}}
  \PY{p}{\PYZob{}}
    \PY{k}{enum} \PY{p}{\PYZob{}} \PY{n}{opts} \PY{o}{=} \PY{n}{opts}\PY{o}{:}\PY{o}{:}\PY{n}{fct} \PY{o}{|} \PY{n}{opts}\PY{o}{:}\PY{o}{:}\PY{n}{tot} \PY{p}{\PYZcb{}}\PY{p}{;}
    \PY{k}{enum} \PY{p}{\PYZob{}} \PY{n}{opts\PYZus{}iters} \PY{o}{=} \PY{l+m+mi}{3}\PY{p}{\PYZcb{}}\PY{p}{;}
    \PY{n}{test}\PY{o}{\PYZlt{}}\PY{n}{opts}\PY{p}{,} \PY{n}{opts\PYZus{}iters}\PY{o}{\PYZgt{}}\PY{p}{(}\PY{l+s}{\PYZdq{}}\PY{l+s}{iters3\PYZus{}tot\PYZus{}fct}\PY{l+s}{\PYZdq{}}\PY{p}{)}\PY{p}{;}
  \PY{p}{\PYZcb{}}
  \PY{p}{\PYZob{}}
    \PY{k}{enum} \PY{p}{\PYZob{}} \PY{n}{opts} \PY{o}{=} \PY{n}{opts}\PY{o}{:}\PY{o}{:}\PY{n}{iga} \PY{o}{|} \PY{n}{opts}\PY{o}{:}\PY{o}{:}\PY{n}{fct}\PY{p}{\PYZcb{}}\PY{p}{;}
    \PY{k}{enum} \PY{p}{\PYZob{}} \PY{n}{opts\PYZus{}iters} \PY{o}{=} \PY{l+m+mi}{2}\PY{p}{\PYZcb{}}\PY{p}{;}
    \PY{n}{test}\PY{o}{\PYZlt{}}\PY{n}{opts}\PY{p}{,} \PY{n}{opts\PYZus{}iters}\PY{o}{\PYZgt{}}\PY{p}{(}\PY{l+s}{\PYZdq{}}\PY{l+s}{iga\PYZus{}fct}\PY{l+s}{\PYZdq{}}\PY{p}{)}\PY{p}{;}
  \PY{p}{\PYZcb{}}
  \PY{p}{\PYZob{}}
    \PY{k}{enum} \PY{p}{\PYZob{}} \PY{n}{opts} \PY{o}{=} \PY{n}{opts}\PY{o}{:}\PY{o}{:}\PY{n}{iga} \PY{o}{|} \PY{n}{opts}\PY{o}{:}\PY{o}{:}\PY{n}{tot} \PY{o}{|} \PY{n}{opts}\PY{o}{:}\PY{o}{:}\PY{n}{fct} \PY{p}{\PYZcb{}}\PY{p}{;}
    \PY{k}{enum} \PY{p}{\PYZob{}} \PY{n}{opts\PYZus{}iters} \PY{o}{=} \PY{l+m+mi}{2}\PY{p}{\PYZcb{}}\PY{p}{;}
    \PY{n}{test}\PY{o}{\PYZlt{}}\PY{n}{opts}\PY{p}{,} \PY{n}{opts\PYZus{}iters}\PY{o}{\PYZgt{}}\PY{p}{(}\PY{l+s}{\PYZdq{}}\PY{l+s}{iga\PYZus{}tot\PYZus{}fct}\PY{l+s}{\PYZdq{}}\PY{p}{)}\PY{p}{;}
  \PY{p}{\PYZcb{}}
\PY{p}{\PYZcb{}}
\end{Verbatim}

%% file: rotating_sphere.cpp.tex
\begin{Verbatim}[commandchars=\\\{\}]
\PY{c+cm}{/**}
\PY{c+cm}{ * @file}
\PY{c+cm}{ * @copyright University of Warsaw}
\PY{c+cm}{ * @section LICENSE}
\PY{c+cm}{ * GPLv3+ (see the COPYING file or http://www.gnu.org/licenses/)}
\PY{c+cm}{ */}

\PY{c+cp}{\PYZsh{}}\PY{c+cp}{include \PYZlt{}boost}\PY{c+cp}{/}\PY{c+cp}{math}\PY{c+cp}{/}\PY{c+cp}{constants}\PY{c+cp}{/}\PY{c+cp}{constants.hpp\PYZgt{}}
\PY{k}{using} \PY{n}{boost}\PY{o}{:}\PY{o}{:}\PY{n}{math}\PY{o}{:}\PY{o}{:}\PY{n}{constants}\PY{o}{:}\PY{o}{:}\PY{n}{pi}\PY{p}{;}

\PY{c+cp}{\PYZsh{}}\PY{c+cp}{include \PYZlt{}libmpdata++}\PY{c+cp}{/}\PY{c+cp}{solvers}\PY{c+cp}{/}\PY{c+cp}{mpdata.hpp\PYZgt{}}
\PY{c+cp}{\PYZsh{}}\PY{c+cp}{include \PYZlt{}libmpdata++}\PY{c+cp}{/}\PY{c+cp}{concurr}\PY{c+cp}{/}\PY{c+cp}{threads.hpp\PYZgt{}}
\PY{c+cp}{\PYZsh{}}\PY{c+cp}{include \PYZlt{}libmpdata++}\PY{c+cp}{/}\PY{c+cp}{output}\PY{c+cp}{/}\PY{c+cp}{hdf5\PYZus{}xdmf.hpp\PYZgt{}}
\PY{k}{using} \PY{k}{namespace} \PY{n}{libmpdataxx}\PY{p}{;}

\PY{k}{enum} \PY{p}{\PYZob{}}\PY{n}{x}\PY{p}{,} \PY{n}{y}\PY{p}{,} \PY{n}{z}\PY{p}{\PYZcb{}}\PY{p}{;}
\PY{k}{struct} \PY{k+kt}{ct\PYZus{}params\PYZus{}t} \PY{o}{:} \PY{k+kt}{ct\PYZus{}params\PYZus{}default\PYZus{}t}
\PY{p}{\PYZob{}}
  \PY{k}{using} \PY{k+kt}{real\PYZus{}t} \PY{o}{=} \PY{k+kt}{double}\PY{p}{;}
\PY{c+c1}{//\PYZlt{}listing\PYZhy{}1\PYZgt{}}
  \PY{k}{enum} \PY{p}{\PYZob{}} \PY{n}{n\PYZus{}dims} \PY{o}{=} \PY{l+m+mi}{3} \PY{p}{\PYZcb{}}\PY{p}{;}
\PY{c+c1}{//\PYZlt{}/listing\PYZhy{}1\PYZgt{}}
  \PY{k}{enum} \PY{p}{\PYZob{}} \PY{n}{n\PYZus{}eqns} \PY{o}{=} \PY{l+m+mi}{1} \PY{p}{\PYZcb{}}\PY{p}{;}
  \PY{k}{enum} \PY{p}{\PYZob{}} \PY{n}{opts} \PY{o}{=} \PY{n}{opts}\PY{o}{:}\PY{o}{:}\PY{n}{abs} \PY{p}{\PYZcb{}}\PY{p}{;}
\PY{p}{\PYZcb{}}\PY{p}{;}

\PY{k}{template}\PY{o}{\PYZlt{}}\PY{k}{class} \PY{n+nc}{T}\PY{o}{\PYZgt{}}
\PY{k+kt}{void} \PY{n}{setup}\PY{p}{(}\PY{n}{T} \PY{o}{\PYZam{}}\PY{n}{solver}\PY{p}{)}
\PY{p}{\PYZob{}}
  \PY{k}{const} \PY{k+kt}{ct\PYZus{}params\PYZus{}t}\PY{o}{:}\PY{o}{:}\PY{k+kt}{real\PYZus{}t}
    \PY{n}{dt} \PY{o}{=} \PY{l+m+mf}{0.2}\PY{p}{,}
    \PY{n}{dx} \PY{o}{=} \PY{l+m+mf}{2.5}\PY{p}{,}
    \PY{n}{dy} \PY{o}{=} \PY{l+m+mf}{2.5}\PY{p}{,}
    \PY{n}{dz} \PY{o}{=} \PY{l+m+mf}{2.5}\PY{p}{,}
    \PY{n}{h} \PY{o}{=} \PY{l+m+mf}{4.}\PY{p}{,}
    \PY{n}{r} \PY{o}{=} \PY{l+m+mi}{7} \PY{o}{*} \PY{n}{dx}\PY{p}{,}
    \PY{n}{x0} \PY{o}{=} \PY{p}{(}\PY{l+m+mi}{20} \PY{o}{\PYZhy{}} \PY{l+m+mi}{7} \PY{o}{*} \PY{n}{pow}\PY{p}{(}\PY{l+m+mi}{6}\PY{p}{,} \PY{o}{\PYZhy{}}\PY{l+m+mf}{0.5}\PY{p}{)}\PY{p}{)} \PY{o}{*} \PY{n}{dx}\PY{p}{,}
    \PY{n}{y0} \PY{o}{=} \PY{p}{(}\PY{l+m+mi}{20} \PY{o}{\PYZhy{}} \PY{l+m+mi}{7} \PY{o}{*} \PY{n}{pow}\PY{p}{(}\PY{l+m+mi}{6}\PY{p}{,} \PY{o}{\PYZhy{}}\PY{l+m+mf}{0.5}\PY{p}{)}\PY{p}{)} \PY{o}{*} \PY{n}{dy}\PY{p}{,}
    \PY{n}{z0} \PY{o}{=} \PY{p}{(}\PY{l+m+mi}{20} \PY{o}{+} \PY{l+m+mi}{14} \PY{o}{*} \PY{n}{pow}\PY{p}{(}\PY{l+m+mi}{6}\PY{p}{,} \PY{o}{\PYZhy{}}\PY{l+m+mf}{0.5}\PY{p}{)}\PY{p}{)} \PY{o}{*} \PY{n}{dz}\PY{p}{;}

  \PY{n}{blitz}\PY{o}{:}\PY{o}{:}\PY{n}{firstIndex} \PY{n}{i}\PY{p}{;}
  \PY{n}{blitz}\PY{o}{:}\PY{o}{:}\PY{n}{secondIndex} \PY{n}{j}\PY{p}{;}
  \PY{n}{blitz}\PY{o}{:}\PY{o}{:}\PY{n}{thirdIndex} \PY{n}{k}\PY{p}{;}

  \PY{c+c1}{// sphere shape}
  \PY{n}{decltype}\PY{p}{(}\PY{n}{solver}\PY{p}{.}\PY{n}{advectee}\PY{p}{(}\PY{p}{)}\PY{p}{)} \PY{n}{tmp}\PY{p}{(}\PY{n}{solver}\PY{p}{.}\PY{n}{advectee}\PY{p}{(}\PY{p}{)}\PY{p}{.}\PY{n}{extent}\PY{p}{(}\PY{p}{)}\PY{p}{)}\PY{p}{;}
  \PY{n}{tmp} \PY{o}{=}   \PY{n}{blitz}\PY{o}{:}\PY{o}{:}\PY{n}{pow}\PY{p}{(}\PY{n}{i} \PY{o}{*} \PY{n}{dx} \PY{o}{\PYZhy{}} \PY{n}{x0}\PY{p}{,} \PY{l+m+mi}{2}\PY{p}{)}
        \PY{o}{+} \PY{n}{blitz}\PY{o}{:}\PY{o}{:}\PY{n}{pow}\PY{p}{(}\PY{n}{j} \PY{o}{*} \PY{n}{dx} \PY{o}{\PYZhy{}} \PY{n}{y0}\PY{p}{,} \PY{l+m+mi}{2}\PY{p}{)}
        \PY{o}{+} \PY{n}{blitz}\PY{o}{:}\PY{o}{:}\PY{n}{pow}\PY{p}{(}\PY{n}{k} \PY{o}{*} \PY{n}{dx} \PY{o}{\PYZhy{}} \PY{n}{z0}\PY{p}{,} \PY{l+m+mi}{2}\PY{p}{)}\PY{p}{;}
  \PY{n}{solver}\PY{p}{.}\PY{n}{advectee}\PY{p}{(}\PY{p}{)} \PY{o}{=} \PY{n}{where}\PY{p}{(}\PY{n}{tmp} \PY{o}{\PYZhy{}} \PY{n}{pow}\PY{p}{(}\PY{n}{r}\PY{p}{,} \PY{l+m+mi}{2}\PY{p}{)} \PY{o}{\PYZlt{}}\PY{o}{=} \PY{l+m+mi}{0}\PY{p}{,} \PY{n}{h} \PY{o}{*} \PY{p}{(}\PY{l+m+mi}{1} \PY{o}{\PYZhy{}} \PY{n}{blitz}\PY{o}{:}\PY{o}{:}\PY{n}{sqrt}\PY{p}{(}\PY{n}{tmp}\PY{p}{)} \PY{o}{/} \PY{n}{r}\PY{p}{)} \PY{p}{,} \PY{l+m+mi}{0}\PY{p}{)}\PY{p}{;}

  \PY{k}{const} \PY{k+kt}{ct\PYZus{}params\PYZus{}t}\PY{o}{:}\PY{o}{:}\PY{k+kt}{real\PYZus{}t}
    \PY{n}{omega} \PY{o}{=} \PY{l+m+mf}{0.1}\PY{p}{,}
    \PY{n}{xc} \PY{o}{=} \PY{l+m+mi}{20} \PY{o}{*} \PY{n}{dx}\PY{p}{,}
    \PY{n}{yc} \PY{o}{=} \PY{l+m+mi}{20} \PY{o}{*} \PY{n}{dy}\PY{p}{,}
    \PY{n}{zc} \PY{o}{=} \PY{l+m+mi}{20} \PY{o}{*} \PY{n}{dz}\PY{p}{;}
  \PY{c+c1}{// constant angular velocity rotational field}
  \PY{n}{solver}\PY{p}{.}\PY{n}{advector}\PY{p}{(}\PY{n}{x}\PY{p}{)} \PY{o}{=} \PY{n}{omega} \PY{o}{/} \PY{n}{sqrt}\PY{p}{(}\PY{l+m+mi}{3}\PY{p}{)} \PY{o}{*} \PY{p}{(}\PY{o}{\PYZhy{}}\PY{p}{(}\PY{n}{j} \PY{o}{*} \PY{n}{dy} \PY{o}{\PYZhy{}} \PY{n}{yc}\PY{p}{)} \PY{o}{+} \PY{p}{(}\PY{n}{k} \PY{o}{*} \PY{n}{dz} \PY{o}{\PYZhy{}} \PY{n}{zc}\PY{p}{)}\PY{p}{)} \PY{o}{*} \PY{n}{dt} \PY{o}{/} \PY{n}{dx}\PY{p}{;}
  \PY{n}{solver}\PY{p}{.}\PY{n}{advector}\PY{p}{(}\PY{n}{y}\PY{p}{)} \PY{o}{=} \PY{n}{omega} \PY{o}{/} \PY{n}{sqrt}\PY{p}{(}\PY{l+m+mi}{3}\PY{p}{)} \PY{o}{*} \PY{p}{(} \PY{p}{(}\PY{n}{i} \PY{o}{*} \PY{n}{dx} \PY{o}{\PYZhy{}} \PY{n}{xc}\PY{p}{)} \PY{o}{\PYZhy{}} \PY{p}{(}\PY{n}{k} \PY{o}{*} \PY{n}{dz} \PY{o}{\PYZhy{}} \PY{n}{zc}\PY{p}{)}\PY{p}{)} \PY{o}{*} \PY{n}{dt} \PY{o}{/} \PY{n}{dy}\PY{p}{;}
  \PY{n}{solver}\PY{p}{.}\PY{n}{advector}\PY{p}{(}\PY{n}{z}\PY{p}{)} \PY{o}{=} \PY{n}{omega} \PY{o}{/} \PY{n}{sqrt}\PY{p}{(}\PY{l+m+mi}{3}\PY{p}{)} \PY{o}{*} \PY{p}{(}\PY{o}{\PYZhy{}}\PY{p}{(}\PY{n}{i} \PY{o}{*} \PY{n}{dx} \PY{o}{\PYZhy{}} \PY{n}{xc}\PY{p}{)} \PY{o}{+} \PY{p}{(}\PY{n}{j} \PY{o}{*} \PY{n}{dy} \PY{o}{\PYZhy{}} \PY{n}{yc}\PY{p}{)}\PY{p}{)} \PY{o}{*} \PY{n}{dt} \PY{o}{/} \PY{n}{dz}\PY{p}{;}
\PY{p}{\PYZcb{}}

\PY{k+kt}{int} \PY{n}{main}\PY{p}{(}\PY{p}{)}
\PY{p}{\PYZob{}}
  \PY{k+kt}{int} \PY{n}{nt} \PY{o}{=} \PY{l+m+mi}{5} \PY{o}{*} \PY{l+m+mi}{314}\PY{p}{;}
  \PY{k}{using} \PY{k+kt}{slv\PYZus{}t} \PY{o}{=} \PY{n}{solvers}\PY{o}{:}\PY{o}{:}\PY{n}{mpdata}\PY{o}{\PYZlt{}}\PY{k+kt}{ct\PYZus{}params\PYZus{}t}\PY{o}{\PYZgt{}}\PY{p}{;}
\PY{c+c1}{//\PYZlt{}listing\PYZhy{}2\PYZgt{}}
  \PY{k}{using} \PY{k+kt}{slv\PYZus{}out\PYZus{}t} \PY{o}{=} \PY{n}{output}\PY{o}{:}\PY{o}{:}\PY{n}{hdf5\PYZus{}xdmf}\PY{o}{\PYZlt{}}\PY{k+kt}{slv\PYZus{}t}\PY{o}{\PYZgt{}}\PY{p}{;}
\PY{c+c1}{//\PYZlt{}/listing\PYZhy{}2\PYZgt{}}
  \PY{k+kt}{slv\PYZus{}out\PYZus{}t}\PY{o}{:}\PY{o}{:}\PY{k+kt}{rt\PYZus{}params\PYZus{}t} \PY{n}{p}\PY{p}{;}

  \PY{c+c1}{// pre instantation}
  \PY{n}{p}\PY{p}{.}\PY{n}{n\PYZus{}iters} \PY{o}{=} \PY{l+m+mi}{4}\PY{p}{;}
  \PY{n}{p}\PY{p}{.}\PY{n}{grid\PYZus{}size} \PY{o}{=} \PY{p}{\PYZob{}}\PY{l+m+mi}{41}\PY{p}{,} \PY{l+m+mi}{41}\PY{p}{,} \PY{l+m+mi}{41}\PY{p}{\PYZcb{}}\PY{p}{;}

  \PY{n}{p}\PY{p}{.}\PY{n}{outfreq} \PY{o}{=} \PY{n}{nt}\PY{p}{;}
  \PY{n}{p}\PY{p}{.}\PY{n}{outvars}\PY{p}{[}\PY{l+m+mi}{0}\PY{p}{]}\PY{p}{.}\PY{n}{name} \PY{o}{=} \PY{l+s}{\PYZdq{}}\PY{l+s}{psi}\PY{l+s}{\PYZdq{}}\PY{p}{;}
\PY{c+c1}{//\PYZlt{}listing\PYZhy{}3\PYZgt{}}
  \PY{n}{p}\PY{p}{.}\PY{n}{outdir} \PY{o}{=} \PY{l+s}{\PYZdq{}}\PY{l+s}{rotating\PYZus{}sphere\PYZus{}3d}\PY{l+s}{\PYZdq{}}\PY{p}{;}
\PY{c+c1}{//\PYZlt{}/listing\PYZhy{}3\PYZgt{}}

  \PY{c+c1}{// instantation}
  \PY{n}{concurr}\PY{o}{:}\PY{o}{:}\PY{n}{threads}\PY{o}{\PYZlt{}}
  \PY{k+kt}{slv\PYZus{}out\PYZus{}t}\PY{p}{,}
  \PY{n}{bcond}\PY{o}{:}\PY{o}{:}\PY{n}{open}\PY{p}{,} \PY{n}{bcond}\PY{o}{:}\PY{o}{:}\PY{n}{open}\PY{p}{,}
  \PY{n}{bcond}\PY{o}{:}\PY{o}{:}\PY{n}{open}\PY{p}{,} \PY{n}{bcond}\PY{o}{:}\PY{o}{:}\PY{n}{open}\PY{p}{,}
  \PY{n}{bcond}\PY{o}{:}\PY{o}{:}\PY{n}{open}\PY{p}{,} \PY{n}{bcond}\PY{o}{:}\PY{o}{:}\PY{n}{open}
  \PY{o}{\PYZgt{}} \PY{n}{slv}\PY{p}{(}\PY{n}{p}\PY{p}{)}\PY{p}{;}

  \PY{c+c1}{// post instantation}
  \PY{n}{setup}\PY{p}{(}\PY{n}{slv}\PY{p}{)}\PY{p}{;}

  \PY{c+c1}{// time stepping}
  \PY{n}{slv}\PY{p}{.}\PY{n}{advance}\PY{p}{(}\PY{n}{nt}\PY{p}{)}\PY{p}{;}
\PY{p}{\PYZcb{}}
\end{Verbatim}

%% file: over_the_pole.cpp.tex
\begin{Verbatim}[commandchars=\\\{\}]
\PY{c+cm}{/* }
\PY{c+cm}{ * @file}
\PY{c+cm}{ * @copyright University of Warsaw}
\PY{c+cm}{ * @section LICENSE}
\PY{c+cm}{ * GPLv3+ (see the COPYING file or http://www.gnu.org/licenses/)}
\PY{c+cm}{ */}

\PY{c+cp}{\PYZsh{}}\PY{c+cp}{include \PYZlt{}cmath\PYZgt{}}

\PY{c+cp}{\PYZsh{}}\PY{c+cp}{include \PYZlt{}boost}\PY{c+cp}{/}\PY{c+cp}{math}\PY{c+cp}{/}\PY{c+cp}{constants}\PY{c+cp}{/}\PY{c+cp}{constants.hpp\PYZgt{}}
\PY{k}{using} \PY{n}{boost}\PY{o}{:}\PY{o}{:}\PY{n}{math}\PY{o}{:}\PY{o}{:}\PY{n}{constants}\PY{o}{:}\PY{o}{:}\PY{n}{pi}\PY{p}{;}

\PY{c+cp}{\PYZsh{}}\PY{c+cp}{include \PYZlt{}libmpdata++}\PY{c+cp}{/}\PY{c+cp}{solvers}\PY{c+cp}{/}\PY{c+cp}{mpdata.hpp\PYZgt{}}
\PY{c+cp}{\PYZsh{}}\PY{c+cp}{include \PYZlt{}libmpdata++}\PY{c+cp}{/}\PY{c+cp}{concurr}\PY{c+cp}{/}\PY{c+cp}{threads.hpp\PYZgt{}}
\PY{c+cp}{\PYZsh{}}\PY{c+cp}{include \PYZlt{}libmpdata++}\PY{c+cp}{/}\PY{c+cp}{output}\PY{c+cp}{/}\PY{c+cp}{hdf5\PYZus{}xdmf.hpp\PYZgt{}}
\PY{k}{using} \PY{k}{namespace} \PY{n}{libmpdataxx}\PY{p}{;}

\PY{k}{template} \PY{o}{\PYZlt{}}\PY{k+kt}{int} \PY{n}{opts\PYZus{}arg}\PY{p}{,} \PY{k+kt}{int} \PY{n}{opts\PYZus{}iters}\PY{o}{\PYZgt{}}
\PY{k+kt}{void} \PY{n}{test}\PY{p}{(}\PY{k}{const} \PY{n}{std}\PY{o}{:}\PY{o}{:}\PY{n}{string} \PY{n}{filename}\PY{p}{)}
\PY{p}{\PYZob{}}
  \PY{k}{enum} \PY{p}{\PYZob{}}\PY{n}{x}\PY{p}{,} \PY{n}{y}\PY{p}{\PYZcb{}}\PY{p}{;}
  \PY{k}{struct} \PY{k+kt}{ct\PYZus{}params\PYZus{}t} \PY{o}{:} \PY{k+kt}{ct\PYZus{}params\PYZus{}default\PYZus{}t}
  \PY{p}{\PYZob{}}
    \PY{k}{using} \PY{k+kt}{real\PYZus{}t} \PY{o}{=} \PY{k+kt}{double}\PY{p}{;}
    \PY{k}{enum} \PY{p}{\PYZob{}} \PY{n}{n\PYZus{}dims} \PY{o}{=} \PY{l+m+mi}{2} \PY{p}{\PYZcb{}}\PY{p}{;}
    \PY{k}{enum} \PY{p}{\PYZob{}} \PY{n}{n\PYZus{}eqns} \PY{o}{=} \PY{l+m+mi}{1} \PY{p}{\PYZcb{}}\PY{p}{;}
    \PY{k}{enum} \PY{p}{\PYZob{}} \PY{n}{opts} \PY{o}{=} \PY{n}{opts\PYZus{}arg} \PY{p}{\PYZcb{}}\PY{p}{;}
  \PY{p}{\PYZcb{}}\PY{p}{;}
  
  \PY{k+kt}{int}
    \PY{n}{nlon} \PY{o}{=} \PY{l+m+mi}{129}\PY{p}{,}
    \PY{n}{nlat} \PY{o}{=} \PY{l+m+mi}{64}\PY{p}{,}
    \PY{n}{nt} \PY{o}{=} \PY{l+m+mi}{5120}\PY{p}{;}
  
  \PY{k}{typename} \PY{k+kt}{ct\PYZus{}params\PYZus{}t}\PY{o}{:}\PY{o}{:}\PY{k+kt}{real\PYZus{}t} 
    \PY{n}{pi} \PY{o}{=} \PY{n}{boost}\PY{o}{:}\PY{o}{:}\PY{n}{math}\PY{o}{:}\PY{o}{:}\PY{n}{constants}\PY{o}{:}\PY{o}{:}\PY{n}{pi}\PY{o}{\PYZlt{}}\PY{k}{typename} \PY{k+kt}{ct\PYZus{}params\PYZus{}t}\PY{o}{:}\PY{o}{:}\PY{k+kt}{real\PYZus{}t}\PY{o}{\PYZgt{}}\PY{p}{(}\PY{p}{)}\PY{p}{,}
    \PY{n}{dlmb} \PY{o}{=} \PY{l+m+mi}{2} \PY{o}{*} \PY{n}{pi} \PY{o}{/} \PY{p}{(}\PY{n}{nlon} \PY{o}{\PYZhy{}} \PY{l+m+mi}{1}\PY{p}{)}\PY{p}{,}
    \PY{n}{dphi} \PY{o}{=} \PY{n}{pi} \PY{o}{/} \PY{n}{nlat}\PY{p}{;}

  \PY{k}{using} \PY{k+kt}{slv\PYZus{}out\PYZus{}t} \PY{o}{=} \PY{n}{output}\PY{o}{:}\PY{o}{:}\PY{n}{hdf5\PYZus{}xdmf}\PY{o}{\PYZlt{}}\PY{n}{solvers}\PY{o}{:}\PY{o}{:}\PY{n}{mpdata}\PY{o}{\PYZlt{}}\PY{k+kt}{ct\PYZus{}params\PYZus{}t}\PY{o}{\PYZgt{}}\PY{o}{\PYZgt{}}\PY{p}{;}
  \PY{k}{typename} \PY{k+kt}{slv\PYZus{}out\PYZus{}t}\PY{o}{:}\PY{o}{:}\PY{k+kt}{rt\PYZus{}params\PYZus{}t} \PY{n}{p}\PY{p}{;}

  \PY{n}{p}\PY{p}{.}\PY{n}{n\PYZus{}iters} \PY{o}{=} \PY{n}{opts\PYZus{}iters}\PY{p}{;} 
  \PY{n}{p}\PY{p}{.}\PY{n}{grid\PYZus{}size} \PY{o}{=} \PY{p}{\PYZob{}}\PY{n}{nlon}\PY{p}{,} \PY{n}{nlat}\PY{p}{\PYZcb{}}\PY{p}{;}

  \PY{n}{p}\PY{p}{.}\PY{n}{outfreq} \PY{o}{=} \PY{n}{nt}\PY{p}{;} 
  \PY{n}{p}\PY{p}{.}\PY{n}{outvars}\PY{p}{[}\PY{l+m+mi}{0}\PY{p}{]}\PY{p}{.}\PY{n}{name} \PY{o}{=} \PY{l+s}{\PYZdq{}}\PY{l+s}{psi}\PY{l+s}{\PYZdq{}}\PY{p}{;}
  \PY{n}{p}\PY{p}{.}\PY{n}{outdir} \PY{o}{=} \PY{n}{filename}\PY{p}{;}

\PY{c+c1}{//\PYZlt{}listing\PYZhy{}2\PYZgt{}}
  \PY{n}{concurr}\PY{o}{:}\PY{o}{:}\PY{n}{threads}\PY{o}{\PYZlt{}}
    \PY{k+kt}{slv\PYZus{}out\PYZus{}t}\PY{p}{,} 
    \PY{n}{bcond}\PY{o}{:}\PY{o}{:}\PY{n}{cyclic}\PY{p}{,} \PY{n}{bcond}\PY{o}{:}\PY{o}{:}\PY{n}{cyclic}\PY{p}{,}
    \PY{n}{bcond}\PY{o}{:}\PY{o}{:}\PY{n}{polar}\PY{p}{,} \PY{n}{bcond}\PY{o}{:}\PY{o}{:}\PY{n}{polar}
  \PY{o}{\PYZgt{}} \PY{n}{run}\PY{p}{(}\PY{n}{p}\PY{p}{)}\PY{p}{;} 
\PY{c+c1}{//\PYZlt{}/listing\PYZhy{}2\PYZgt{}}
  
  \PY{k}{typename} \PY{k+kt}{ct\PYZus{}params\PYZus{}t}\PY{o}{:}\PY{o}{:}\PY{k+kt}{real\PYZus{}t}
    \PY{n}{r} \PY{o}{=} \PY{l+m+mi}{7} \PY{o}{*} \PY{n}{dlmb}\PY{p}{,}
    \PY{n}{x0} \PY{o}{=} \PY{l+m+mi}{3} \PY{o}{*} \PY{n}{pi} \PY{o}{/} \PY{l+m+mi}{2}\PY{p}{,}
    \PY{n}{y0} \PY{o}{=} \PY{l+m+mi}{0}\PY{p}{;}

  \PY{n}{blitz}\PY{o}{:}\PY{o}{:}\PY{n}{firstIndex} \PY{n}{i}\PY{p}{;}
  \PY{n}{blitz}\PY{o}{:}\PY{o}{:}\PY{n}{secondIndex} \PY{n}{j}\PY{p}{;}

  \PY{n}{decltype}\PY{p}{(}\PY{n}{run}\PY{p}{.}\PY{n}{advectee}\PY{p}{(}\PY{p}{)}\PY{p}{)} 
    \PY{n}{tmp}\PY{p}{(}\PY{n}{run}\PY{p}{.}\PY{n}{advectee}\PY{p}{(}\PY{p}{)}\PY{p}{.}\PY{n}{extent}\PY{p}{(}\PY{p}{)}\PY{p}{)}\PY{p}{;}

  \PY{n}{tmp} \PY{o}{=} \PY{l+m+mi}{2} \PY{o}{*} \PY{p}{(}  \PY{n}{blitz}\PY{o}{:}\PY{o}{:}\PY{n}{pow2}\PY{p}{(}\PY{n}{blitz}\PY{o}{:}\PY{o}{:}\PY{n}{cos}\PY{p}{(}\PY{n}{dphi} \PY{o}{*} \PY{p}{(}\PY{n}{j} \PY{o}{+} \PY{l+m+mf}{0.5}\PY{p}{)} \PY{o}{\PYZhy{}} \PY{n}{pi} \PY{o}{/} \PY{l+m+mi}{2}\PY{p}{)} \PY{o}{*} \PY{n}{blitz}\PY{o}{:}\PY{o}{:}\PY{n}{sin}\PY{p}{(}\PY{p}{(}\PY{n}{dlmb} \PY{o}{*} \PY{n}{i} \PY{o}{\PYZhy{}} \PY{n}{x0}\PY{p}{)} \PY{o}{/} \PY{l+m+mi}{2}\PY{p}{)}\PY{p}{)}
             \PY{o}{+} \PY{n}{blitz}\PY{o}{:}\PY{o}{:}\PY{n}{pow2}\PY{p}{(}\PY{n}{blitz}\PY{o}{:}\PY{o}{:}\PY{n}{sin}\PY{p}{(}\PY{p}{(}\PY{n}{dphi} \PY{o}{*} \PY{p}{(}\PY{n}{j} \PY{o}{+} \PY{l+m+mf}{0.5}\PY{p}{)} \PY{o}{\PYZhy{}} \PY{n}{pi} \PY{o}{/} \PY{l+m+mi}{2} \PY{o}{\PYZhy{}} \PY{n}{y0}\PY{p}{)} \PY{o}{/} \PY{l+m+mi}{2}\PY{p}{)}\PY{p}{)}                     \PY{p}{)}\PY{p}{;}

  \PY{n}{run}\PY{p}{.}\PY{n}{advectee}\PY{p}{(}\PY{p}{)} \PY{o}{=} \PY{n}{where}\PY{p}{(}
    \PY{n}{tmp} \PY{o}{\PYZhy{}} \PY{n}{pow}\PY{p}{(}\PY{n}{r}\PY{p}{,} \PY{l+m+mi}{2}\PY{p}{)} \PY{o}{\PYZlt{}}\PY{o}{=} \PY{l+m+mi}{0}\PY{p}{,}                  \PY{c+c1}{//if}
    \PY{l+m+mi}{1} \PY{o}{\PYZhy{}} \PY{n}{sqrt}\PY{p}{(}\PY{n}{tmp}\PY{p}{)} \PY{o}{/} \PY{n}{r}\PY{p}{,}   \PY{c+c1}{//then}
    \PY{l+m+mf}{0.}                                     \PY{c+c1}{//else}
  \PY{p}{)}\PY{p}{;}
  
  \PY{k}{typename} \PY{k+kt}{ct\PYZus{}params\PYZus{}t}\PY{o}{:}\PY{o}{:}\PY{k+kt}{real\PYZus{}t}
    \PY{n}{udt} \PY{o}{=} \PY{l+m+mi}{2} \PY{o}{*} \PY{n}{pi} \PY{o}{/} \PY{n}{nt}\PY{p}{,}
    \PY{n}{b} \PY{o}{=} \PY{o}{\PYZhy{}}\PY{n}{pi} \PY{o}{/} \PY{l+m+mi}{2}\PY{p}{;}

  \PY{n}{run}\PY{p}{.}\PY{n}{advector}\PY{p}{(}\PY{n}{x}\PY{p}{)} \PY{o}{=} \PY{n}{dlmb} \PY{o}{*} \PY{n}{udt} \PY{o}{*} \PY{p}{(}\PY{n}{cos}\PY{p}{(}\PY{n}{b}\PY{p}{)} \PY{o}{*} \PY{n}{blitz}\PY{o}{:}\PY{o}{:}\PY{n}{cos}\PY{p}{(}\PY{p}{(}\PY{n}{j}\PY{o}{+}\PY{l+m+mf}{0.5}\PY{p}{)} \PY{o}{*} \PY{n}{dphi} \PY{o}{\PYZhy{}} \PY{n}{pi} \PY{o}{/} \PY{l+m+mi}{2}\PY{p}{)} \PY{o}{+} \PY{n}{sin}\PY{p}{(}\PY{n}{b}\PY{p}{)} \PY{o}{*} \PY{n}{blitz}\PY{o}{:}\PY{o}{:}\PY{n}{sin}\PY{p}{(}\PY{p}{(}\PY{n}{j}\PY{o}{+}\PY{l+m+mf}{0.5}\PY{p}{)} \PY{o}{*} \PY{n}{dphi} \PY{o}{\PYZhy{}} \PY{n}{pi} \PY{o}{/} \PY{l+m+mi}{2}\PY{p}{)} \PY{o}{*} \PY{n}{blitz}\PY{o}{:}\PY{o}{:}\PY{n}{cos}\PY{p}{(}\PY{p}{(}\PY{n}{i}\PY{o}{+}\PY{l+m+mf}{0.5}\PY{p}{)} \PY{o}{*} \PY{n}{dlmb}\PY{p}{)}\PY{p}{)}\PY{p}{;}
  
  \PY{n}{run}\PY{p}{.}\PY{n}{advector}\PY{p}{(}\PY{n}{y}\PY{p}{)} \PY{o}{=} \PY{o}{\PYZhy{}}\PY{n}{dlmb} \PY{o}{*} \PY{n}{udt} \PY{o}{*} \PY{n}{sin}\PY{p}{(}\PY{n}{b}\PY{p}{)} \PY{o}{*} \PY{n}{blitz}\PY{o}{:}\PY{o}{:}\PY{n}{sin}\PY{p}{(}\PY{p}{(}\PY{n}{i}\PY{p}{)} \PY{o}{*} \PY{n}{dlmb}\PY{p}{)}\PY{o}{*} \PY{n}{blitz}\PY{o}{:}\PY{o}{:}\PY{n}{cos}\PY{p}{(}\PY{p}{(}\PY{n}{j}\PY{o}{+}\PY{l+m+mi}{1}\PY{p}{)} \PY{o}{*} \PY{n}{dphi} \PY{o}{\PYZhy{}} \PY{n}{pi} \PY{o}{/} \PY{l+m+mi}{2}\PY{p}{)}\PY{p}{;}
  
\PY{c+c1}{//\PYZlt{}listing\PYZhy{}3\PYZgt{}}
  \PY{n}{run}\PY{p}{.}\PY{n}{g\PYZus{}factor}\PY{p}{(}\PY{p}{)} \PY{o}{=} \PY{n}{dlmb} \PY{o}{*} \PY{n}{dphi} \PY{o}{*}
    \PY{n}{blitz}\PY{o}{:}\PY{o}{:}\PY{n}{cos}\PY{p}{(}\PY{n}{dphi} \PY{o}{*} \PY{p}{(}\PY{n}{j} \PY{o}{+} \PY{l+m+mf}{0.5}\PY{p}{)} \PY{o}{\PYZhy{}} \PY{n}{pi} \PY{o}{/} \PY{l+m+mi}{2}\PY{p}{)}\PY{p}{;}
\PY{c+c1}{//\PYZlt{}/listing\PYZhy{}3\PYZgt{}}

  \PY{n}{run}\PY{p}{.}\PY{n}{advance}\PY{p}{(}\PY{n}{nt}\PY{p}{)}\PY{p}{;}
\PY{p}{\PYZcb{}}

\PY{k+kt}{int} \PY{n}{main}\PY{p}{(}\PY{p}{)}
\PY{p}{\PYZob{}}
\PY{c+c1}{//\PYZlt{}listing\PYZhy{}1\PYZgt{}}
    \PY{k}{enum} \PY{p}{\PYZob{}} \PY{n}{opts} \PY{o}{=} \PY{n}{opts}\PY{o}{:}\PY{o}{:}\PY{n}{nug} \PY{p}{\PYZcb{}}\PY{p}{;}
\PY{c+c1}{//\PYZlt{}/listing\PYZhy{}1\PYZgt{}}
  \PY{p}{\PYZob{}}
    \PY{k}{enum} \PY{p}{\PYZob{}} \PY{n}{opts} \PY{o}{=} \PY{n}{opts}\PY{o}{:}\PY{o}{:}\PY{n}{nug} \PY{o}{|} \PY{n}{opts}\PY{o}{:}\PY{o}{:}\PY{n}{iga} \PY{o}{|} \PY{n}{opts}\PY{o}{:}\PY{o}{:}\PY{n}{fct} \PY{p}{\PYZcb{}}\PY{p}{;}
    \PY{k}{const} \PY{k+kt}{int} \PY{n}{opts\PYZus{}iters} \PY{o}{=} \PY{l+m+mi}{2}\PY{p}{;}
    \PY{n}{test}\PY{o}{\PYZlt{}}\PY{n}{opts}\PY{p}{,} \PY{n}{opts\PYZus{}iters}\PY{o}{\PYZgt{}}\PY{p}{(}\PY{l+s}{\PYZdq{}}\PY{l+s}{default}\PY{l+s}{\PYZdq{}}\PY{p}{)}\PY{p}{;}
  \PY{p}{\PYZcb{}}
  \PY{p}{\PYZob{}}  
    \PY{k}{enum} \PY{p}{\PYZob{}} \PY{n}{opts} \PY{o}{=} \PY{n}{opts}\PY{o}{:}\PY{o}{:}\PY{n}{nug} \PY{o}{|} \PY{n}{opts}\PY{o}{:}\PY{o}{:}\PY{n}{tot} \PY{o}{|} \PY{n}{opts}\PY{o}{:}\PY{o}{:}\PY{n}{fct} \PY{p}{\PYZcb{}}\PY{p}{;}
    \PY{k}{const} \PY{k+kt}{int} \PY{n}{opts\PYZus{}iters} \PY{o}{=} \PY{l+m+mi}{3}\PY{p}{;}
    \PY{n}{test}\PY{o}{\PYZlt{}}\PY{n}{opts}\PY{p}{,} \PY{n}{opts\PYZus{}iters}\PY{o}{\PYZgt{}}\PY{p}{(}\PY{l+s}{\PYZdq{}}\PY{l+s}{best}\PY{l+s}{\PYZdq{}}\PY{p}{)}\PY{p}{;}
  \PY{p}{\PYZcb{}}
\PY{p}{\PYZcb{}}
\end{Verbatim}

%% file: mpdata_rhs.cpp.tex
\begin{Verbatim}[commandchars=\\\{\}]
\PY{c+cm}{/** }
\PY{c+cm}{ * @file}
\PY{c+cm}{ * @copyright University of Warsaw}
\PY{c+cm}{ * @section LICENSE}
\PY{c+cm}{ * GPLv3+ (see the COPYING file or http://www.gnu.org/licenses/)}
\PY{c+cm}{ *}
\PY{c+cm}{ * @brief improved Euler inhomogeneous solver  }
\PY{c+cm}{ *        (cf. eq. 32 in Smolarkiewicz 1998) // TODO cite}
\PY{c+cm}{ */}

\PY{c+cp}{\PYZsh{}}\PY{c+cp}{include \PYZlt{}libmpdata++}\PY{c+cp}{/}\PY{c+cp}{solvers}\PY{c+cp}{/}\PY{c+cp}{mpdata.hpp\PYZgt{}}

\PY{c+cp}{\PYZsh{}}\PY{c+cp}{pragma once}

\PY{k}{namespace} \PY{n}{libmpdataxx}
\PY{p}{\PYZob{}}
  \PY{k}{namespace} \PY{n}{solvers}
  \PY{p}{\PYZob{}}
    \PY{k}{enum} \PY{k+kt}{rhs\PYZus{}scheme\PYZus{}t} 
    \PY{p}{\PYZob{}} 
      \PY{n}{euler\PYZus{}a}\PY{p}{,} \PY{c+c1}{// Euler\PYZsq{}s method, Eulerian spirit:        psi\PYZca{}n+1 = ADV(psi\PYZca{}n) + R\PYZca{}n}
      \PY{n}{euler\PYZus{}b}\PY{p}{,} \PY{c+c1}{// Euler\PYZsq{}s method, semi\PYZhy{}Lagrangian spirit: psi\PYZca{}n+1 = ADV(psi\PYZca{}n + R\PYZca{}n)}
      \PY{n}{trapez}   \PY{c+c1}{// paraphrase of trapezoidal rule:         psi\PYZca{}n+1 = ADV(psi\PYZca{}n + 1/2 * R\PYZca{}n) + 1/2 * R\PYZca{}n+1 }
    \PY{p}{\PYZcb{}}\PY{p}{;}

    \PY{k}{template} \PY{o}{\PYZlt{}}\PY{k}{class} \PY{n+nc}{ct\PYZus{}params\PYZus{}t}\PY{o}{\PYZgt{}}
    \PY{k}{class} \PY{n+nc}{mpdata\PYZus{}rhs} \PY{o}{:} \PY{k}{public} \PY{n}{mpdata}\PY{o}{\PYZlt{}}\PY{k+kt}{ct\PYZus{}params\PYZus{}t}\PY{o}{\PYZgt{}}
    \PY{p}{\PYZob{}}
      \PY{k}{using} \PY{k+kt}{parent\PYZus{}t} \PY{o}{=} \PY{n}{mpdata}\PY{o}{\PYZlt{}}\PY{k+kt}{ct\PYZus{}params\PYZus{}t}\PY{o}{\PYZgt{}}\PY{p}{;}

      \PY{k}{enum} \PY{p}{\PYZob{}} \PY{n}{n} \PY{o}{=} \PY{l+m+mi}{0} \PY{p}{\PYZcb{}}\PY{p}{;} \PY{c+c1}{// just to make n, n+1 look nice :)}

\PY{c+cp}{\PYZsh{}}\PY{c+cp}{if !defined(NDEBUG)}
      \PY{k+kt}{bool} \PY{n}{update\PYZus{}rhs\PYZus{}called} \PY{o}{=} \PY{n+nb}{true}\PY{p}{;} \PY{c+c1}{// so that it nt=0 there\PYZsq{}s no complain}
\PY{c+cp}{\PYZsh{}}\PY{c+cp}{endif}

      \PY{n+nl}{protected:}

      \PY{c+c1}{// member fields}
      \PY{k}{typename} \PY{k+kt}{parent\PYZus{}t}\PY{o}{:}\PY{o}{:}\PY{k+kt}{real\PYZus{}t} \PY{n}{dt}\PY{p}{;}
      \PY{k+kt}{arrvec\PYZus{}t}\PY{o}{\PYZlt{}}\PY{k}{typename} \PY{k+kt}{parent\PYZus{}t}\PY{o}{:}\PY{o}{:}\PY{k+kt}{arr\PYZus{}t}\PY{o}{\PYZgt{}} \PY{o}{\PYZam{}}\PY{n}{rhs}\PY{p}{;}

\PY{c+c1}{//\PYZlt{}listing\PYZhy{}1\PYZgt{}}
      \PY{k}{virtual} \PY{k+kt}{void} \PY{n+nf}{update\PYZus{}rhs}\PY{p}{(}
        \PY{k+kt}{arrvec\PYZus{}t}\PY{o}{\PYZlt{}}\PY{k}{typename} \PY{k+kt}{parent\PYZus{}t}\PY{o}{:}\PY{o}{:}\PY{k+kt}{arr\PYZus{}t}\PY{o}{\PYZgt{}} \PY{o}{\PYZam{}}\PY{n}{rhs}\PY{p}{,} 
        \PY{k}{const} \PY{k}{typename} \PY{k+kt}{parent\PYZus{}t}\PY{o}{:}\PY{o}{:}\PY{k+kt}{real\PYZus{}t} \PY{o}{\PYZam{}}\PY{n}{dt}\PY{p}{,}
        \PY{k}{const} \PY{k+kt}{int} \PY{o}{\PYZam{}}\PY{n}{at}
      \PY{p}{)} 
\PY{c+c1}{//\PYZlt{}/listing\PYZhy{}1\PYZgt{}}
      \PY{p}{\PYZob{}}
        \PY{n}{assert}\PY{p}{(}\PY{n}{at} \PY{o}{=}\PY{o}{=} \PY{n}{n} \PY{o}{|}\PY{o}{|} \PY{n}{at} \PY{o}{=}\PY{o}{=} \PY{n}{n}\PY{o}{+}\PY{l+m+mi}{1}\PY{p}{)}\PY{p}{;}
\PY{c+cp}{\PYZsh{}}\PY{c+cp}{if !defined(NDEBUG)}
        \PY{n}{update\PYZus{}rhs\PYZus{}called} \PY{o}{=} \PY{n+nb}{true}\PY{p}{;}
\PY{c+cp}{\PYZsh{}}\PY{c+cp}{endif}
        \PY{c+c1}{// zero\PYZhy{}out all rhs arrays}
	\PY{k}{for} \PY{p}{(}\PY{k+kt}{int} \PY{n}{e} \PY{o}{=} \PY{l+m+mi}{0}\PY{p}{;} \PY{n}{e} \PY{o}{\PYZlt{}} \PY{k+kt}{parent\PYZus{}t}\PY{o}{:}\PY{o}{:}\PY{n}{n\PYZus{}eqns}\PY{p}{;} \PY{o}{+}\PY{o}{+}\PY{n}{e}\PY{p}{)} 
        \PY{p}{\PYZob{}}
          \PY{c+c1}{// do nothing for equations with no rhs}
          \PY{k}{if} \PY{p}{(}\PY{n}{opts}\PY{o}{:}\PY{o}{:}\PY{n}{isset}\PY{p}{(}\PY{k+kt}{ct\PYZus{}params\PYZus{}t}\PY{o}{:}\PY{o}{:}\PY{n}{hint\PYZus{}norhs}\PY{p}{,} \PY{n}{opts}\PY{o}{:}\PY{o}{:}\PY{n}{bit}\PY{p}{(}\PY{n}{e}\PY{p}{)}\PY{p}{)}\PY{p}{)} \PY{k}{continue}\PY{p}{;}

          \PY{c+c1}{// otherwise zero out the rhs}
          \PY{n}{rhs}\PY{p}{.}\PY{n}{at}\PY{p}{(}\PY{n}{e}\PY{p}{)}\PY{p}{(}\PY{k}{this}\PY{o}{\PYZhy{}}\PY{o}{\PYZgt{}}\PY{n}{ijk}\PY{p}{)} \PY{o}{=} \PY{l+m+mi}{0}\PY{p}{;}
        \PY{p}{\PYZcb{}}

        \PY{c+c1}{// fill halos with data (e.g. for computing gradients)}
        \PY{k}{for} \PY{p}{(}\PY{k+kt}{int} \PY{n}{e} \PY{o}{=} \PY{l+m+mi}{0}\PY{p}{;} \PY{n}{e} \PY{o}{\PYZlt{}} \PY{k+kt}{parent\PYZus{}t}\PY{o}{:}\PY{o}{:}\PY{n}{n\PYZus{}eqns}\PY{p}{;} \PY{o}{+}\PY{o}{+}\PY{n}{e}\PY{p}{)} \PY{k}{this}\PY{o}{\PYZhy{}}\PY{o}{\PYZgt{}}\PY{n}{xchng}\PY{p}{(}\PY{n}{e}\PY{p}{)}\PY{p}{;}
      \PY{p}{\PYZcb{}}

      \PY{k}{virtual} \PY{k+kt}{void} \PY{n}{apply\PYZus{}rhs}\PY{p}{(}
        \PY{k}{const} \PY{k}{typename} \PY{k+kt}{parent\PYZus{}t}\PY{o}{:}\PY{o}{:}\PY{k+kt}{real\PYZus{}t} \PY{o}{\PYZam{}}\PY{n}{dt}
      \PY{p}{)} \PY{n}{final}
      \PY{p}{\PYZob{}}
        \PY{k}{for} \PY{p}{(}\PY{k+kt}{int} \PY{n}{e} \PY{o}{=} \PY{l+m+mi}{0}\PY{p}{;} \PY{n}{e} \PY{o}{\PYZlt{}} \PY{k+kt}{parent\PYZus{}t}\PY{o}{:}\PY{o}{:}\PY{n}{n\PYZus{}eqns}\PY{p}{;} \PY{o}{+}\PY{o}{+}\PY{n}{e}\PY{p}{)} 
        \PY{p}{\PYZob{}}
          \PY{c+c1}{// do nothing for equations with no rhs}
          \PY{k}{if} \PY{p}{(}\PY{n}{opts}\PY{o}{:}\PY{o}{:}\PY{n}{isset}\PY{p}{(}\PY{k+kt}{ct\PYZus{}params\PYZus{}t}\PY{o}{:}\PY{o}{:}\PY{n}{hint\PYZus{}norhs}\PY{p}{,} \PY{n}{opts}\PY{o}{:}\PY{o}{:}\PY{n}{bit}\PY{p}{(}\PY{n}{e}\PY{p}{)}\PY{p}{)}\PY{p}{)} \PY{k}{continue}\PY{p}{;}

          \PY{c+c1}{// otherwise apply the rhs}
          \PY{k}{this}\PY{o}{\PYZhy{}}\PY{o}{\PYZgt{}}\PY{n}{state}\PY{p}{(}\PY{n}{e}\PY{p}{)}\PY{p}{(}\PY{k}{this}\PY{o}{\PYZhy{}}\PY{o}{\PYZgt{}}\PY{n}{ijk}\PY{p}{)} \PY{o}{+}\PY{o}{=} \PY{n}{dt} \PY{o}{*} \PY{n}{rhs}\PY{p}{.}\PY{n}{at}\PY{p}{(}\PY{n}{e}\PY{p}{)}\PY{p}{(}\PY{k}{this}\PY{o}{\PYZhy{}}\PY{o}{\PYZgt{}}\PY{n}{ijk}\PY{p}{)}\PY{p}{;}
        \PY{p}{\PYZcb{}}
      \PY{p}{\PYZcb{}}

      \PY{n+nl}{public:}

      \PY{k}{struct} \PY{k+kt}{rt\PYZus{}params\PYZus{}t} \PY{o}{:} \PY{k+kt}{parent\PYZus{}t}\PY{o}{:}\PY{o}{:}\PY{k+kt}{rt\PYZus{}params\PYZus{}t} 
      \PY{p}{\PYZob{}} 
        \PY{k}{typename} \PY{k+kt}{parent\PYZus{}t}\PY{o}{:}\PY{o}{:}\PY{k+kt}{real\PYZus{}t} \PY{n}{dt} \PY{o}{=} \PY{l+m+mi}{0}\PY{p}{;} 
      \PY{p}{\PYZcb{}}\PY{p}{;}
      
      \PY{c+c1}{// ctor}
      \PY{n}{mpdata\PYZus{}rhs}\PY{p}{(}
	\PY{k}{typename} \PY{k+kt}{parent\PYZus{}t}\PY{o}{:}\PY{o}{:}\PY{k+kt}{ctor\PYZus{}args\PYZus{}t} \PY{n}{args}\PY{p}{,} 
	\PY{k}{const} \PY{k+kt}{rt\PYZus{}params\PYZus{}t} \PY{o}{\PYZam{}}\PY{n}{p}
      \PY{p}{)} \PY{o}{:}
	\PY{k+kt}{parent\PYZus{}t}\PY{p}{(}\PY{n}{args}\PY{p}{,} \PY{n}{p}\PY{p}{)}\PY{p}{,} 
        \PY{n}{dt}\PY{p}{(}\PY{n}{p}\PY{p}{.}\PY{n}{dt}\PY{p}{)}\PY{p}{,}
        \PY{n}{rhs}\PY{p}{(}\PY{n}{args}\PY{p}{.}\PY{n}{mem}\PY{o}{\PYZhy{}}\PY{o}{\PYZgt{}}\PY{n}{tmp}\PY{p}{[}\PY{n}{\PYZus{}\PYZus{}FILE\PYZus{}\PYZus{}}\PY{p}{]}\PY{p}{[}\PY{l+m+mi}{0}\PY{p}{]}\PY{p}{)}
      \PY{p}{\PYZob{}}
        \PY{n}{assert}\PY{p}{(}\PY{n}{dt} \PY{o}{!}\PY{o}{=} \PY{l+m+mi}{0}\PY{p}{)}\PY{p}{;}
      \PY{p}{\PYZcb{}}

      \PY{c+c1}{// dtor}
      \PY{o}{\PYZti{}}\PY{n}{mpdata\PYZus{}rhs}\PY{p}{(}\PY{p}{)}
      \PY{p}{\PYZob{}}
\PY{c+cp}{\PYZsh{}}\PY{c+cp}{if !defined(NDEBUG)}
        \PY{n}{assert}\PY{p}{(}\PY{n}{update\PYZus{}rhs\PYZus{}called} \PY{o}{\PYZam{}}\PY{o}{\PYZam{}} \PY{l+s}{\PYZdq{}}\PY{l+s}{any overriding update\PYZus{}rhs() must call parent\PYZus{}t::update\PYZus{}rhs()}\PY{l+s}{\PYZdq{}}\PY{p}{)}\PY{p}{;}
\PY{c+cp}{\PYZsh{}}\PY{c+cp}{endif}
      \PY{p}{\PYZcb{}}

      \PY{k+kt}{void} \PY{n}{hook\PYZus{}ante\PYZus{}loop}\PY{p}{(}\PY{k+kt}{int} \PY{n}{nt}\PY{p}{)}
      \PY{p}{\PYZob{}}
        \PY{k+kt}{parent\PYZus{}t}\PY{o}{:}\PY{o}{:}\PY{n}{hook\PYZus{}ante\PYZus{}loop}\PY{p}{(}\PY{n}{nt}\PY{p}{)}\PY{p}{;}

        \PY{k}{switch} \PY{p}{(}\PY{p}{(}\PY{k+kt}{rhs\PYZus{}scheme\PYZus{}t}\PY{p}{)}\PY{k+kt}{ct\PYZus{}params\PYZus{}t}\PY{o}{:}\PY{o}{:}\PY{n}{rhs\PYZus{}scheme}\PY{p}{)}
        \PY{p}{\PYZob{}}
          \PY{k}{case} \PY{k+kt}{rhs\PYZus{}scheme\PYZus{}t}:\PY{o}{:}\PY{n}{euler\PYZus{}a}\PY{o}{:} 
          \PY{k}{case} \PY{k+kt}{rhs\PYZus{}scheme\PYZus{}t}:\PY{o}{:}\PY{n}{euler\PYZus{}b}\PY{o}{:} 
            \PY{k}{break}\PY{p}{;}
          \PY{k}{case} \PY{k+kt}{rhs\PYZus{}scheme\PYZus{}t}:\PY{o}{:}\PY{n}{trapez}\PY{o}{:}
            \PY{n}{update\PYZus{}rhs}\PY{p}{(}\PY{n}{rhs}\PY{p}{,} \PY{n}{dt} \PY{o}{/} \PY{l+m+mi}{2}\PY{p}{,} \PY{n}{n}\PY{p}{)}\PY{p}{;}
            \PY{k}{break}\PY{p}{;}
          \PY{n+nl}{default:} 
            \PY{n}{assert}\PY{p}{(}\PY{n+nb}{false}\PY{p}{)}\PY{p}{;}
        \PY{p}{\PYZcb{}}
      \PY{p}{\PYZcb{}}

      \PY{k+kt}{void} \PY{n}{hook\PYZus{}ante\PYZus{}step}\PY{p}{(}\PY{p}{)}
      \PY{p}{\PYZob{}}
        \PY{k+kt}{parent\PYZus{}t}\PY{o}{:}\PY{o}{:}\PY{n}{hook\PYZus{}ante\PYZus{}step}\PY{p}{(}\PY{p}{)}\PY{p}{;}

\PY{c+cp}{\PYZsh{}}\PY{c+cp}{if !defined(NDEBUG)}
        \PY{n}{update\PYZus{}rhs\PYZus{}called} \PY{o}{=} \PY{n+nb}{false}\PY{p}{;}
\PY{c+cp}{\PYZsh{}}\PY{c+cp}{endif}

        \PY{k}{switch} \PY{p}{(}\PY{p}{(}\PY{k+kt}{rhs\PYZus{}scheme\PYZus{}t}\PY{p}{)}\PY{k+kt}{ct\PYZus{}params\PYZus{}t}\PY{o}{:}\PY{o}{:}\PY{n}{rhs\PYZus{}scheme}\PY{p}{)}
        \PY{p}{\PYZob{}}
          \PY{k}{case} \PY{k+kt}{rhs\PYZus{}scheme\PYZus{}t}:\PY{o}{:}\PY{n}{euler\PYZus{}a}\PY{o}{:} 
            \PY{n}{update\PYZus{}rhs}\PY{p}{(}\PY{n}{rhs}\PY{p}{,} \PY{n}{dt}\PY{p}{,} \PY{n}{n}\PY{p}{)}\PY{p}{;}
            \PY{k}{break}\PY{p}{;}
          \PY{k}{case} \PY{k+kt}{rhs\PYZus{}scheme\PYZus{}t}:\PY{o}{:}\PY{n}{euler\PYZus{}b}\PY{o}{:} 
            \PY{n}{update\PYZus{}rhs}\PY{p}{(}\PY{n}{rhs}\PY{p}{,} \PY{n}{dt}\PY{p}{,} \PY{n}{n}\PY{p}{)}\PY{p}{;}
            \PY{n}{apply\PYZus{}rhs}\PY{p}{(}\PY{n}{dt}\PY{p}{)}\PY{p}{;} 
            \PY{k}{break}\PY{p}{;}
          \PY{k}{case} \PY{k+kt}{rhs\PYZus{}scheme\PYZus{}t}:\PY{o}{:}\PY{n}{trapez}\PY{o}{:} 
            \PY{n}{apply\PYZus{}rhs}\PY{p}{(}\PY{n}{dt} \PY{o}{/} \PY{l+m+mi}{2}\PY{p}{)}\PY{p}{;} 
            \PY{k}{break}\PY{p}{;}
          \PY{n+nl}{default:} 
            \PY{n}{assert}\PY{p}{(}\PY{n+nb}{false}\PY{p}{)}\PY{p}{;}
        \PY{p}{\PYZcb{}}
      \PY{p}{\PYZcb{}}

      \PY{k+kt}{void} \PY{n}{hook\PYZus{}post\PYZus{}step}\PY{p}{(}\PY{p}{)}
      \PY{p}{\PYZob{}}
        \PY{k+kt}{parent\PYZus{}t}\PY{o}{:}\PY{o}{:}\PY{n}{hook\PYZus{}post\PYZus{}step}\PY{p}{(}\PY{p}{)}\PY{p}{;}
        \PY{k}{switch} \PY{p}{(}\PY{p}{(}\PY{k+kt}{rhs\PYZus{}scheme\PYZus{}t}\PY{p}{)}\PY{k+kt}{ct\PYZus{}params\PYZus{}t}\PY{o}{:}\PY{o}{:}\PY{n}{rhs\PYZus{}scheme}\PY{p}{)}
        \PY{p}{\PYZob{}}
          \PY{k}{case} \PY{k+kt}{rhs\PYZus{}scheme\PYZus{}t}:\PY{o}{:}\PY{n}{euler\PYZus{}a}\PY{o}{:} 
            \PY{n}{apply\PYZus{}rhs}\PY{p}{(}\PY{n}{dt}\PY{p}{)}\PY{p}{;}
            \PY{k}{break}\PY{p}{;}
          \PY{k}{case} \PY{k+kt}{rhs\PYZus{}scheme\PYZus{}t}:\PY{o}{:}\PY{n}{euler\PYZus{}b}\PY{o}{:} 
            \PY{k}{break}\PY{p}{;}
          \PY{k}{case} \PY{k+kt}{rhs\PYZus{}scheme\PYZus{}t}:\PY{o}{:}\PY{n}{trapez}\PY{o}{:} 
            \PY{n}{update\PYZus{}rhs}\PY{p}{(}\PY{n}{rhs}\PY{p}{,} \PY{n}{dt} \PY{o}{/} \PY{l+m+mi}{2}\PY{p}{,} \PY{n}{n}\PY{o}{+}\PY{l+m+mi}{1}\PY{p}{)}\PY{p}{;}
            \PY{n}{apply\PYZus{}rhs}\PY{p}{(}\PY{n}{dt} \PY{o}{/} \PY{l+m+mi}{2}\PY{p}{)}\PY{p}{;}
            \PY{k}{break}\PY{p}{;}
          \PY{n+nl}{default:}
            \PY{n}{assert}\PY{p}{(}\PY{n+nb}{false}\PY{p}{)}\PY{p}{;}
        \PY{p}{\PYZcb{}}
      \PY{p}{\PYZcb{}} 

      \PY{k}{static} \PY{k+kt}{void} \PY{n}{alloc}\PY{p}{(}\PY{k}{typename} \PY{k+kt}{parent\PYZus{}t}\PY{o}{:}\PY{o}{:}\PY{k+kt}{mem\PYZus{}t} \PY{o}{*}\PY{n}{mem}\PY{p}{,} \PY{k}{const} \PY{k+kt}{rt\PYZus{}params\PYZus{}t} \PY{o}{\PYZam{}}\PY{n}{p}\PY{p}{)}
      \PY{p}{\PYZob{}}
        \PY{c+c1}{// TODO: optimise to skip allocs for equations with no rhs}
	\PY{k+kt}{parent\PYZus{}t}\PY{o}{:}\PY{o}{:}\PY{n}{alloc}\PY{p}{(}\PY{n}{mem}\PY{p}{,} \PY{n}{p}\PY{p}{)}\PY{p}{;}
        \PY{k+kt}{parent\PYZus{}t}\PY{o}{:}\PY{o}{:}\PY{n}{alloc\PYZus{}tmp\PYZus{}sclr}\PY{p}{(}\PY{n}{mem}\PY{p}{,} \PY{n}{p}\PY{p}{.}\PY{n}{grid\PYZus{}size}\PY{p}{,} \PY{n}{\PYZus{}\PYZus{}FILE\PYZus{}\PYZus{}}\PY{p}{,} \PY{k+kt}{parent\PYZus{}t}\PY{o}{:}\PY{o}{:}\PY{n}{n\PYZus{}eqns}\PY{p}{)}\PY{p}{;} \PY{c+c1}{// rhs array for each equation}
      \PY{p}{\PYZcb{}}
    \PY{p}{\PYZcb{}}\PY{p}{;}
  \PY{p}{\PYZcb{}}\PY{p}{;} \PY{c+c1}{// namespace solvers}
\PY{p}{\PYZcb{}}\PY{p}{;} \PY{c+c1}{// namespace libmpdataxx}
\end{Verbatim}

%% file: example_7.cpp.tex
\begin{Verbatim}[commandchars=\\\{\}]
\PY{c+cm}{/** }
\PY{c+cm}{ * @file}
\PY{c+cm}{ * @copyright University of Warsaw}
\PY{c+cm}{ * @section LICENSE}
\PY{c+cm}{ * GPLv3+ (see the COPYING file or http://www.gnu.org/licenses/)}
\PY{c+cm}{ *}
\PY{c+cm}{ * @brief a minimalistic model of a harmonic oscillator}
\PY{c+cm}{ * (consult eq. 28\PYZhy{}30 in Smolarkiewicz 2006, IJNMF)}
\PY{c+cm}{ *}
\PY{c+cm}{ */}
\PY{c+c1}{//}

\PY{c+c1}{//\PYZlt{}listing\PYZhy{}1\PYZgt{}}
\PY{c+cp}{\PYZsh{}}\PY{c+cp}{include \PYZlt{}libmpdata++}\PY{c+cp}{/}\PY{c+cp}{solvers}\PY{c+cp}{/}\PY{c+cp}{mpdata\PYZus{}rhs.hpp\PYZgt{}}

\PY{k}{template} \PY{o}{\PYZlt{}}\PY{k}{class} \PY{n+nc}{ct\PYZus{}params\PYZus{}t}\PY{o}{\PYZgt{}}
\PY{k}{struct} \PY{n}{coupled\PYZus{}harmosc} \PY{o}{:} \PY{k}{public} 
  \PY{n}{libmpdataxx}\PY{o}{:}\PY{o}{:}\PY{n}{solvers}\PY{o}{:}\PY{o}{:}\PY{n}{mpdata\PYZus{}rhs}\PY{o}{\PYZlt{}}\PY{k+kt}{ct\PYZus{}params\PYZus{}t}\PY{o}{\PYZgt{}}
\PY{p}{\PYZob{}} \PY{c+c1}{// aliases}
  \PY{k}{using} \PY{k+kt}{parent\PYZus{}t} \PY{o}{=} 
    \PY{n}{libmpdataxx}\PY{o}{:}\PY{o}{:}\PY{n}{solvers}\PY{o}{:}\PY{o}{:}\PY{n}{mpdata\PYZus{}rhs}\PY{o}{\PYZlt{}}\PY{k+kt}{ct\PYZus{}params\PYZus{}t}\PY{o}{\PYZgt{}}\PY{p}{;}
  \PY{k}{using} \PY{n}{ix} \PY{o}{=} \PY{k}{typename} \PY{k+kt}{ct\PYZus{}params\PYZus{}t}\PY{o}{:}\PY{o}{:}\PY{n}{ix}\PY{p}{;}
  \PY{c+c1}{// member fields}
  \PY{k}{typename} \PY{k+kt}{ct\PYZus{}params\PYZus{}t}\PY{o}{:}\PY{o}{:}\PY{k+kt}{real\PYZus{}t} \PY{n}{omega}\PY{p}{;}

  \PY{c+c1}{// method called by mpdata\PYZus{}rhs}
  \PY{k+kt}{void} \PY{n+nf}{update\PYZus{}rhs}\PY{p}{(}
    \PY{n}{libmpdataxx}\PY{o}{:}\PY{o}{:}\PY{k+kt}{arrvec\PYZus{}t}\PY{o}{\PYZlt{}}
      \PY{k}{typename} \PY{k+kt}{parent\PYZus{}t}\PY{o}{:}\PY{o}{:}\PY{k+kt}{arr\PYZus{}t}
    \PY{o}{\PYZgt{}} \PY{o}{\PYZam{}}\PY{n}{rhs}\PY{p}{,} 
    \PY{k}{const} \PY{k}{typename} \PY{k+kt}{parent\PYZus{}t}\PY{o}{:}\PY{o}{:}\PY{k+kt}{real\PYZus{}t} \PY{o}{\PYZam{}}\PY{n}{dt}\PY{p}{,}
    \PY{k}{const} \PY{k+kt}{int} \PY{o}{\PYZam{}}\PY{n}{at}
  \PY{p}{)} \PY{p}{\PYZob{}}
    \PY{k+kt}{parent\PYZus{}t}\PY{o}{:}\PY{o}{:}\PY{n}{update\PYZus{}rhs}\PY{p}{(}\PY{n}{rhs}\PY{p}{,} \PY{n}{dt}\PY{p}{,} \PY{n}{at}\PY{p}{)}\PY{p}{;}

    \PY{c+c1}{// just to shorten code}
    \PY{k}{const} \PY{k}{auto} \PY{o}{\PYZam{}}\PY{n}{psi} \PY{o}{=} \PY{k}{this}\PY{o}{\PYZhy{}}\PY{o}{\PYZgt{}}\PY{n}{state}\PY{p}{(}\PY{n}{ix}\PY{o}{:}\PY{o}{:}\PY{n}{psi}\PY{p}{)}\PY{p}{;}
    \PY{k}{const} \PY{k}{auto} \PY{o}{\PYZam{}}\PY{n}{phi} \PY{o}{=} \PY{k}{this}\PY{o}{\PYZhy{}}\PY{o}{\PYZgt{}}\PY{n}{state}\PY{p}{(}\PY{n}{ix}\PY{o}{:}\PY{o}{:}\PY{n}{phi}\PY{p}{)}\PY{p}{;}
    \PY{k}{const} \PY{k}{auto} \PY{o}{\PYZam{}}\PY{n}{i} \PY{o}{=} \PY{k}{this}\PY{o}{\PYZhy{}}\PY{o}{\PYZgt{}}\PY{n}{i}\PY{p}{;}

    \PY{k}{switch} \PY{p}{(}\PY{n}{at}\PY{p}{)} 
    \PY{p}{\PYZob{}} \PY{c+c1}{// explicit solution for R\PYZca{}\PYZob{}n\PYZcb{} }
      \PY{c+c1}{// (note: with trapez used only at t=0)}
      \PY{k}{case} \PY{p}{(}\PY{l+m+mi}{0}\PY{p}{)}: 
      \PY{n}{rhs}\PY{p}{.}\PY{n}{at}\PY{p}{(}\PY{n}{ix}\PY{o}{:}\PY{o}{:}\PY{n}{psi}\PY{p}{)}\PY{p}{(}\PY{n}{i}\PY{p}{)} \PY{o}{+}\PY{o}{=} \PY{n}{omega} \PY{o}{*} \PY{n}{phi}\PY{p}{(}\PY{n}{i}\PY{p}{)}\PY{p}{;}
      \PY{n}{rhs}\PY{p}{.}\PY{n}{at}\PY{p}{(}\PY{n}{ix}\PY{o}{:}\PY{o}{:}\PY{n}{phi}\PY{p}{)}\PY{p}{(}\PY{n}{i}\PY{p}{)} \PY{o}{\PYZhy{}}\PY{o}{=} \PY{n}{omega} \PY{o}{*} \PY{n}{psi}\PY{p}{(}\PY{n}{i}\PY{p}{)}\PY{p}{;}
      \PY{k}{break}\PY{p}{;}
   
      \PY{c+c1}{// implicit solution for R\PYZca{}\PYZob{}n+1\PYZcb{}}
      \PY{k}{case} \PY{p}{(}\PY{l+m+mi}{1}\PY{p}{)}: 
      \PY{n}{rhs}\PY{p}{.}\PY{n}{at}\PY{p}{(}\PY{n}{ix}\PY{o}{:}\PY{o}{:}\PY{n}{psi}\PY{p}{)}\PY{p}{(}\PY{n}{i}\PY{p}{)} \PY{o}{+}\PY{o}{=} \PY{p}{(}
	\PY{p}{(}\PY{n}{psi}\PY{p}{(}\PY{n}{i}\PY{p}{)} \PY{o}{+} \PY{n}{dt} \PY{o}{*} \PY{n}{omega} \PY{o}{*} \PY{n}{phi}\PY{p}{(}\PY{n}{i}\PY{p}{)}\PY{p}{)} 
        \PY{o}{/} \PY{p}{(}\PY{l+m+mi}{1} \PY{o}{+} \PY{n}{pow}\PY{p}{(}\PY{n}{dt} \PY{o}{*} \PY{n}{omega}\PY{p}{,} \PY{l+m+mi}{2}\PY{p}{)}\PY{p}{)}
	\PY{o}{\PYZhy{}} \PY{n}{psi}\PY{p}{(}\PY{n}{i}\PY{p}{)}
      \PY{p}{)} \PY{o}{/} \PY{n}{dt}\PY{p}{;}
      \PY{n}{rhs}\PY{p}{.}\PY{n}{at}\PY{p}{(}\PY{n}{ix}\PY{o}{:}\PY{o}{:}\PY{n}{phi}\PY{p}{)}\PY{p}{(}\PY{n}{i}\PY{p}{)} \PY{o}{+}\PY{o}{=} \PY{p}{(}
        \PY{p}{(}\PY{n}{phi}\PY{p}{(}\PY{n}{i}\PY{p}{)} \PY{o}{\PYZhy{}} \PY{n}{dt} \PY{o}{*} \PY{n}{omega} \PY{o}{*} \PY{n}{psi}\PY{p}{(}\PY{n}{i}\PY{p}{)}\PY{p}{)} 
        \PY{o}{/} \PY{p}{(}\PY{l+m+mi}{1} \PY{o}{+} \PY{n}{pow}\PY{p}{(}\PY{n}{dt} \PY{o}{*} \PY{n}{omega}\PY{p}{,} \PY{l+m+mi}{2}\PY{p}{)}\PY{p}{)}
        \PY{o}{\PYZhy{}} \PY{n}{phi}\PY{p}{(}\PY{n}{i}\PY{p}{)}
      \PY{p}{)} \PY{o}{/} \PY{n}{dt}\PY{p}{;}
      \PY{k}{break}\PY{p}{;}
    \PY{p}{\PYZcb{}}
  \PY{p}{\PYZcb{}}
  \PY{c+c1}{// run\PYZhy{}time parameters}
  \PY{k}{struct} \PY{k+kt}{rt\PYZus{}params\PYZus{}t} \PY{o}{:} \PY{k+kt}{parent\PYZus{}t}\PY{o}{:}\PY{o}{:}\PY{k+kt}{rt\PYZus{}params\PYZus{}t} \PY{p}{\PYZob{}} 
    \PY{k}{typename} \PY{k+kt}{ct\PYZus{}params\PYZus{}t}\PY{o}{:}\PY{o}{:}\PY{k+kt}{real\PYZus{}t} \PY{n}{omega} \PY{o}{=} \PY{l+m+mi}{0}\PY{p}{;} 
  \PY{p}{\PYZcb{}}\PY{p}{;}
  \PY{c+c1}{// ctor}
  \PY{n}{coupled\PYZus{}harmosc}\PY{p}{(}
    \PY{k}{typename} \PY{k+kt}{parent\PYZus{}t}\PY{o}{:}\PY{o}{:}\PY{k+kt}{ctor\PYZus{}args\PYZus{}t} \PY{n}{args}\PY{p}{,}
    \PY{k}{const} \PY{k+kt}{rt\PYZus{}params\PYZus{}t} \PY{o}{\PYZam{}}\PY{n}{p}
  \PY{p}{)} \PY{o}{:} \PY{k+kt}{parent\PYZus{}t}\PY{p}{(}\PY{n}{args}\PY{p}{,} \PY{n}{p}\PY{p}{)}\PY{p}{,} \PY{n}{omega}\PY{p}{(}\PY{n}{p}\PY{p}{.}\PY{n}{omega}\PY{p}{)}
  \PY{p}{\PYZob{}} \PY{n}{assert}\PY{p}{(}\PY{n}{omega} \PY{o}{!}\PY{o}{=} \PY{l+m+mi}{0}\PY{p}{)}\PY{p}{;} \PY{p}{\PYZcb{}}
\PY{p}{\PYZcb{}}\PY{p}{;}
\PY{c+c1}{//\PYZlt{}/listing\PYZhy{}1\PYZgt{}}
\end{Verbatim}

%% file: example_71.cpp.tex
\begin{Verbatim}[commandchars=\\\{\}]
\PY{c+cm}{/** }
\PY{c+cm}{ * @file}
\PY{c+cm}{ * @copyright University of Warsaw}
\PY{c+cm}{ * @section LICENSE}
\PY{c+cm}{ * GPLv3+ (see the COPYING file or http://www.gnu.org/licenses/)}
\PY{c+cm}{ *}
\PY{c+cm}{ * @brief a minimalistic model of a harmonic oscillator}
\PY{c+cm}{ * (consult eq. 28 in Smolarkiewicz 2006, IJNMF)}
\PY{c+cm}{ */}

\PY{c+cp}{\PYZsh{}}\PY{c+cp}{include \PYZdq{}coupled\PYZus{}harmosc.hpp\PYZdq{}}
\PY{c+cp}{\PYZsh{}}\PY{c+cp}{include \PYZlt{}libmpdata++}\PY{c+cp}{/}\PY{c+cp}{concurr}\PY{c+cp}{/}\PY{c+cp}{threads.hpp\PYZgt{}}
\PY{c+cp}{\PYZsh{}}\PY{c+cp}{include \PYZlt{}libmpdata++}\PY{c+cp}{/}\PY{c+cp}{output}\PY{c+cp}{/}\PY{c+cp}{gnuplot.hpp\PYZgt{}}
\PY{k}{using} \PY{k}{namespace} \PY{n}{libmpdataxx}\PY{p}{;}

\PY{c+cp}{\PYZsh{}}\PY{c+cp}{include \PYZlt{}boost}\PY{c+cp}{/}\PY{c+cp}{math}\PY{c+cp}{/}\PY{c+cp}{constants}\PY{c+cp}{/}\PY{c+cp}{constants.hpp\PYZgt{}}

\PY{k}{const} \PY{k+kt}{int} \PY{n}{nt} \PY{o}{=} \PY{l+m+mi}{1400}\PY{p}{;}

\PY{k+kt}{int} \PY{n+nf}{main}\PY{p}{(}\PY{p}{)} 
\PY{p}{\PYZob{}}
\PY{c+c1}{//\PYZlt{}listing\PYZhy{}1\PYZgt{}}
  \PY{k}{struct} \PY{k+kt}{ct\PYZus{}params\PYZus{}t} \PY{o}{:} \PY{k+kt}{ct\PYZus{}params\PYZus{}default\PYZus{}t}
  \PY{p}{\PYZob{}}
    \PY{k}{using} \PY{k+kt}{real\PYZus{}t} \PY{o}{=} \PY{k+kt}{double}\PY{p}{;}
    \PY{k}{enum} \PY{p}{\PYZob{}} \PY{n}{n\PYZus{}dims} \PY{o}{=} \PY{l+m+mi}{1} \PY{p}{\PYZcb{}}\PY{p}{;}
    \PY{k}{enum} \PY{p}{\PYZob{}} \PY{n}{n\PYZus{}eqns} \PY{o}{=} \PY{l+m+mi}{2} \PY{p}{\PYZcb{}}\PY{p}{;}
    \PY{k}{enum} \PY{p}{\PYZob{}} \PY{n}{rhs\PYZus{}scheme} \PY{o}{=} 
      \PY{n}{solvers}\PY{o}{:}\PY{o}{:}\PY{k+kt}{rhs\PYZus{}scheme\PYZus{}t}\PY{o}{:}\PY{o}{:}\PY{n}{trapez} \PY{p}{\PYZcb{}}\PY{p}{;}
    \PY{k}{struct} \PY{n}{ix} \PY{p}{\PYZob{}} \PY{k}{enum} \PY{p}{\PYZob{}}\PY{n}{psi}\PY{p}{,} \PY{n}{phi}\PY{p}{\PYZcb{}}\PY{p}{;} \PY{p}{\PYZcb{}}\PY{p}{;}
  \PY{p}{\PYZcb{}}\PY{p}{;}
\PY{c+c1}{//\PYZlt{}/listing\PYZhy{}1\PYZgt{}}
  \PY{k}{using} \PY{k+kt}{real\PYZus{}t} \PY{o}{=} \PY{k}{typename} \PY{k+kt}{ct\PYZus{}params\PYZus{}t}\PY{o}{:}\PY{o}{:}\PY{k+kt}{real\PYZus{}t}\PY{p}{;}
  
  \PY{k}{using} \PY{k+kt}{sim\PYZus{}t} \PY{o}{=} \PY{n}{output}\PY{o}{:}\PY{o}{:}\PY{n}{gnuplot}\PY{o}{\PYZlt{}}
    \PY{n}{coupled\PYZus{}harmosc}\PY{o}{\PYZlt{}}\PY{k+kt}{ct\PYZus{}params\PYZus{}t}\PY{o}{\PYZgt{}}
  \PY{o}{\PYZgt{}}\PY{p}{;}
  \PY{k}{typename} \PY{k+kt}{sim\PYZus{}t}\PY{o}{:}\PY{o}{:}\PY{k+kt}{rt\PYZus{}params\PYZus{}t} \PY{n}{p}\PY{p}{;} 

\PY{c+c1}{//\PYZlt{}listing\PYZhy{}2\PYZgt{}}
  \PY{c+c1}{// run\PYZhy{}time parameters}
  \PY{k}{using} \PY{n}{boost}\PY{o}{:}\PY{o}{:}\PY{n}{math}\PY{o}{:}\PY{o}{:}\PY{n}{constants}\PY{o}{:}\PY{o}{:}\PY{n}{pi}\PY{p}{;}
  \PY{n}{p}\PY{p}{.}\PY{n}{dt} \PY{o}{=} \PY{l+m+mi}{1}\PY{p}{;}
  \PY{n}{p}\PY{p}{.}\PY{n}{omega} \PY{o}{=} \PY{l+m+mi}{2} \PY{o}{*} \PY{n}{pi}\PY{o}{\PYZlt{}}\PY{k+kt}{real\PYZus{}t}\PY{o}{\PYZgt{}}\PY{p}{(}\PY{p}{)} \PY{o}{/} \PY{n}{p}\PY{p}{.}\PY{n}{dt} \PY{o}{/} \PY{l+m+mi}{400}\PY{p}{;}
\PY{c+c1}{//\PYZlt{}/listing\PYZhy{}2\PYZgt{}}
  \PY{n}{p}\PY{p}{.}\PY{n}{grid\PYZus{}size} \PY{o}{=} \PY{p}{\PYZob{}}\PY{l+m+mi}{1001}\PY{p}{\PYZcb{}}\PY{p}{;}
  \PY{n}{p}\PY{p}{.}\PY{n}{outfreq} \PY{o}{=} \PY{l+m+mi}{10}\PY{p}{;}

  \PY{k}{using} \PY{n}{ix} \PY{o}{=} \PY{k}{typename} \PY{k+kt}{ct\PYZus{}params\PYZus{}t}\PY{o}{:}\PY{o}{:}\PY{n}{ix}\PY{p}{;}
  \PY{n}{p}\PY{p}{.}\PY{n}{outvars} \PY{o}{=} \PY{p}{\PYZob{}}
    \PY{p}{\PYZob{}}\PY{n}{ix}\PY{o}{:}\PY{o}{:}\PY{n}{psi}\PY{p}{,} \PY{p}{\PYZob{}}\PY{p}{.}\PY{n}{name} \PY{o}{=} \PY{l+s}{\PYZdq{}}\PY{l+s}{psi}\PY{l+s}{\PYZdq{}}\PY{p}{,} \PY{p}{.}\PY{n}{unit} \PY{o}{=} \PY{l+s}{\PYZdq{}}\PY{l+s}{1}\PY{l+s}{\PYZdq{}}\PY{p}{\PYZcb{}}\PY{p}{\PYZcb{}}\PY{p}{,}
    \PY{p}{\PYZob{}}\PY{n}{ix}\PY{o}{:}\PY{o}{:}\PY{n}{phi}\PY{p}{,} \PY{p}{\PYZob{}}\PY{p}{.}\PY{n}{name} \PY{o}{=} \PY{l+s}{\PYZdq{}}\PY{l+s}{phi}\PY{l+s}{\PYZdq{}}\PY{p}{,} \PY{p}{.}\PY{n}{unit} \PY{o}{=} \PY{l+s}{\PYZdq{}}\PY{l+s}{1}\PY{l+s}{\PYZdq{}}\PY{p}{\PYZcb{}}\PY{p}{\PYZcb{}}
  \PY{p}{\PYZcb{}}\PY{p}{;}
  \PY{n}{p}\PY{p}{.}\PY{n}{gnuplot\PYZus{}command} \PY{o}{=} \PY{l+s}{\PYZdq{}}\PY{l+s}{plot}\PY{l+s}{\PYZdq{}}\PY{p}{;}

  \PY{c+c1}{// instantiation}
  \PY{n}{concurr}\PY{o}{:}\PY{o}{:}\PY{n}{threads}\PY{o}{\PYZlt{}}\PY{k+kt}{sim\PYZus{}t}\PY{p}{,} \PY{n}{bcond}\PY{o}{:}\PY{o}{:}\PY{n}{cyclic}\PY{p}{,} \PY{n}{bcond}\PY{o}{:}\PY{o}{:}\PY{n}{cyclic}\PY{o}{\PYZgt{}} \PY{n}{run}\PY{p}{(}\PY{n}{p}\PY{p}{)}\PY{p}{;}

  \PY{c+c1}{// initial condition}
  \PY{p}{\PYZob{}}
    \PY{n}{blitz}\PY{o}{:}\PY{o}{:}\PY{n}{firstIndex} \PY{n}{i}\PY{p}{;}
    \PY{n}{run}\PY{p}{.}\PY{n}{advectee}\PY{p}{(}\PY{n}{ix}\PY{o}{:}\PY{o}{:}\PY{n}{psi}\PY{p}{)} \PY{o}{=} \PY{n}{where}\PY{p}{(}        
      \PY{n}{i}\PY{o}{\PYZlt{}}\PY{o}{=} \PY{l+m+mi}{50} \PY{o}{|}\PY{o}{|} \PY{n}{i}\PY{o}{\PYZgt{}}\PY{o}{=} \PY{l+m+mi}{150}\PY{p}{,}                            \PY{c+c1}{// if}
      \PY{l+m+mi}{0}\PY{p}{,}                                            \PY{c+c1}{// then}
      \PY{l+m+mf}{0.5} \PY{o}{*} \PY{p}{(}\PY{l+m+mi}{1} \PY{o}{+} \PY{n}{cos}\PY{p}{(}\PY{l+m+mi}{2} \PY{o}{*} \PY{n}{pi}\PY{o}{\PYZlt{}}\PY{k+kt}{real\PYZus{}t}\PY{o}{\PYZgt{}}\PY{p}{(}\PY{p}{)} \PY{o}{*} \PY{n}{i} \PY{o}{/} \PY{l+m+mi}{100}\PY{p}{)}\PY{p}{)}   \PY{c+c1}{// else}
    \PY{p}{)}\PY{p}{;}
    \PY{n}{run}\PY{p}{.}\PY{n}{advectee}\PY{p}{(}\PY{n}{ix}\PY{o}{:}\PY{o}{:}\PY{n}{phi}\PY{p}{)} \PY{o}{=} \PY{k+kt}{real\PYZus{}t}\PY{p}{(}\PY{l+m+mi}{0}\PY{p}{)}\PY{p}{;}
  \PY{p}{\PYZcb{}}
  \PY{n}{run}\PY{p}{.}\PY{n}{advector}\PY{p}{(}\PY{p}{)} \PY{o}{=} \PY{l+m+mf}{.5}\PY{p}{;}

  \PY{c+c1}{// integration}
  \PY{n}{run}\PY{p}{.}\PY{n}{advance}\PY{p}{(}\PY{n}{nt}\PY{p}{)}\PY{p}{;}
\PY{p}{\PYZcb{}}
\end{Verbatim}

%% file: shallow_water.cpp.tex
\begin{Verbatim}[commandchars=\\\{\}]
\PY{c+cm}{/** }
\PY{c+cm}{ * @file}
\PY{c+cm}{ * @copyright University of Warsaw}
\PY{c+cm}{ * @section LICENSE}
\PY{c+cm}{ * GPLv3+ (see the COPYING file or http://www.gnu.org/licenses/)}
\PY{c+cm}{ */}

\PY{c+cp}{\PYZsh{}}\PY{c+cp}{include \PYZlt{}libmpdata++}\PY{c+cp}{/}\PY{c+cp}{solvers}\PY{c+cp}{/}\PY{c+cp}{mpdata\PYZus{}rhs\PYZus{}vip.hpp\PYZgt{} }
\PY{c+cp}{\PYZsh{}}\PY{c+cp}{include \PYZlt{}libmpdata++}\PY{c+cp}{/}\PY{c+cp}{formulae}\PY{c+cp}{/}\PY{c+cp}{nabla\PYZus{}formulae.hpp\PYZgt{}}

\PY{c+cm}{/** @brief the 2D shallow\PYZhy{}water equations system}
\PY{c+cm}{  *}
\PY{c+cm}{  * Consult chapter 3 in Vallis 2008 for a detailed derivation.}
\PY{c+cm}{  *}
\PY{c+cm}{  * The key assumptions are:}
\PY{c+cm}{  * \PYZhy{} horizontal scale is much larger than the vertical scale (\PYZbs{}f\PYZdl{} u \PYZbs{}approx u(x) \PYZbs{}f\PYZdl{})}
\PY{c+cm}{  * \PYZhy{} hydrostatic equillibrium}
\PY{c+cm}{  * \PYZhy{} constant density}
\PY{c+cm}{  *}
\PY{c+cm}{  * Nomenclature:}
\PY{c+cm}{  * \PYZhy{} \PYZbs{}f\PYZdl{} \PYZbs{}eta(x,y) \PYZbs{}f\PYZdl{} \PYZhy{} (absolute) height of the fluid surface}
\PY{c+cm}{  * \PYZhy{} \PYZbs{}f\PYZdl{} \PYZbs{}eta\PYZus{}0(x,y) \PYZbs{}f\PYZdl{} \PYZhy{} bathymetry}
\PY{c+cm}{  * \PYZhy{} \PYZbs{}f\PYZdl{} h = \PYZbs{}eta \PYZhy{} \PYZbs{}eta\PYZus{}0 \PYZbs{}f\PYZdl{} \PYZhy{} thickness of the fluid layer}
\PY{c+cm}{  * \PYZhy{} \PYZbs{}f\PYZdl{} \PYZbs{}vec\PYZob{}u\PYZcb{} = (u,v) \PYZbs{}f\PYZdl{}}
\PY{c+cm}{  * \PYZhy{} \PYZbs{}f\PYZdl{} \PYZbs{}nabla\PYZus{}z = (\PYZbs{}partial\PYZus{}x, \PYZbs{}partial\PYZus{}y) \PYZbs{}f\PYZdl{} }
\PY{c+cm}{  *}
\PY{c+cm}{  * momentum equation:}
\PY{c+cm}{  * \PYZbs{}f\PYZdl{} \PYZbs{}partial\PYZus{}t u + u \PYZbs{}cdot \PYZbs{}nabla\PYZus{}z u = \PYZhy{} \PYZbs{}frac\PYZob{}1\PYZcb{}\PYZob{}\PYZbs{}rho\PYZcb{} \PYZbs{}nabla\PYZus{}z p \PYZbs{}f\PYZdl{}}
\PY{c+cm}{  *}
\PY{c+cm}{  * pressure in a column of the constant\PYZhy{}density fluid:}
\PY{c+cm}{  * \PYZbs{}f\PYZdl{} p = p\PYZus{}0 \PYZhy{} \PYZbs{}rho g z = p\PYZus{}0 + \PYZbs{}rho g \PYZbs{}cdot (\PYZbs{}eta(x) \PYZhy{} z) \PYZbs{}f\PYZdl{}}
\PY{c+cm}{  *}
\PY{c+cm}{  * mass continuity equation:}
\PY{c+cm}{  * \PYZbs{}f\PYZdl{} \PYZbs{}partial\PYZus{}t h + \PYZbs{}nabla\PYZus{}z (h \PYZbs{}cdot u) = 0 \PYZbs{}f\PYZdl{}}
\PY{c+cm}{  *}
\PY{c+cm}{  * h times momentum eq. plus u times mass continuity equation:}
\PY{c+cm}{  * \PYZbs{}f\PYZdl{} \PYZbs{}partial\PYZus{}t (uh) + \PYZbs{}nabla\PYZus{}z (u \PYZbs{}cdot uh) = \PYZhy{}g h \PYZbs{}nabla\PYZus{}z \PYZbs{}eta \PYZbs{}f\PYZdl{}}
\PY{c+cm}{  */}
\PY{k}{template} \PY{o}{\PYZlt{}}\PY{k}{typename} \PY{k+kt}{ct\PYZus{}params\PYZus{}t}\PY{p}{,} \PY{k}{class} \PY{n+nc}{enableif} \PY{o}{=} \PY{k+kt}{void}\PY{o}{\PYZgt{}}
\PY{k}{class} \PY{n+nc}{shallow\PYZus{}water} 
\PY{p}{\PYZob{}}\PY{p}{\PYZcb{}}\PY{p}{;}

\PY{k}{template} \PY{o}{\PYZlt{}}\PY{k}{class} \PY{n+nc}{ct\PYZus{}params\PYZus{}t}\PY{o}{\PYZgt{}}
\PY{k}{class} \PY{n+nc}{shallow\PYZus{}water\PYZus{}common} \PY{o}{:} \PY{k}{public} \PY{n}{libmpdataxx}\PY{o}{:}\PY{o}{:}\PY{n}{solvers}\PY{o}{:}\PY{o}{:}\PY{n}{mpdata\PYZus{}rhs\PYZus{}vip}\PY{o}{\PYZlt{}}\PY{k+kt}{ct\PYZus{}params\PYZus{}t}\PY{o}{\PYZgt{}}
\PY{p}{\PYZob{}}
  \PY{k}{using} \PY{k+kt}{parent\PYZus{}t} \PY{o}{=} \PY{n}{libmpdataxx}\PY{o}{:}\PY{o}{:}\PY{n}{solvers}\PY{o}{:}\PY{o}{:}\PY{n}{mpdata\PYZus{}rhs\PYZus{}vip}\PY{o}{\PYZlt{}}\PY{k+kt}{ct\PYZus{}params\PYZus{}t}\PY{o}{\PYZgt{}}\PY{p}{;}

  \PY{n+nl}{protected:}

  \PY{c+c1}{// member fields}
  \PY{k}{const} \PY{k}{typename} \PY{k+kt}{ct\PYZus{}params\PYZus{}t}\PY{o}{:}\PY{o}{:}\PY{k+kt}{real\PYZus{}t} \PY{n}{g}\PY{p}{;}

  \PY{c+c1}{// }
  \PY{k+kt}{void} \PY{n+nf}{update\PYZus{}rhs}\PY{p}{(}
    \PY{n}{libmpdataxx}\PY{o}{:}\PY{o}{:}\PY{k+kt}{arrvec\PYZus{}t}\PY{o}{\PYZlt{}}\PY{k}{typename} \PY{k+kt}{parent\PYZus{}t}\PY{o}{:}\PY{o}{:}\PY{k+kt}{arr\PYZus{}t}\PY{o}{\PYZgt{}} \PY{o}{\PYZam{}}\PY{n}{rhs}\PY{p}{,}
    \PY{k}{const} \PY{k}{typename} \PY{k+kt}{parent\PYZus{}t}\PY{o}{:}\PY{o}{:}\PY{k+kt}{real\PYZus{}t} \PY{o}{\PYZam{}}\PY{n}{dt}\PY{p}{,}
    \PY{k}{const} \PY{k+kt}{int} \PY{o}{\PYZam{}}\PY{n}{at}
  \PY{p}{)} \PY{p}{\PYZob{}}
    \PY{k+kt}{parent\PYZus{}t}\PY{o}{:}\PY{o}{:}\PY{n}{update\PYZus{}rhs}\PY{p}{(}\PY{n}{rhs}\PY{p}{,} \PY{n}{dt}\PY{p}{,} \PY{n}{at}\PY{p}{)}\PY{p}{;}
    \PY{k}{enum} \PY{p}{\PYZob{}} \PY{n}{n} \PY{o}{=} \PY{l+m+mi}{0} \PY{p}{\PYZcb{}}\PY{p}{;}    \PY{c+c1}{// just to make n, n+1 look nice :)}
    \PY{n}{assert}\PY{p}{(}
      \PY{k}{this}\PY{o}{\PYZhy{}}\PY{o}{\PYZgt{}}\PY{n}{timestep} \PY{o}{=}\PY{o}{=} \PY{l+m+mi}{0} \PY{o}{\PYZam{}}\PY{o}{\PYZam{}} \PY{n}{at} \PY{o}{=}\PY{o}{=} \PY{n}{n} 
      \PY{o}{|}\PY{o}{|}
      \PY{k}{this}\PY{o}{\PYZhy{}}\PY{o}{\PYZgt{}}\PY{n}{timestep}  \PY{o}{\PYZgt{}} \PY{l+m+mi}{0} \PY{o}{\PYZam{}}\PY{o}{\PYZam{}} \PY{n}{at} \PY{o}{=}\PY{o}{=} \PY{n}{n}\PY{o}{+}\PY{l+m+mi}{1}
    \PY{p}{)}\PY{p}{;} \PY{c+c1}{// note: we know only how to calculate R\PYZca{}\PYZob{}n+1\PYZcb{}}
       \PY{c+c1}{//       thus allowing to treat R\PYZca{}\PYZob{}n+1\PYZcb{} as R\PYZca{}\PYZob{}n\PYZcb{}}
       \PY{c+c1}{//       only in the first timestep}
  \PY{p}{\PYZcb{}}

  \PY{k+kt}{void} \PY{n+nf}{hook\PYZus{}post\PYZus{}step}\PY{p}{(}\PY{p}{)}
  \PY{p}{\PYZob{}}
    \PY{k+kt}{parent\PYZus{}t}\PY{o}{:}\PY{o}{:}\PY{n}{hook\PYZus{}post\PYZus{}step}\PY{p}{(}\PY{p}{)}\PY{p}{;}
    \PY{n}{assert}\PY{p}{(}\PY{n}{min}\PY{p}{(}\PY{k}{this}\PY{o}{\PYZhy{}}\PY{o}{\PYZgt{}}\PY{n}{state}\PY{p}{(}\PY{k+kt}{ct\PYZus{}params\PYZus{}t}\PY{o}{:}\PY{o}{:}\PY{n}{ix}\PY{o}{:}\PY{o}{:}\PY{n}{h}\PY{p}{)}\PY{p}{(}\PY{k}{this}\PY{o}{\PYZhy{}}\PY{o}{\PYZgt{}}\PY{n}{ijk}\PY{p}{)}\PY{p}{)} \PY{o}{\PYZgt{}}\PY{o}{=} \PY{l+m+mi}{0}\PY{p}{)}\PY{p}{;}  
\PY{c+c1}{//   std::cerr\PYZlt{}\PYZlt{}min(this\PYZhy{}\PYZgt{}state(ct\PYZus{}params\PYZus{}t::ix::h)(this\PYZhy{}\PYZgt{}ijk))\PYZlt{}\PYZlt{}std::endl;}
  \PY{p}{\PYZcb{}}

  \PY{k+kt}{void} \PY{n+nf}{hook\PYZus{}ante\PYZus{}step}\PY{p}{(}\PY{p}{)}
  \PY{p}{\PYZob{}}
    \PY{k+kt}{parent\PYZus{}t}\PY{o}{:}\PY{o}{:}\PY{n}{hook\PYZus{}ante\PYZus{}step}\PY{p}{(}\PY{p}{)}\PY{p}{;}
    \PY{n}{assert}\PY{p}{(}\PY{n}{min}\PY{p}{(}\PY{k}{this}\PY{o}{\PYZhy{}}\PY{o}{\PYZgt{}}\PY{n}{state}\PY{p}{(}\PY{k+kt}{ct\PYZus{}params\PYZus{}t}\PY{o}{:}\PY{o}{:}\PY{n}{ix}\PY{o}{:}\PY{o}{:}\PY{n}{h}\PY{p}{)}\PY{p}{(}\PY{k}{this}\PY{o}{\PYZhy{}}\PY{o}{\PYZgt{}}\PY{n}{ijk}\PY{p}{)}\PY{p}{)} \PY{o}{\PYZgt{}}\PY{o}{=} \PY{l+m+mi}{0}\PY{p}{)}\PY{p}{;}  
  \PY{p}{\PYZcb{}}

  \PY{n+nl}{public:}

  \PY{c+c1}{// run\PYZhy{}time parameters}
  \PY{k}{struct} \PY{k+kt}{rt\PYZus{}params\PYZus{}t} \PY{o}{:} \PY{k+kt}{parent\PYZus{}t}\PY{o}{:}\PY{o}{:}\PY{k+kt}{rt\PYZus{}params\PYZus{}t} 
  \PY{p}{\PYZob{}}   
    \PY{k}{typename} \PY{k+kt}{parent\PYZus{}t}\PY{o}{:}\PY{o}{:}\PY{k+kt}{real\PYZus{}t} \PY{n}{g} \PY{o}{=} \PY{l+m+mf}{9.81}\PY{p}{;} \PY{c+c1}{// default value }
  \PY{p}{\PYZcb{}}\PY{p}{;}

  \PY{c+c1}{// ctor}
  \PY{n}{shallow\PYZus{}water\PYZus{}common}\PY{p}{(} 
    \PY{k}{typename} \PY{k+kt}{parent\PYZus{}t}\PY{o}{:}\PY{o}{:}\PY{k+kt}{ctor\PYZus{}args\PYZus{}t} \PY{n}{args}\PY{p}{,} 
    \PY{k}{const} \PY{k+kt}{rt\PYZus{}params\PYZus{}t} \PY{o}{\PYZam{}}\PY{n}{p}
  \PY{p}{)} \PY{o}{:}
    \PY{k+kt}{parent\PYZus{}t}\PY{p}{(}\PY{n}{args}\PY{p}{,} \PY{n}{p}\PY{p}{)}\PY{p}{,} 
    \PY{n}{g}\PY{p}{(}\PY{n}{p}\PY{p}{.}\PY{n}{g}\PY{p}{)}
  \PY{p}{\PYZob{}}\PY{p}{\PYZcb{}}
\PY{p}{\PYZcb{}}\PY{p}{;}

\PY{c+c1}{// 1D version}
\PY{k}{template} \PY{o}{\PYZlt{}}\PY{k}{typename} \PY{k+kt}{ct\PYZus{}params\PYZus{}t}\PY{o}{\PYZgt{}}
\PY{k}{class} \PY{n+nc}{shallow\PYZus{}water}\PY{o}{\PYZlt{}}
  \PY{k+kt}{ct\PYZus{}params\PYZus{}t}\PY{p}{,} 
  \PY{k}{typename} \PY{n}{std}\PY{o}{:}\PY{o}{:}\PY{n}{enable\PYZus{}if}\PY{o}{\PYZlt{}}\PY{k+kt}{ct\PYZus{}params\PYZus{}t}\PY{o}{:}\PY{o}{:}\PY{n}{n\PYZus{}dims} \PY{o}{=}\PY{o}{=} \PY{l+m+mi}{1}\PY{o}{\PYZgt{}}\PY{o}{:}\PY{o}{:}\PY{n}{type}
\PY{o}{\PYZgt{}} \PY{o}{:} \PY{k}{public} \PY{n}{shallow\PYZus{}water\PYZus{}common}\PY{o}{\PYZlt{}}\PY{k+kt}{ct\PYZus{}params\PYZus{}t}\PY{o}{\PYZgt{}}
\PY{p}{\PYZob{}}
  \PY{n}{static\PYZus{}assert}\PY{p}{(}\PY{k+kt}{ct\PYZus{}params\PYZus{}t}\PY{o}{:}\PY{o}{:}\PY{n}{n\PYZus{}eqns} \PY{o}{=}\PY{o}{=} \PY{l+m+mi}{2}\PY{p}{,} \PY{l+s}{\PYZdq{}}\PY{l+s}{\PYZob{}qx, h\PYZcb{} in 1D}\PY{l+s}{\PYZdq{}}\PY{p}{)}\PY{p}{;}
  \PY{k}{using} \PY{k+kt}{parent\PYZus{}t} \PY{o}{=} \PY{n}{shallow\PYZus{}water\PYZus{}common}\PY{o}{\PYZlt{}}\PY{k+kt}{ct\PYZus{}params\PYZus{}t}\PY{o}{\PYZgt{}}\PY{p}{;}
  \PY{k}{using} \PY{k+kt}{parent\PYZus{}t}\PY{o}{:}\PY{o}{:}\PY{k+kt}{parent\PYZus{}t}\PY{p}{;} \PY{c+c1}{// inheriting ctors}
  \PY{k}{using} \PY{n}{ix} \PY{o}{=} \PY{k}{typename} \PY{k+kt}{ct\PYZus{}params\PYZus{}t}\PY{o}{:}\PY{o}{:}\PY{n}{ix}\PY{p}{;}

\PY{c+c1}{//\PYZlt{}listing\PYZhy{}1\PYZgt{}}
  \PY{k+kt}{void} \PY{n+nf}{update\PYZus{}rhs}\PY{p}{(}
    \PY{n}{libmpdataxx}\PY{o}{:}\PY{o}{:}\PY{k+kt}{arrvec\PYZus{}t}\PY{o}{\PYZlt{}}
      \PY{k}{typename} \PY{k+kt}{parent\PYZus{}t}\PY{o}{:}\PY{o}{:}\PY{k+kt}{arr\PYZus{}t}
    \PY{o}{\PYZgt{}} \PY{o}{\PYZam{}}\PY{n}{rhs}\PY{p}{,}
    \PY{k}{const} \PY{k}{typename} \PY{k+kt}{parent\PYZus{}t}\PY{o}{:}\PY{o}{:}\PY{k+kt}{real\PYZus{}t} \PY{o}{\PYZam{}}\PY{n}{dt}\PY{p}{,}
    \PY{k}{const} \PY{k+kt}{int} \PY{o}{\PYZam{}}\PY{n}{at}
  \PY{p}{)} \PY{p}{\PYZob{}}
    \PY{k}{using} \PY{k}{namespace} \PY{n}{libmpdataxx}\PY{o}{:}\PY{o}{:}\PY{n}{formulae}\PY{o}{:}\PY{o}{:}\PY{n}{nabla}\PY{p}{;}

    \PY{k+kt}{parent\PYZus{}t}\PY{o}{:}\PY{o}{:}\PY{n}{update\PYZus{}rhs}\PY{p}{(}\PY{n}{rhs}\PY{p}{,} \PY{n}{dt}\PY{p}{,} \PY{n}{at}\PY{p}{)}\PY{p}{;}

    \PY{n}{rhs}\PY{p}{.}\PY{n}{at}\PY{p}{(}\PY{n}{ix}\PY{o}{:}\PY{o}{:}\PY{n}{qx}\PY{p}{)}\PY{p}{(}\PY{k}{this}\PY{o}{\PYZhy{}}\PY{o}{\PYZgt{}}\PY{n}{i}\PY{p}{)} \PY{o}{\PYZhy{}}\PY{o}{=} 
      \PY{k}{this}\PY{o}{\PYZhy{}}\PY{o}{\PYZgt{}}\PY{n}{g} 
      \PY{o}{*} \PY{k}{this}\PY{o}{\PYZhy{}}\PY{o}{\PYZgt{}}\PY{n}{state}\PY{p}{(}\PY{n}{ix}\PY{o}{:}\PY{o}{:}\PY{n}{h}\PY{p}{)}\PY{p}{(}\PY{k}{this}\PY{o}{\PYZhy{}}\PY{o}{\PYZgt{}}\PY{n}{i}\PY{p}{)} 
      \PY{o}{*} \PY{n}{grad}\PY{p}{(}\PY{k}{this}\PY{o}{\PYZhy{}}\PY{o}{\PYZgt{}}\PY{n}{state}\PY{p}{(}\PY{n}{ix}\PY{o}{:}\PY{o}{:}\PY{n}{h}\PY{p}{)}\PY{p}{,} \PY{k}{this}\PY{o}{\PYZhy{}}\PY{o}{\PYZgt{}}\PY{n}{i}\PY{p}{,} \PY{k}{this}\PY{o}{\PYZhy{}}\PY{o}{\PYZgt{}}\PY{n}{di}\PY{p}{)}\PY{p}{;} 
  \PY{p}{\PYZcb{}}
\PY{c+c1}{//\PYZlt{}/listing\PYZhy{}1\PYZgt{}}
\PY{p}{\PYZcb{}}\PY{p}{;}

\PY{c+c1}{// 2D version}
\PY{k}{template} \PY{o}{\PYZlt{}}\PY{k}{typename} \PY{k+kt}{ct\PYZus{}params\PYZus{}t}\PY{o}{\PYZgt{}}
\PY{k}{class} \PY{n+nc}{shallow\PYZus{}water}\PY{o}{\PYZlt{}}
  \PY{k+kt}{ct\PYZus{}params\PYZus{}t}\PY{p}{,} 
  \PY{k}{typename} \PY{n}{std}\PY{o}{:}\PY{o}{:}\PY{n}{enable\PYZus{}if}\PY{o}{\PYZlt{}}\PY{k+kt}{ct\PYZus{}params\PYZus{}t}\PY{o}{:}\PY{o}{:}\PY{n}{n\PYZus{}dims} \PY{o}{=}\PY{o}{=} \PY{l+m+mi}{2}\PY{o}{\PYZgt{}}\PY{o}{:}\PY{o}{:}\PY{n}{type}
\PY{o}{\PYZgt{}} \PY{o}{:} \PY{k}{public} \PY{n}{shallow\PYZus{}water\PYZus{}common}\PY{o}{\PYZlt{}}\PY{k+kt}{ct\PYZus{}params\PYZus{}t}\PY{o}{\PYZgt{}}
\PY{p}{\PYZob{}}
  \PY{n}{static\PYZus{}assert}\PY{p}{(}\PY{k+kt}{ct\PYZus{}params\PYZus{}t}\PY{o}{:}\PY{o}{:}\PY{n}{n\PYZus{}eqns} \PY{o}{=}\PY{o}{=} \PY{l+m+mi}{3}\PY{p}{,} \PY{l+s}{\PYZdq{}}\PY{l+s}{\PYZob{}qx, qy, h\PYZcb{} in 2D}\PY{l+s}{\PYZdq{}}\PY{p}{)}\PY{p}{;}
  \PY{k}{using} \PY{k+kt}{parent\PYZus{}t} \PY{o}{=} \PY{n}{shallow\PYZus{}water\PYZus{}common}\PY{o}{\PYZlt{}}\PY{k+kt}{ct\PYZus{}params\PYZus{}t}\PY{o}{\PYZgt{}}\PY{p}{;}
  \PY{k}{using} \PY{k+kt}{parent\PYZus{}t}\PY{o}{:}\PY{o}{:}\PY{k+kt}{parent\PYZus{}t}\PY{p}{;} \PY{c+c1}{// inheriting ctors}
  \PY{k}{using} \PY{n}{ix} \PY{o}{=} \PY{k}{typename} \PY{k+kt}{ct\PYZus{}params\PYZus{}t}\PY{o}{:}\PY{o}{:}\PY{n}{ix}\PY{p}{;}

  \PY{k}{template} \PY{o}{\PYZlt{}}\PY{k+kt}{int} \PY{n}{d}\PY{p}{,} \PY{k}{class} \PY{n+nc}{arr\PYZus{}t}\PY{o}{\PYZgt{}}
  \PY{k+kt}{void} \PY{n}{forcings\PYZus{}helper}\PY{p}{(}
    \PY{k+kt}{arr\PYZus{}t} \PY{n}{rhs}\PY{p}{,} \PY{c+c1}{// TODO: ref?}
    \PY{k}{const} \PY{n}{libmpdataxx}\PY{o}{:}\PY{o}{:}\PY{k+kt}{rng\PYZus{}t} \PY{o}{\PYZam{}}\PY{n}{i}\PY{p}{,}
    \PY{k}{const} \PY{n}{libmpdataxx}\PY{o}{:}\PY{o}{:}\PY{k+kt}{rng\PYZus{}t} \PY{o}{\PYZam{}}\PY{n}{j}\PY{p}{,}
    \PY{k}{const} \PY{k}{typename} \PY{k+kt}{ct\PYZus{}params\PYZus{}t}\PY{o}{:}\PY{o}{:}\PY{k+kt}{real\PYZus{}t} \PY{o}{\PYZam{}}\PY{n}{di}
  \PY{p}{)}
  \PY{p}{\PYZob{}}
    \PY{k}{using} \PY{k}{namespace} \PY{n}{libmpdataxx}\PY{o}{:}\PY{o}{:}\PY{n}{formulae}\PY{o}{:}\PY{o}{:}\PY{n}{nabla}\PY{p}{;}
    \PY{n}{rhs}\PY{p}{(}\PY{n}{pi}\PY{o}{\PYZlt{}}\PY{n}{d}\PY{o}{\PYZgt{}}\PY{p}{(}\PY{n}{i}\PY{p}{,}\PY{n}{j}\PY{p}{)}\PY{p}{)} \PY{o}{\PYZhy{}}\PY{o}{=} \PY{k}{this}\PY{o}{\PYZhy{}}\PY{o}{\PYZgt{}}\PY{n}{g} \PY{o}{*} \PY{k}{this}\PY{o}{\PYZhy{}}\PY{o}{\PYZgt{}}\PY{n}{state}\PY{p}{(}\PY{n}{ix}\PY{o}{:}\PY{o}{:}\PY{n}{h}\PY{p}{)}\PY{p}{(}\PY{n}{pi}\PY{o}{\PYZlt{}}\PY{n}{d}\PY{o}{\PYZgt{}}\PY{p}{(}\PY{n}{i}\PY{p}{,}\PY{n}{j}\PY{p}{)}\PY{p}{)} \PY{o}{*} \PY{n}{grad}\PY{o}{\PYZlt{}}\PY{n}{d}\PY{o}{\PYZgt{}}\PY{p}{(}\PY{k}{this}\PY{o}{\PYZhy{}}\PY{o}{\PYZgt{}}\PY{n}{state}\PY{p}{(}\PY{n}{ix}\PY{o}{:}\PY{o}{:}\PY{n}{h}\PY{p}{)}\PY{p}{,} \PY{n}{i}\PY{p}{,} \PY{n}{j}\PY{p}{,} \PY{n}{di}\PY{p}{)}\PY{p}{;} 
  \PY{p}{\PYZcb{}}

  \PY{c+c1}{/// @brief Shallow Water Equations: Momentum forcings for the X and Y coordinates}
  \PY{k+kt}{void} \PY{n}{update\PYZus{}rhs}\PY{p}{(}
    \PY{n}{libmpdataxx}\PY{o}{:}\PY{o}{:}\PY{k+kt}{arrvec\PYZus{}t}\PY{o}{\PYZlt{}}\PY{k}{typename} \PY{k+kt}{parent\PYZus{}t}\PY{o}{:}\PY{o}{:}\PY{k+kt}{arr\PYZus{}t}\PY{o}{\PYZgt{}} \PY{o}{\PYZam{}}\PY{n}{rhs}\PY{p}{,}
    \PY{k}{const} \PY{k}{typename} \PY{k+kt}{parent\PYZus{}t}\PY{o}{:}\PY{o}{:}\PY{k+kt}{real\PYZus{}t} \PY{o}{\PYZam{}}\PY{n}{dt}\PY{p}{,}
    \PY{k}{const} \PY{k+kt}{int} \PY{o}{\PYZam{}}\PY{n}{at}
  \PY{p}{)} \PY{p}{\PYZob{}}
    \PY{c+c1}{//}
    \PY{k+kt}{parent\PYZus{}t}\PY{o}{:}\PY{o}{:}\PY{n}{update\PYZus{}rhs}\PY{p}{(}\PY{n}{rhs}\PY{p}{,} \PY{n}{dt}\PY{p}{,} \PY{n}{at}\PY{p}{)}\PY{p}{;}

    \PY{c+c1}{//}
    \PY{n}{forcings\PYZus{}helper}\PY{o}{\PYZlt{}}\PY{l+m+mi}{0}\PY{o}{\PYZgt{}}\PY{p}{(}\PY{n}{rhs}\PY{p}{.}\PY{n}{at}\PY{p}{(}\PY{n}{ix}\PY{o}{:}\PY{o}{:}\PY{n}{qx}\PY{p}{)}\PY{p}{,} \PY{k}{this}\PY{o}{\PYZhy{}}\PY{o}{\PYZgt{}}\PY{n}{i}\PY{p}{,} \PY{k}{this}\PY{o}{\PYZhy{}}\PY{o}{\PYZgt{}}\PY{n}{j}\PY{p}{,} \PY{k}{this}\PY{o}{\PYZhy{}}\PY{o}{\PYZgt{}}\PY{n}{di}\PY{p}{)}\PY{p}{;}
    \PY{n}{forcings\PYZus{}helper}\PY{o}{\PYZlt{}}\PY{l+m+mi}{1}\PY{o}{\PYZgt{}}\PY{p}{(}\PY{n}{rhs}\PY{p}{.}\PY{n}{at}\PY{p}{(}\PY{n}{ix}\PY{o}{:}\PY{o}{:}\PY{n}{qy}\PY{p}{)}\PY{p}{,} \PY{k}{this}\PY{o}{\PYZhy{}}\PY{o}{\PYZgt{}}\PY{n}{j}\PY{p}{,} \PY{k}{this}\PY{o}{\PYZhy{}}\PY{o}{\PYZgt{}}\PY{n}{i}\PY{p}{,} \PY{k}{this}\PY{o}{\PYZhy{}}\PY{o}{\PYZgt{}}\PY{n}{dj}\PY{p}{)}\PY{p}{;} 
  \PY{p}{\PYZcb{}}
\PY{p}{\PYZcb{}}\PY{p}{;}
\end{Verbatim}

%% file: shallow_water_params.cpp.tex
\begin{Verbatim}[commandchars=\\\{\}]
\PY{c+cm}{/** }
\PY{c+cm}{ * @file}
\PY{c+cm}{ * @copyright University of Warsaw}
\PY{c+cm}{ * @section LICENSE}
\PY{c+cm}{ * GPLv3+ (see the COPYING file or http://www.gnu.org/licenses/)}
\PY{c+cm}{ */}

\PY{c+cp}{\PYZsh{}}\PY{c+cp}{include \PYZdq{}shallow\PYZus{}water.hpp\PYZdq{}}
\PY{c+cp}{\PYZsh{}}\PY{c+cp}{include \PYZlt{}libmpdata++}\PY{c+cp}{/}\PY{c+cp}{concurr}\PY{c+cp}{/}\PY{c+cp}{serial.hpp\PYZgt{}}
\PY{k}{using} \PY{k}{namespace} \PY{n}{libmpdataxx}\PY{p}{;} 

\PY{c+cp}{\PYZsh{}}\PY{c+cp}{include \PYZlt{}boost}\PY{c+cp}{/}\PY{c+cp}{math}\PY{c+cp}{/}\PY{c+cp}{constants}\PY{c+cp}{/}\PY{c+cp}{constants.hpp\PYZgt{}}
\PY{k}{using} \PY{n}{boost}\PY{o}{:}\PY{o}{:}\PY{n}{math}\PY{o}{:}\PY{o}{:}\PY{n}{constants}\PY{o}{:}\PY{o}{:}\PY{n}{pi}\PY{p}{;}

\PY{c+cp}{\PYZsh{}}\PY{c+cp}{include \PYZlt{}fstream\PYZgt{}}

\PY{k}{const} \PY{k+kt}{int} 
  \PY{n}{nt} \PY{o}{=} \PY{l+m+mi}{300}\PY{p}{,}
  \PY{n}{outfreq} \PY{o}{=} \PY{l+m+mi}{1}\PY{p}{;}

\PY{k}{using} \PY{k+kt}{real\PYZus{}t} \PY{o}{=} \PY{k+kt}{double}\PY{p}{;}

\PY{c+c1}{// compile\PYZhy{}time parameters}
\PY{c+c1}{// enum \PYZob{} hint\PYZus{}noneg = opts::bit(ix::h) \PYZcb{};  // TODO: reconsider?}
\PY{c+c1}{//\PYZlt{}listing\PYZhy{}1\PYZgt{}}
\PY{k}{template} \PY{o}{\PYZlt{}}\PY{k+kt}{int} \PY{n}{opts\PYZus{}arg}\PY{o}{\PYZgt{}}
\PY{k}{struct} \PY{k+kt}{ct\PYZus{}params\PYZus{}t} \PY{o}{:} \PY{k+kt}{ct\PYZus{}params\PYZus{}default\PYZus{}t}
\PY{p}{\PYZob{}}
  \PY{k}{using} \PY{k+kt}{real\PYZus{}t} \PY{o}{=} \PY{o}{:}\PY{o}{:}\PY{k+kt}{real\PYZus{}t}\PY{p}{;}
  \PY{k}{enum} \PY{p}{\PYZob{}} \PY{n}{n\PYZus{}dims} \PY{o}{=} \PY{l+m+mi}{1} \PY{p}{\PYZcb{}}\PY{p}{;}
  \PY{k}{enum} \PY{p}{\PYZob{}} \PY{n}{n\PYZus{}eqns} \PY{o}{=} \PY{l+m+mi}{2} \PY{p}{\PYZcb{}}\PY{p}{;}
  
  \PY{c+c1}{// options}
  \PY{k}{enum} \PY{p}{\PYZob{}} \PY{n}{opts} \PY{o}{=} \PY{n}{opts\PYZus{}arg} \PY{o}{|} \PY{n}{opts}\PY{o}{:}\PY{o}{:}\PY{n}{dfl} \PY{p}{\PYZcb{}}\PY{p}{;}
  \PY{k}{enum} \PY{p}{\PYZob{}} \PY{n}{rhs\PYZus{}scheme} \PY{o}{=} \PY{n}{solvers}\PY{o}{:}\PY{o}{:}\PY{n}{trapez} \PY{p}{\PYZcb{}}\PY{p}{;}
  
  \PY{c+c1}{// indices}
  \PY{k}{struct} \PY{n}{ix} \PY{p}{\PYZob{}} 
    \PY{k}{enum} \PY{p}{\PYZob{}} \PY{n}{qx}\PY{p}{,} \PY{n}{h} \PY{p}{\PYZcb{}}\PY{p}{;}
    \PY{k}{enum} \PY{p}{\PYZob{}} \PY{n}{vip\PYZus{}i}\PY{o}{=}\PY{n}{qx}\PY{p}{,} \PY{n}{vip\PYZus{}den}\PY{o}{=}\PY{n}{h} \PY{p}{\PYZcb{}}\PY{p}{;}
  \PY{p}{\PYZcb{}}\PY{p}{;} 
  
  \PY{c+c1}{// hints}
  \PY{k}{enum} \PY{p}{\PYZob{}} \PY{n}{hint\PYZus{}norhs} \PY{o}{=} \PY{n}{opts}\PY{o}{:}\PY{o}{:}\PY{n}{bit}\PY{p}{(}\PY{n}{ix}\PY{o}{:}\PY{o}{:}\PY{n}{h}\PY{p}{)} \PY{p}{\PYZcb{}}\PY{p}{;} 
\PY{p}{\PYZcb{}}\PY{p}{;}
\PY{c+c1}{//\PYZlt{}/listing\PYZhy{}1\PYZgt{}}

\PY{k}{struct} \PY{n}{intcond}
\PY{p}{\PYZob{}}
  \PY{k+kt}{real\PYZus{}t} \PY{k}{operator}\PY{p}{(}\PY{p}{)}\PY{p}{(}\PY{k}{const} \PY{k+kt}{real\PYZus{}t} \PY{o}{\PYZam{}}\PY{n}{x}\PY{p}{)} \PY{k}{const}
  \PY{p}{\PYZob{}}
    \PY{k}{return} 
      \PY{n}{std}\PY{o}{:}\PY{o}{:}\PY{n}{abs}\PY{p}{(}\PY{n}{x}\PY{p}{)} \PY{o}{\PYZlt{}}\PY{o}{=} \PY{l+m+mi}{1} \PY{c+c1}{// if}
      \PY{o}{?} \PY{l+m+mi}{1} \PY{o}{\PYZhy{}} \PY{n}{x}\PY{o}{*}\PY{n}{x}        \PY{c+c1}{// then}
      \PY{o}{:} \PY{l+m+mi}{0}\PY{p}{;}             \PY{c+c1}{// else}
  \PY{p}{\PYZcb{}}
  \PY{n}{BZ\PYZus{}DECLARE\PYZus{}FUNCTOR}\PY{p}{(}\PY{n}{intcond}\PY{p}{)}\PY{p}{;}
\PY{p}{\PYZcb{}}\PY{p}{;}

\PY{c+c1}{// TODO: all this plotting logic should be done with a new libmpdataxx::output}
\PY{k}{struct} \PY{k+kt}{out\PYZus{}t}
\PY{p}{\PYZob{}}
  \PY{n}{std}\PY{o}{:}\PY{o}{:}\PY{n}{ofstream} \PY{n}{x}\PY{p}{,} \PY{n}{h}\PY{p}{,} \PY{n}{q}\PY{p}{,} \PY{n}{t}\PY{p}{;}
  \PY{k+kt}{out\PYZus{}t}\PY{p}{(}\PY{k}{const} \PY{n}{std}\PY{o}{:}\PY{o}{:}\PY{n}{string} \PY{o}{\PYZam{}}\PY{n}{pfx}\PY{p}{)} \PY{o}{:} 
    \PY{n}{x}\PY{p}{(}\PY{n}{pfx} \PY{o}{+} \PY{l+s}{\PYZdq{}}\PY{l+s}{.x}\PY{l+s}{\PYZdq{}}\PY{p}{)}\PY{p}{,} 
    \PY{n}{h}\PY{p}{(}\PY{n}{pfx} \PY{o}{+} \PY{l+s}{\PYZdq{}}\PY{l+s}{.h}\PY{l+s}{\PYZdq{}}\PY{p}{)}\PY{p}{,} 
    \PY{n}{q}\PY{p}{(}\PY{n}{pfx} \PY{o}{+} \PY{l+s}{\PYZdq{}}\PY{l+s}{.q}\PY{l+s}{\PYZdq{}}\PY{p}{)}\PY{p}{,} 
    \PY{n}{t}\PY{p}{(}\PY{n}{pfx} \PY{o}{+} \PY{l+s}{\PYZdq{}}\PY{l+s}{.t}\PY{l+s}{\PYZdq{}}\PY{p}{)}
  \PY{p}{\PYZob{}}\PY{p}{\PYZcb{}}
\PY{p}{\PYZcb{}}\PY{p}{;}

\PY{k}{template} \PY{o}{\PYZlt{}}\PY{k}{class} \PY{n+nc}{ix}\PY{p}{,} \PY{k}{class} \PY{n+nc}{run\PYZus{}t}\PY{o}{\PYZgt{}}
\PY{k+kt}{void} \PY{n}{output}\PY{p}{(}\PY{k+kt}{run\PYZus{}t} \PY{o}{\PYZam{}}\PY{n}{run}\PY{p}{,} \PY{k}{const} \PY{k+kt}{int} \PY{o}{\PYZam{}}\PY{n}{t}\PY{p}{,} \PY{k}{const} \PY{k+kt}{real\PYZus{}t} \PY{o}{\PYZam{}}\PY{n}{dx}\PY{p}{,} \PY{k}{const} \PY{k+kt}{real\PYZus{}t} \PY{o}{\PYZam{}}\PY{n}{dt}\PY{p}{,} \PY{k+kt}{out\PYZus{}t} \PY{o}{\PYZam{}}\PY{n}{out}\PY{p}{)}
\PY{p}{\PYZob{}}
  \PY{c+c1}{// x coordinate (once)}
  \PY{k}{if} \PY{p}{(}\PY{n}{t} \PY{o}{=}\PY{o}{=} \PY{l+m+mi}{0}\PY{p}{)}
  \PY{p}{\PYZob{}}
    \PY{k}{for} \PY{p}{(}\PY{k+kt}{int} \PY{n}{i} \PY{o}{=} \PY{l+m+mi}{0}\PY{p}{;} \PY{n}{i} \PY{o}{\PYZlt{}} \PY{n}{run}\PY{p}{.}\PY{n}{advectee}\PY{p}{(}\PY{p}{)}\PY{p}{.}\PY{n}{extent}\PY{p}{(}\PY{l+m+mi}{0}\PY{p}{)}\PY{p}{;} \PY{o}{+}\PY{o}{+}\PY{n}{i}\PY{p}{)} 
      \PY{n}{out}\PY{p}{.}\PY{n}{x} \PY{o}{\PYZlt{}}\PY{o}{\PYZlt{}} \PY{n}{i} \PY{o}{*} \PY{n}{dx} \PY{o}{\PYZlt{}}\PY{o}{\PYZlt{}} \PY{l+s}{\PYZdq{}}\PY{l+s+se}{\PYZbs{}t}\PY{l+s}{\PYZdq{}}\PY{p}{;}
    \PY{n}{out}\PY{p}{.}\PY{n}{x} \PY{o}{\PYZlt{}}\PY{o}{\PYZlt{}} \PY{l+s}{\PYZdq{}}\PY{l+s+se}{\PYZbs{}n}\PY{l+s}{\PYZdq{}}\PY{p}{;}
  \PY{p}{\PYZcb{}} 

  \PY{c+c1}{// time}
  \PY{n}{out}\PY{p}{.}\PY{n}{t} \PY{o}{\PYZlt{}}\PY{o}{\PYZlt{}} \PY{n}{t} \PY{o}{*} \PY{n}{dt} \PY{o}{\PYZlt{}}\PY{o}{\PYZlt{}} \PY{l+s}{\PYZdq{}}\PY{l+s+se}{\PYZbs{}t}\PY{l+s}{\PYZdq{}} \PY{o}{\PYZlt{}}\PY{o}{\PYZlt{}} \PY{l+s}{\PYZdq{}}\PY{l+s+se}{\PYZbs{}n}\PY{l+s}{\PYZdq{}}\PY{p}{;} 
  
  \PY{c+c1}{// layer depth}
  \PY{k}{for} \PY{p}{(}\PY{k}{auto} \PY{o}{\PYZam{}}\PY{n}{it} \PY{o}{:} \PY{n}{run}\PY{p}{.}\PY{n}{advectee}\PY{p}{(}\PY{n}{ix}\PY{o}{:}\PY{o}{:}\PY{n}{h}\PY{p}{)}\PY{p}{)} \PY{n}{out}\PY{p}{.}\PY{n}{h} \PY{o}{\PYZlt{}}\PY{o}{\PYZlt{}} \PY{n}{it} \PY{o}{\PYZlt{}}\PY{o}{\PYZlt{}} \PY{l+s}{\PYZdq{}}\PY{l+s+se}{\PYZbs{}t}\PY{l+s}{\PYZdq{}}\PY{p}{;}
  \PY{n}{out}\PY{p}{.}\PY{n}{h} \PY{o}{\PYZlt{}}\PY{o}{\PYZlt{}} \PY{l+s}{\PYZdq{}}\PY{l+s+se}{\PYZbs{}n}\PY{l+s}{\PYZdq{}}\PY{p}{;}

  \PY{c+c1}{// momentum}
  \PY{k}{for} \PY{p}{(}\PY{k}{auto} \PY{o}{\PYZam{}}\PY{n}{it} \PY{o}{:} \PY{n}{run}\PY{p}{.}\PY{n}{advectee}\PY{p}{(}\PY{n}{ix}\PY{o}{:}\PY{o}{:}\PY{n}{qx}\PY{p}{)}\PY{p}{)} \PY{n}{out}\PY{p}{.}\PY{n}{q} \PY{o}{\PYZlt{}}\PY{o}{\PYZlt{}} \PY{n}{it} \PY{o}{\PYZlt{}}\PY{o}{\PYZlt{}} \PY{l+s}{\PYZdq{}}\PY{l+s+se}{\PYZbs{}t}\PY{l+s}{\PYZdq{}}\PY{p}{;}
  \PY{n}{out}\PY{p}{.}\PY{n}{q} \PY{o}{\PYZlt{}}\PY{o}{\PYZlt{}} \PY{l+s}{\PYZdq{}}\PY{l+s+se}{\PYZbs{}n}\PY{l+s}{\PYZdq{}}\PY{p}{;}
\PY{p}{\PYZcb{}}

\PY{k}{template}\PY{o}{\PYZlt{}}\PY{k+kt}{int} \PY{n}{opts}\PY{o}{\PYZgt{}}
\PY{k+kt}{void} \PY{n}{test}\PY{p}{(}\PY{k}{const} \PY{n}{std}\PY{o}{:}\PY{o}{:}\PY{n}{string} \PY{o}{\PYZam{}}\PY{n}{pfx}\PY{p}{)} 
\PY{p}{\PYZob{}}
  \PY{k}{using} \PY{n}{ix} \PY{o}{=} \PY{k}{typename} \PY{k+kt}{ct\PYZus{}params\PYZus{}t}\PY{o}{\PYZlt{}}\PY{n}{opts}\PY{o}{\PYZgt{}}\PY{o}{:}\PY{o}{:}\PY{n}{ix}\PY{p}{;}

  \PY{c+c1}{// solver choice}
  \PY{k}{using} \PY{k+kt}{solver\PYZus{}t} \PY{o}{=} \PY{n}{shallow\PYZus{}water}\PY{o}{\PYZlt{}}\PY{k+kt}{ct\PYZus{}params\PYZus{}t}\PY{o}{\PYZlt{}}\PY{n}{opts}\PY{o}{\PYZgt{}}\PY{o}{\PYZgt{}}\PY{p}{;}

\PY{c+c1}{//\PYZlt{}listing\PYZhy{}2\PYZgt{}}
  \PY{c+c1}{// run\PYZhy{}time parameters}
  \PY{k}{typename} \PY{k+kt}{solver\PYZus{}t}\PY{o}{:}\PY{o}{:}\PY{k+kt}{rt\PYZus{}params\PYZus{}t} \PY{n}{p}\PY{p}{;} 
  \PY{n}{p}\PY{p}{.}\PY{n}{dt} \PY{o}{=} \PY{l+m+mf}{.01}\PY{p}{;}
  \PY{n}{p}\PY{p}{.}\PY{n}{di} \PY{o}{=} \PY{l+m+mf}{.05}\PY{p}{;}
  \PY{n}{p}\PY{p}{.}\PY{n}{grid\PYZus{}size} \PY{o}{=} \PY{p}{\PYZob{}} \PY{k+kt}{int}\PY{p}{(}\PY{l+m+mi}{16} \PY{o}{/} \PY{n}{p}\PY{p}{.}\PY{n}{di}\PY{p}{)} \PY{p}{\PYZcb{}}\PY{p}{;}
  \PY{n}{p}\PY{p}{.}\PY{n}{g} \PY{o}{=} \PY{l+m+mi}{1}\PY{p}{;}
  \PY{n}{p}\PY{p}{.}\PY{n}{vip\PYZus{}eps} \PY{o}{=} \PY{l+m+mf}{1e\PYZhy{}8}\PY{p}{;} 
\PY{c+c1}{//\PYZlt{}/listing\PYZhy{}2\PYZgt{}}

  \PY{c+c1}{// instantiation}
  \PY{n}{concurr}\PY{o}{:}\PY{o}{:}\PY{n}{serial}\PY{o}{\PYZlt{}}
    \PY{k+kt}{solver\PYZus{}t}\PY{p}{,} 
    \PY{n}{bcond}\PY{o}{:}\PY{o}{:}\PY{n}{cyclic}\PY{p}{,} \PY{n}{bcond}\PY{o}{:}\PY{o}{:}\PY{n}{cyclic}
  \PY{o}{\PYZgt{}} \PY{n}{run}\PY{p}{(}\PY{n}{p}\PY{p}{)}\PY{p}{;} \PY{c+c1}{// TODO: change into open bc}

  \PY{c+c1}{// initial condition}
  \PY{p}{\PYZob{}}
    \PY{n}{blitz}\PY{o}{:}\PY{o}{:}\PY{n}{firstIndex} \PY{n}{i}\PY{p}{;}
    \PY{n}{run}\PY{p}{.}\PY{n}{advectee}\PY{p}{(}\PY{n}{ix}\PY{o}{:}\PY{o}{:}\PY{n}{h}\PY{p}{)} \PY{o}{=} \PY{n}{intcond}\PY{p}{(}\PY{p}{)}\PY{p}{(}\PY{n}{p}\PY{p}{.}\PY{n}{di} \PY{o}{*} \PY{n}{i} \PY{o}{\PYZhy{}} \PY{p}{(}\PY{n}{p}\PY{p}{.}\PY{n}{grid\PYZus{}size}\PY{p}{[}\PY{l+m+mi}{0}\PY{p}{]}\PY{o}{\PYZhy{}}\PY{l+m+mi}{1}\PY{p}{)} \PY{o}{*} \PY{n}{p}\PY{p}{.}\PY{n}{di} \PY{o}{/} \PY{l+m+mi}{2}\PY{p}{)}\PY{p}{;}
  \PY{p}{\PYZcb{}}
  \PY{n}{run}\PY{p}{.}\PY{n}{advectee}\PY{p}{(}\PY{n}{ix}\PY{o}{:}\PY{o}{:}\PY{n}{qx}\PY{p}{)} \PY{o}{=} \PY{l+m+mi}{0}\PY{p}{;}

  \PY{c+c1}{// integration}
  \PY{k+kt}{double} \PY{n}{tmp} \PY{o}{=} \PY{l+m+mi}{0}\PY{p}{;}
  \PY{k+kt}{out\PYZus{}t} \PY{n+nf}{out}\PY{p}{(}\PY{n}{pfx}\PY{p}{)}\PY{p}{;}
  \PY{n}{output}\PY{o}{\PYZlt{}}\PY{n}{ix}\PY{o}{\PYZgt{}}\PY{p}{(}\PY{n}{run}\PY{p}{,} \PY{l+m+mi}{0}\PY{p}{,} \PY{n}{p}\PY{p}{.}\PY{n}{di}\PY{p}{,} \PY{n}{p}\PY{p}{.}\PY{n}{dt}\PY{p}{,} \PY{n}{out}\PY{p}{)}\PY{p}{;}
  \PY{k}{for} \PY{p}{(}\PY{k+kt}{int} \PY{n}{t} \PY{o}{=} \PY{l+m+mi}{0}\PY{p}{;} \PY{n}{t} \PY{o}{\PYZlt{}} \PY{n}{nt}\PY{p}{;} \PY{n}{t} \PY{o}{+}\PY{o}{=} \PY{n}{outfreq}\PY{p}{)}
  \PY{p}{\PYZob{}}
    \PY{n}{run}\PY{p}{.}\PY{n}{advance}\PY{p}{(}\PY{n}{outfreq}\PY{p}{)}\PY{p}{;} 
    \PY{n}{output}\PY{o}{\PYZlt{}}\PY{n}{ix}\PY{o}{\PYZgt{}}\PY{p}{(}\PY{n}{run}\PY{p}{,} \PY{n}{t} \PY{o}{+} \PY{n}{outfreq}\PY{p}{,} \PY{n}{p}\PY{p}{.}\PY{n}{di}\PY{p}{,} \PY{n}{p}\PY{p}{.}\PY{n}{dt}\PY{p}{,} \PY{n}{out}\PY{p}{)}\PY{p}{;}
   
    \PY{k}{if} \PY{p}{(}\PY{n}{max}\PY{p}{(}\PY{n}{run}\PY{p}{.}\PY{n}{advector}\PY{p}{(}\PY{p}{)}\PY{p}{)} \PY{o}{\PYZgt{}} \PY{n}{tmp}\PY{p}{)} \PY{n}{tmp} \PY{o}{=} \PY{n}{max}\PY{p}{(}\PY{n}{run}\PY{p}{.}\PY{n}{advector}\PY{p}{(}\PY{p}{)}\PY{p}{)}\PY{p}{;}
    \PY{n}{std}\PY{o}{:}\PY{o}{:}\PY{n}{cerr}\PY{o}{\PYZlt{}}\PY{o}{\PYZlt{}}\PY{l+s}{\PYZdq{}}\PY{l+s}{max advector = }\PY{l+s}{\PYZdq{}}\PY{o}{\PYZlt{}}\PY{o}{\PYZlt{}} \PY{n}{tmp}  \PY{o}{\PYZlt{}}\PY{o}{\PYZlt{}}\PY{n}{std}\PY{o}{:}\PY{o}{:}\PY{n}{endl}\PY{p}{;}
  \PY{p}{\PYZcb{}}
\PY{p}{\PYZcb{}}

\PY{k+kt}{int} \PY{n}{main}\PY{p}{(}\PY{p}{)}
\PY{p}{\PYZob{}}
  \PY{n}{test}\PY{o}{\PYZlt{}}\PY{n}{opts}\PY{o}{:}\PY{o}{:}\PY{n}{abs} \PY{o}{|} \PY{n}{opts}\PY{o}{:}\PY{o}{:}\PY{n}{fct} \PY{o}{\PYZgt{}}\PY{p}{(}\PY{l+s}{\PYZdq{}}\PY{l+s}{fct+abs}\PY{l+s}{\PYZdq{}}\PY{p}{)}\PY{p}{;}
  \PY{n}{test}\PY{o}{\PYZlt{}}\PY{n}{opts}\PY{o}{:}\PY{o}{:}\PY{n}{iga} \PY{o}{|} \PY{n}{opts}\PY{o}{:}\PY{o}{:}\PY{n}{fct} \PY{o}{\PYZgt{}}\PY{p}{(}\PY{l+s}{\PYZdq{}}\PY{l+s}{fct+iga}\PY{l+s}{\PYZdq{}}\PY{p}{)}\PY{p}{;}
  \PY{c+c1}{//plotting model results and analitic solution; }
  \PY{c+c1}{//python uses sys.argv[1:0] for choosing model outputs}
  \PY{n}{system}\PY{p}{(}\PY{l+s}{\PYZdq{}}\PY{l+s}{python ../../../../tests/tutorial/7\PYZus{}shallow\PYZus{}water/plot.py fct+abs fct+iga }\PY{l+s}{\PYZdq{}}\PY{p}{)}\PY{p}{;}
\PY{p}{\PYZcb{}}
\end{Verbatim}

%% file: bombel_params.cpp.tex
\begin{Verbatim}[commandchars=\\\{\}]
\PY{c+cm}{/** }
\PY{c+cm}{ * @file}
\PY{c+cm}{ * @copyright University of Warsaw}
\PY{c+cm}{ * @section LICENSE}
\PY{c+cm}{ * GPLv3+ (see the COPYING file or http://www.gnu.org/licenses/)}
\PY{c+cm}{ *}
\PY{c+cm}{ * @brief test for pressure solvers (as in Smolarkiewicz \PYZam{} Pudykiewicz 1992, fig.3) }
\PY{c+cm}{ * buoyant convection in Boussinesq flow}
\PY{c+cm}{ */}

\PY{c+cp}{\PYZsh{}}\PY{c+cp}{include \PYZdq{}boussinesq.hpp\PYZdq{}}
\PY{c+cp}{\PYZsh{}}\PY{c+cp}{include \PYZlt{}libmpdata++}\PY{c+cp}{/}\PY{c+cp}{concurr}\PY{c+cp}{/}\PY{c+cp}{threads.hpp\PYZgt{}}
\PY{c+cp}{\PYZsh{}}\PY{c+cp}{include \PYZlt{}libmpdata++}\PY{c+cp}{/}\PY{c+cp}{output}\PY{c+cp}{/}\PY{c+cp}{gnuplot.hpp\PYZgt{}}
\PY{k}{using} \PY{k}{namespace} \PY{n}{libmpdataxx}\PY{p}{;}

\PY{k+kt}{int} \PY{n+nf}{main}\PY{p}{(}\PY{p}{)} 
\PY{p}{\PYZob{}}
  \PY{c+c1}{// compile\PYZhy{}time parameters}
\PY{c+c1}{//\PYZlt{}listing\PYZhy{}1\PYZgt{}}
  \PY{k}{struct} \PY{k+kt}{ct\PYZus{}params\PYZus{}t} \PY{o}{:} \PY{k+kt}{ct\PYZus{}params\PYZus{}default\PYZus{}t}
  \PY{p}{\PYZob{}}
    \PY{k}{using} \PY{k+kt}{real\PYZus{}t} \PY{o}{=} \PY{k+kt}{double}\PY{p}{;}
    \PY{k}{enum} \PY{p}{\PYZob{}} \PY{n}{n\PYZus{}dims} \PY{o}{=} \PY{l+m+mi}{2} \PY{p}{\PYZcb{}}\PY{p}{;}
    \PY{k}{enum} \PY{p}{\PYZob{}} \PY{n}{n\PYZus{}eqns} \PY{o}{=} \PY{l+m+mi}{3} \PY{p}{\PYZcb{}}\PY{p}{;}
    \PY{k}{enum} \PY{p}{\PYZob{}} \PY{n}{rhs\PYZus{}scheme} \PY{o}{=} \PY{n}{solvers}\PY{o}{:}\PY{o}{:}\PY{n}{trapez} \PY{p}{\PYZcb{}}\PY{p}{;}
    \PY{k}{enum} \PY{p}{\PYZob{}} \PY{n}{prs\PYZus{}scheme} \PY{o}{=} \PY{n}{solvers}\PY{o}{:}\PY{o}{:}\PY{n}{cr} \PY{p}{\PYZcb{}}\PY{p}{;}
    \PY{k}{struct} \PY{n}{ix} \PY{p}{\PYZob{}} \PY{k}{enum} \PY{p}{\PYZob{}}
      \PY{n}{u}\PY{p}{,} \PY{n}{w}\PY{p}{,} \PY{n}{tht}\PY{p}{,} 
      \PY{n}{vip\PYZus{}i}\PY{o}{=}\PY{n}{u}\PY{p}{,} \PY{n}{vip\PYZus{}j}\PY{o}{=}\PY{n}{w}\PY{p}{,} \PY{n}{vip\PYZus{}den}\PY{o}{=}\PY{o}{\PYZhy{}}\PY{l+m+mi}{1}
    \PY{p}{\PYZcb{}}\PY{p}{;} \PY{p}{\PYZcb{}}\PY{p}{;}
  \PY{p}{\PYZcb{}}\PY{p}{;} 
\PY{c+c1}{//\PYZlt{}/listing\PYZhy{}1\PYZgt{}}
  \PY{k}{using} \PY{n}{ix} \PY{o}{=} \PY{k}{typename} \PY{k+kt}{ct\PYZus{}params\PYZus{}t}\PY{o}{:}\PY{o}{:}\PY{n}{ix}\PY{p}{;}
  \PY{k}{using} \PY{k+kt}{real\PYZus{}t} \PY{o}{=} \PY{k}{typename} \PY{k+kt}{ct\PYZus{}params\PYZus{}t}\PY{o}{:}\PY{o}{:}\PY{k+kt}{real\PYZus{}t}\PY{p}{;}

  \PY{k}{const} \PY{k+kt}{int} \PY{n}{r0} \PY{o}{=} \PY{l+m+mi}{250}\PY{p}{;} 
  \PY{k}{const} \PY{k+kt}{int} \PY{n}{nx} \PY{o}{=} \PY{l+m+mi}{201}\PY{p}{,} \PY{n}{ny} \PY{o}{=} \PY{l+m+mi}{201}\PY{p}{,} \PY{n}{nt} \PY{o}{=} \PY{l+m+mi}{800}\PY{p}{;}
  \PY{k}{typename} \PY{k+kt}{ct\PYZus{}params\PYZus{}t}\PY{o}{:}\PY{o}{:}\PY{k+kt}{real\PYZus{}t} \PY{n}{Tht\PYZus{}ref} \PY{o}{=} \PY{l+m+mi}{300}\PY{p}{;} \PY{c+c1}{//1; // reference state (constant throughout the domain)}

  \PY{c+c1}{// conjugate residual}
  \PY{k}{using} \PY{k+kt}{solver\PYZus{}t} \PY{o}{=} \PY{n}{output}\PY{o}{:}\PY{o}{:}\PY{n}{gnuplot}\PY{o}{\PYZlt{}}\PY{n}{boussinesq}\PY{o}{\PYZlt{}}\PY{k+kt}{ct\PYZus{}params\PYZus{}t}\PY{o}{\PYZgt{}}\PY{o}{\PYZgt{}}\PY{p}{;}

  \PY{c+c1}{// run\PYZhy{}time parameters}
  \PY{k+kt}{solver\PYZus{}t}\PY{o}{:}\PY{o}{:}\PY{k+kt}{rt\PYZus{}params\PYZus{}t} \PY{n}{p}\PY{p}{;}

  \PY{n}{p}\PY{p}{.}\PY{n}{dt} \PY{o}{=} \PY{l+m+mf}{.75}\PY{p}{;}
  \PY{n}{p}\PY{p}{.}\PY{n}{di} \PY{o}{=} \PY{n}{p}\PY{p}{.}\PY{n}{dj} \PY{o}{=} \PY{l+m+mf}{10.}\PY{p}{;} 
  \PY{n}{p}\PY{p}{.}\PY{n}{Tht\PYZus{}ref} \PY{o}{=} \PY{n}{Tht\PYZus{}ref}\PY{p}{;} 

  \PY{n}{p}\PY{p}{.}\PY{n}{outfreq} \PY{o}{=} \PY{l+m+mi}{100}\PY{p}{;} \PY{c+c1}{//12;}
  \PY{n}{p}\PY{p}{.}\PY{n}{outvars} \PY{o}{=} \PY{p}{\PYZob{}}
\PY{c+c1}{//    \PYZob{}ix::u,   \PYZob{}.name = \PYZdq{}u\PYZdq{},   .unit = \PYZdq{}m/s\PYZdq{}\PYZcb{}\PYZcb{}, }
\PY{c+c1}{//    \PYZob{}ix::w,   \PYZob{}.name = \PYZdq{}w\PYZdq{},   .unit = \PYZdq{}m/s\PYZdq{}\PYZcb{}\PYZcb{}, }
    \PY{p}{\PYZob{}}\PY{n}{ix}\PY{o}{:}\PY{o}{:}\PY{n}{tht}\PY{p}{,} \PY{p}{\PYZob{}}\PY{p}{.}\PY{n}{name} \PY{o}{=} \PY{l+s}{\PYZdq{}}\PY{l+s}{tht}\PY{l+s}{\PYZdq{}}\PY{p}{,} \PY{p}{.}\PY{n}{unit} \PY{o}{=} \PY{l+s}{\PYZdq{}}\PY{l+s}{K}\PY{l+s}{\PYZdq{}}  \PY{p}{\PYZcb{}}\PY{p}{\PYZcb{}}
  \PY{p}{\PYZcb{}}\PY{p}{;}
  \PY{n}{p}\PY{p}{.}\PY{n}{gnuplot\PYZus{}view} \PY{o}{=} \PY{l+s}{\PYZdq{}}\PY{l+s}{map}\PY{l+s}{\PYZdq{}}\PY{p}{;}
  \PY{n}{p}\PY{p}{.}\PY{n}{gnuplot\PYZus{}output} \PY{o}{=} \PY{l+s}{\PYZdq{}}\PY{l+s}{figure\PYZus{}\PYZpc{}s\PYZus{}\PYZpc{}04d.svg}\PY{l+s}{\PYZdq{}}\PY{p}{;}
  \PY{n}{p}\PY{p}{.}\PY{n}{gnuplot\PYZus{}with} \PY{o}{=} \PY{l+s}{\PYZdq{}}\PY{l+s}{lines}\PY{l+s}{\PYZdq{}}\PY{p}{;}
  \PY{n}{p}\PY{p}{.}\PY{n}{gnuplot\PYZus{}surface} \PY{o}{=} \PY{n+nb}{false}\PY{p}{;}
  \PY{n}{p}\PY{p}{.}\PY{n}{gnuplot\PYZus{}contour} \PY{o}{=} \PY{n+nb}{true}\PY{p}{;}

  \PY{k+kt}{real\PYZus{}t} \PY{n}{eps} \PY{o}{=} \PY{l+m+mf}{.01}\PY{p}{;}

  \PY{k}{if} \PY{p}{(}\PY{n+nb}{false}\PY{p}{)} \PY{c+c1}{// TODO}
  \PY{p}{\PYZob{}}
    \PY{c+c1}{// physics\PYZhy{}oriented plot}
    \PY{n}{p}\PY{p}{.}\PY{n}{gnuplot\PYZus{}cntrparam} \PY{o}{=} \PY{l+s}{\PYZdq{}}\PY{l+s}{levels incremental 299.95, 0.1, 300.55}\PY{l+s}{\PYZdq{}}\PY{p}{;}
    \PY{n}{p}\PY{p}{.}\PY{n}{gnuplot\PYZus{}cbrange} \PY{o}{=} \PY{l+s}{\PYZdq{}}\PY{l+s}{[299.95 : 300.55]}\PY{l+s}{\PYZdq{}}\PY{p}{;}
    \PY{n}{p}\PY{p}{.}\PY{n}{gnuplot\PYZus{}cbtics} \PY{o}{=} \PY{l+s}{\PYZdq{}}\PY{l+s}{300.05, 0.1, 300.45}\PY{l+s}{\PYZdq{}}\PY{p}{;}
    \PY{n}{p}\PY{p}{.}\PY{n}{gnuplot\PYZus{}palette} \PY{o}{=} \PY{l+s}{\PYZdq{}}\PY{l+s}{defined (}\PY{l+s}{\PYZdq{}}
      \PY{l+s}{\PYZdq{}}\PY{l+s}{299.95 \PYZsq{}\PYZsh{}ffffff\PYZsq{}, }\PY{l+s}{\PYZdq{}}
      \PY{l+s}{\PYZdq{}}\PY{l+s}{300.05 \PYZsq{}\PYZsh{}ffffff\PYZsq{}, 300.05 \PYZsq{}\PYZsh{}993399\PYZsq{}, }\PY{l+s}{\PYZdq{}}
      \PY{l+s}{\PYZdq{}}\PY{l+s}{300.15 \PYZsq{}\PYZsh{}993399\PYZsq{}, 300.15 \PYZsq{}\PYZsh{}00CCFF\PYZsq{}, }\PY{l+s}{\PYZdq{}}
      \PY{l+s}{\PYZdq{}}\PY{l+s}{300.25 \PYZsq{}\PYZsh{}00CCFF\PYZsq{}, 300.25 \PYZsq{}\PYZsh{}66CC00\PYZsq{}, }\PY{l+s}{\PYZdq{}}
      \PY{l+s}{\PYZdq{}}\PY{l+s}{300.35 \PYZsq{}\PYZsh{}66CC00\PYZsq{}, 300.35 \PYZsq{}\PYZsh{}FC8727\PYZsq{}, }\PY{l+s}{\PYZdq{}}
      \PY{l+s}{\PYZdq{}}\PY{l+s}{300.45 \PYZsq{}\PYZsh{}FC8727\PYZsq{}, 300.45 \PYZsq{}\PYZsh{}FFFF00\PYZsq{}, }\PY{l+s}{\PYZdq{}}
      \PY{l+s}{\PYZdq{}}\PY{l+s}{300.55 \PYZsq{}\PYZsh{}FFFF00\PYZsq{}}\PY{l+s}{\PYZdq{}}
    \PY{l+s}{\PYZdq{}}\PY{l+s}{)}\PY{l+s}{\PYZdq{}}\PY{p}{;}
  \PY{p}{\PYZcb{}} 
  \PY{k}{else}
  \PY{p}{\PYZob{}}
    \PY{c+c1}{// numerics\PYZhy{}oriented plot}
    \PY{n}{p}\PY{p}{.}\PY{n}{gnuplot\PYZus{}cntrparam} \PY{o}{=} \PY{l+s}{\PYZdq{}}\PY{l+s}{levels discrete 299.99, 300.00, 300.05, 300.50, 300.51}\PY{l+s}{\PYZdq{}}\PY{p}{;}
    \PY{n}{p}\PY{p}{.}\PY{n}{gnuplot\PYZus{}cbrange} \PY{o}{=} \PY{l+s}{\PYZdq{}}\PY{l+s}{[299.95 : 300.55]}\PY{l+s}{\PYZdq{}}\PY{p}{;}
    \PY{n}{p}\PY{p}{.}\PY{n}{gnuplot\PYZus{}cbtics} \PY{o}{=} \PY{l+s}{\PYZdq{}}\PY{l+s}{(\PYZsq{}299.99\PYZsq{} 299.95, \PYZsq{}300.00\PYZsq{} 300.00, \PYZsq{}300.05\PYZsq{} 300.05, \PYZsq{}300.50\PYZsq{} 300.50, \PYZsq{}300.51\PYZsq{} 300.55)}\PY{l+s}{\PYZdq{}}\PY{p}{;} \PY{c+c1}{// note: intentionally non\PYZhy{}linear!!! TODO: use eps to construct the strings!}
    \PY{n}{p}\PY{p}{.}\PY{n}{gnuplot\PYZus{}palette} \PY{o}{=} \PY{l+s}{\PYZdq{}}\PY{l+s}{defined (}\PY{l+s}{\PYZdq{}}
      \PY{l+s}{\PYZdq{}}\PY{l+s}{299.95 \PYZsq{}\PYZsh{}BACA66\PYZsq{}, }\PY{l+s}{\PYZdq{}}
      \PY{l+s}{\PYZdq{}}\PY{l+s}{300.00 \PYZsq{}\PYZsh{}BACA66\PYZsq{}, 300.00 \PYZsq{}\PYZsh{}ffffff\PYZsq{}, }\PY{l+s}{\PYZdq{}}
      \PY{l+s}{\PYZdq{}}\PY{l+s}{300.05 \PYZsq{}\PYZsh{}ffffff\PYZsq{}, 300.05 \PYZsq{}\PYZsh{}cccccc\PYZsq{}, }\PY{l+s}{\PYZdq{}}
      \PY{l+s}{\PYZdq{}}\PY{l+s}{300.50 \PYZsq{}\PYZsh{}cccccc\PYZsq{},}\PY{l+s}{\PYZdq{}}
      \PY{l+s}{\PYZdq{}}\PY{l+s}{300.50 \PYZsq{}\PYZsh{}ff0000\PYZsq{}, 300.55 \PYZsq{}\PYZsh{}ff0000\PYZsq{}}\PY{l+s}{\PYZdq{}}
    \PY{l+s}{\PYZdq{}}\PY{l+s}{)}\PY{l+s}{\PYZdq{}}\PY{p}{;}
  \PY{p}{\PYZcb{}}

  \PY{n}{p}\PY{p}{.}\PY{n}{gnuplot\PYZus{}term} \PY{o}{=} \PY{l+s}{\PYZdq{}}\PY{l+s}{svg}\PY{l+s}{\PYZdq{}}\PY{p}{;}
\PY{c+c1}{//\PYZlt{}listing\PYZhy{}2\PYZgt{}}
  \PY{n}{p}\PY{p}{.}\PY{n}{prs\PYZus{}tol} \PY{o}{=} \PY{l+m+mf}{1e\PYZhy{}7}\PY{p}{;}
\PY{c+c1}{//\PYZlt{}/listing\PYZhy{}2\PYZgt{}}
  \PY{n}{p}\PY{p}{.}\PY{n}{grid\PYZus{}size} \PY{o}{=} \PY{p}{\PYZob{}}\PY{n}{nx}\PY{p}{,} \PY{n}{ny}\PY{p}{\PYZcb{}}\PY{p}{;}

  \PY{n}{libmpdataxx}\PY{o}{:}\PY{o}{:}\PY{n}{concurr}\PY{o}{:}\PY{o}{:}\PY{n}{threads}\PY{o}{\PYZlt{}}
    \PY{k+kt}{solver\PYZus{}t}\PY{p}{,} 
    \PY{n}{bcond}\PY{o}{:}\PY{o}{:}\PY{n}{cyclic}\PY{p}{,} \PY{n}{bcond}\PY{o}{:}\PY{o}{:}\PY{n}{cyclic}\PY{p}{,}
    \PY{n}{bcond}\PY{o}{:}\PY{o}{:}\PY{n}{cyclic}\PY{p}{,} \PY{n}{bcond}\PY{o}{:}\PY{o}{:}\PY{n}{cyclic}
  \PY{o}{\PYZgt{}} \PY{n}{slv}\PY{p}{(}\PY{n}{p}\PY{p}{)}\PY{p}{;}

  \PY{p}{\PYZob{}}
    \PY{c+c1}{// initial condition}
    \PY{n}{blitz}\PY{o}{:}\PY{o}{:}\PY{n}{firstIndex} \PY{n}{i}\PY{p}{;}
    \PY{n}{blitz}\PY{o}{:}\PY{o}{:}\PY{n}{secondIndex} \PY{n}{j}\PY{p}{;}

    \PY{n}{slv}\PY{p}{.}\PY{n}{advectee}\PY{p}{(}\PY{n}{ix}\PY{o}{:}\PY{o}{:}\PY{n}{tht}\PY{p}{)} \PY{o}{=} \PY{n}{Tht\PYZus{}ref} \PY{o}{+} \PY{n}{where}\PY{p}{(}
      \PY{c+c1}{// if}
      \PY{n}{pow}\PY{p}{(}\PY{n}{i} \PY{o}{*} \PY{n}{p}\PY{p}{.}\PY{n}{di} \PY{o}{\PYZhy{}} \PY{l+m+mi}{4}    \PY{o}{*} \PY{n}{r0} \PY{p}{,} \PY{l+m+mi}{2}\PY{p}{)} \PY{o}{+} 
      \PY{n}{pow}\PY{p}{(}\PY{n}{j} \PY{o}{*} \PY{n}{p}\PY{p}{.}\PY{n}{dj} \PY{o}{\PYZhy{}} \PY{l+m+mf}{1.04} \PY{o}{*} \PY{n}{r0} \PY{p}{,} \PY{l+m+mi}{2}\PY{p}{)} \PY{o}{\PYZlt{}}\PY{o}{=} \PY{n}{pow}\PY{p}{(}\PY{n}{r0}\PY{p}{,} \PY{l+m+mi}{2}\PY{p}{)}\PY{p}{,} 
      \PY{c+c1}{// then}
      \PY{l+m+mf}{.5}\PY{p}{,} 
      \PY{c+c1}{// else}
      \PY{l+m+mi}{0}
    \PY{p}{)}\PY{p}{;}
\PY{n}{std}\PY{o}{:}\PY{o}{:}\PY{n}{cerr} \PY{o}{\PYZlt{}}\PY{o}{\PYZlt{}} \PY{l+s}{\PYZdq{}}\PY{l+s}{min(psi) = }\PY{l+s}{\PYZdq{}} \PY{o}{\PYZlt{}}\PY{o}{\PYZlt{}} \PY{n}{min}\PY{p}{(}\PY{n}{slv}\PY{p}{.}\PY{n}{advectee}\PY{p}{(}\PY{n}{ix}\PY{o}{:}\PY{o}{:}\PY{n}{tht}\PY{p}{)}\PY{p}{)} \PY{o}{\PYZlt{}}\PY{o}{\PYZlt{}} \PY{l+s}{\PYZdq{}}\PY{l+s+se}{\PYZbs{}n}\PY{l+s}{\PYZdq{}}\PY{p}{;}
\PY{n}{std}\PY{o}{:}\PY{o}{:}\PY{n}{cerr} \PY{o}{\PYZlt{}}\PY{o}{\PYZlt{}} \PY{l+s}{\PYZdq{}}\PY{l+s}{max(u)\PYZca{}2 + max(w)\PYZca{}2 = }\PY{l+s}{\PYZdq{}} \PY{o}{\PYZlt{}}\PY{o}{\PYZlt{}} \PY{n}{max}\PY{p}{(}\PY{n}{pow}\PY{p}{(}\PY{n}{slv}\PY{p}{.}\PY{n}{advectee}\PY{p}{(}\PY{n}{ix}\PY{o}{:}\PY{o}{:}\PY{n}{u}\PY{p}{)}\PY{p}{,}\PY{l+m+mi}{2}\PY{p}{)} \PY{o}{+} \PY{n}{pow}\PY{p}{(}\PY{n}{slv}\PY{p}{.}\PY{n}{advectee}\PY{p}{(}\PY{n}{ix}\PY{o}{:}\PY{o}{:}\PY{n}{w}\PY{p}{)}\PY{p}{,}\PY{l+m+mi}{2}\PY{p}{)}\PY{p}{)} \PY{o}{\PYZlt{}}\PY{o}{\PYZlt{}} \PY{l+s}{\PYZdq{}}\PY{l+s+se}{\PYZbs{}n}\PY{l+s}{\PYZdq{}}\PY{p}{;}
    \PY{n}{slv}\PY{p}{.}\PY{n}{advectee}\PY{p}{(}\PY{n}{ix}\PY{o}{:}\PY{o}{:}\PY{n}{u}\PY{p}{)} \PY{o}{=} \PY{l+m+mi}{0}\PY{p}{;} 
    \PY{n}{slv}\PY{p}{.}\PY{n}{advectee}\PY{p}{(}\PY{n}{ix}\PY{o}{:}\PY{o}{:}\PY{n}{w}\PY{p}{)} \PY{o}{=} \PY{l+m+mi}{0}\PY{p}{;} 
  \PY{p}{\PYZcb{}}

  \PY{c+c1}{// integration}
  \PY{n}{slv}\PY{p}{.}\PY{n}{advance}\PY{p}{(}\PY{n}{nt}\PY{p}{)}\PY{p}{;} 
\PY{n}{std}\PY{o}{:}\PY{o}{:}\PY{n}{cerr} \PY{o}{\PYZlt{}}\PY{o}{\PYZlt{}} \PY{l+s}{\PYZdq{}}\PY{l+s}{min(psi)\PYZhy{}300 = }\PY{l+s}{\PYZdq{}} \PY{o}{\PYZlt{}}\PY{o}{\PYZlt{}} \PY{n}{min}\PY{p}{(}\PY{n}{slv}\PY{p}{.}\PY{n}{advectee}\PY{p}{(}\PY{n}{ix}\PY{o}{:}\PY{o}{:}\PY{n}{tht}\PY{p}{)}\PY{p}{)}\PY{o}{\PYZhy{}}\PY{l+m+mf}{300.0} \PY{o}{\PYZlt{}}\PY{o}{\PYZlt{}} \PY{l+s}{\PYZdq{}}\PY{l+s+se}{\PYZbs{}n}\PY{l+s}{\PYZdq{}}\PY{p}{;}
\PY{n}{std}\PY{o}{:}\PY{o}{:}\PY{n}{cerr} \PY{o}{\PYZlt{}}\PY{o}{\PYZlt{}} \PY{l+s}{\PYZdq{}}\PY{l+s}{max(psi)\PYZhy{}300 = }\PY{l+s}{\PYZdq{}} \PY{o}{\PYZlt{}}\PY{o}{\PYZlt{}} \PY{n}{max}\PY{p}{(}\PY{n}{slv}\PY{p}{.}\PY{n}{advectee}\PY{p}{(}\PY{n}{ix}\PY{o}{:}\PY{o}{:}\PY{n}{tht}\PY{p}{)}\PY{p}{)}\PY{o}{\PYZhy{}}\PY{l+m+mf}{300.5} \PY{o}{\PYZlt{}}\PY{o}{\PYZlt{}} \PY{l+s}{\PYZdq{}}\PY{l+s+se}{\PYZbs{}n}\PY{l+s}{\PYZdq{}}\PY{p}{;}
\PY{n}{std}\PY{o}{:}\PY{o}{:}\PY{n}{cerr} \PY{o}{\PYZlt{}}\PY{o}{\PYZlt{}} \PY{l+s}{\PYZdq{}}\PY{l+s}{max(u)\PYZca{}2 + max(w)\PYZca{}2 = }\PY{l+s}{\PYZdq{}} \PY{o}{\PYZlt{}}\PY{o}{\PYZlt{}} \PY{n}{max}\PY{p}{(}\PY{n}{pow}\PY{p}{(}\PY{n}{slv}\PY{p}{.}\PY{n}{advectee}\PY{p}{(}\PY{n}{ix}\PY{o}{:}\PY{o}{:}\PY{n}{u}\PY{p}{)}\PY{p}{,}\PY{l+m+mi}{2}\PY{p}{)} \PY{o}{+} \PY{n}{pow}\PY{p}{(}\PY{n}{slv}\PY{p}{.}\PY{n}{advectee}\PY{p}{(}\PY{n}{ix}\PY{o}{:}\PY{o}{:}\PY{n}{w}\PY{p}{)}\PY{p}{,}\PY{l+m+mi}{2}\PY{p}{)}\PY{p}{)} \PY{o}{\PYZlt{}}\PY{o}{\PYZlt{}} \PY{l+s}{\PYZdq{}}\PY{l+s+se}{\PYZbs{}n}\PY{l+s}{\PYZdq{}}\PY{p}{;}
  \PY{k}{if} \PY{p}{(}\PY{n}{min}\PY{p}{(}\PY{n}{slv}\PY{p}{.}\PY{n}{advectee}\PY{p}{(}\PY{n}{ix}\PY{o}{:}\PY{o}{:}\PY{n}{tht}\PY{p}{)}\PY{p}{)} \PY{o}{\PYZlt{}} \PY{l+m+mi}{300}\PY{o}{\PYZhy{}}\PY{n}{eps} \PY{o}{|}\PY{o}{|} \PY{n}{max}\PY{p}{(}\PY{n}{slv}\PY{p}{.}\PY{n}{advectee}\PY{p}{(}\PY{n}{ix}\PY{o}{:}\PY{o}{:}\PY{n}{tht}\PY{p}{)}\PY{p}{)} \PY{o}{\PYZgt{}} \PY{l+m+mf}{300.5}\PY{o}{+}\PY{n}{eps}\PY{p}{)}
    \PY{k}{throw} \PY{n}{std}\PY{o}{:}\PY{o}{:}\PY{n}{runtime\PYZus{}error}\PY{p}{(}\PY{l+s}{\PYZdq{}}\PY{l+s}{too big under\PYZhy{} or over\PYZhy{}shots :(}\PY{l+s}{\PYZdq{}}\PY{p}{)}\PY{p}{;} 
\PY{p}{\PYZcb{}}\PY{p}{;}
\end{Verbatim}

%% file: bombel_rhs.cpp.tex
\begin{Verbatim}[commandchars=\\\{\}]
\PY{c+cm}{/** }
\PY{c+cm}{ * @file}
\PY{c+cm}{ * @copyright University of Warsaw}
\PY{c+cm}{ * @section LICENSE}
\PY{c+cm}{ * GPLv3+ (see the COPYING file or http://www.gnu.org/licenses/)}
\PY{c+cm}{ */}

\PY{c+cp}{\PYZsh{}}\PY{c+cp}{pragma once}

\PY{c+cp}{\PYZsh{}}\PY{c+cp}{include \PYZlt{}libmpdata++}\PY{c+cp}{/}\PY{c+cp}{solvers}\PY{c+cp}{/}\PY{c+cp}{mpdata\PYZus{}rhs\PYZus{}vip\PYZus{}prs.hpp\PYZgt{}}

\PY{k}{template} \PY{o}{\PYZlt{}}\PY{k}{class} \PY{n+nc}{ct\PYZus{}params\PYZus{}t}\PY{o}{\PYZgt{}}
\PY{k}{class} \PY{n+nc}{boussinesq} \PY{o}{:} \PY{k}{public} \PY{n}{libmpdataxx}\PY{o}{:}\PY{o}{:}\PY{n}{solvers}\PY{o}{:}\PY{o}{:}\PY{n}{mpdata\PYZus{}rhs\PYZus{}vip\PYZus{}prs}\PY{o}{\PYZlt{}}\PY{k+kt}{ct\PYZus{}params\PYZus{}t}\PY{o}{\PYZgt{}}
\PY{p}{\PYZob{}}
  \PY{k}{using} \PY{k+kt}{parent\PYZus{}t} \PY{o}{=} \PY{n}{libmpdataxx}\PY{o}{:}\PY{o}{:}\PY{n}{solvers}\PY{o}{:}\PY{o}{:}\PY{n}{mpdata\PYZus{}rhs\PYZus{}vip\PYZus{}prs}\PY{o}{\PYZlt{}}\PY{k+kt}{ct\PYZus{}params\PYZus{}t}\PY{o}{\PYZgt{}}\PY{p}{;}
  \PY{k}{using} \PY{n}{ix} \PY{o}{=} \PY{k}{typename} \PY{k+kt}{ct\PYZus{}params\PYZus{}t}\PY{o}{:}\PY{o}{:}\PY{n}{ix}\PY{p}{;}

  \PY{n+nl}{public:}
  \PY{k}{using} \PY{k+kt}{real\PYZus{}t} \PY{o}{=} \PY{k}{typename} \PY{k+kt}{ct\PYZus{}params\PYZus{}t}\PY{o}{:}\PY{o}{:}\PY{k+kt}{real\PYZus{}t}\PY{p}{;}

  \PY{n+nl}{private:}
  \PY{c+c1}{// member fields}
  \PY{k+kt}{real\PYZus{}t} \PY{n}{g}\PY{p}{,} \PY{n}{Tht\PYZus{}ref}\PY{p}{;}

\PY{c+c1}{//\PYZlt{}listing\PYZhy{}1\PYZgt{}}
  \PY{c+c1}{// explicit forcings }
  \PY{k+kt}{void} \PY{n+nf}{update\PYZus{}rhs}\PY{p}{(}
    \PY{n}{libmpdataxx}\PY{o}{:}\PY{o}{:}\PY{k+kt}{arrvec\PYZus{}t}\PY{o}{\PYZlt{}}
      \PY{k}{typename} \PY{k+kt}{parent\PYZus{}t}\PY{o}{:}\PY{o}{:}\PY{k+kt}{arr\PYZus{}t}
    \PY{o}{\PYZgt{}} \PY{o}{\PYZam{}}\PY{n}{rhs}\PY{p}{,} 
    \PY{k}{const} \PY{k+kt}{real\PYZus{}t} \PY{o}{\PYZam{}}\PY{n}{dt}\PY{p}{,} 
    \PY{k}{const} \PY{k+kt}{int} \PY{o}{\PYZam{}}\PY{n}{at} 
  \PY{p}{)} \PY{p}{\PYZob{}}
    \PY{k+kt}{parent\PYZus{}t}\PY{o}{:}\PY{o}{:}\PY{n}{update\PYZus{}rhs}\PY{p}{(}\PY{n}{rhs}\PY{p}{,} \PY{n}{dt}\PY{p}{,} \PY{n}{at}\PY{p}{)}\PY{p}{;} 

    \PY{k}{const} \PY{k}{auto} \PY{o}{\PYZam{}}\PY{n}{Tht} \PY{o}{=} \PY{k}{this}\PY{o}{\PYZhy{}}\PY{o}{\PYZgt{}}\PY{n}{state}\PY{p}{(}\PY{n}{ix}\PY{o}{:}\PY{o}{:}\PY{n}{tht}\PY{p}{)}\PY{p}{;} 
    \PY{k}{const} \PY{k}{auto} \PY{o}{\PYZam{}}\PY{n}{ijk} \PY{o}{=} \PY{k}{this}\PY{o}{\PYZhy{}}\PY{o}{\PYZgt{}}\PY{n}{ijk}\PY{p}{;}

    \PY{n}{rhs}\PY{p}{.}\PY{n}{at}\PY{p}{(}\PY{n}{ix}\PY{o}{:}\PY{o}{:}\PY{n}{w}\PY{p}{)}\PY{p}{(}\PY{n}{ijk}\PY{p}{)} \PY{o}{+}\PY{o}{=} 
      \PY{n}{g} \PY{o}{*} \PY{p}{(}\PY{n}{Tht}\PY{p}{(}\PY{n}{ijk}\PY{p}{)} \PY{o}{\PYZhy{}} \PY{n}{Tht\PYZus{}ref}\PY{p}{)} \PY{o}{/} \PY{n}{Tht\PYZus{}ref}\PY{p}{;} 
  \PY{p}{\PYZcb{}}
\PY{c+c1}{//\PYZlt{}/listing\PYZhy{}1\PYZgt{}}
\PY{c+c1}{//    rhs.at(ix::w)(ijk) += g /*/ 300*/ * (Tht(ijk) \PYZhy{} Tht\PYZus{}ref) / Tht\PYZus{}ref; }

  \PY{n+nl}{public:}

  \PY{k}{struct} \PY{k+kt}{rt\PYZus{}params\PYZus{}t} \PY{o}{:} \PY{k+kt}{parent\PYZus{}t}\PY{o}{:}\PY{o}{:}\PY{k+kt}{rt\PYZus{}params\PYZus{}t} 
  \PY{p}{\PYZob{}} 
    \PY{k+kt}{real\PYZus{}t} \PY{n}{g} \PY{o}{=} \PY{l+m+mf}{9.81}\PY{p}{,} \PY{n}{Tht\PYZus{}ref} \PY{o}{=} \PY{l+m+mi}{0}\PY{p}{;} 
  \PY{p}{\PYZcb{}}\PY{p}{;}

  \PY{c+c1}{// ctor}
  \PY{n}{boussinesq}\PY{p}{(} 
    \PY{k}{typename} \PY{k+kt}{parent\PYZus{}t}\PY{o}{:}\PY{o}{:}\PY{k+kt}{ctor\PYZus{}args\PYZus{}t} \PY{n}{args}\PY{p}{,} 
    \PY{k}{const} \PY{k+kt}{rt\PYZus{}params\PYZus{}t} \PY{o}{\PYZam{}}\PY{n}{p}
  \PY{p}{)} \PY{o}{:}
    \PY{k+kt}{parent\PYZus{}t}\PY{p}{(}\PY{n}{args}\PY{p}{,} \PY{n}{p}\PY{p}{)}\PY{p}{,}
    \PY{n}{g}\PY{p}{(}\PY{n}{p}\PY{p}{.}\PY{n}{g}\PY{p}{)}\PY{p}{,}
    \PY{n}{Tht\PYZus{}ref}\PY{p}{(}\PY{n}{p}\PY{p}{.}\PY{n}{Tht\PYZus{}ref}\PY{p}{)}
  \PY{p}{\PYZob{}}
    \PY{n}{assert}\PY{p}{(}\PY{n}{Tht\PYZus{}ref} \PY{o}{!}\PY{o}{=} \PY{l+m+mi}{0}\PY{p}{)}\PY{p}{;}
  \PY{p}{\PYZcb{}}
\PY{p}{\PYZcb{}}\PY{p}{;}
\end{Verbatim}